\documentclass[12pt,vi,oneside]{mitthesis}

\usepackage{cite}

\usepackage[dvipdfm]{graphicx} 
\usepackage{bmpsize}

\def\L{{\cal L}}
\usepackage{amsmath}
\usepackage{amssymb}
\usepackage{mathrsfs}

\usepackage{balance}

\usepackage[subrefformat=parens,labelformat=parens]{subfig}

\usepackage{amsfonts}
\usepackage{graphicx}

\usepackage{amssymb}
\usepackage{multirow}
\usepackage{color}
\usepackage{amsthm}
\usepackage{xfrac}
\usepackage{adjustbox}
\makeatletter
\newcommand*{\rom}[1]{\expandafter\@slowromancap\romannumeral #1@}
\makeatother

  \usepackage{bm}
\usepackage{accents}
\newlength{\dhatheight}
\newcommand{\doublehat}[1]{%
	\settoheight{\dhatheight}{\ensuremath{\hat{#1}}}%
	\addtolength{\dhatheight}{-0.35ex}%
	\hat{\vphantom{\rule{1pt}{\dhatheight}}%
		\smash{\hat{#1}}}}

\usepackage{cite}
\usepackage{balance}

\usepackage{epstopdf}

\usepackage{pgfplots}

\usepackage{amsmath,stmaryrd,pict2e,picture}
\usepackage{float}

\usepackage{amsmath,amssymb,amsthm}
\usepackage{mathrsfs}
\usepackage{algorithm, algorithmic}

\newtheorem{prop}{Proposition}
\usepackage{lgrind}

\usepackage{cmap}
\usepackage[T1]{fontenc}
\usepackage{amssymb}
\usepackage{sectsty}
\usepackage{amsmath}
\usepackage{epstopdf}
\usepackage{multirow}
\usepackage{graphicx}
\usepackage{caption}
\usepackage{color}
\usepackage{float}
\usepackage{flushend}
\usepackage{url}
\usepackage[table]{}

\usepackage{titlesec}
\titlespacing*{\chapter}{0pt}{-0.25in}{12pt}
\titleformat{\chapter}[display]
{\normalfont\large\centering}{\chaptertitlename\ \thechapter}{0pt}{\large\uppercase}

\usepackage{titletoc}
\titlecontents{chapter}
[0pt]
{}
{\chaptername\ \thecontentslabel:\quad}
{}
{\hfill\contentspage}

\renewcommand*\listtablename{LIST OF TABLES}
\renewcommand*\listfigurename{LIST OF FIGURES}

\pgfplotsset{compat=1.16}

\begin{document}
%
%
%
%
%
%
%
%
%
%
%

\begin{titlepage}
\textbf{Sparse Array Beamformer Design for Active and Passive Sensing} \\
\vspace{10mm}
by \\
\vspace{10mm}
{Syed Ali Hamza}\\
\vspace{10mm}
\vspace{10mm}
\vspace{10mm}
{College of Engineering, Villanova University}\\
\vspace{10mm}
\vspace{10mm}
{DOCTORATE OF PHILOSOPHY} \\
\vspace{10mm}
{Advisor: Moeness Amin}\\

\cleardoublepage
\end{titlepage}





\chapter*{}
\vspace*{\fill}
\hspace*{\fill}
 Copyright © 2020, Syed Ali Hamza 
\hspace*{\fill}
\\
\hspace*{\fill}
 All Rights Reserved
\hspace*{\fill}
\vspace*{\fill}

\pagestyle{plain}
\pagenumbering{roman} 
\chapter*{STATEMENT BY AUTHOR}

\addcontentsline{toc}{chapter}{STATEMENT BY AUTHOR}

This dissertation has been submitted in partial fulfillment of requirements for an advanced degree at the Villanova University. 

Brief quotations from this dissertation are allowable without special permission, provided that accurate acknowledgment of source is made. Requests for permission for extended quotation from or reproduction of this manuscript in whole or in part may be granted by the head of the major department or the Associate Dean for Graduate Studies and Research of the College of Engineering when in his or her judgment the proposed use of the material is in the interests of scholarship. In all other instances, however, permission must be obtained from the author.

\clearpage 
\chapter*{ACKNOWLEDGEMENTS}

\addcontentsline{toc}{chapter}{ACKNOWLEDGEMENTS}

This dissertation is the result of my research at Villanova University. The project described in this thesis was supported in part by National Science Foundation and Office of Naval Research. Its contents are solely the responsibility of the author and do not necessarily represent the official views of the sponsors.

First and foremost, I would like to thank my family.  My parents and my siblings made everything seem so easy, and generously supported me along the way. A very special gratitude goes out to my friends who made sure I had a life outside my research.

Many people who I cannot mention individually deserve thanks and appreciation for their support in the preparation of my dissertation. I would like to express my gratitude to my advisor, Prof. Moeness Amin, who provided me an opportunity to work on timely and exciting topics. It would not have been possible without his unfettered support, guidance and encouragement.  Besides my advisor, I would like to thank Prof. Aboulnasr Hassanien for sharing his experiences and providing insightful comments during my research. Special gratitude goes to the rest of my thesis committee, Prof. Bijan Mobasseri and Prof. Ahmad Hoorfar, for their invaluable support. Also, I am grateful to Janice Moughan and to the university staff at the International Student Office for their help in managing administrative tasks, which are always complicated for me. I thank my fellow labmates for the stimulating discussions inside and outside the lab.  

Thanks to everyone who supported me.

\clearpage
\chapter*{DEDICATION}

\addcontentsline{toc}{chapter}{DEDICATION}

\hfill \break
\begin{center}
This dissertation is lovingly dedicated to my parents and  siblings, Maryam, Samrat and Usama!\\
\end{center}
\clearpage
\tableofcontents
\addcontentsline{toc}{chapter}{\listtablename}
{%
	\let\oldnumberline\numberline%
	\renewcommand{\numberline}{\tablename~\oldnumberline}%
	\listoftables%
}
\newpage 
\addcontentsline{toc}{chapter}{\listfigurename}
{%
	\let\oldnumberline\numberline%
	\renewcommand{\numberline}{\figurename~\oldnumberline}%
	\listoffigures%
}
\chapter*{NOMENCLATURE}

\addcontentsline{toc}{chapter}{NOMENCLATURE}

\begin{tabular}{ll}

    CRB & Cramer Rao bound  \\
        CPI & Coherent processing interval  \\
	DFT & Discrete Fourier transform \\
	DNN & Deep Neural Network \\
	DOA & Direction of arrival \\
	FOV & Field of view \\
MLE & Maximum likelihood estimate  \\
	MaxSNR  & Maximizing signal to noise ratio \\
	MaxSINR  & Maximizing signal to interfernce plus noise  ratio \\
	MIMO & Multiple input multiple output \\
    MRA & Minimum redundancy array \\
    
    QCQP & Quadrtically constraint quadratic program \\
	SDR & Semidefinite relaxation \\
    SCA & Successive convex relaxation \\
	TDL & Time delay line \\
	ULA & Uniform linear array \\

\end{tabular}

\chapter*{ABSTRACT}

\addcontentsline{toc}{chapter}{ABSTRACT}

Sparse array design through sensor selection reduces system
transceiver overhead by lowering the hardware costs and processing
complexity. Sparse-based design can potentially  improve active and passive sensing in radar, communications, and underwater acoustics. Sparse array design approaches include coarse sampling in time, frequency, and space at both the transmitter (for active sensing) and the receiver. These approaches present a new paradigm to, for instance, direction-of-arrival (DOA) or Doppler estimation and allow dealing with more sources than physical sensors. 

Sparse sensor placement, with various design objectives, has successfully been employed in diverse application areas, particularly for enhanced parameter estimation and receiver performance. The
sparse array design criteria are generally categorized into
environment-independent and environment-dependent performance
metrics. The former are largely benign to the
underlying environment and, in principle, seek to maximize
the spatial degrees of freedom by extending the coarray
aperture. This enables high resolution DOA estimation possibly involving more sources
than the available physical sensors. Environment-dependent
objectives, on the other hand, consider the operating
conditions characterized by emitters and targets
in the array field of view, in addition to receiver noise.
In this regard, applying such objectives renders the array
configuration as well as the array weights time-varying in
response to dynamic and changing environment.

 This work is geared towards designing environment-dependent sparse array beamformer to improve the output signal-to-interference and noise ratio using both narrowband and wideband signal platforms. This is achieved through low latency sensor selection technology that enables cost effective sparse array design that can be readily configured to meet
environment-dependent performance metrics. In this case, the
system cost can be considerably reduced by multiplexing the
expensive transceiver chains to serve many more perspective
sensor locations. However, at any given time, only a few
sensor locations are operational which correspond to the
active connections to the transceiver chains.

One key challenge in implementing the data-dependent approaches is the lack of knowledge of
exact or estimated values of the data autocorrelation function
across the full sparse array aperture. The sparse array design can
only have few active sensors at a time, in essence making
it difficult to furnish the correlation values corresponding
to the inactive sensor locations. At the core of this work is to address the aforementioned issues by devising innovative solutions using convex optimization and machine learning tools, structured sparsity concepts, low rank matrix completion schemes and fusing the environment-dependent and environment-independent deigns by developing a hybrid approach.   Sparse array design is also proposed using the deep neural network (DNN). The DNN based design  paves the way for a successful real-time nimble implementation of data driven sparse beamformer that efficiently overcomes the bottlenecks specific to sparse implementations.

Finally, active sparse array design is considered  which is critical to implementing an
efficient receiver design for adaptive radar operations. A desirable
transmit beampattern design is achieved which focuses the transmitted power towards
the perceived target locations and, as such, suppresses the
transmission towards undesired directions. In so doing, it can
steer clear of heavy clutter environment or avoid probing
towards an adversary location in covert situations. This work also addresses another
critical design objective for the transmitter, i.e.,  to minimize the
cross correlation of the returns from different targets to
enable effective adaptive receiver processing. High target
cross correlation can severely deteriorate performance and impede the performance of  adaptive radar operations. This is achieved by  lending a separate beamformer for each target location, such that all  beamformers share a common sparse configuration. 
\newpage

\pagenumbering{arabic} 

\chapter{INTRODUCTION AND MOTIVATION}

\section{Sparse Arrays}
\begin{figure}[!tbp]
  \centering
  \subfloat[]{\includegraphics[width=0.63\textwidth]{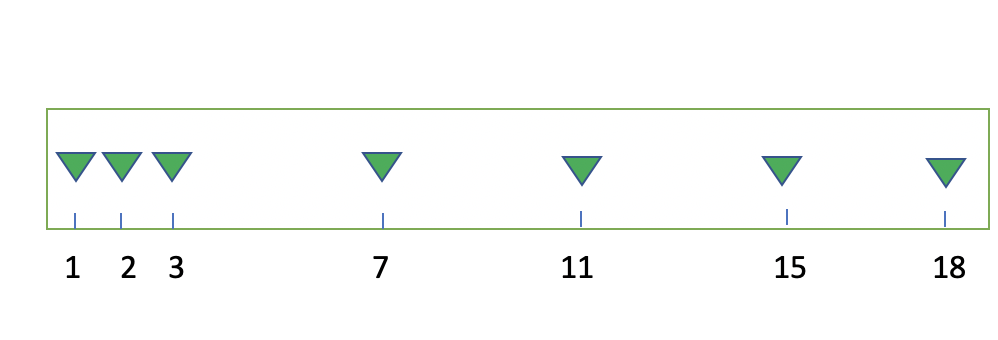}\label{wo_g_ch1}}
  \hfill
  \subfloat[]{\includegraphics[width=0.93\textwidth]{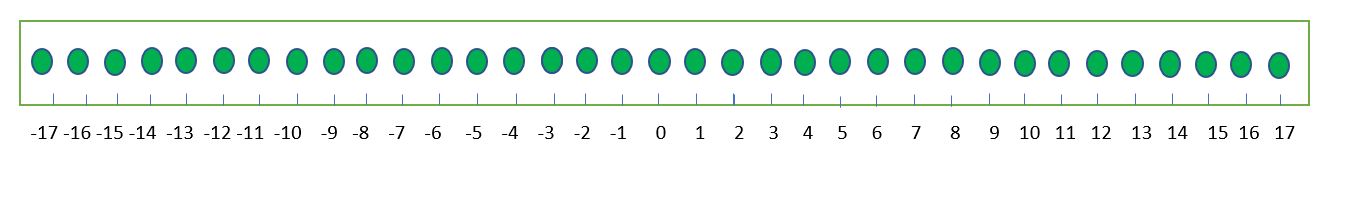}\label{sop_g1_ch1}}
 \hfill
  \caption{(a) Minimum redundancy  array  (7 elements);  (b) Corresponding coarray}
\label{sop_g_wo_g_ch1}
\end{figure}

\begin{figure}[!tbp]
  \centering
  \subfloat[]{\includegraphics[width=0.83\textwidth]{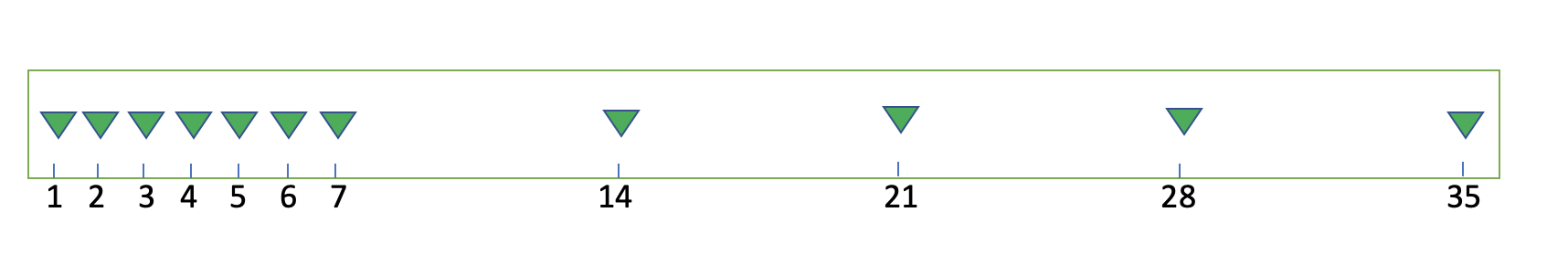}\label{wo_g3_ch1}}
  \hfill
  \subfloat[]{\includegraphics[width=0.93\textwidth]{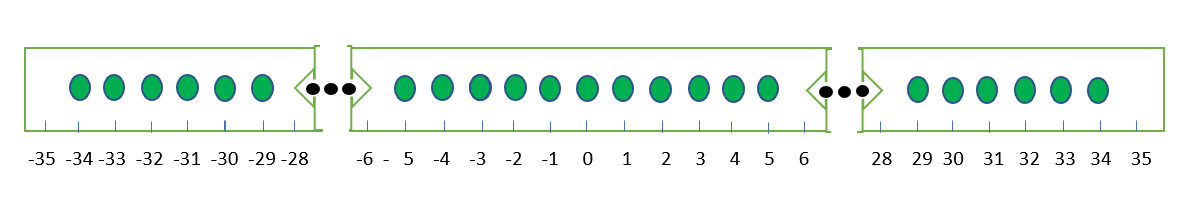}\label{sop_g4_ch1}}
 \hfill
  \caption{(a) Nested array  (11 elements);  (b) Corresponding coarray}
\label{sop_g_wo_g5_ch1}
\end{figure}

Planning sensor locations can potentially economize the receiver cost by  curtailing the valuable hardware and
computational resources. Sparse sensor placement, with
various design objectives, has successfully been employed
in diverse application areas, particularly for enhanced parameter
estimation and receiver performance \cite{710573, 6477161, 1428743, 4058261,  4663892,   6031934}. The
sparse array design criteria are generally categorized into
environment-independent and environment-dependent performance
metrics. The former are largely benign to the
underlying environment and, in principle, seek to maximize
the spatial degrees of freedom by extending the  coarray
aperture.The difference coarray of a given array configuration
demonstrates the unique correlation lags that can be exploited to maximize the degrees-of-freedom (DOF). This enables high resolution direction of
arrival (DOA) estimation possibly involving more sources
than the available physical sensors. 
Towards this goal, sparse array is interpreted in the coarray domain to yield a higher number of degrees-of-freedom (DOFs). For instance, for a
given number of physical sensors, the minimum redundancy array (MRA \cite{1139138}) (Fig. \ref{wo_g_ch1})   maximizes the number of
consecutive virtual sensors in the resulting difference coarray as shown in Fig. \ref{sop_g1_ch1}.
However, for an arbitrary number of sensors, the MRA  configurations are hard to optimize since they don't follow any general expressions. 
Alternatively, several  structured array configurations have recently been proposed to maximize the contiguous difference coarray including the coprime arrays and nested arrays \cite{5456168, 5739227, 7012090}. Nested array is easy to construct configuration obtained by combining two uniform
linear subarrays with different interelement and  result in a larger number of
consecutive virtual sensors under the coarray equivalence as shown in Fig. \ref{sop_g_wo_g5_ch1}. 

\begin{figure}[!t]
\centering
\includegraphics[width=6.0in]{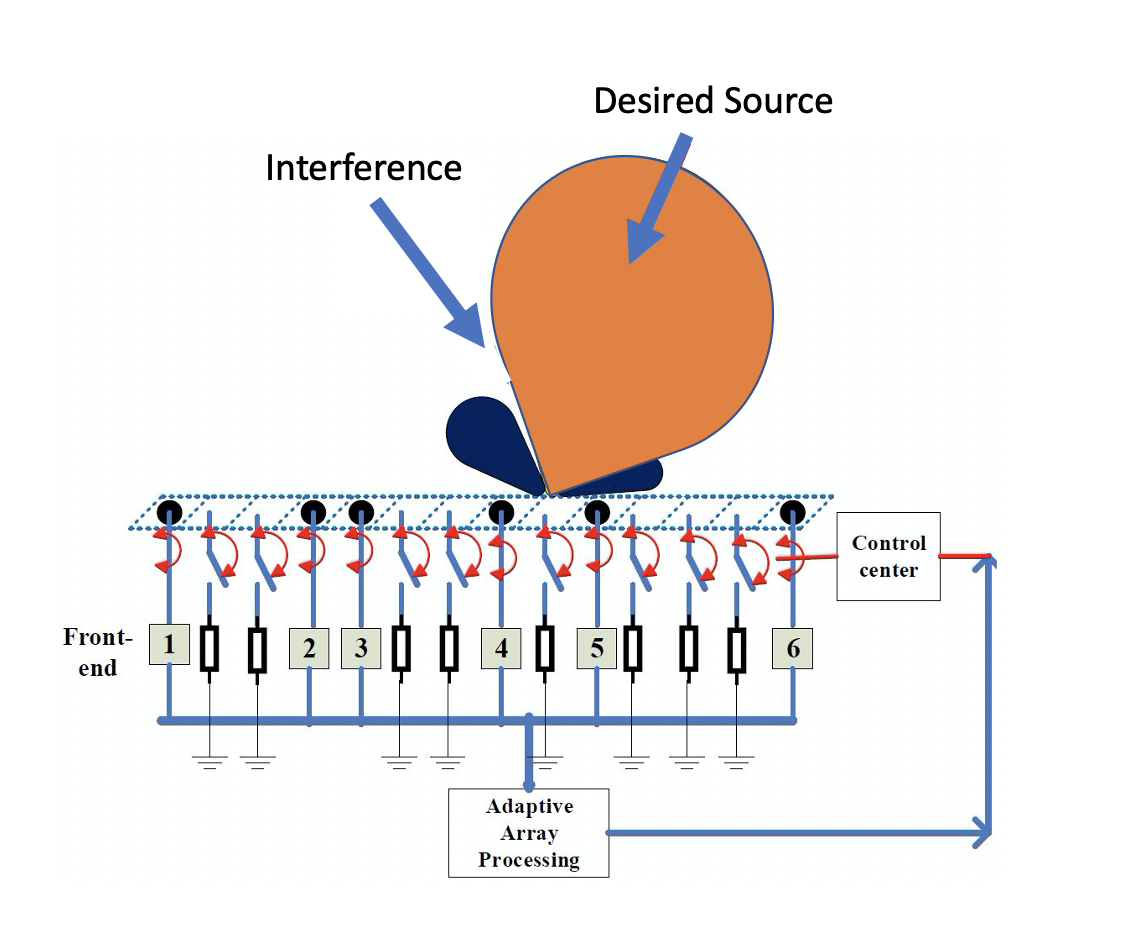}
\caption{Reconfigurable array through sensor switching}
\label{cir_sweep_ch1}
\end{figure}

Environment dependent 
objectives, on the other hand, consider the operating
conditions characterized by  the underlying constellation of emitters and targets or interferences in the array field of view, in addition to receiver noise.
In this regard, applying such objectives renders the array
configuration as well as the array weights time-varying in
response to dynamic and changing environment.
In this thesis, we focus on environment dependent optimum sparse array beamformer design
that maximizes the  signal to-
interference and noise ratio (MaxSINR). Sparse array design typically involves the selection of a
subset of uniform grid points for sensor placements. For a
given number of sensors, it is  assumed that the number of
grid points, spaced by half wavelength, is limited due to the constraint on the physical aperture.  In this approach, antenna positions are selected from uniformly spaced locations
that are served by a limited number of transceiver chains. The environment-sensitive design
objectives have recently become more realizable due to advances in efficient sensor switching technologies that readily
activates a subset of sensors on a predefined grid points resulting in reconfigurable arrays \cite{6774934} depicted in Fig. \ref{cir_sweep_ch1}.
Thereby, the system cost can significantly be reduced by
limiting the number of expensive transceivers chains.

\section{Main Contributions and   Thesis Organization}

The work presented in this thesis is geared towards designing environment-dependent sparse array beamformer to improve the output signal-to-interference and noise ratio using both narrowband and wideband signal platforms. It addresses key challenges in implementing the data-dependent approaches including the lack of knowledge of exact or estimated values of the data autocorrelation function across the full sparse array aperture. This is because the sparse array design can only have few active sensors at a time, in essence making it difficult to furnish the correlation values corresponding to the inactive sensor locations. At the core of this work is to address the aforementioned issues by devising innovative solutions using convex optimization and machine learning tools, structured sparsity concepts, low rank matrix completion schemes and fusing the environment-dependent and environment-independent deigns by developing a hybrid approach. 

The main contributions of the thesis are organized as follows. Chapters 2 deals with sparse array design to optimally process the source seen through the local scatterers. Chapter 3 and 4 focus on the MaxSINR narrowband sparse receive beamformer through the hybrid and matrix completion  approaches. Chapter 5 develops the design of wideband sparse array design through TDL and DFT implementations. Sparse array active design  is investigated in Chapters 6 and Chapter 7 presents DNN based sparse beamformer design, followed by conclusion.

Chapter 2 examines the MaxSNR problem in presence of
local scatterings which may follow specific deterministic or
statistical scattering models. It is shown that if the scatterers
assume a Gaussian distribution centered around the source
angular position, the optimum array configuration for maximizing
the SNR is the commonly used uniform linear array
(ULA). If the scatterers circulate the source, the array design is optimized to harness sparse array topologies with superior performance over ULAs. Simulation results are presented to show
the effectiveness of array configurability in the case of both
Gaussian and circularly spread sources.

In Chapter 3, we consider sparse array design for receive
beamforming achieving maximum signal-to-interference
plus noise ratio (MaxSINR) for both single point source and
multiple point sources, operating in an interference active environment. Unlike existing sparse design methods which either
deal with structured, environment-independent, or non-structured,
environment-dependent arrays, our method is a hybrid approach
and seeks a full augumentable array that optimizes
beamformer performance. Thereby, one can in essence reap the benefits of structured and
non-structured arrays. This paradigm calls for a new aperture
design approach that strives to provide filled co-arrays and, at
the same time, be environment-sensitive. This approach proves important for
limited aperture that constrains the number of possible uniform
grid points for sensor placements. The problem is formulated
as quadratically constraint quadratic program (QCQP), with
the cost function penalized with weighted $l_{1}$-norm squared of
the beamformer weight vector.  Performance comparisons among
the proposed sparse array, the commonly used uniform arrays,
arrays obtained by other design methods, and arrays designed
without the augmentability constraint are provided.

Chapter 4 address the sparse array optimization design
requiring to estimate the data autocorrelations at all spatial lags across the array aperture. Towards this end, we adopt
low rank matrix completion under the semidefinite Toeplitz constraint for interpolating those autocorrelation values
corresponding to the missing lags. We compare the performance of matrix completion approach with that of the
fully augmentable sparse array design acting on the same objective function. The optimization tool employed is the
regularized $l_1$-norm successive convex approximation (SCA). Design examples with simulated data are presented using
different operating scenarios, along with performance comparisons among various configurations.

In Chapter 5, we develop sparse array receive beamformer design
methods achieving maximum signal-to-interference plus noise ratio
(MaxSINR) for wideband sources and jammers. Both tapped
delay line (TDL) filtering and the DFT realizations to wideband
array processing are considered.  The sparse array configuration design problem
is formulated as a quadratically constraint quadratic program
(QCQP) and solved by using SDR (semidefinite relaxation). A
computationally viable approach through SCA (successive convex
relaxation) is also pursued. It is assumed that the sensor
configuration remains the same within the observation time. In
radar, this assumption amounts to configuring the same sparse configuration
over the coherent processing interval (CPI). The notion of mixed $l_{1-q}$ norm regularization
is exploited to thrive the group sparsity to conceive a  wideband sparse beamformer. In order to realize an implementable
design, in presence of missing autocorrelation lags, we propose
parameter-free block Toeplitz matrix completion to estimate
the received data correlation matrix across the entire array
aperture.

Chapter 6 examines the  sparse
array design for transmit beamforming. The main task is to
design the beamformer to direct the bulk of the transmitted
power to target locations. This, in turn, enables improved receiver
performance which strives to maximize the signal-to-interference
plus noise ratio (MaxSINR) of the radar returns. In order
to realize an environmental-dependent adaptive design, it is
incumbent that the directed signals towards different targets
are maximally mutually independent. This is achieved by  lending a separate beamformer for each target location, such that all  beamformers share a common sparse configuration. It is shown that the proposed approach can efficiently utilize the available array aperture to achieve the desired transmitted beampattern characteristics. 

In Chapter 7, we consider sparse array design for receive beamforming achieving maximum signal-to-interference plus noise ratio (MaxSINR) for a desired point source, operating in a narrowband interference active environment. The sparse array design methods developed thus far, are either data driven or rely entirely on the prior knowledge of the interference DOAs and respective powers. In this chapter, we develop a sparse beamformer  spectral analysis    (SBSA) approach, exploiting the prior information of the interference parameters for implementing a MaxSINR beamformer with superior performance and low computational complexity. The data-driven design is essentially conceived by integrating the proposed SBSA design and the Deep Neural Network (DNN). Towards this goal, the training scenarios are generated either through enumeration or simulated more expediently by SBSA methodology for DNN, to learn an effective representation that can perform well in a downstream task. The DNN effectively approximates the unknown mapping from the input received correlation matrix and outputs the sparse configuration with superior interference mitigation capability. The performance of the DNN is evaluated by the ability to learn the enumerated and SBSA designs. It is shown that DNN effectively learns the proposed algorithms and, as such, paves the way for efficient real-time implementation.


 Conclusion and recommendations follow at the end.


\section{List of publications}

The following publications are the result of this research.

\subsection{Journal articles}

\begin{itemize}
\item \textbf{S.  Hamza} and M. Amin, \textit{\textquotedblleft  Sparse Array Design for Maximizing the Signal-to-Interference-plus-Noise-Ratio by
	Matrix Completion,\textquotedblright} Digital Signal Processing, 2020.
	\item \textbf{S.  Hamza} and  M. G. Amin, \textit{\textquotedblleft Hybrid Sparse Array Beamforming Design for General Rank Signal Models,\textquotedblright} IEEE Transactions on Signal Processing, 2019.
	\item \textbf{S.  Hamza} and  M. G. Amin, \textit{\textquotedblleft  Sparse Array  Beamforming for Wideband Signal Models,\textquotedblright} IEEE Transactions on Aerospace and Electronic Systems (Under review).
		\item \textbf{S.  Hamza} and  M. G. Amin, \textit{\textquotedblleft  Sparse Array  Beamformer Design Availing Deep Learning Approach,\textquotedblright} IEEE Transactions on Signal Processing (Ready for submission).
\end{itemize}

\subsection{Conference papers}

\begin{itemize}
\item \textbf{S. Hamza} and   M. Amin, \textit{\textquotedblleft Sparse Array Receiver Beamformer Design  for  Multi-Functional Antenna,\textquotedblright} submitted to EUSIPCO, 2020. 
\item \textbf{S. Hamza} and  M. G. Amin, \textit{\textquotedblleft Sparse Array Design  for Transmit Beamforming,\textquotedblright}  IEEE Radar Conference (RadarConf20), 2020.
\item \textbf{S. Hamza}  and   M. G. Amin, \textit{\textquotedblleft Planar Sparse Array Design for Transmit Beamforming,\textquotedblright}  SPIE Defense and Commercial Sensing, 2020.
\item \textbf{S. Hamza} and  M. G. Amin, \textit{\textquotedblleft A Method for Optimum Sparse Array Beamforming Design,\textquotedblright}  SPIE Defense and Commercial Sensing, 2020.
\item \textbf{S. Hamza} and    M. G. Amin \textit{\textquotedblleft Hybrid Sparse Array Design for Under-determined Models,\textquotedblright}  ICASSP 2019 - 2019 IEEE International Conference on Acoustics, Speech and Signal Processing (ICASSP), Brighton, United Kingdom, 2019.
\item \textbf{S.  Hamza} and  M. G. Amin \textit{\textquotedblleft Sparse Array Design Utilizing Matix Completion,\textquotedblright} Asilomar Conference on Signals, Systems, and Computers, Pacific Grove, CA, USA, 2019.
\item \textbf{S.  Hamza} and   M. Amin, \textit{\textquotedblleft Sparse Array DFT Beamformers for Wideband Sources,\textquotedblright} 2019 IEEE Radar Conference (RadarConf19), Boston, MA, USA.
\item \textbf{S. Hamza} and   M. Amin, \textit{\textquotedblleft Optimum Sparse Array Beamforming for General Rank Signal Models,\textquotedblright} 2018 IEEE Radar Conference (RadarConf18), Oklahoma City, OK, 2018.
\item \textbf{S. Hamza}, M. G. Amin and G. Fabrizio \textit{\textquotedblleft Optimum Sparse Array Design for Maximizing Signal- to-Noise Ratio in Presence of Local Scatterings,\textquotedblright} 2018 IEEE International Conference on Acoustics, Speech and Signal Processing (ICASSP), Calgary, AB, 2018.
\item \textbf{S.  Hamza} and    M. G. Amin \textit{\textquotedblleft Optimum Sparse Array Receive Beamforming for Wideband Signal Model,\textquotedblright} 2018 Asilomar Conference on Signals, Systems, and Computers, Pacific Grove, CA, USA, 2018.
\item \textbf{S.  Hamza} and   M. G. Amin \textit{\textquotedblleft Two Dimensional Sparse Array Design of Wideband Signals for Data Reduction,\textquotedblright}  SPIE Defense and Commercial Sensing,  2019
\end{itemize}





%

\chapter{Optimum Sparse Array Design for Maximizing Signal-to-Noise Ratio in Presence of Local Scatterings}

\section{Introduction}
%
%
%
%
Sparse arrays have recently attracted much attention that is motivated by switched antenna technologies and advances in constrained minimization and convex optimization techniques. There are several metrics to design sparse arrays and decide on optimum array configurations.  Among those metrics, maximum signal-to-noise ratio (MaxSNR), maximum signal-to-interference and noise ratio (MaxSINR), and reduced Cramer-Rao bound (CRB) for direction-of-arrival (DOA) estimation, yield improved beamforming and direction finding performance \cite{6477161, 6774934, 8036237, 1634819, 6714077, 5680595}. 

Designing optimum sparse arrays for MaxSNR strives to maximize the SNR at the array output for a given source direction-of-arrival (DOA). Depending on the number of antennas and permissible antenna locations, it has been shown that a significant performance improvement  can be achieved over other commonly used sparse arrays, including nested and coprime structured array \cite{5456168, 5739227, 7012090}. In  past contributions, point sources have typically been assumed where each source is characterized by a steering vector and provides a rank-one spatial covariance matrix at the array receiver. However, depending on the multipath environments and source signal bandwidth, these steering vectors, along with the corresponding covariance matrix rank, can  significantly  change \cite{768200, 983319,  5592379, 1631777, 8061016}.  

In this chapter,  the effect of the spatial channel on optimum sparse array beamforming for MaxSNR is examined for the first time.   Two models for local scatterings, namely, the Gaussian model  and  circular model,  are considered. These scattering models are most suitable for  dense urban environment, in which the signal encounter rich  scattering prior to its arrival at the array receiver \cite{656151, 471582, 995514, 1002493}. It is shown that the optimum sparse array always elects a configuration that seeks the highest spatial correlation values across the antennas. On the other hand for a spatial correlation that is monotonically decreasing, the sparse array would assume the minimum possible antenna separation, so as to collect the highest correlation values. This is accomplished by   configuring a ULA. For those scattering models in which the correlation rises and falls with increased antenna spacing, the optimum MaxSNR sparse array configuration deviates from a ULA, and positions the antennas such that their separations are consistent with the highest sensor correlation values. \vspace{-.15mm}

We pose the problem as optimally selecting $K$ antennas out of $N$ possible equally spaced locations on a uniform grid. The antenna selection problem  for maximizing SNR amounts to maximizing the principal eigenvalue of the source correlation matrix \cite{1223538}. In order to realize convex relaxation for this NP hard optimization problem and avoid computational burden of singular value decomposition for each possible configuration, we maximize the lower bound of output SNR instead. The lower bound optimization objective is  approximated by Taylor series  and  formulated as an iterative linear  program. \vspace{-.15mm}
%
%

The rest of the chapter is organized as follows:  In  section \ref {Problem Formulation_ch2}, we  formulate  the problem  for  maximizing the  output SNR. The role of  array configurability for MaxSNR for Gaussian and circular scattering models is  explained in section \ref {Spatial spread vector channel models_ch2}. Section \ref{Optimum sparse array design_ch2} deals with the iterative solution of finding optimum array design. Simulations  and conclusion follows at the end. 
\setlength{\parskip}{0mm}
\section{Problem Formulation} \label{Problem Formulation_ch2}
Consider a spatially spread source with $P$ independently scattered components impinging on a  linear array with $N$ uniformly placed antennas. Then, the signal received at the array  at  time instant $t$ is given by: 
\begin {equation} \label{a_ch2}
\mathbf{x}(t)=   \sum_{k=1}^{P} (\alpha _k(t)) \mathbf{a}( \theta_k)  + \mathbf{n}(t),
\end {equation}
where
\begin {equation*}  \label{b_ch2}
\mathbf{a} ({\theta_k})=[1 \,  \,  \, e^{j (2 \pi / \lambda) d cos(\theta_k)  } \,  . \,   . \,  . \, e^{j (2 \pi / \lambda) d (N-1) cos(\theta_k)  }]^T,
\end {equation*}
is the steering vector  corresponding to the $ k$th scatterer with the direction-of-arrival $\theta_k$,  $d$ is the inter-element spacing in terms of  wavelengths $\lambda$, $\alpha _k(t)$ $\in \mathbb{C}$  represents the complex amplitude of $k$th scatterer and $\mathbf{n}(t)$ $\in \mathbb{C}^N$ represents the additive Gaussian noise with variance $\sigma_n^2$ at the receiver output. The received signal $\mathbf{x}(t)$  is  linearly combined by the $N$-antenna  beamformer that strives to maximize the output SNR. The output signal $y(t)$ of the optimum beamformer for maximum SNR is given by \cite{1223538},
\vspace{1mm}
\begin {equation}  \label{c_ch2}
y(t) = \mathbf{w_0}^H \mathbf{x}(t),
\end {equation}
 with \vspace{-2mm}
\begin {equation}  \label{d_ch2}
\mathbf{w_0} = \mathscr{P} \{  \mathbf{R_n}^{-1} \mathbf{R}  \}.
\end {equation}
The operator $\mathscr{P} \{. \}$  computes the principal eigenvector, $\mathbf{R}=\mathbf{UR_sU}^H$ is the received  source correlation matrix with $\mathbf{U}=[\mathbf{a} ({\theta_1}) \,  . \,   . \,  . \,\mathbf{a}
({\theta_P})]$ and $k$, $l$th entry of $\mathbf{R_s}(k,l)=E\{\alpha _k(t)\alpha _k^H(t)\}$ for $k=l$ and zero otherwise. For spatially uncorrelated noise, $\mathbf{R_n}=\sigma_n^2\mathbf{I}$, we
obtain the corresponding optimum output $SNR_o$ as follows: \vspace{.5mm}
\begin {equation}  \label{f_ch2}
\mathbf{w_0} = \mathscr{P} \{  \mathbf{R}  \},
\end {equation}
\begin {equation}  \label{g_ch2}
SNR_o=\frac {\mathbf{w_0}^H \mathbf{R} \mathbf{w_0}} { \mathbf{w_0}^H \mathbf{R_n} \mathbf{w_0}} = \frac {||\mathbf{R}||_2} {\sigma_n^2}.
\end {equation}
Here, $||.||_2$ denotes the spectral norm or the maximum eigenvalue of the matrix. Equations~(\ref{f_ch2}) and~(\ref{g_ch2}) show that the optimum beamformer for MaxSNR is directly tied to the eigen-structure of the correlation matrix. As such, there is a need to analyze the correlation matrix under the scattering models. 
\section{Gaussian and circular scattering models} \label{Spatial spread vector channel models_ch2}
Gaussian model assumes that  the directions of arrival of the scatterers  are Gaussianly distributed, $\mathscr{N}(\theta_{0},\sigma_\theta)$, having mean direction of arrival $\theta_{0}$ and variance $\sigma_\theta$. Consequently, each element of the steering vector would be jointly Gaussian with  zero mean and covariance matrix given by \cite{656151}, \vspace{1mm}
\begin {equation} \label{a2_ch2}
\mathbf{R_g}_{(\theta_0,\sigma_0)} \approx (\mathbf{a}({\theta_0})\mathbf{a}^H({\theta_0})) \circ  \mathbf{B}_{(\theta_0,\sigma_0)},
\end {equation}
 where `$\circ$' denotes Hadamard product and \vspace{1mm}
\begin {equation} \label{b2_ch2}
\mathbf{B}_{(\theta_0,\sigma_0)}(k,l) = e^{-2(\pi\delta (k-l))^2\sigma_0^2 cos^2(\theta_0)}.
\end {equation}
 Denote $\mathbf{z} \in \{0,1 \}^N $ a selection vector  whose entries $0$'s and $1$'s represents the presence and absence  of corresponding antennas respectively. The steering vector  corresponding to antenna selection vector  $\mathbf{z}$  is given by $\mathbf{a} _{{\theta_0}}(\mathbf{z})=\mathbf{a} {{(\theta_0)}}\odot \mathbf{z}$. Here `$\odot$'  is the element-wise product operator which allows the selection of antenna elements according to $\mathbf{z}$. Similarly, $\mathbf{B}_{(\theta_0,\sigma_0)}(\mathbf{z}) =   \mathbf{B}_{(\theta_0,\sigma_0)}  \odot \mathbf{Z} $ with $\mathbf{Z}=\mathbf{z}\mathbf{z}^T$ being the corresponding antenna selection  matrix. Equation (\ref{a2_ch2}) with selected antennas can be re-written as follows: \vspace{1mm}
\begin {equation} \label{c2_ch2}
\mathbf{R_g}_{(\theta_0,\sigma_0)}(\mathbf{z})\approx  (\mathbf{a} _{{\theta_0}}(\mathbf{z})\mathbf{a}^H _{{\theta_0}}(\mathbf{z}))\circ  \mathbf{B}_{(\theta_0,\sigma_0)}(\mathbf{z}).
\end {equation}
We  note that the trace of $ \mathbf{R_g}_{(\theta_0,\sigma_0)}(\mathbf{z})$ is constant since the
input source power remains the same irrespective of the array configuration.
Accordingly, the  sum of eigenvalues is constant  for all possible correlation matrices associated with the $K$ antenna selection problem. To be more explicit, the problem formulated  in Eq.~(\ref{g_ch2}) can  be expressed as:  \vspace{1mm}
\begin{equation} \label{d2_ch2}
\begin{aligned}
 \underset{\mathbf{z}}{\text{max}} &  \quad \quad  {||\mathbf{R_g}_{(\theta_0,\sigma_0)}(\mathbf{z})||_2}, \\ 
\text{given}; &  \,      \sum_{k=1}^{K} v_\mathbf{z}(k) =  \mathbf{Tr} \, ( \mathbf{R_g}_{(\theta_0,\sigma_0)}(\mathbf{z})) = K\mathbf{Tr}(\mathbf{R_s}),
\end{aligned}
\end{equation}
 where  $\mathbf{Tr}$ $(.)$ denotes the trace of the matrix, $v_\mathbf{z}(k)$ is the $k$th eigenvalue of correlation matrix $ \mathbf{R_g}_{(\theta_0,\sigma_0)}(\mathbf{z})$. 
Equations~(\ref{a2_ch2})  and~(\ref{b2_ch2}) show that the correlation drops monotonically with increased correlation lag. As shown below, this property compels the optimum sparse array to assume a ULA with minimum inter-element spacing; hence, the solution  does not require any iteration-based method or enumeration. \par
Let  $ \mathbf{R_g}_{(\theta_0,\sigma_0)}(\mathbf{u})$  be the correlation matrix for ULA `$\mathbf{u}$' and  $ \mathbf{R_g}_{(\theta_0,\sigma_0)}(\mathbf{s})$ be  the correlation  matrix associated with the  sparse  `$\mathbf{s}$'  configuration with the same number of antennas, $K$. The $k$th eigenvalue $v_{\mathbf{z}M}(k)$ of $\mathbf{R}^M_{\mathbf{g}(\theta_0,\sigma_0)}(\mathbf{z})$ is  related to its corresponding eigenvlaue $v_\mathbf{z}(k)$ of $ \mathbf{R_g}_{(\theta_0,\sigma_0)}(\mathbf{z})$ by \cite{CIT-006}, \vspace{1mm}
\begin {equation} \label{f2_ch2}
v_{\mathbf{z}M}(k)= v^M_\mathbf{z}(k),   \, \, \, \,  \forall \, M \geq 0.
\end {equation}
For the Gaussian spatial channel,   $\mathbf{Tr}$ $(\mathbf{R}^M_{\mathbf{g}(\theta_0,\sigma_0)}(\mathbf{u})) >  \mathbf{Tr}    \, (\mathbf{R}^M_{\mathbf{g}(\theta_0,\sigma_0)}(\mathbf{s}))$ (proof in section \ref{Appendix_ch2}) along with Eq.~(\ref{f2_ch2}), yields
 \vspace{-1.5mm}
\begin {equation} \label{g2_ch2}
\sum_{k=1}^{K} v^M_\mathbf{u}(k) > \sum_{k=1}^{K} v^M_\mathbf{s}(k), \, \, \, \, \,   \forall \, M \geq 2.
\end {equation}
\vspace{-.65mm}
From Eq.~(\ref{g2_ch2}), it can be readily shown that $max \,(v_\mathbf{u}(k)) > max \, (v_\mathbf{s}(k))$. Here, we make use of the fact that all the  eigenvalues of the  correlation matrices are greater or equal to zero. 
\begin{figure}[!tbp]
  \centering
  \subfloat[]{\includegraphics[width=0.63\textwidth]{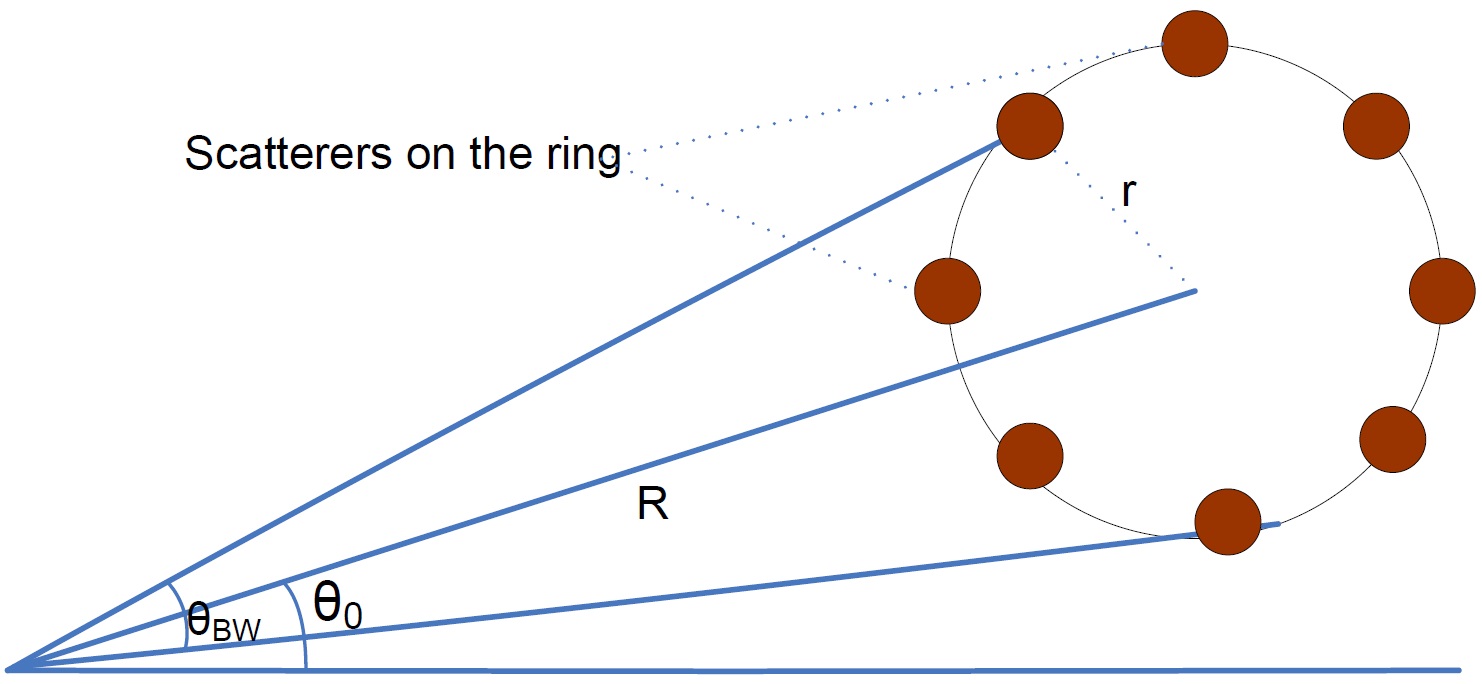}\label{circular_f}}
  \hfill
  \subfloat[]{\includegraphics[width=0.83\textwidth]{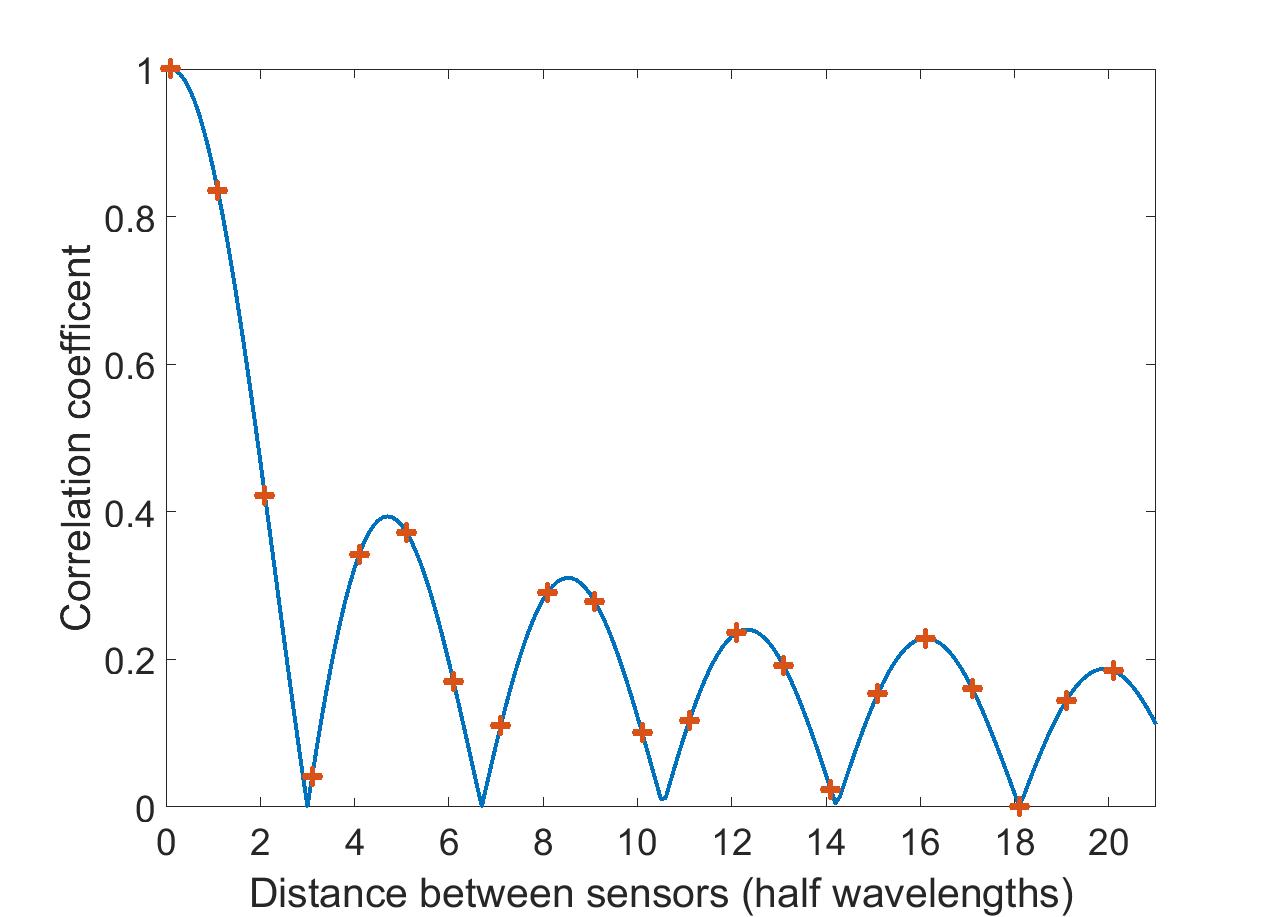}\label{plotcorr}}
 \hfill
  \caption{(a) Circular model;  (b) Correlation coefficient as a function of lags }
\label{circular_fplotcorr}
\end{figure}
Therefore, for the  Gaussian scattering model where the correlation function is monotonically decreasing with sensors spacing, the optimum sparse array would always cluster with minimum spatial separation, configuring a  ULA. \vspace{.5mm}

The correlation between consecutive sensors for the circular model is given by \cite{Lee:1997:MCE:522758}, 
\begin {equation}  \label{i2_ch2}
r_{c} (\theta_0,\sigma_0)=\frac{1}{P}\sum_{i=0}^{P-1}e^{-j2 \pi (\delta (k-l))cos(\theta_0+\theta_i)},
\end {equation}
where $r_{c} (\theta_0,\sigma_0)$ is the $k,l$th element of the correlation matrix $\mathbf{R_{c}}_{ (\theta_0,\sigma_0)}$  and $\theta_i$'s are circularly distributed around the source as shown in Fig. \ref{circular_f}. Contrary to  Gaussian model, the sensor data correlation in circular case shows oscillatory behaviour (Fig. \ref{plotcorr}) as a function of lags  (Eq.~(\ref{i2_ch2})). We can, however, bound the upper and lower limits of the optimum SNR as follows:
\begin {equation} \label{e2_ch2}
\frac {\mathbf{Tr}\,( \mathbf{R_c}_{(\theta_0,\sigma_0)}(\mathbf{w}))}{K}    \leq SNR_o  \leq \mathbf{Tr}\,(\mathbf{R_c}_{(\theta_0,\sigma_0)}(\mathbf{o})).
\end {equation}
$\mathbf{R_c}_{(\theta_0,\sigma_0)}(\mathbf{o})$  and $\mathbf{R_c}_{(\theta_0,\sigma_0)}(\mathbf{w})$ are the optimum  `$\mathbf{o}$'  and worst  `$\mathbf{w}$'  array correlation matrices, respectively. In the optimal  case, the eigenvalues are maximally spread, whereas the worst possibility for MaxSNR arises when all the eigenvalues are equal. We also note that   Eq.~(\ref{g2_ch2}) in fact determines the eigen-spread of the correlation matrix asymptotically. It can be shown  that this equation  remains valid for circular correlation matrix for some $\zeta$ sufficiently large, such that $\sum_{k=1}^{K} v^M_\mathbf{o}(k) > \sum_{k=1}^{K} v^M_\mathbf{s}(k), \, \, \, \, \,   \forall \, M \geq \zeta$. Therefore, finding the optimum configuration amounts to maximizing $\sum_{k=1}^{K} v^M_\mathbf{z}(k)$ or equivalently, Tr  ($\mathbf{R}^M_{\mathbf{c}(\theta_0,\sigma_0)} (\mathbf{z})$) for any $M \geq \zeta$, over all possible configurations. Though maximizing Tr  ($\mathbf{R}^M_{\mathbf{c}(\theta_0,\sigma_0)} (\mathbf{z})$)   is computationally less expensive than maximizing the  principal eigenvalue,  yet it is highly computationally involved. Therefore, in the next section we resort to  lower bound relaxation  to design the optimum  configuration. 
 \section{Optimum sparse array design}  \label{Optimum sparse array design_ch2}
Following the approach in \cite{8061016}, we assume that  we have an estimate of  $\mathbf{ R_{(\theta_0, \sigma_0)}}$, which is the full antenna array source correlation matrix. Then, the problem in  Eq.~(\ref{d2_ch2}) can be rewritten as follows: \vspace{-1.5mm}
\begin{equation} \label{a3_ch2}
\begin{aligned}
 \underset{\mathbf{z}}{\text{max}} &&  ||\mathbf{ R_{(\theta_0, \sigma_0)}(z)} ||_2,\\
\text{s.t.} &&    ||\mathbf{z}||_0=K.
\end{aligned}
\end{equation}
Here,  $||.||_0$ determines the cardinality of selection vector $\mathbf{z}$. Given $\mathbf{e_0}$, the principal eigenvector corresponding to the full antenna array, we approximate the  thinned vector $\mathbf{e_0}(\mathbf{z})/||\mathbf{e_0}(\mathbf{z})||_2$ as the principal eigenvector of the selected $K$-element subarray $\mathbf{z}$ \cite{7944303}. The problem in Eq.~(\ref{a3_ch2})  can then be approximated by: \vspace{-1mm}
\begin{equation} \label{b3_ch2}
\begin{aligned}
 \underset{\mathbf{z}}{\text{max}} \quad &  \frac{\mathbf{ e_0}^H(\mathbf{z})\mathbf{R_{(\theta_0, \sigma_0)}}(\mathbf{z})\mathbf{e_0}(\mathbf{z})}{||\mathbf{e_0}(\mathbf{z})||_2^2},\\
\text{s.t.} \quad &   ||\mathbf{z}||_0=K.
\end{aligned}
\end{equation}
This approximation represents a lower bound of optimum SNR. Define, $\tilde{\mathbf{R}}_{\theta, \sigma_0} = \mathbf{R}_{(\theta_0, \sigma_0)} \circ  (\mathbf{e_0}\mathbf{e_0}^H)$ and $\tilde{\mathbf{e_o}}=\mathbf{e^*_o} \circ \mathbf{e_o}$.  Equation~(\ref{b3_ch2})  can be rephrased as follows:  \vspace{.5mm}
\begin{equation} \label{c3_ch2}
\begin{aligned}
 \underset{\mathbf{z}}{\text{max}} && \, \, \, \, \frac{\mathbf{ z}^T\mathbf{\tilde{R}_{\theta, \sigma_0}z}}{\mathbf{z}^T\mathbf{\tilde{e_0}}},\\
\text{s.t.} &&  \, \,  ||\mathbf{z}||_1=K, \\
 &&  \, \,  0 \leq \mathbf{z} \leq 1.
\end{aligned}
\end{equation}
The  constraint in  Eq.~(\ref{b3_ch2}) is relaxed to affine equality constraint ($||.||_1$ denotes $l^1$-norm) and a  box  constraint, but the objective function still  remains non-convex. Therefore, we resort to iterative first order Taylor approximation as follows:
\begin{equation} \label{d3_ch2}
\begin{aligned}
 \underset{\mathbf{z}}{\text{max}} & \quad \quad \frac{-\mathbf{ z}^{iT}\mathbf{\tilde{R}_{\theta, \sigma_0}z}^i + 2\mathbf{ z}^{iT}\mathbf{\tilde{R}_{\theta, \sigma_0}z}}{\mathbf{z}^T\mathbf{\tilde{e_0}}}, \\
\text{s.t.} &  \quad \quad ||\mathbf{z}||_1=K, \\
 &   \quad \quad  0 \leq \mathbf{z} \leq 1.
\end{aligned}
\end{equation}
 Here, $i$ is the iteration number. This linear fractional program (LFP) can be turned into LP by simple change of variables, $\alpha=1/\mathbf{z}^T\mathbf{\tilde{e_0}}$  and $\mathbf{v}=\bf{z}/\mathbf{z}^T\mathbf{\tilde{e_0}}$ as follows \cite{Boyd:2004:CO:993483},  \vspace{1.5mm}
\begin{equation} \label{e3_ch2}
\begin{aligned}
 \underset{\mathbf{v},\alpha}{\text{max}} &  \quad \quad  {-\mathbf{ z}^{iT}\mathbf{\tilde{R}_{\theta, \sigma_0}z}^i\alpha + 2\mathbf{ z}^{iT}\mathbf{\tilde{R}_{\theta, \sigma_0}v}},\\
\text{s.t.} &   \quad \quad  \mathbf{1}^T\mathbf{v}-K\alpha=0, \, \, \mathbf{\tilde{e_0}}^T\mathbf{v}=1,\\
 &   \quad \quad  \mathbf{v} \geq 0, \, \, \mathbf{v}-\alpha \leq 0, \, \, \alpha \geq 0.
\end{aligned}
\end{equation}
yielding, the estimate of the selection vector $\mathbf{z}=\mathbf{v}/\alpha$.
 \section{Simulations} \label{Simulations_ch2}
For  both  scattering models, we select $K=9$ sensors from $N=21$  possible equally spaced locations with  inter-element spacing of $\lambda/2$. The source SNR   is  $0$ dB.
 \vspace{-1mm}
\begin{figure}[!tbp]
  \centering
  \subfloat[]{\includegraphics[width=0.5\textwidth]{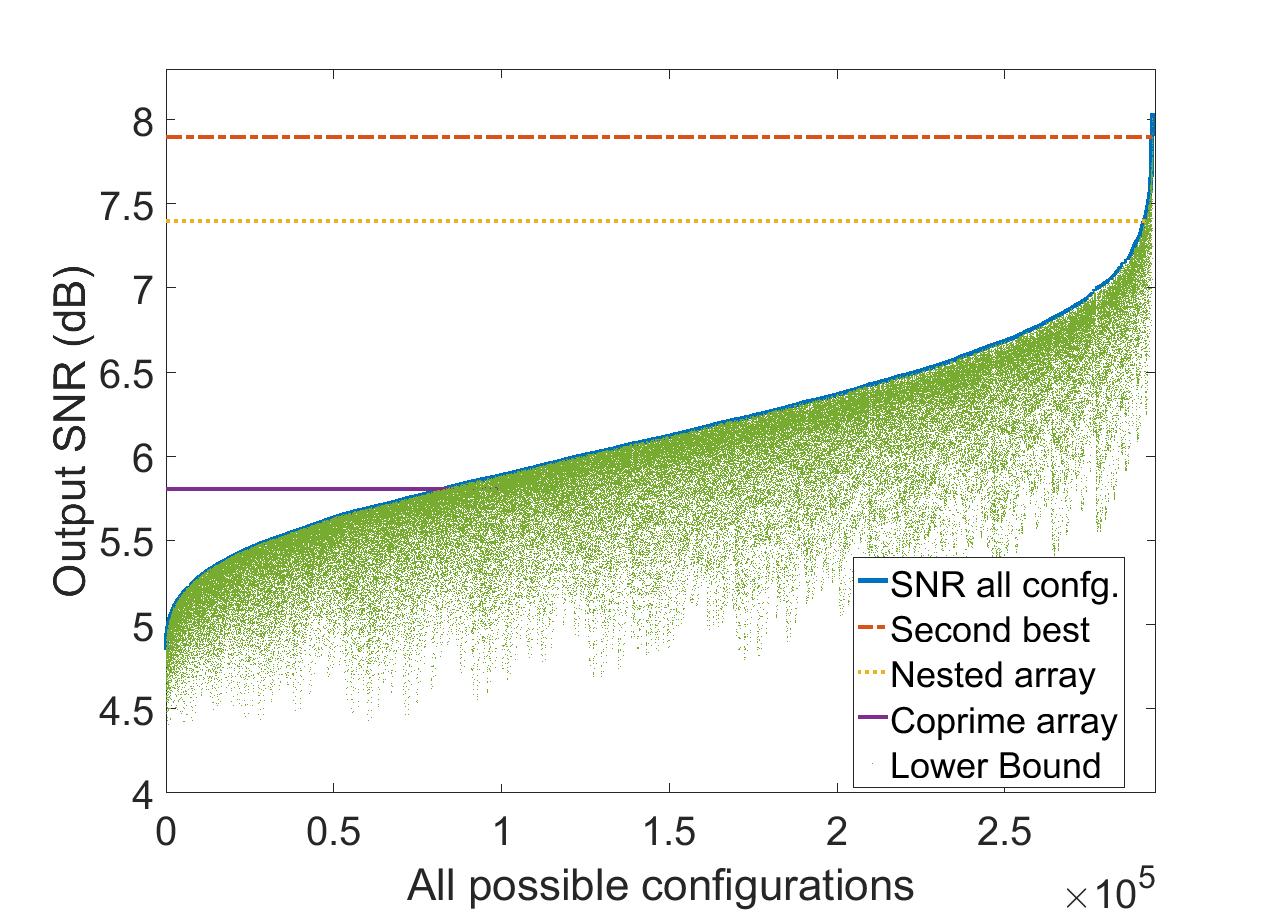}\label{gauss_90_ch2}}
 \hspace*{\fill}
  \subfloat[]{\includegraphics[width=0.5\textwidth]{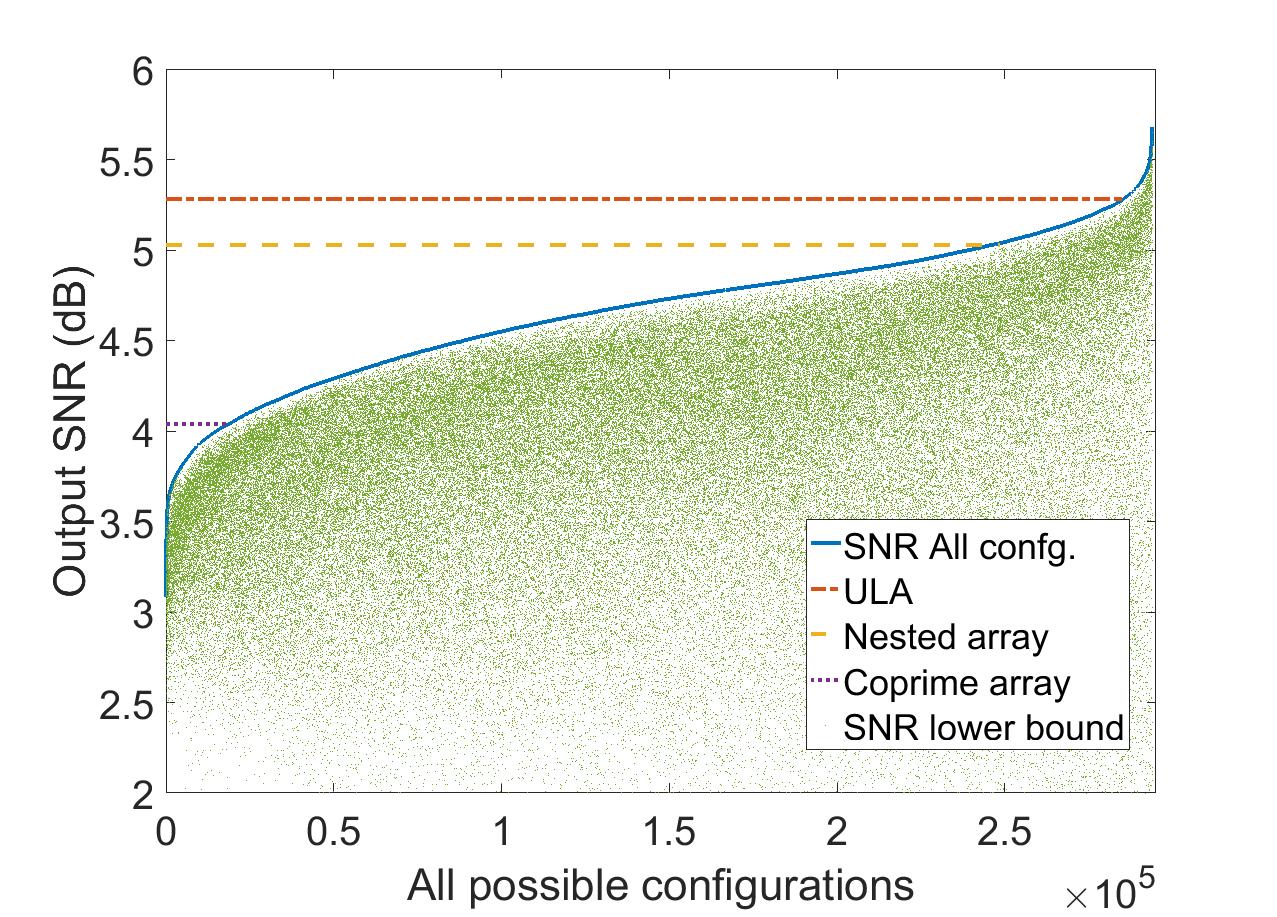}\label{cir_90_ch2}}
 \vspace*{\fill}
  \subfloat[]{\includegraphics[width=0.5\textwidth]{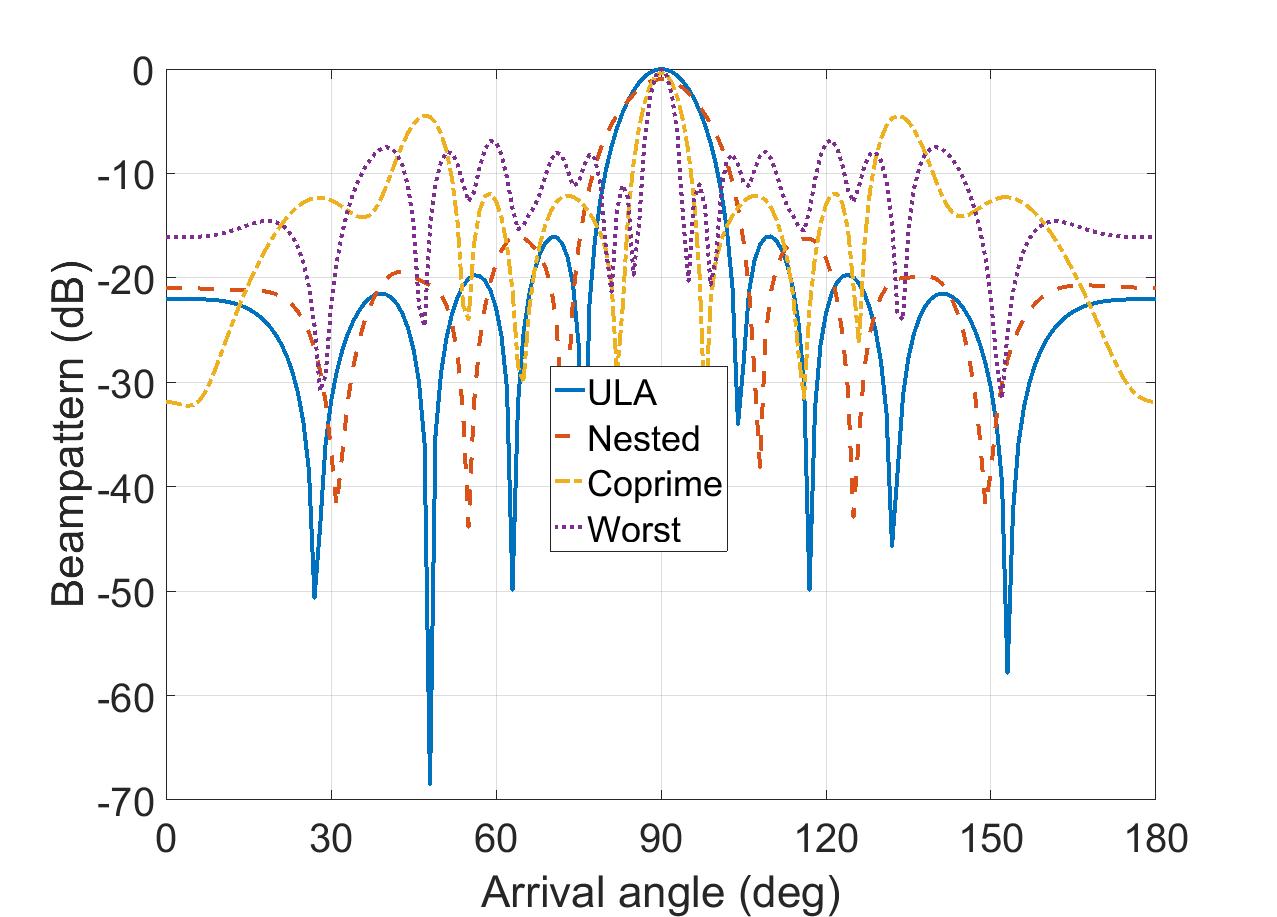}\label{bp_g_ch2}}
 \hspace*{\fill}
  \subfloat[]{\includegraphics[width=0.5\textwidth]{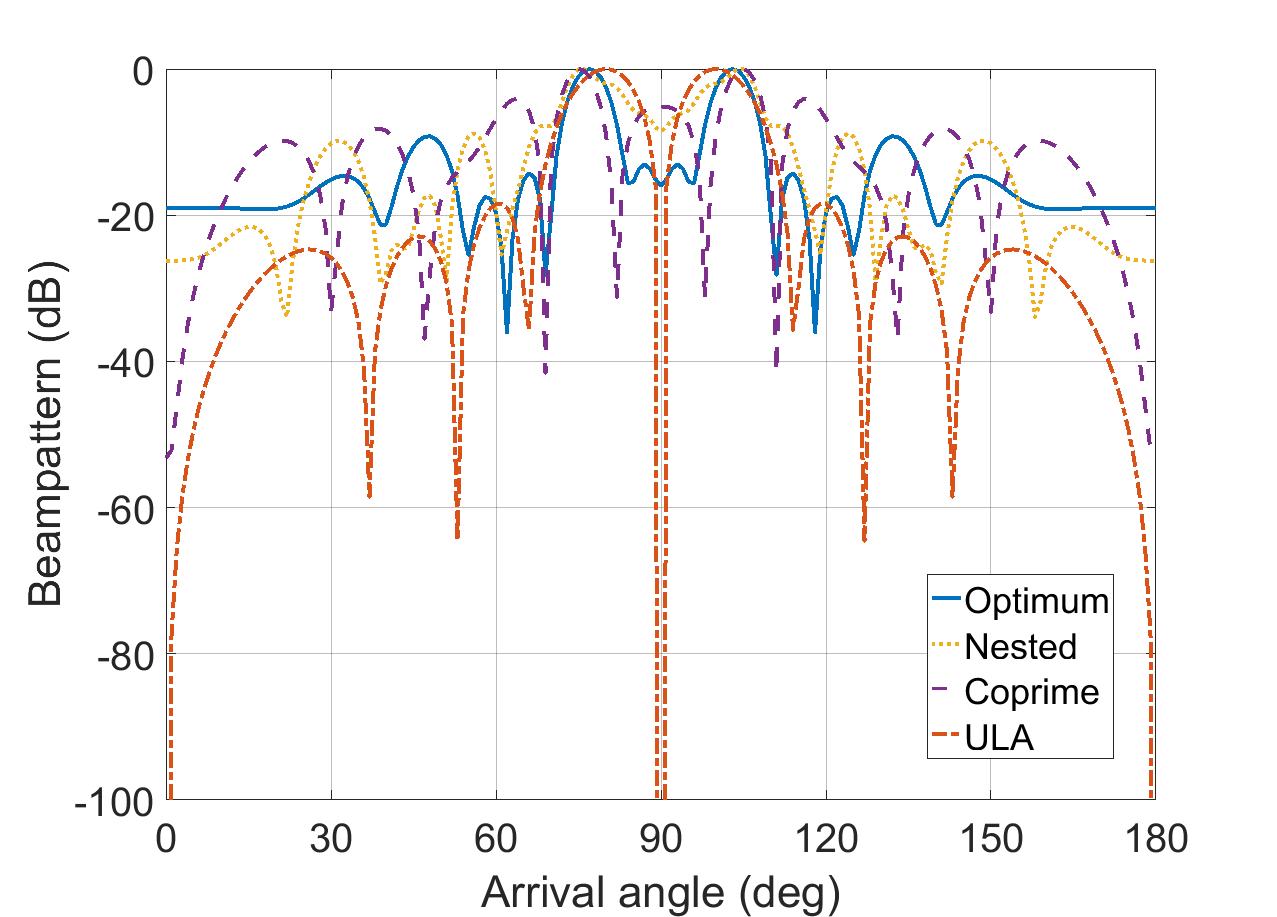}\label{bp_c_ch2}}
 \hspace*{\fill}
  \caption{Output  SNR comparison for various array configurations; (a) Gaussian model; (b) circular model. Beampattern of optimum array; (c) Gaussian model; (d) circular model.}
\end{figure}
\subsection{Gaussian model}
Figure~\ref{gauss_90_ch2}  plots the optimum SNR for all possible array configurations in ascending order of output SNR for Gaussian spread source with centre angle of  90$^0$ and variance of  5$^0$. As expected, the optimum array emerges as a ULA with output SNR of 8 dB. The lower bound relaxation  is also depicted in Fig.~\ref{gauss_90_ch2} and is shown to offer a good metric in the underlying case. The ULA has more than 3 dB advantage over the worst array configuration having less than 5 dB output SNR. The worst array configuration is shown in the Fig.~\ref{wo_g_ch2} where  `$.$'  and  `$\times$'  represent the presence and absence of sensor respectively. We also observe that if  sensor malfunction prevents a contiguous uniformly spaced antenna array configuration, the optimum sparse array opts to minimally stretch, incorporating the next closest antenna position, as shown   in the Fig.~\ref{sop_g_ch2}. In order to simulate this case, we set antenna 9 as a faulty antenna. The output SNR corresponding to this new configuration is 7.9 dB which is 0.1 dB less than the optimum array configuration. This lower SNR is  the price paid by including a smaller correlation value corresponding to antenna position 10 compared to that of antenna position 9.  The output SNR corresponding to coprime and nested sparse arrays, as depicted in  Fig.~\ref{gauss_90_ch2}, are significantly less as compared to the ULA. Figure~\ref{bp_g_ch2} shows the optimum beampattern   corresponding to the different configurations with ULA giving the  widest main lobe with the highest gain possible at the scatterer's centre. Moreover, the sidelobes for ULA are significantly lower as compared to other sparse arrays which is highly desirable. \vspace{-2.5mm}

\subsection{Circular model}
\begin{figure}[!tbp]
  \centering
  \subfloat[]{\includegraphics[width=0.83\textwidth]{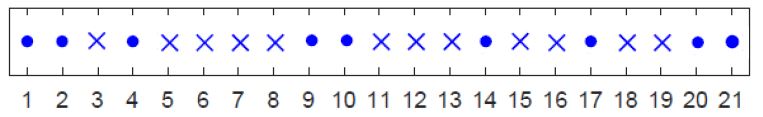}\label{wo_g_ch2}}
  \hfill
  \subfloat[]{\includegraphics[width=0.83\textwidth]{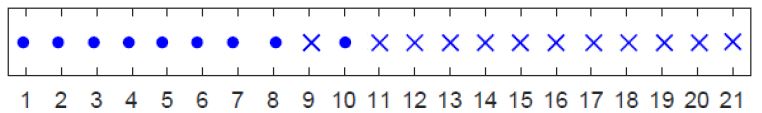}\label{sop_g_ch2}}
 \vfill
\subfloat[]{\includegraphics[width=0.83\textwidth]{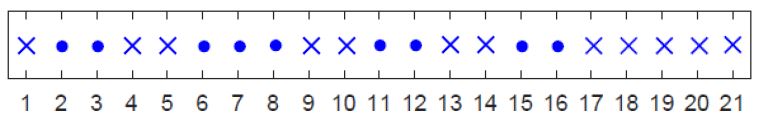}\label{op_c_ch2}}
 \hfill
  \subfloat[]{\includegraphics[width=0.83\textwidth]{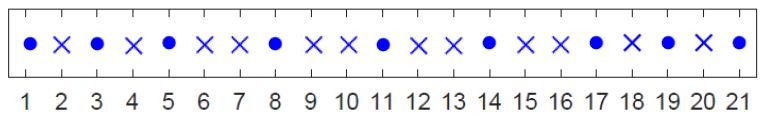}\label{wo_c_ch2}}
 \vfill
\subfloat[]{\includegraphics[width=0.83\textwidth]{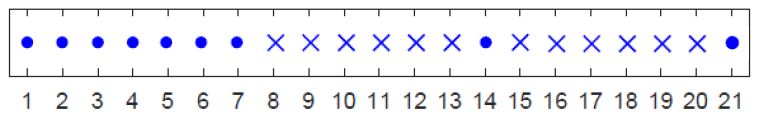}\label{n_c_ch2}}
 \hfill
  \subfloat[]{\includegraphics[width=0.83\textwidth]{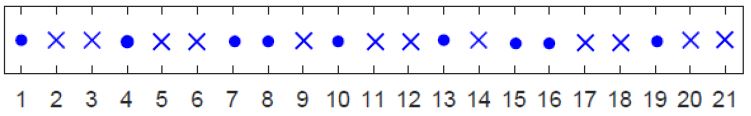}\label{co_c_ch2}}
  \caption{(a) Worst sparse array  (Gaussian);  (b) Second best (Gaussian); (c) Optimum array (circular);  (d) Worst  sparse array  (circular); (e) Nested array;   (f) Coprime array }
\label{sop_g_wo_g}
\end{figure}

\begin{figure}[!t]
\centering
\includegraphics[width=6.0in]{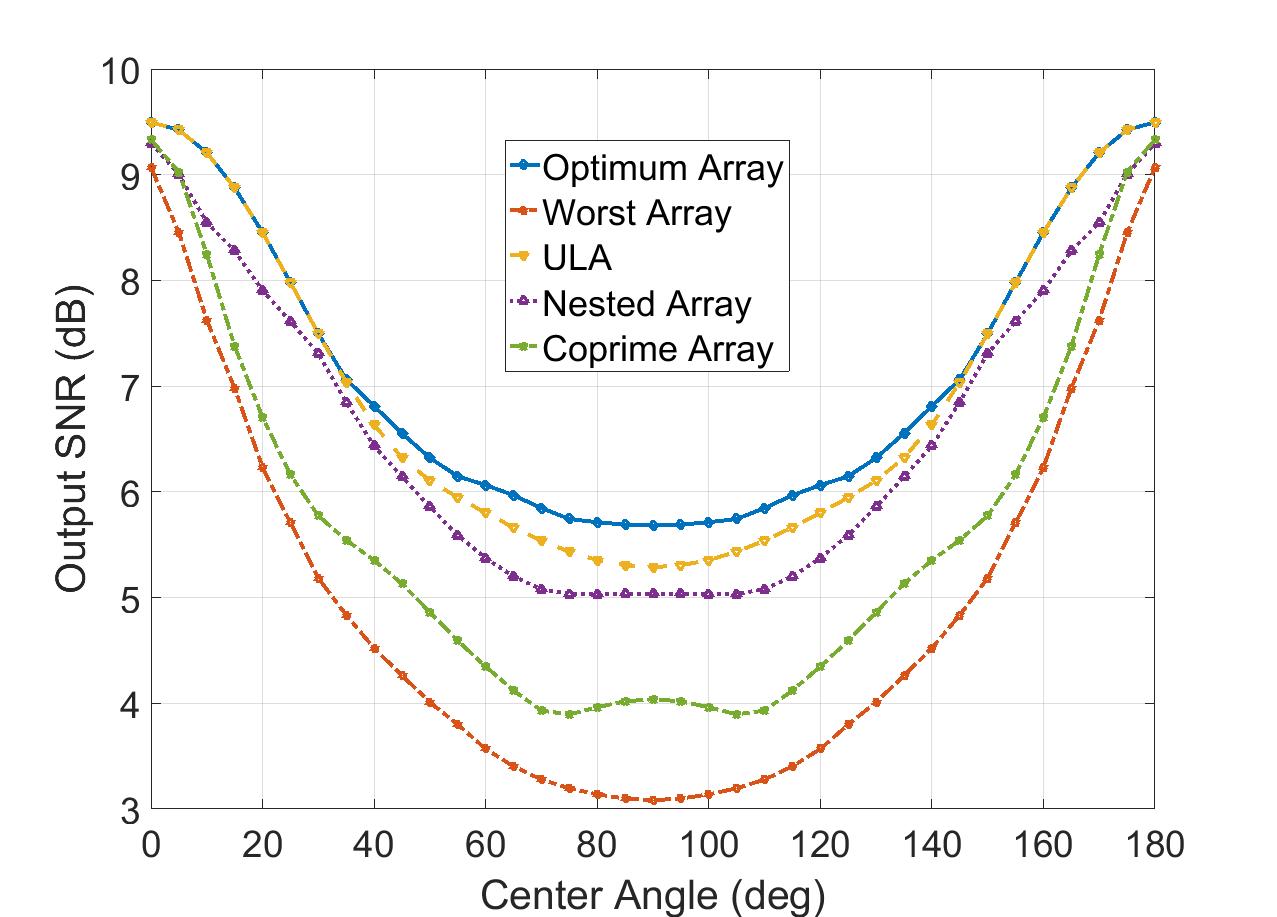}
\caption{Output  SNR for different  arrays  vs centre angle.}
\label{cir_sweep_ch2}
\end{figure}

Figure~\ref{cir_90_ch2} shows the scenario for  circular scattering model with a single source  at the broadside with the spatial spread of  $30^0$.    It  shows that the optimum sparse  array (shown in Fig.~\ref{op_c_ch2}) has SNR of $5.68$ dB, whereas the worst sparse array (shown in Fig.~\ref{wo_c_ch2}) has output SNR of $3$ dB, which is   more than  2.5 dB  down as compared to the  optimum array topology. 
It is informative  to compare  the performance of optimum array with the ULA,  which is a de facto array configuration in many applications. The ULA gives output SNR of 5.28 dB  which is  0.4   dB  lower  than the optimum performance. The significant difference however  lies in the beampattern of the two arrays, as shown in Fig.~\ref{bp_c_ch2}. Contrary to the optimum configuration, the ULA attempts to maximize the output SNR by placing  a null exactly in the centre of the scatterer beam which is highly  undesirable. Figure~\ref{cir_90_ch2} also shows that the optimum sparse  array has a clear advantage over coprime and nested array topologies in terms of output SNR. This is beacuse the optimum array manages  wider mainlobe with higher gain where the scatterers are most dense (Fig.~\ref{bp_c_ch2}). Figure~\ref{cir_sweep_ch2} shows that  better performance of the optimum sparse array is more pronounced at the broadside, whereas the ULA is the optimum array configuration for DOAs near the array  end-fire  location. It can be seen that the  performance of the optimum array and sub optimum sparse arrays differ significantly over a wide  field of view. \vspace{-1mm}
\section{Conclusion}
This chapter considered optimum sparse array configuration for maximizing the beamformer output SNR for a single source that is seen to the receiver through its local scatterers.  It is shown that for the Gaussian local scattering model,  the correlation is weakened monotonically across the receiver antennas. As such, the optimum configuration, in seeking to capture the highest spatial correlation values, becomes the ULA. We showed that for a circular local scattering model, the optimum sparse array loses uniformity in quest of including  the high correlation values corresponding to large antenna separations. In both cases, we solved the optimization problem both iteratively and by enumerations and showed strong agreement between the two methods. \vspace{-1mm}
\section{Appendix} \label{Appendix_ch2}
\renewenvironment{proof}{{ Proof of the result used in Eq.~(\ref{g2_ch2}):}}{\qed}
\begin{proof}\\
 \vspace{2mm}
\begin{equation} \label{a4_ch2}
\begin{aligned}
  \mathbf{Tr} \, (\mathbf{R}^M_{\mathbf{g}(\theta_0,\sigma_0)}(\mathbf{z})) & \approx    \mathbf{Tr} \, ( ( \mathbf{a} _{{\theta_0}}(\mathbf{z})\mathbf{a}^H _{{\theta_0}}(\mathbf{z})\circ  \mathbf{B}_{(\theta_0,\sigma_0)}(\mathbf{z}))^M)  \\
& =    \mathbf{Tr}\,(  \mathbf{a} _{{\theta_0}}(\mathbf{z})\mathbf{a}^H _{{\theta_0}}(\mathbf{z})\circ  \mathbf{B}^M_{(\theta_0,\sigma_0)}(\mathbf{z}))\\
 & =  \mathbf{Tr}\,( \mathbf{B}^M_{(\theta_0,\sigma_0)}(\mathbf{z})).
\end{aligned}
\end{equation}
For Gaussian model, $ \mathbf{B}_{(\theta_0,\sigma_0)}(\mathbf{u}) \geq    \mathbf{B}_{(\theta_0,\sigma_0)}(\mathbf{s})  >   0$. Here, `$\geq$' means element wise comparison and strict equality holds only for  diagonal entries. This implies, 
 \vspace{1mm}
\begin {equation} \label{b4_ch2}
\begin{aligned}
 \mathbf{B}^M_{(\theta_0,\sigma_0)}(\mathbf{u})  &>  \mathbf{B}^M_{(\theta_0,\sigma_0)}(\mathbf{s}), \, \, \, \, \quad \quad \forall \, M \geq 2\\
\mathbf{Tr} \, (\mathbf{B}^M_{(\theta_0,\sigma_0)}(\mathbf{u}) ) &>  \mathbf{Tr}\, (\mathbf{B}^M_{(\theta_0,\sigma_0)}(\mathbf{s})).
\end{aligned}
\end {equation}
\balance 
Combining  (\ref{a4_ch2}) and (\ref{b4_ch2}), we have,
\vspace{1mm}
\begin {equation*} \label{d4_ch2}
\mathbf{Tr} \,( \mathbf{R}^M_{\mathbf{g}(\theta_0,\sigma_0)}(\mathbf{u})) >  \mathbf{Tr} \, ( \mathbf{R}^M_{\mathbf{g}(\theta_0,\sigma_0)}(\mathbf{s}))   \quad 
\forall \, M \geq 2
\end {equation*}
\end{proof}
\renewenvironment{proof}{}{\qed}

\chapter{Hybrid Sparse Array  Beamforming Design for General Rank Signal Models}

\section{Introduction}
Sparse array design through sensor selection reduces system receiver overhead by lowering the hardware costs and processing complexity. It finds applications in sensor signal processing for communications, radar, sonar, satellite navigation,  radio telescopes, speech enhancement  and ultrasonic imaging \cite{710573, 6477161, 1428743, 4058261,  4663892,   6031934}. One primary goal in these applications is to determine sensor locations to achieve optimality for some pre-determined performance criteria. This optimality includes minimizing the mean radius of the confidence ellipsoid associated with the estimation error covariance matrix  \cite{4663892}, and lowering the  Cramer Rao bound (CRB)   for angle estimation in direction finding problem \cite{6981988}. The receiver performance then depends largely on the operating environment, which may change according to the source and interference signals and locations.  This is in contrast to sparse arrays whose configurations follow certain formulas and seek to attain high extended aperture co-arrays. The driving objective, in this case,  is to enable direction of arrival (DOA) estimation of more sources than physical sensors. Common examples are structured arrays such as nested and coprime arrays \cite{1139138, 5456168, 7012090}.
 
Sparse array design typically involves the selection of a subset of uniform grid points for sensor placements. For a given number of sensors, it is often assumed that the number of grid points, spaced by half wavelength,  is unlimited. However, in many applications, there is a constraint on  the spatial extent of the system aperture. In this case,  a structured array, in seeking to maximize the number of spatial autocorrelation lags,  may find itself placing  sensors beyond the available physical aperture. The problem then becomes that of dual constraints, one relates to the number of sensors, and the other to the number of grid-points.

 With a limited aperture constraint invoked, few sensors may in fact be sufficient to produce a desirable filled structured co-array, even with narrowband assumption and without needing wideband or multiple frequencies \cite{7106511}. In this case, any additional sensors, constitute a surplus that  can be utilized to meet an environment-dependent performance criterion, such as maximum signal-to-interference and noise ratio (SINR).  Thereby, one   can in essence reap the benefits of structured and non-structured arrays. This paradigm calls for a new aperture design approach that strives to provide filled co-arrays and, at the same time, be environment-sensitive.  This hybrid design approach is the core contribution of this chapter.

Sparse sensor  design has  thoroughly been studied  to economize the  receive beamformer \cite{1144829, 1144961, 1137831, Lo1966ASO, 1534, 97353, 5680595, doi:10.1029/2010RS004522,  299602, 775291, 6663667, article1, 6145602, 6822510}.  However, in contrast to MaxSINR design,  the main focus of  the efforts, therein, was in achieving  desirable beampattern characteristics with nominal sidelobe levels, since the sparse beamformer  is susceptible to high  sidelobe levels. For example, an array thinning design was proposed for sidelobe minimization in \cite{1534}  by starting from a fully populated array and sequentially  removing sensors in a systematic manner. Instead, the sparse array design presented in \cite{97353} to  optimize the peak sidelobe level   involves a  joint design of sensor locations and their corresponding beamforming weights. A beampattern matching design explained in \cite{5680595} can  effectively recover  sparse topologies  through an iterative cyclic approach.  Additionally, global optimization tools such as Genetic Algorithms/Simulated Annealing and  convex relaxation schemes based on re-weighted $l_1$-norm minimization have  been  rigorously exploited in sensor selection problem  for synthesizing a user-specified receive beampattern response \cite{299602, 775291, 6663667, article1, 6145602, 6822510}.

In environment-dependent array design, signal power estimation and enhancement in an interference active environment  has a direct bearing on improving   target detection and localization for radar   signal processing,   increasing   throughput or  channel capacity for MIMO wireless communication systems, and enhancing   resolution capability in  medical imaging \cite{Goldsmith:2005:WC:993515, Trees:1992:DEM:530789, 1206680}. It is noted that with sparse array, the commonly used Capon beamforming must not only find the optimum weights but also the optimum array configuration. This is clearly an entwined optimization problem, and requires finding maximum SINR over all possible sparse array configurations. Maximum signal to noise ratio (MaxSNR) and MaxSINR have been shown to yield significantly efficient beamforming with  performance depending mainly on the positions of the sensors as well as the locations of sources in the field of view (FOV) \cite{6774934,  1634819, 6714077}.      \par

In this chapter, we  consider a bi-objective optimization problem, namely achieving the filled co-array  and maximizing the SINR. The proposed technique enjoys  key advantages as compared to state-of-the-art sparse aperture design, namely, (a) It does not require any \textit{a priori} knowledge of the  jammers directions of arrival and their respective power which is implicitly assumed in  previous contributions \cite{8061016, 8313065, 8036237}. As such, it is possible to directly work  on the received data   correlation matrix   (b) It extends to spatial spread sources in a straightforward way.  \par
\begin{figure}[!t]
	\centering
	\includegraphics[width=4.35 in, height=2.3 in]{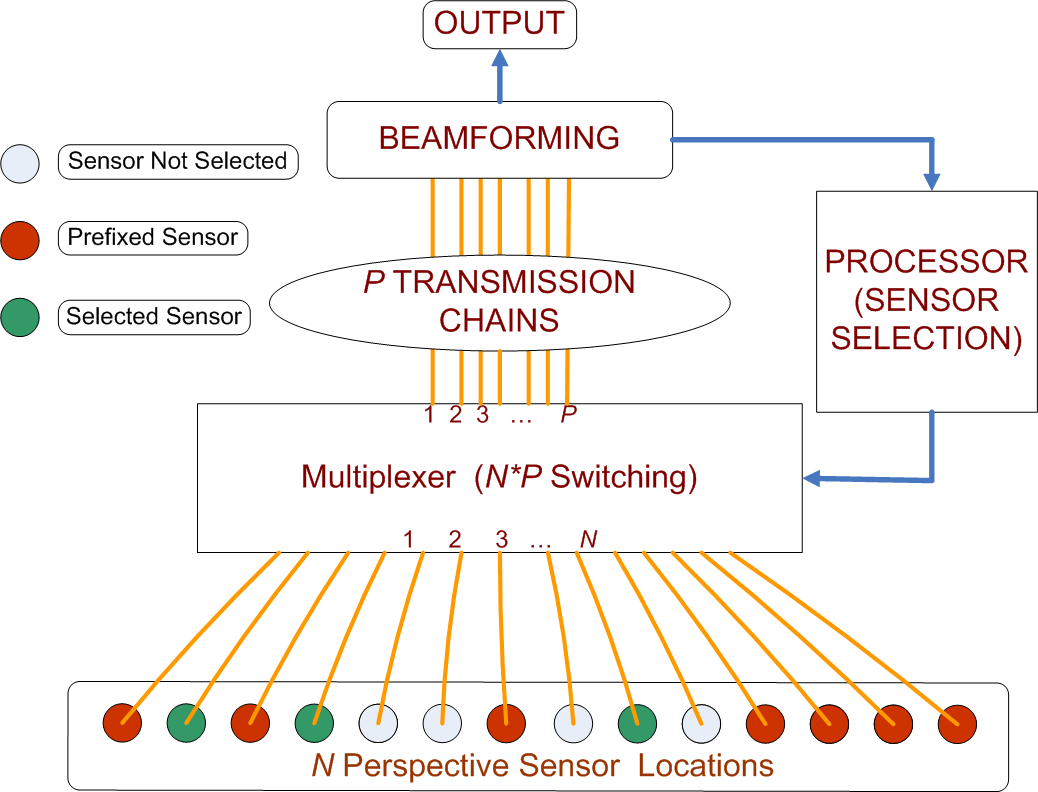}
	\caption{Block diagram of adaptive switched sensor beamformer}
	\label{Block_diagram_ch3}
\end{figure}
 The proposed hybrid approach first determines a prefixed sparse array that results in a filled   co-array with minimum number of sensors. This prefixed configuration could be a minimum redundancy array (MRA) \cite{1139138}, nested or coprime array configuration that fills the aperture under consideration with minimal sensors, allowing maximum degrees of freedom for SINR maximization. This prefixed sensor configuration can  be  achieved by an optimization problem involving the minimum number of sensors spanning a pre-determined aperture. However, for the scope of this chapter, the prefixed configuration is  set by MRA or other structured arrays.   The remaining sensors after forming the prefixed array are  utilized to maximize the  SINR.  The cascade nature of the proposed hybrid approach is relatively simpler than the  ultimate design approach that produces the optimum filled sparse array that maximizes SINR. Environment-dependent array design lowers the hardware complexity by reducing the expensive transmission chains through sensor switching as shown in the  block diagram   in  Fig. \ref{Block_diagram_ch3}. The proposed hybrid approach, however,  has an added advantage of offering a simplified sensor switching in time-varying environment.  This is attributed to   large number of fixed location sensors which would always remain  non-switched, irrespective of the sources and interferences in the FOV. 
 
  The proposed hybrid approach is  particularly permissive as the number $N$ of possible sensor locations increases.  To further clarify, it is noted that sparse arrays having $N$ available sensors can typically span a filled array aperture of the order of $\mathcal{O}(N(N-1)/2)$ \cite{5456168}; conversely, given an aperture spanning  $N$ possible sensor locations, only  $\mathcal{O}(N^{1/2})$ sensors are sufficient to synthesize a fully augmentable array design. This emphasizes the fact that as the possible aperture size increases, then relatively  few sensors are required to meet the full augmentability condition,  leaving more degrees of freedom to optimize  for SINR enhancement. The hybrid  approach also lends itself to  more desirable  beampattern characteristics  by  maintaining  minimum spacing between sensor elements. It is important to note  that having fully augmentable arrays not only provide    the benefits of simplified sensor switching and improved identifiability of large number of sources,  but also they ensure  the availability of full array data covariance matrix essential to carry  optimized SINR configuration \cite{709534}, \cite{Abramovich:1999:PTC:2197953.2201686}.  Therefore, the proposed  simplified hybrid  sensor switching architecture ensures the knowledge of global data statistics at all times,   in contrast to  previous efforts in \cite{7096707, 6333987, 8234698} that sort to optimize data dependent microphone placement viz a viz transmission power. The  proposed methodology therein   targets a different objective function and  primarily relies  on  local heuristics. In this  case,  sensor switching comes with an  additional  implementation overhead, in an  attempt to recursively match the  performance offered by the knowledge of global statistics.     \par

We consider the problem of MaxSINR sparse arrays with limited aperture for both single and higher rank signal correlation matrices. The case of single rank correlation matrix arises when there is one desired source signal in the FOV, whereas the case of higher rank signal model occurs for  spatially spread source.   The   problem is posed as optimally selecting $P$ sensors out of $N$ possible equally spaced grid points. Maximizing SINR amounts to maximizing the  principal eigenvalue of the product of  the inverse of data correlation matrix and the desired source correlation matrix \cite{1223538}. Since it   is an NP hard optimization problem, we pose this problem as QCQP with weighted $l_1$-norm squared to promote sparsity. The re-weighted $l_{1}$-norm squared relaxation is effective for reducing the required sensors and   minimizing the transmit power for multicast beamforming  \cite{ 6477161}. We propose a modified re-weighting matrix  based iterative approach to control the sparsity of the optimum weight vector so that  $P$ sensor fully augmentable hybrid array is finally selected. This modified regularization re-weighting matrix  based approach  incorporates the prefixed structured array   assumption in our design and works by minimizing the objective function   around the presumed prefixed array.   \par  
The rest of the chapter is organized as follows:  In the next section, we  state the problem formulation for  maximizing the  output SINR under general rank signal correlation matrix. Section \ref{Optimum sparse array design_ch3} deals with the optimum sparse array design by  semidefinite relaxation (SDR) and proposed  modified re-weighting  based iterative algorithm of finding $P$ sensor fully augmentable  hybrid  sparse array design. In section \ref{Simulations_ch3}, with the aid of number of  design examples,  we demonstrate the usefulness of fully augmentable arrays achieving MaxSINR and highlight the effectiveness of the proposed methodology  for sparse array design. Concluding remarks follow at the end.
\section{Problem Formulation} \label{Problem Formulation_ch3}
Consider  $K$  desired sources and $L$ independent interfering source signals impinging on a  linear array with $N$ uniformly placed sensors. The baseband signal received at the array  at  time instant $t$ is then given by; 
\begin {equation} \label{a_ch3}
\mathbf{x}(t)=   \sum_{k=1}^{K} (\alpha _k(t)) \mathbf{s}( \theta_k)  + \sum_{l=1}^{L} (\beta _l(t)) \mathbf{v}( \theta_l)  + \mathbf{n}(t),
\end {equation}
where, $\mathbf{s} ({\theta_k})$  and $\mathbf{v} ({\theta_l})$ $\in \mathbb{C}^N$ are  the corresponding steering vectors    respective to directions  of arrival, $\theta_k$ or $\theta_l$, and are defined  as follows;
\vspace{+2mm}
\begin {equation}  \label{b_ch3}
\mathbf{s} ({\theta_k})=[1 \,  \,  \, e^{j (2 \pi / \lambda) d cos(\theta_k)  } \,  . \,   . \,  . \, e^{j (2 \pi / \lambda) d (N-1) cos(\theta_k)  }]^T.
\end {equation}
The inter-element spacing  is denoted  by $d$, ($\alpha _k(t)$, $\beta _l(t))$ $\in \mathbb{C}$  denote the complex amplitudes of the incoming baseband signals \cite{trove.nla.gov.au/work/15617720}. The additive Gaussian noise $\mathbf{n}(t)$ $\in \mathbb{C}^N$ has a  variance of  $\sigma_n^2$ at the receiver output.
The received signal vector $\mathbf{x}(t)$   is   combined linearly by the $N$-sensor  beamformer that strives to maximize the output SINR. The output signal $y(t)$ of the optimum beamformer for maximum SINR is given by \cite{1223538}, 
\begin {equation}  \label{c_ch3}
y(t) = \mathbf{w}_o^H \mathbf{x}(t),
\end {equation}
where $\mathbf{w}_o$ is the solution of the  optimization problem given below;
\begin{equation} \label{d_ch3}
\begin{aligned}
\underset{\mathbf{w} \in \mathbb{C}^N}{\text{minimize}} & \quad   \mathbf{w}^H\mathbf{R}_{s^{'}}\mathbf{w},\\
\text{s.t.} & \quad     \mathbf{ w}^H\mathbf{R}_{s}\mathbf{ w}=1.
\end{aligned}
\end{equation}
For statistically independent signals, the desired source correlation matrix is given by, $\mathbf{R}_s= \sum_{k=1}^{K} \sigma^2_k \mathbf{s}( \theta_k)\mathbf{s}^H( \theta_k)$, where, $ \sigma^2_k =E\{\alpha _k(t)\alpha _k^H(t)\}$. Likewise,  we have the  interference and noise correlation matrix $\mathbf{R}_{s^{'}}= \sum_{l=1}^{L} (\sigma^2_l \mathbf{v}( \theta_l)\mathbf{v}^H( \theta_l)$) + $\sigma_n^2\mathbf{I}_{N\times N}$,    with $ \sigma^2_l =E\{\beta _l(t)\beta_l^H(t)\}$ being the power of the $l$th interfering source. The problem in (\ref{d_ch3}) can  be written equivalently by replacing $\mathbf{R}_{s^{'}}$ with the received data covariance matrix, $\mathbf{R_{xx}}=\mathbf{R}_s+ \mathbf{R}_{s^{'}}$  as follows \cite{1223538},
\begin{equation} \label{e_ch3}
\begin{aligned}
\underset{\mathbf{w} \in \mathbb{C}^N }{\text{minimize}} & \quad   \mathbf{ w}^H\mathbf{R_{xx}}\mathbf{ w},\\
\text{s.t.} & \quad     \mathbf{ w}^H\mathbf{R}_{s}\mathbf{ w} \geq 1.
\end{aligned}
\end{equation}
It is noted that the  equality constraint in (4) is  relaxed in (5) due to  the inclusion of  the constraint as part of the objective function, and as such, (5)  converges to  the equality constraint. Additionally, the optimal solution in (5) is invariant up to  uncertainty of the absolute powers of the sources of interest.  Accordingly, the relative power profile of the sources of interest would suffice.  For a single desired point source, this implies that only the knowledge of the DOA of the desired source is sufficient  rather than the exact knowledge of the desired source correlation matrix. Similarly, neither the source power nor the average power of the scatterers is required in (5) for spatially spread sources when the spatial channel model, such as the Gaussian or circular, is assumed \cite{656151}.  However, in practice, these assumptions can deviate from the actual received data statistics and hence the discrepancy is typically mitigated, to an extent,  by preprocessing  the received data correlation matrix through diagonal loading or tapering the  correlation matrix \cite{1206680}.

     There exists a  closed form solution of the above  optimization problem  and  is given by $\mathbf{w}_o = \mathscr{P} \{  \mathbf{R}_{s^{'}}^{-1} \mathbf{R}_s  \}=\mathscr{P} \{  \mathbf{R_{xx}}^{-1} \mathbf{R}_s  \}$. The operator $\mathscr{P} \{. \}$  computes the principal eigenvector of the input matrix. Substituting $\mathbf{w}_o$ into  (\ref{c_ch3})  yields the corresponding optimum output SINR$_o$;
\begin{equation}  \label{f_ch3}
\text{SINR}_o=\frac {\mathbf{w}_o^H \mathbf{R}_s \mathbf{w}_o} { \mathbf{w}_o^H \mathbf{R}_{s^{'}} \mathbf{w}_o} = \Lambda_{max}\{\mathbf{R}^{-1}_{s^{'}} \mathbf{R}_s\}.
\end{equation}
 This shows that the optimum output SINR$_o$ is given by the maximum eigenvalue ($\Lambda_{max}$) associated with   the product of  the inverse of interference plus noise correlation matrix  and the desired source correlation matrix. Therefore, the performance of the optimum beamformer for maximizing the output SINR is directly related to the desired and interference plus noise correlation matrix. It is to be noted that the  rank of the desired source signal correlation matrix equals $K$, i.e. the cardinality of the desired sources.

\section{Optimum sparse array design}  \label{Optimum sparse array design_ch3}
The problem of locating the maximum  principal eigenvalue among all the   correlation matrices associated with $P$ sensor selection is a combinatorial optimization problem. The  constraint optimization (\ref{e_ch3}) can be re-formulated for optimum sparse array design by incorporating  an additional constraint on the cardinality of  the weight vector;   
\begin{equation} \label{a2_ch3}
\begin{aligned}
\underset{\mathbf{w \in \mathbb{C}}^{N}}{\text{minimize}} & \quad  \mathbf{ w}^H\mathbf{R_{xx}}\mathbf{w},\\
\text{s.t.} & \quad   \mathbf{ w}^H\mathbf{R}_{s}\mathbf{ w}\geq1,  \\
& \quad   ||\mathbf{w}||_0=P.
\end{aligned}
\end{equation}
Here,  $||.||_0$ determines the cardinality of the weight   vector $\mathbf{w}$.  We  assume that we have an estimate of all the filled co-array  correlation lags corresponding to the correlation matrix of the  full  aperture array. The problem expressed in  (\ref{a2_ch3})  can  be relaxed  to induce the sparsity in the beamforming weight vector $\mathbf{w}$ without placing a hard constraint on the specific cardinality of $\mathbf{w}$, as follows \cite{doi:10.1002/cpa.20132};
\begin{equation} \label{b2_ch3}
\begin{aligned}
\underset{\mathbf{w \in \mathbb{C}}^{N}}{\text{minimize}} & \quad   \mathbf{ w}^H\mathbf{R_{xx}}\mathbf{w} + \mu(||\mathbf{w}||_1),\\
\text{s.t.} & \quad    \mathbf{ w}^H\mathbf{R}_{s}\mathbf{ w} \geq1.  \\
\end{aligned}
\end{equation}
Here, $||.||_1$  is the  sparsity inducing $l_1$-norm and $\mu$ is a parameter to control the desired sparsity in the solution. Even though the relaxed problem expressed in  (\ref{b2_ch3}) is not exactly similar to that of  (\ref{a2_ch3}), yet it is well known that  $l_1$-norm  regularization has been an effective tool for recovering sparse solutions in many diverse formulations \cite{Bruckstein:2009:SSS:1654028.1654037,5651522}. The  problem in $(\ref{b2_ch3})$  can be  penalized instead  by the weighted $l_1$-norm function  which is a well known sparsity promoting formulation \cite{Candes2008},
\begin{equation} \label{c2_ch3}
\begin{aligned}
\underset{\mathbf{w \in \mathbb{C}}^{N}}{\text{minimize}} & \quad   \mathbf{ w}^H\mathbf{R_{xx}}\mathbf{w} + \mu(||(\mathbf{b}^i\circ|\mathbf{w}|)||_1),\\
\text{s.t.} & \quad    \mathbf{ w}^H\mathbf{R_{s}}\mathbf{ w} \geq1.  \\
\end{aligned}
\end{equation}
where, \textquotedblleft$\circ$\textquotedblright \, denotes the element wise product, \textquotedblleft$|.|$\textquotedblright \, is the modulus operator and $\mathbf{b}^i \in \mathbb{R}^N$  is the regularization re-weighting vector  at  the $i$th iteration. Therefore, $(\ref{c2_ch3})$ is the sequential optimization methodology, where the regularization re-weighting vector $\mathbf{b}^i$ is typically chosen as an inverse function of the beamforming weight vector obtained at the previous iteration. This, in turn,  suppresses the sensors corresponding to  smaller beamforming weights, thereby encouraging sparsity in an iterative fashion.    The weighted $l_1$-norm function   in $(\ref{c2_ch3})$  is replaced by the  $l_1$-norm  squared  function which does not  alter the regularization property of  the weighted  $l_1$-norm function \cite{6477161},
\begin{equation} \label{d2_ch3}
\begin{aligned}
\underset{\mathbf{w \in \mathbb{C}}^{N}}{\text{minimize}} & \quad   \mathbf{ w}^H\mathbf{R_{xx}}\mathbf{w} + \mu(||(\mathbf{b}^i\circ|\mathbf{w}|)||^2_1),\\
\text{s.t.} & \quad    \mathbf{ w}^H\mathbf{R}_{s}\mathbf{ w} \geq1.  \\
\end{aligned}
\end{equation}
The semidefinite  formulation (SDP) of the above problem can then be realized  by re-expressing the quadratic form, $ \mathbf{ w}^H\mathbf{R_{xx}}\mathbf{w}= $Tr$(\mathbf{ w}^H\mathbf{R_{xx}}\mathbf{w})= $Tr$(\mathbf{R_{xx}}\mathbf{w}\mathbf{ w}^H)=$ Tr$(\mathbf{R_{xx}}\mathbf{W})$, where Tr(.) is the trace of the matrix. Similarly, the regularization term $||(\mathbf{b}^i\circ|\mathbf{w}|)||^2_1=(|\mathbf{w}|^T\mathbf{b}^i)( (\mathbf{b}^i)^T|\mathbf{w}|)=\mathbf{|w|}^T\mathbf{B}^i\mathbf{|w|}=$Tr$(\mathbf{B}^i\mathbf{|W|})$. Here,  $\mathbf{W}=\mathbf{w}\mathbf{w}^H$ and $\mathbf{B}^i=\mathbf{b}^i(\mathbf{b}^i)^T$ is the regularization re-weighting matrix at the $i$th iteration. Utilizing these quadratic expressions in (\ref{d2_ch3})  yields the following problem \cite{Bengtsson99optimaldownlink, 8378759, 6477161},
\begin{equation} \label{e2_ch3}
\begin{aligned}
\underset{\mathbf{W \in \mathbb{C}}^{N\times N}, \mathbf{\tilde{W} \in \mathbb{R}}^{N\times N}}{\text{minimize}} & \quad   \text{Tr}(\mathbf{R_{xx}}\mathbf{W}) + \mu \text{Tr}(\mathbf{B}^i\mathbf{\tilde{W}}),\\
\text{s.t.} & \quad    \text{Tr}(\mathbf{R}_s\mathbf{W}) \geq 1,  \\
& \quad \mathbf{\tilde{W}} \geq |\mathbf{W}|,\\
& \quad   \mathbf{W} \succeq 0, \, \text{Rank}(\mathbf{W})=1.
\end{aligned}
\end{equation}

The function \textquotedblleft$|.|$\textquotedblright \, returns the  absolute values of the entries of the matrix, \textquotedblleft$\geq$\textquotedblright \, is the element wise comparison and \textquotedblleft$\succeq$\textquotedblright \, denotes the  generalized  matrix inequality. The auxiliary matrix $\mathbf{\tilde{W}} \in \mathbb{R}^{N\times N}$ implements the weighted  $l_1$-norm  squared regularization along with the re-weighting matrix $\mathbf{B}^i$.   The  rank constraint in  (\ref{e2_ch3}) is non convex and therefore need to be removed. The rank relaxed approximation works well for the underlying problem. In case, the solution matrix is not rank $1$, we can resort to randomization to harness  rank $1$ approximate solutions \cite{5447068}. Alternatively, one could  minimize the nuclear norm of $\mathbf{W}$,  as a  surrogate for $l_1 $-norm in the case of   matrices, to induce sparsity in the eigenvalues of $\mathbf{W} $ and promote rank one solutions  \cite{Recht:2010:GMS:1958515.1958520, Mohan:2012:IRA:2503308.2503351}. The  resulting rank relaxed semidefinite program (SDR) is  given  by;
\begin{equation} \label{f2_ch3}
\begin{aligned}
\underset{\mathbf{W \in \mathbb{C}}^{N\times N}, \mathbf{\tilde{W} \in \mathbb{R}}^{N\times N}}{\text{minimize}} & \quad   \text{Tr}(\mathbf{R_{xx}}\mathbf{W}) + \mu \text{Tr}(\mathbf{B}^i\mathbf{\tilde{W}}),\\
\text{s.t.} & \quad    \text{Tr}(\mathbf{R}_s\mathbf{W}) \geq 1,  \\
& \quad \mathbf{\tilde{W}} \geq |\mathbf{W}|,\\
& \quad   \mathbf{W} \succeq 0.
\end{aligned}
\end{equation}
 In general, QCQP  is NP hard and cannot  be solved in polynomial time.  The formulation in  (\ref{f2_ch3})  is clearly convex, in terms of unknown matrices, as all the  other correlation matrices involved  are guaranteed to be positive semidefinite. The sparsity parameter $\mu$ largely determines the cardinality of the solution beamforming weight vector.   To ensure $P$ sensor selection, appropriate value of $\mu$ is typically found by carrying a binary search over the probable range of $\mu$. After achieving the desired cardinality, the reduced size thinned correlation matrix $\mathbf{R_{xx}}$ is formed corresponding to the non-zero values of $\mathbf{\tilde{W}}$. The reduced dimension SDR is now solved with setting $\mu=0$, yielding optimum beamformer $\mathbf{w}_o=\mathscr{P} \{  \mathbf{W} \}$. 
\subsection{Fair gain beamforming}
The optimization in   (\ref{f2_ch3}) strives to incorporate the signal from all the directions of interest while optimally removing the interfering signals. To achieve this objective, the optimum sparse array may  show leaning towards a certain source of interest, consequently, not offering fair gain towards all sources.  In an effort to promote equal gain towards all sources, we put a separate constraint on the power towards all desired sources as follows;
\begin{equation} \label{g2_ch3}
\begin{aligned}
\underset{\mathbf{W \in \mathbb{C}}^{N\times N}, \mathbf{\tilde{W} \in \mathbb{R}}^{N\times N}}{\text{minimize}} & \quad   \text{Tr}(\mathbf{R_{xx}}\mathbf{W}) + \mu \text{Tr}(\mathbf{B}^i\mathbf{\tilde{W}}),\\
\text{s.t.} & \quad    \text{Tr}(\mathbf{R}_k\mathbf{W}) \geq 1, \, \, \, \, \forall k\in (1,2,3 ... K)   \\
& \quad \mathbf{\tilde{W}} \geq |\mathbf{W}|,\\
& \quad   \mathbf{W} \succeq 0.
\end{aligned}
\end{equation}
 Here,  $\mathbf{R}_k= \mathbf{s}( \theta_k)\mathbf{s}^H( \theta_k)$ is the rank $1$ covariance matrix associated with the source at DOA $( \theta_k)$.   However, the above SDR can   be solved to an arbitrary small accuracy  $\zeta$, by employing  interior point methods  involving   the   worst case complexity of  $\mathcal{O}\{$max$(K,N)^4N^{(1/2)}\log(1/\zeta)\}$   \cite{5447068}.
\begin{table}[t!] 
\noindent\rule{16.5cm}{0.6pt}
	\caption{Proposed algorithm to achieve desired cardinality of optimal weight vector}
\noindent\rule{16.5cm}{0.6pt}

	\begin{algorithmic}
		
		\renewcommand{\algorithmicrequire}{\textbf{Input:}}
		
		\renewcommand{\algorithmicensure}{\textbf{Output:}}
		
		\REQUIRE Data correlation matrix $\mathbf{R_{xx}}$, $N$, $P$, look direction DOA's $\theta_k$,  hybrid selection vector $\mathbf{z}$. \\
		
		\ENSURE  $P$ sensor beamforming weight vector $\mathbf{w}_o$,   \\
				    Initialize  $\epsilon$.\\
				 	Initialize $\mu_{lower}$, $\mu_{upper}$ (Initializing lower and upper limits of sparsity parameter range for	binary search for desired cardinality $P$)\\

		   \textbf{FSDR:} Initialize  $\mathbf{B=zz}^T$. \\
		   \textbf{NFSDR:} For optimum array design without the augmentability constraint, initialize  $\mathbf{z}$  to be all ones vector, $\mathbf{B=zz}^T$ (all ones matrix). \\
		   \textbf{Perturbed-NFSDR:} Locate the sensor $i$ such that, if not selected, results in the minimum compromise of the objective function. Perturb $\mathbf{z}$ at position $i$, $\mathbf{z}(i)=\mathbf{z}(i)+\gamma$, afterwards calculating $\mathbf{B=zz}^T$. \\
		 \WHILE {(Cardinality of $\mathbf{w}_o$ $\neq$ $P$)}
		\STATE  Update $\mu$ through binary search.
		\FOR {(Typically requires five to six iterations)}
			\STATE   Run the  SDR of  (\ref{f2_ch3}) or (\ref{g2_ch3}) (Fair gain case).
		   \STATE   Update the regularization weighting matrix $\mathbf{B}$ according to (\ref{i2_ch3}). \\ 
		\ENDFOR
         	\ENDWHILE
		
		\STATE After achieving the desired cardinality, run SDR for reduced size correlation matrix corresponding to nonzero values of $\mathbf{\tilde{W}}$  and $\mu=0$,  yielding, $\mathbf{w}_o=\mathscr{P} \{  \mathbf{W} \} $. 
		\RETURN $\mathbf{w}_o$
		
	\end{algorithmic}
	\label{algorithm_ch3}
	\noindent\rule{16.5cm}{0.6pt}
\end{table}
\subsection{Modified re-weighting for fully augmentable hybrid array}\label{Modified Re-Weighting_ch3}
 For the case without the full augmentability constraint the   regularization re-weighting matrix  $\mathbf{B}$ is initialized unweighted i.e. by all ones matrix and the $m,n$th element of \textbf{B} is iteratively updated as follows \cite{Candes2008},
\begin{equation} \label{h2_ch3}
\mathbf{B}_{m,n}^{i+1}=\frac{1}{|\mathbf{W}_{m,n}^i|+\epsilon}.
\end{equation}
The parameter $\epsilon$ avoids the unwanted case of division by zero, though its choice is fairly independent to the performance of the iterative algorithm but at times very small values of $\epsilon$ can result in the algorithm getting trapped in the local minima.  For the hybrid array design, we initialize the re-weighting matrix instead as an outer product of hybrid selection vector $\mathbf{z}$.  The hybrid selection vector $\mathbf{z}$ is an $N$ dimensional vector containing binary entries of zero and one,  where, zeros correspond to the pre-selected sensors and ones correspond to the remaining sensors to be selected. Hence, the  cardinality  of $\mathbf{z}$ is equal to the  difference of the total number of available sensors and the number of pre-selected sensors. This modified re-weighting approach ensures  that the sensors corresponding to the pre-selected configuration is  not penalized as part of the regularization, hence,  $\mathbf{B=zz}^T$, thrives solutions that incorporate the pre-selected array topology. The modified penalizing weight update for the hybrid array design can be expressed as;
\begin{equation} \label{i2_ch3}
\mathbf{B}^{i+1}=(\mathbf{zz}^T) \varoslash ({|\mathbf{W}^i|+\epsilon}).
\end{equation}
The symbol \textquotedblleft$\varoslash$\textquotedblright \, denotes element wise division. For the hybrid design, (15) is proposed with appropriate selection of  $\mathbf{z}$, as explained above, and hereafter referred to as the Fixed SDR (FSDR).  The array designed without the augmentability consideration is the special case of (15) with $\mathbf{z}$ being an all ones vector and  the algorithm is subsequently regarded as the Non-Fixed SDR (NFSDR).   The pseudo-code  for controlling the sparsity of the optimal weight vector $\mathbf{w}_o$ is summarized in  Table \ref{algorithm_ch3}.
\subsection{Symmetric arrays}\label{Symmetric arrays_ch3}
 The solution of the  NFSDR formulation is penchant for symmetric arrays in the case of  symmetric initialization vector   $\mathbf{z}$.  The plausible explanation is  as follows.  We first show that the  beamforming weights which maximizes the output SINR for symmetric sparse array topologies are conjugate symmetric w.r.t. the array center.
\begin{prop}
The conjugate symmetry of the optimal weight vector   holds for centro-symmetric sparse array configurations in case of the general rank desired source  model.
\end{prop}
\begin{proof}
(Refer to the section \ref{Appendix_ch3} for the proof.) 
\end{proof}
      We observe that the regularized cost function does not  invoke sparsity until after the first  few initial iterations. Consequently,   the initial solutions of the semidefinite program has symmetric coefficients as the NFSDR seeks near optimal solutions which  are  analytically shown to be conjugate symmetric.  Moreover, the iterative sparsity enhancing formulation introduces sparsity by penalizing the beamforming weight vector according to  (\ref{i2_ch3}), where, it only accounts the magnitude of the beamforming weights. Therefore, at each iteration the regularization re-weighting matrix $\mathbf{B}$ happens to penalize the solution weight vector in a symmetric fashion around the array center.  Thus, the   iterative NFSDR sparse solution favors   symmetric configurations by discarding corresponding symmetric  sensors simultaneously. Though, the symmetric configuration can  be suitable for certain applications \cite{8448767}, and can have desirable performance, yet,  it reduces the available degrees of freedom. Therefore, to avoid curtailing  the available degrees of freedom,  we   perturb the re-weighting regularization matrix   $\mathbf{B}$ at the initial iteration, as follows.   From $N$ prospective locations,  find the   sensor position,  which if not selected, results in the least compromise of  the objective function performance. Corresponding to the aforementioned  position,  set the  regularization weight to be relatively high through perturbation by parameter $\gamma$. By so doing, we resolve the issues arising from the  symmetric regularization re-weighting matrix. This modified algorithm is henceforth referred to as the perturbed-NFSDR and is detailed in Table \ref{algorithm_ch3}. 
\begin{figure}[t]
	\centering
	\includegraphics[width=4.38 in, height=3.5 in]{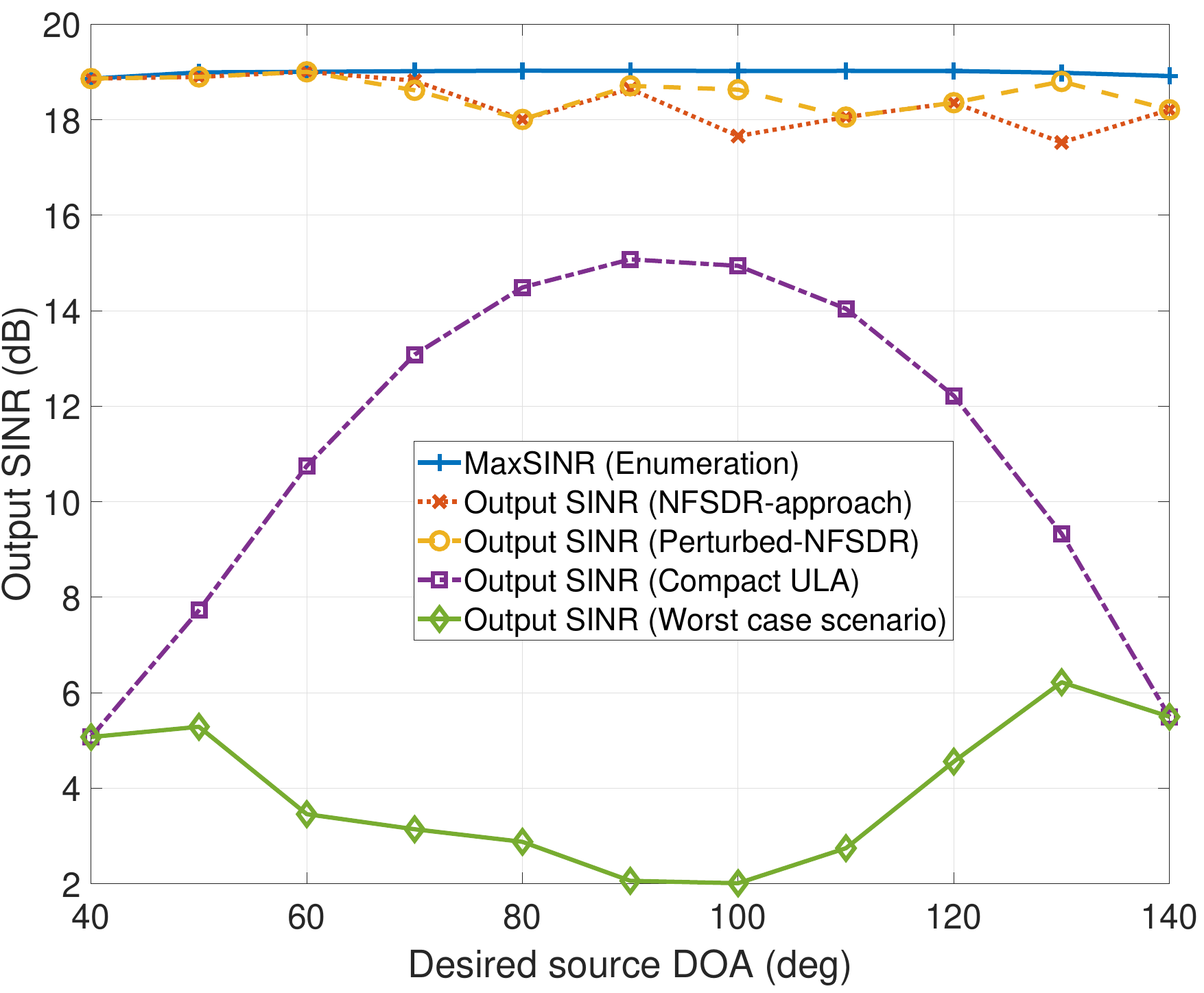}
	\caption{Output SINR for different array topologies}
	\label{scan_ch3}
\end{figure}
\section{Simulations} \label{Simulations_ch3}
In this section, we show the effectiveness of the proposed techniques for the sparse array design for  MaxSINR.  We initially examine the  proposed approach for  array configurability   by considering  arbitrary arrays  without  the augmentability constraint. In the later examples,  we  demonstrate the effectiveness  of fully augmentable  hybrid sparse array design through linear and 2D arrays. We focus on the EM modality, and as such we use antennas for sensors.

\subsection{Single point source} \label{Single point source_ch3}
We select $P=8$ antennas from $N=16$  possible equally spaced locations with  inter-element spacing of $\lambda/2$. Figure \ref{scan_ch3} shows the output SINR for different  array configurations for  the case  of single desired point source with its DOA varying from   $40^0$ to $140^0$. The interfering signals are located at $20^0$ and $\pm10^0$ degree apart  from the desired source angle. To explain this scenario, suppose that the desired source is at $60^0$, we consider the respective directions of arrival of the three interfering signals   at $40^0$, $50^0$ and $70^0$. The SNR of the desired signal is $10$ dB, and the  interference to noise ratio (INR)  is set to $10$ dB for each scenario. The input SINR is $-4.9$ dB. The upper and lower limit of the sparsity parameter $\mu$ is set to $1.5$ and $0.01$ respectively, $\gamma=0.05$  and $\epsilon=0.1$. From  the  Fig. \ref{scan_ch3}, it is evident that the NFSDR-approach performs  close to the performance of the optimum array found by exhaustive search ($12870$ possible configurations), which has very high computational cost attributed to expensive singular value decomposition (SVD) for each enumeration. Moreover, the perturbed-NFSDR algorithm results in comparable or better performance. Except for the slightly lower performance at the desired source of DOA of $70^0$, we observe that for the desired source of DOA at $90^0$, $100^0$ and $130^0$, the perturbed-NFSDR recovers a sparse array with better performance than the NFSDR-approach. For the other DOAs, the perturbed-NFSDR recovers the same symmetric configuration as that recovered by the NFSDR-approach.  This emphasizes that the perturbed-NFSDR does not  eliminate the possibility of  symmetric solutions and optimizes over  both the symmetrical and unsymmetrical array configurations.    On average, the proposed algorithms  takes six to seven  iterations to converge to the optimum antenna locations; hence, offering  considerable savings in the computational cost. It is of interest to compare  the optimum sparse array performance with that of compact uniform linear array (ULA). It can be seen from    Fig. \ref{scan_ch3}, that the optimum sparse array offers considerable SINR advantage over the compact  ULA for all   source  angles of arrival. The  ULA performance degrades  severely when the source of interest is more towards the array end-fire location.  In this case, the   ULA fails to resolve and cancel the strong interferers as they are located close to the desired source. 
\begin{figure}[t]
	\centering
	\includegraphics[width=4.38 in, height=3.50 in]{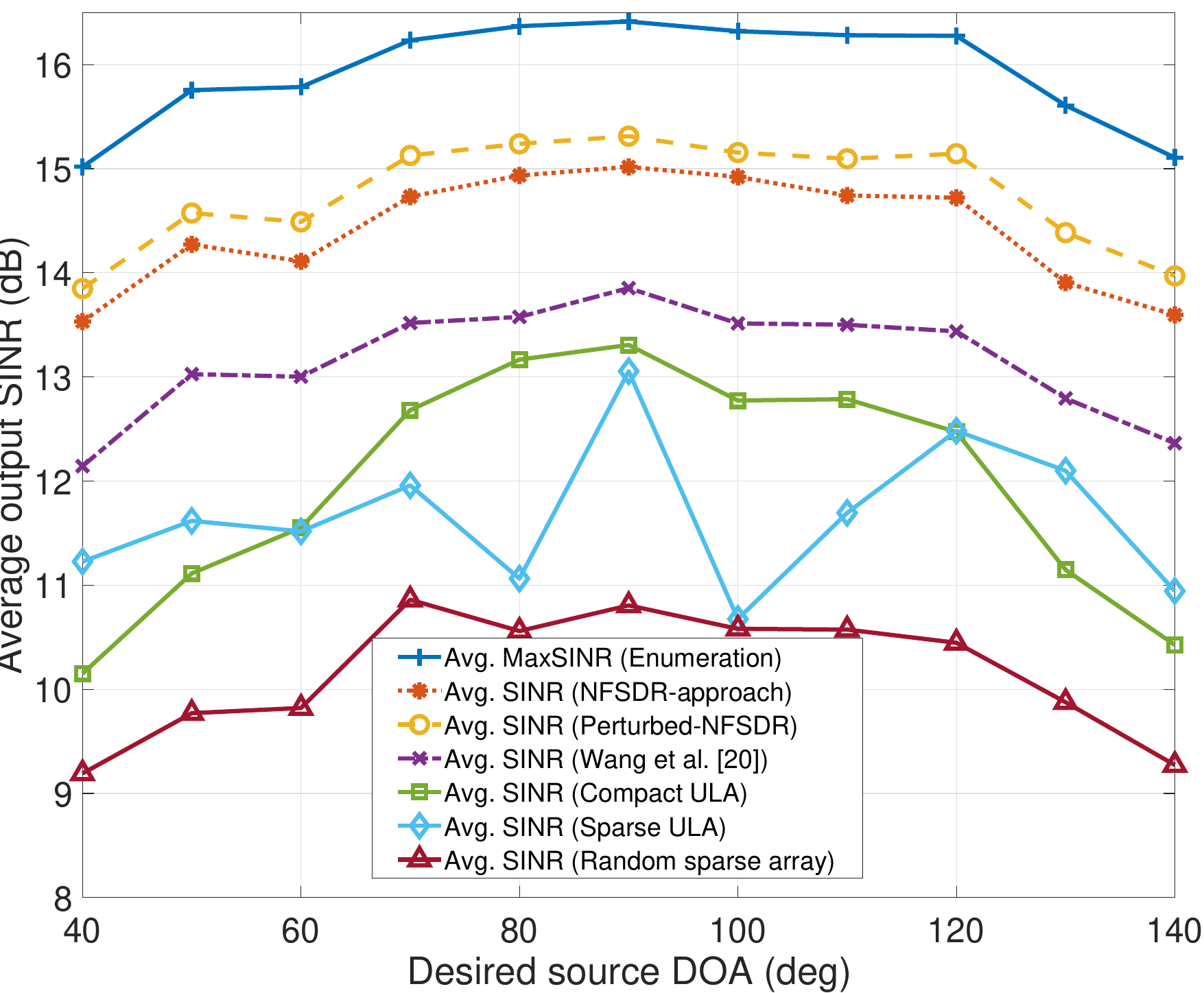}
	\caption{Average Output SINR for different array topologies over $6000$ Monte Carlo trials}
	\label{scan_monte_ch3}
\end{figure}

\begin{figure}[!t]
	\centering
	\subfloat[]{\includegraphics[width=0.83\textwidth]{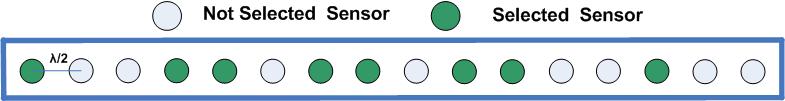}
		\label{best_90_0_ch3}}
		  \hfill
	\subfloat[]{\includegraphics[width=0.83\textwidth]{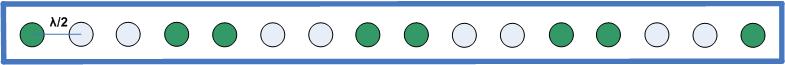}
		\label{op_algo_90_0_ch3}}
		  \hfill
\subfloat[]{\includegraphics[width=0.83\textwidth]{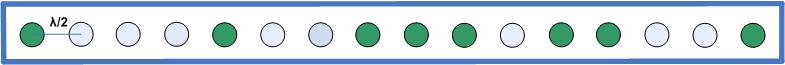}
		\label{op_symalgo_90_0_ch3}}
		  \hfill
	\subfloat[]{\includegraphics[width=0.83\textwidth]{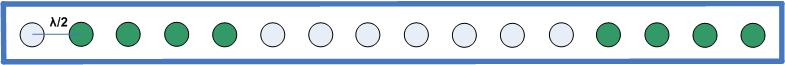}
		\label{worst_90_0_ch3}}
		  \hfill
	\caption{Array configurations obtained  for the  point source at the array broadside (a) Optimum  (Enumeration)  (b) NFSDR-approach  (c) Perturbed-NFSDR  (d) Worst performing array configuration  }
	\label{point_ch3}
\end{figure}

For the case of the desired source  at the array broadside,  the maximum output SINR  of the optimum array found through enumeration (Fig. \ref{best_90_0_ch3})  is $19$ dB.  The optimum array  design obtained through the  NFSDR-approach yields an output SINR of $18.6$ dB, which is  $0.4$ dB less than the corresponding SINR of the optimum array found through exhaustive search.  The broadside source arrays are shown in the  Fig. \ref{point_ch3} (where green-filled circle indicates antenna present whereas gray-filled circle indicates antenna absent).  The sparse array recovered through NFSDR-approach is clearly a symmetric configuration (Fig. \ref{op_algo_90_0_ch3}). Figure  \ref{op_symalgo_90_0_ch3} shows the sparse array found after addressing the symmetry bias by the  approach explained in Section \ref{Symmetric arrays_ch3}. The SINR for this non-symmetric configuration is $18.7$ dB and is suboptimal merely by  $0.3$ dB.   It is worth noticing  that  the worst performing sparse array configuration (Fig. \ref{worst_90_0_ch3}) comparatively engages larger  array aperture than the optimum array found through enumeration (Fig. \ref{best_90_0_ch3}), yet it  has an output SINR as low as  $2.06$ dB. This  emphasizes the fact that if an arbitrary  sparse array structure is employed, it could degrade the performance catastrophically irrespective of the occupied aperture  and could perform far worst than the compact ULA, which offers modest output SINR of  $15.07$ dB  for the  scenario under consideration. 
\subsubsection{Monte Carlo Simulation} \label{Monte1_ch3}
To thoroughly examine the performance of the proposed algorithms under  random  interfering environments, we perform 6000 Monte Carlo simulations.  For this purpose, the desired source DOA is fixed with SNR of $10$ dB, and eight  interferences are generated  which are uniformly distributed anywhere  from $20^0$ to $160^0$.  The INRs of these sources are  uniformly drawn from $10$ dB to $15$ dB. We choose 8 antennas out of 16 possible locations. The upper and lower limit of the sparsity parameter $\mu$ is set to 3 and 0.01 respectively, $\gamma=0.1$  and $\epsilon=0.05$. The performance curves are shown in  Fig. \ref{scan_monte_ch3} for the desired source fixed at $11$ different DOAs varying from $40^0$ to $140^0$. On  average, the proposed perturbed-NFSDR  algorithm consistently provided superior SINR performance. However,  this performance is around $1.2$ dB suboptimal than the average SINR computed through enumeration. The average SINR performance of the perturbed-NFSDR algorithm is around $0.35$ dB better than the  proposed NFSDR-approach. This is because the degrees of freedom are limited by the inherent array symmetry enforced by the re-weighted optimization scheme. The performances of the proposed algorithms are compared with the  design methodology proposed in \cite{8313065}, which relies on the \textit{a priori} knowledge of the interference steering vectors and respective powers. It is noted  that in the underlying scenario the design in \cite{8313065} is more than $1$ dB suboptimal than the proposed algorithms and around $2$ dB suboptimal as compared to the performance upper bound. The algorithm in \cite{8313065} relies on successive linear approximation of the objective function as opposed to the  quadratic implementation of the SDR, thereby suffering  in performance. The SINR performances  for the compact ULA, sparse ULA and randomly employed sparse topology are also shown in the Fig. \ref{scan_monte_ch3}, further highlighting the utility of sparse array design.
\begin{figure}[t]
	\centering
	\subfloat[]{\includegraphics[width=0.83\textwidth]{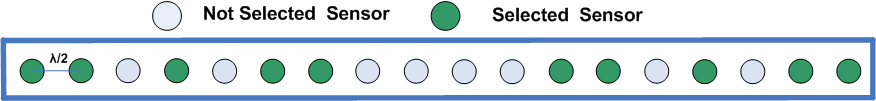}
		\label{op_alg_45_5_ch3}}
		\hfill
\subfloat[]{\includegraphics[width=0.83\textwidth]{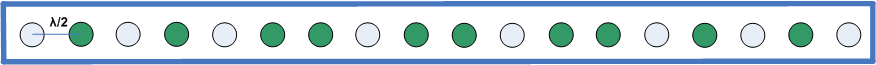}
		\label{op_45_5_ch3}}
		\hfill
\subfloat[]{\includegraphics[width=0.83\textwidth]{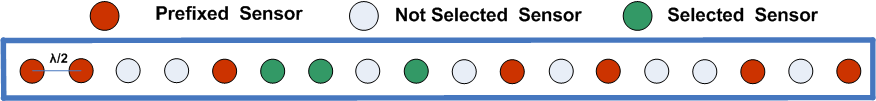}
		\label{op_alg_multi_ch3}}
		\hfill
	\caption{(a) Antenna array multiple sources (NFSDR-approach)   (b)   Fair gain $10$ element antenna array (NFSDR-approach)    (c)  Hybrid  $10$ antenna array  for multiple desired sources (FSDR) }
	\label{op_c_co_c_ch3}
\end{figure}
\begin{figure}[t]
	\centering
	\includegraphics[width=4.38 in, height=3.50 in]{beampatterneune-eps-converted-to.pdf}
	\caption{Beampattern  for  multiple point sources}
	\label{multi_bp_ch3}
\end{figure}
\begin{figure}[t]
	\centering
	\subfloat[]{\includegraphics[width=0.83\textwidth]{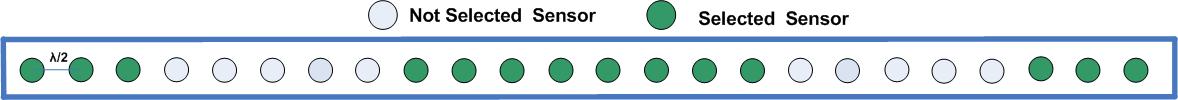}
		\label{op_1D_ch3}}
		\hfill
\subfloat[]{\includegraphics[width=0.83\textwidth]{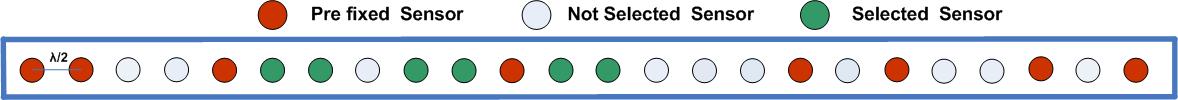}
		\label{mm1_1D-hybrid_ch3}}
		\hfill
\subfloat[]{\includegraphics[width=0.83\textwidth]{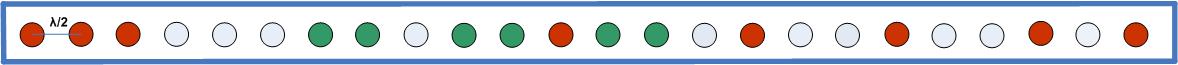}
		\label{mm2_1D-hybrid_ch3}}
		\hfill
	\caption{(a) $14$ element antenna  array (NFSDR-approach)  (b)    Hybrid  $14$ antenna sparse array ($8$ prefixed, $6$ selected through FSDR)   (c)  Hybrid $14$ antenna sparse array ($8$ prefixed, $6$ selected through FSDR)}
	\label{1D-hybrid_ch3}
\end{figure}

\subsection{ Multiple point sources} \label{Multiple point sources_ch3}
For the multiple point sources scenario,  consider three desired signals  impinging from DOAs $40^0$,  $65^0$ and $90^0$ with SNR of $0$ dB each. Unlike the example in \ref{Single point source_ch3}, we set four strong interferers with INR of $30$ dB are operational at DOAs $50^0$,  $60^0$,  $120^0$  and $150^0$. In so doing, we  analyze the robustness of the proposed scheme under very low input SINR of $-36.02$ dB.  We select $10$ antennas out of $18$ available slots. The optimum array recovered through convex relaxation is shown in Fig. \ref{op_alg_45_5_ch3}. This configuration results with an  output SINR of $11.85$ dB  against SINR of $12.1$ dB for the optimum configuration found through enumeration. For the fair gain beamforming, we apply the optimization of   (13) and the  array configuration for MaxSINR for  the fair gain beamforming is shown in Fig \ref{op_45_5_ch3}. The output SINR for the fair beamforming case is $11.6$ dB which is slightly less than the optimum array without the fair gain consideration ($11.85$ dB). However, the advantage of  fair beamforming is well apparent from the beampatterns in both cases as shown in Fig \ref{multi_bp_ch3}, where the  gain towards the source at $65^0$ is around $4.24$ dB higher than the case of optimum array without the fair gain consideration. The maximum gain deviation for the fair gain case is $3.5$ dB vs. $8$ dB variation without the fair gain consideration.   The SINR  of compact ULA is  compromised more than $3$ dB   as compared to the optimum sparse array (Fig. \ref{op_alg_45_5_ch3}) obtained through the proposed methodology. This improved performance is due to  the optimum sparse array smartly engaging  its degrees of freedom to eradicate the interfering signals while maintaining maximum gain towards all  sources of interest.
\subsection{ Fully augmentable linear arrays}
Consider selecting $14$ antennas out of $24$ possible available  locations with   antenna spacing of $\lambda/2$. A desired source is impinging from DOA of $30^0$ and SNR of $10$ dB, whereas  narrowband jammers are operating at $20^0$, $40^0$ and  $120^0$ with INR of $10$ dB each. The range of $\mu$ and other parameters are the same as in \ref{Monte1_ch3}. Optimum array configuration (Fig. \ref{op_1D_ch3}) achieved through convex relaxation (NFSDR-approach) has an  output SINR of $21.29 $ dB   as compared to SINR of  $21.32$ dB of an optimum array recovered through enumeration ($1.96*10^6$ possible configurations). It should be noted that the array recovered without filled co-array constraint is  not essentially fully augmentable as is the case  in the optimum array (Fig. \ref{op_1D_ch3}) which clearly has missing co-array lags. 

In quest of fully augmentable array design we  prefix  8 antennas (red elements in Fig. \ref{mm1_1D-hybrid_ch3}) in a minimum redundancy array (MRA)  configuration over 24 uniform grid points. This provides 24 consecutive autocorrelation lags.   We are, therefore, left with six antennas to be placed in the remaining 16 possible locations ($8008$ possible configurations). We  enumerated the performance of all possible hybrid arrays associated with underlying MRA configuration  and found the output SINR ranges from $18.1$ dB to $21.3$ dB. Figure \ref{mm1_1D-hybrid_ch3} shows  the  configuration recovered through the proposed approach  which has an output SINR of $20.96$ dB.  The proposed approach thus recovers the hybrid sparse array with performance close to the best possible, moreover it  approximately yields $3$ dB advantage over worst fully augmentable hybrid array. As MRAs are not unique we started with   a different 8 element MRA structured  array (red elements in Fig. \ref{mm2_1D-hybrid_ch3}),  to further reinforce the effectiveness of fully augmentable sparse arrays. The dynamic performance range associated with  MRA of Fig. \ref{mm2_1D-hybrid_ch3}, is  from $17.59$ dB to   $21.3$ dB. The performance in this case is very similar to the aforementioned MRA configuration  with the output SINR of  $21.08$ dB for the hybrid array recovered through proposed methodology (Fig. \ref{mm2_1D-hybrid_ch3}). The maximum possible SINR offered by both hybrid arrays is $21.3$ dB which  is extremely close to SINR performance of $21.32$ dB offered by the optimum array without augmentability constraint.  
\begin{figure}[!t]
	\centering
	\includegraphics[width=4.38 in, height=3.50 in]{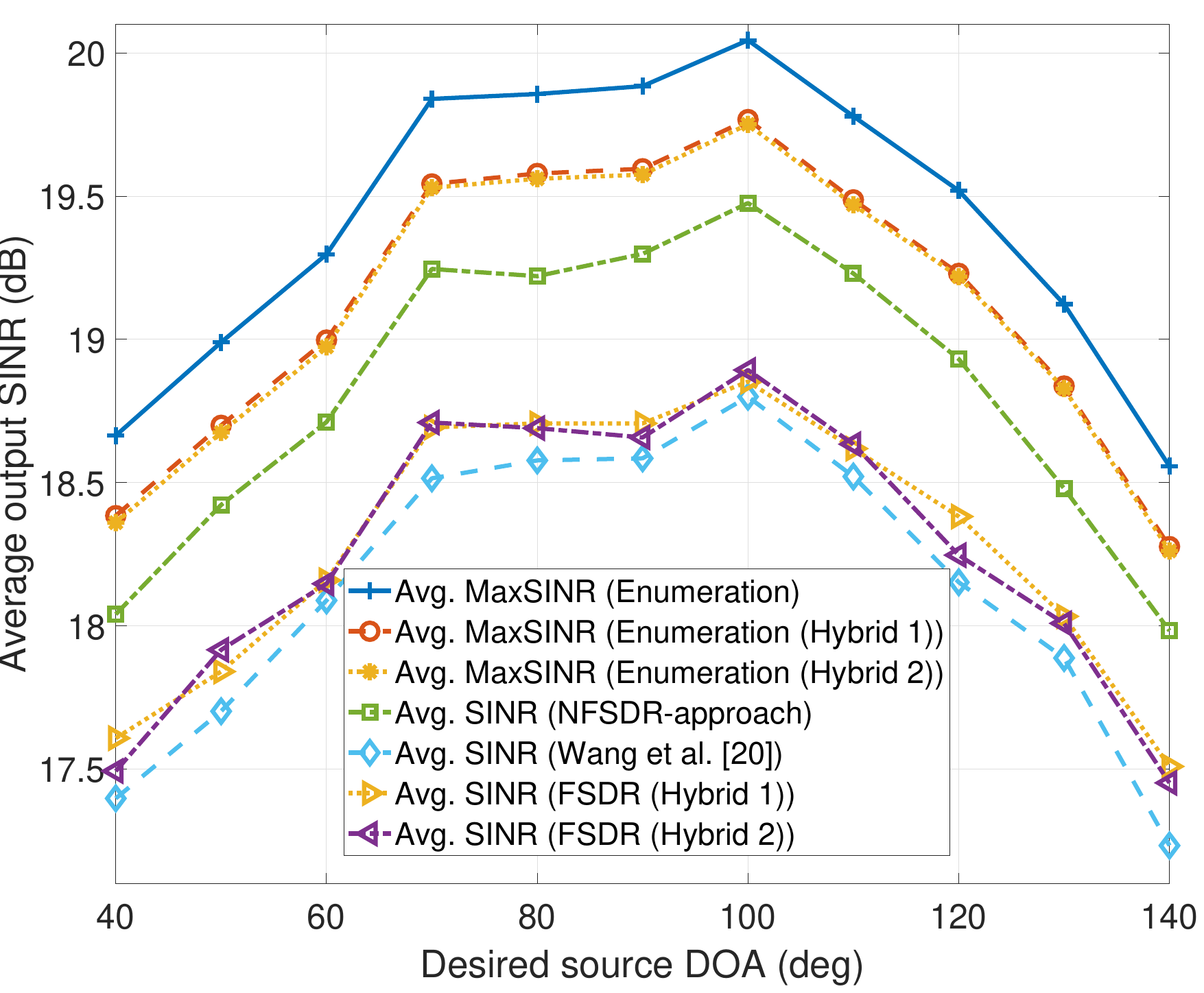}
	\caption{Average Output SINR for different array topologies over $3500$ Monte Carlo trials}
	\label{scan_monte2_ch3}
\end{figure}
\subsubsection{Monte Carlo Simulation}
We generate 3500  Monte Carlo simulations for comparison between  the performance of the sparse arrays that are designed freely and that of sparse array design involving   full augmentability constraint.  We choose $16$ antennas out of $24$ available locations.  The desired source DOA is fixed  with SNR of $10$ dB as in \ref{Monte1_ch3}.  We assume twelve   narrowband interferences drawn uniformly  from $20^0$ to $160^0$ with  respective INRs   uniformly distributed from $10$~dB to $15$~dB.  For binary search, the upper and lower limit of the sparsity parameter $\mu$ is  5 and 0.01 respectively and $\epsilon=0.1$,  for all 3500 scenarios.  Fig. \ref{scan_monte2_ch3} shows the average SINR performance, where the proposed NFSDR-approach  is only $0.57$ dB suboptimal relative to the optimum array found through enumeration (choosing 16 antennas out of 24 involves $735471$ prospective configurations). However,  this performance is achieved by sparse arrays without ensuring the augmentabilty constraint. Therefore, we prefix $8$ antennas in MRA topology, namely Hybrid $1$  and Hybrid $2$ prefix configurations,  shown in red circles in  Figs.  \ref{mm1_1D-hybrid_ch3} and Fig. \ref{mm2_1D-hybrid_ch3} respectively. The MaxSINR performance, found by enumeration, for either of the  underlying hybrid topologies competes very closely as evident in Fig. 8. The average MaxSINR (found by enumeration), under both prefixed configurations, is only compromised by $0.28$ dB relative to the average MaxSINR performance offered without the augmentability constraint. It is noted that in this case, the possible sparse configurations are drastically reduced from $735471$ to  $12870$ (choose the remaining $8$ antennas from the remaining $16$ possible locations due to prefixing $8$ antennas \textit{a~priori}). It is clear from Fig. \ref{scan_monte2_ch3} that the proposed FSDR algorithm successfully  recovers the hybrid sparse array with an average SINR performance loss of $0.8$ dB. We remark that  the performance of the hybrid sparse array  is still slightly better than the optimum sparse array receive beamforming proposed in  \cite{8313065} that assumes the knowledge of jammers' steering vectors and utilizes all the available degrees of freedom, unlike the hybrid sparse array. 

\subsection{ Fully augmentable 2D arrays}
\begin{figure}[!t]
  	\centering
  	\includegraphics[height=2.48in,width=3in]{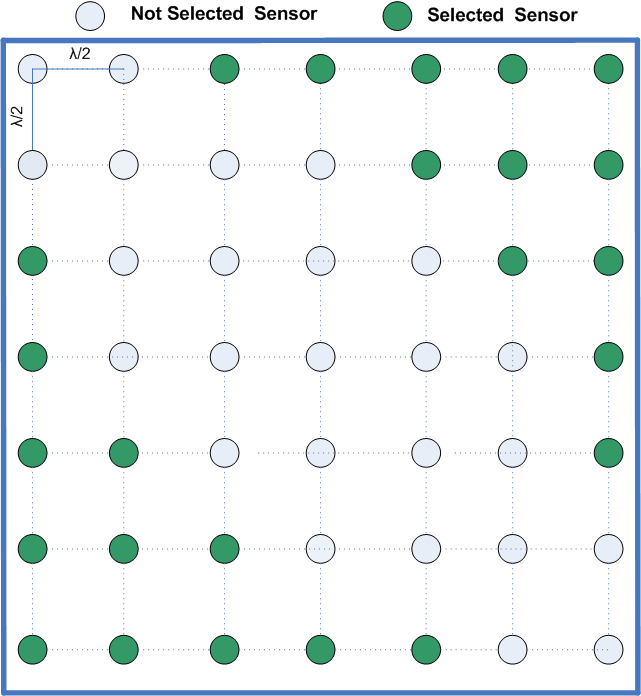}
  	\caption{$24$ element antenna sparse array (NFSDR-approach)}
  	\label{2d_ideal_ch3}
  \end{figure}

 Consider a $7 \times 7$ planar array with grid pacing of $\lambda/2$ where we place 24 antennas at 49 possible positions.  A desired source is impinging from elevation angle $\theta=50^0$  and azimuth angle of  $\phi=90^0$. Here, elevation angle is with respect to the plane carrying the array  rather than reference from the zenith. Four strong interferes are impinging from  ($\theta=20^0$, $\phi=30^0$), ($\theta=40^0$, $\phi=80^0$), ($\theta=120^0$, $\phi=75^0$) and ($\theta=35^0$, $\phi=20^0$). The INR corresponding to each interference is $20 $ dB and SNR is set to $0$ dB. There are of the order of $10^{14}$ possible 24 antenna configurations, hence the problem is prohibitive even by exhaustive search. Therefore, we resort to  the upper bound of performance limits to compare our results. Here, we utilize the fact that the best possible performance  occurs when the interferes are completely canceled  in the array output and the output SINR in that case would equal the array gain offered by the 24 element array which amounts to  $13.8$~dB.  Figure \ref{2d_ideal_ch3} shows the optimum antenna locations recovered by the proposed NFSDR-approach. The output SINR for this configuration is $13.68$ dB which is sufficiently  close to the ideal performance. It should be noted that again the array recovered in the Fig. \ref{2d_ideal_ch3} is not fully augmentable  as it is missing quiet a few correlation lags.
 
\begin{figure}[!t]
  	\centering
  	\includegraphics[height=2.48in,width=3in]{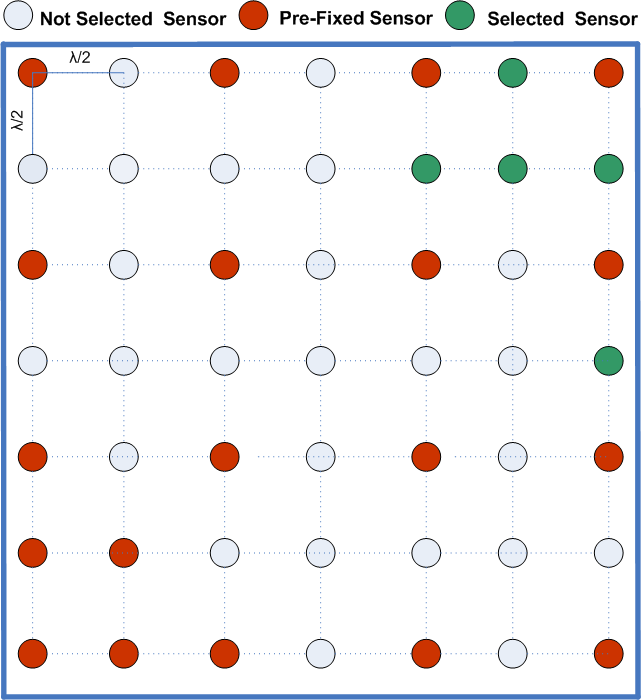}
  	\caption{ $24$ element  hybrid antenna sparse array  ($19$ prefixed, $5$ selected through FSDR)}
  	\label{2d_hybridop_ch3}
  \end{figure}

\begin{figure}[!t]
  	\centering
  	\includegraphics[height=2.48in,width=3in]{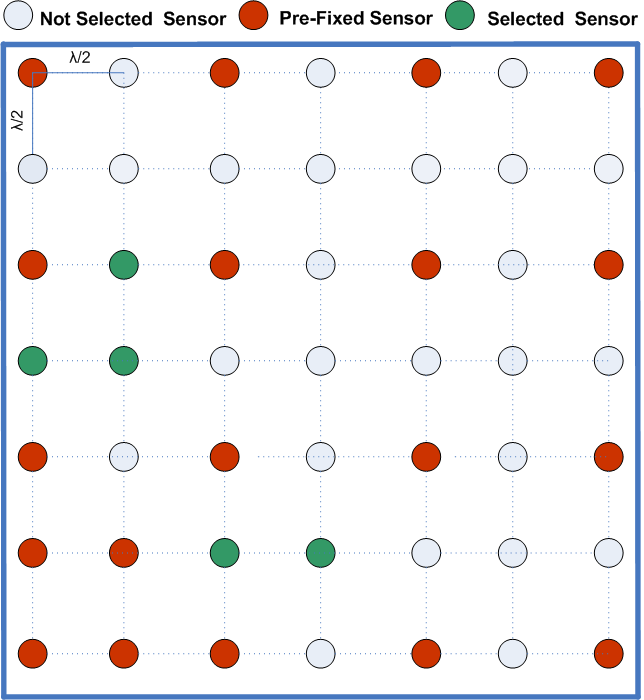}
  	\caption{ $24$ element worst performing hybrid antenna sparse array  ($19$ prefixed, $5$ selected)}
  	\label{2d_hybridwo_ch3}
  \end{figure}

  We now introduce the condition of full augmentability  by placing 19 antennas in nested lattice configuration \cite{6214623}  to form a filled co-array (red elements in Fig.  \ref{2d_hybridop_ch3}). The rest of five available antennas can be placed in the remaining 30 possible locations hence resulting in approximately $1.5 * 10 ^5$ possibilities. Figure \ref{2d_hybridop_ch3} shows the  hybrid sparse geometry recovered by FSDR algorithm and offers SINR of $13.25$ dB which is around $0.4$ dB less than the optimum array. The performance range of the hybrid arrays associated with the structured nested lattice array ranges from $11.4$ dB to $13.38$ dB (found through exhaustive search). In this regard the FSDR algorithm finds the  hybrid sparse  array with the performance degradation  of little more than $0.1$ dB. The worst performing hybrid array (Fig.  \ref{2d_hybridwo_ch3}) has an output SINR of $11.4$ dB and is around $2$ dB lower than the  best performing hybrid sparse array.
  
\begin{figure}[!t]
	\centering
	\includegraphics[width=4.48 in, height=2.95 in]{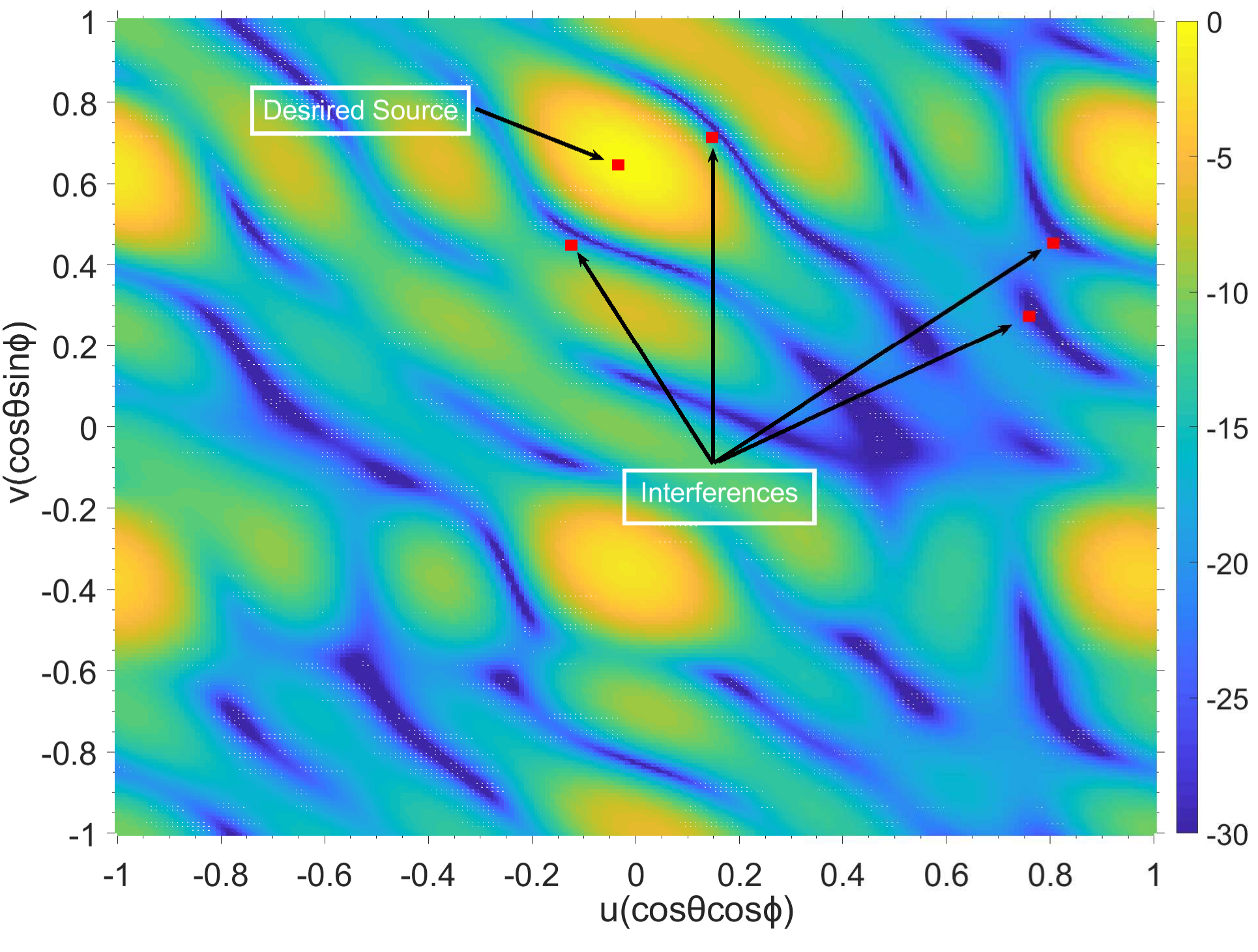}
	\caption{Beampattern  for the antenna array in Fig. \ref{2d_hybridop_ch3}}
	\label{2d_bp_ideal_ch3}
\end{figure}

\begin{figure}[!t]
	\centering
	\includegraphics[width=4.48 in, height=2.95 in]{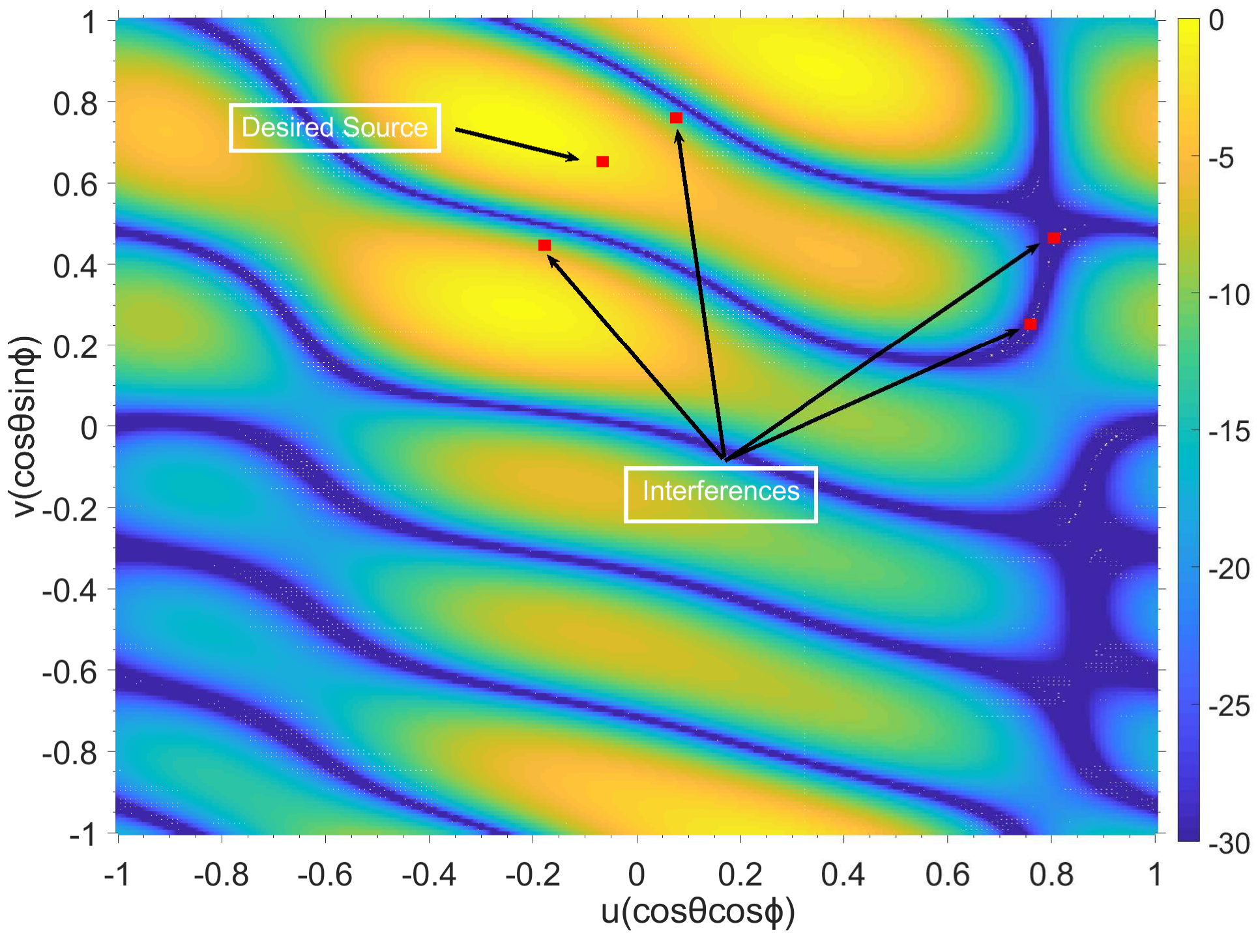}
	\caption{Beampattern  for a $6 \times 4$ compact rectangular  array}
	\label{2d_bp_rec_ch3}
\end{figure}

   It is of interest to compare the performance of aforementioned sparse arrays with a compact $2$D array. For this purpose, we chose a $6 \times 4$ rectangular array. The compact rectangular array performs very poorly in the underlying scenario and has an output SINR of $7.8$ dB which is more than $ 5$ dB down from the  hybrid  sparse array recovered through the semidefinite relaxation. This performance degradation is very clear from  the beampattern of both arrays shown in Figs. \ref{2d_bp_ideal_ch3} and \ref{2d_bp_rec_ch3} (normalized beampattern in dB). In the  case of the  hybrid sparse array recovered through FSDR (Fig. \ref{2d_hybridop_ch3}), the target has the maximum gain towards the direction of interest with minimum gain simultaneously towards all unwanted DOAs (Fig. \ref{2d_bp_ideal_ch3}).  In contrast,  it is clear from Fig. \ref{2d_bp_rec_ch3} that the beampattern of the compact rectangular array  could not manage maximum gain towards the direction of interest while effectively rejecting the interfering signals. Although, the $6 \times 5$  and $6 \times 6$ compact arrays utilize $6$ and $12$ additional sensors, yet the  respective output SINRs of $9.04$ dB and  $11$ dB are considerably suboptimal relative to the proposed solutions. It is noted that adding 18  additional sensors resulting in $7 \times 6$ rectangular array has an output SINR of $12.87$ dB. Still, the 24 element free-design as well as the hybrid design outperform the compact $42$ element  rectangular array. However, a $49$ element fully populated $7 \times 7$ rectangular array has an output SINR of $14.37$ dB, which is marginal improvement given the SINR of 24 element designed topologies.  The hybrid array also appears to be more  robust as it has higher dynamic performance range threshold ($11.4$ dB).  The   performance of arbitrarily designed arrays is more prone  to deteriorate catastrophically  even far worse than  that   of the  compact uniform or  rectangular arrays.

  We also test the fully augmentable array design for the case of  multiple point source scenario described previously (Section \ref{Multiple point sources_ch3}). The hybrid array recovered through proposed methodology is shown in the Fig. \ref{op_alg_multi_ch3} (red elements showing the $7$ element MRA). The output SINR is $11.566$ dB and is sufficiently close to the performance achieved through enumeration. 

\section{Conclusion}
This chapter considered fully augmentable  sparse array configurations for maximizing the beamformer output SINR for general rank desired signal  correlation matrices.  It proposed a  hybrid  sparse array design that  simultaneously considers co-array and environment-dependent objectives.  The proposed array design approach uses a subset of the available antennas to obtain a fully augmentable array while employing the remaining antennas  for achieving the highest SINR. It was shown that the hybrid design  is data driven and  hence practically  viable, as it ensures the availability of the full data correlation matrix  with a reasonable trade off in the SINR performance. We applied the modified re-weighting QCQP  which proved effective     in recovering  superior SINR performance for hybrid sparse arrays in polynomial run times. The proposed approach  was  extended  for    fair gain beamforming towards   multiple   sources. We solved the optimization problem by both the proposed algorithms and  enumeration and showed strong agreement between the two methods. 

\section{Appendix} \label{Appendix_ch3}
\subsection{Proof of the Conjugate symmetric property of optimal weight vector}
\begin{proof}
	The correlation matrix $\mathbf{R}$ for centro-symmetric arrays have a conjugate persymmetric structure such that \cite{86911}: \\
\begin{equation}\label{appendixa_ch3}
\mathbf{TR^{'}T}=\mathbf{R}
\end{equation}

Here  $\{'\}$ is the conjugate operator and  $\mathbf{T}$ is the  transformation matrix which flips the entries of a vector upside down by left multiplication;\\ 
\[
\mathbf{T}=
\begin{bmatrix}
	0       & \dots & 0 &0  & 1 \\
	0       & \dots & 0 & 1 & 0 \\
	\vdots     & \dots   &  &\vdots \\
	1       & \dots & 0 &  & 0
\end{bmatrix}
\]
The  optimal weight vector which maximizes the SINR is given by;
\begin{equation}\label{appendixb_ch3}
\mathbf{w}_o = \mathscr{P} \{  \mathbf{R_{s^{'}}}^{-1}\mathbf{R_s}  \}
\end{equation}
where,
\begin{equation}\label{appendixc_ch3}
\{  \mathbf{R_{s^{'}}}^{-1}\mathbf{R_s} \}  \mathbf{w}_o = \Lambda_{max} \mathbf{w}_o
\end{equation}
Using the relation in  (\ref{appendixa_ch3}),    (\ref{appendixc_ch3}) can be re-expressed as follows,
 \begin{equation}\label{appendixd_ch3}
 \begin{aligned}
 \{  \mathbf{(TR^{'}_{s^{'}}T)^{-1}}\mathbf{(TR^{'}_sT)} \}  \mathbf{w}_o & = \Lambda_{max} \mathbf{w}_o \\
  \{  \mathbf{T^{-1}(R^{'}_{s^{'}})^{-1}T^{-1}}\mathbf{(TR^{'}_sT)} \}  \mathbf{w}_o & = \Lambda_{max} \mathbf{w}_o \\
  \end{aligned}
 \end{equation}
 Multiplying both sides by $\mathbf{T}$ and applying the conjugate operator,
 \begin{equation}\label{appendixe_ch3}
 \{  \mathbf{R_{s^{'}}}^{-1}\mathbf{R_s} \}  \mathbf{T}\mathbf{w}^{'}_o = \Lambda_{max} \mathbf{T}\mathbf{w}^{'}_o
 \end{equation}
 From (\ref{appendixe_ch3}), we note, that $\mathbf{T}\mathbf{w}_o^{'}$ is also the principal eigenvector associated with matrix $   \mathbf{R_{s^{'}}}^{-1}\mathbf{R_s} $. Since the principal eigenvector of the positive definite Hermitian matrix  is unique up to  the scalar complex multiplier, this directly implies that;
  \begin{equation*}\label{appendixf_ch3}
 \mathbf{w}_o= \mathbf{T}\mathbf{w}_o^{'} 
 \end{equation*} 
\end{proof}




\chapter{Sparse Array Design for Maximizing the Signal-to-Interference-plus-Noise-Ratio by Matrix Completion}


\section{Introduction}
Sensor selection schemes strive to optimize various performance metrics while curtailing   valuable hardware  and  computational resources. Sparse sensor placement, with various design objectives, has successfully been  employed   in  diverse application areas, particularly for enhanced parameter estimation and receiver performance \cite{710573, 6477161, 4058261, 6822510, 4663892, 5680595, 6031934, 8448767}.   The sparse array design criteria  are generally  categorized into environment-independent and  environment-dependent performance metrics. The  former are largely benign to the underlying environment and, in principle, seek to maximize the spatial degrees of freedom by extending the co-array aperture. This  enables high resolution direction of arrival (DOA) estimation possibly involving more sources than the available physical sensors \cite{1139138, 5456168, 7012090, 7106511, 8683315}. Environment-dependent objectives, on the other hand, consider the operating conditions characterized by emitters and targets in the array field of view, in addition to receiver noise. In this regard, applying such objectives  renders the array configuration as well as the array weights time-varying in response to dynamic and changing   environment.

In this chapter, we focus on optimum sparse array design for receive beamforming that maximizes the output signal-to-interference and noise ratio (MaxSINR) \cite{10.1117/12.2519879, 1206680, 8378759, hamza2019sparse, 8645552}.   It has been shown that optimum sparse array beamforming involves both array configuration and weights, and  can yield significant dividends in terms of SINR performance in presence of desired and interfering sources    \cite{6774934,  1634819, 6714077, 8061016, 8036237, 8313065, 7944303}.   However, one key challenge in implementing the data-dependent approaches, like Capon beamforming, is the need to have the exact or estimated values of the data autocorrelation function across the full sparse array aperture \cite{1206680, 1223538}. This underlying predicament arises as the sparse array design can only have few active sensors at a time, in essence  making it difficult to  furnish the correlation values corresponding to the inactive sensor locations. 

To address the aforementioned problem, we propose in this chapter a  matrix completion strategy assuming a single desired source and multiple interfering sources.  This strategy permits the interpolation of the missing data correlation lags, thus enabling optimum “thinning” of the array for MaxSINR. The low rank matrix completion has  been utilized successfully in many applications, including the high-resolution direction of arrival estimation. We compare the matrix completion strategy to  the hybrid sparse array design that has been recently introduced and which also provides full spatial autocorrelation function for array thinning \cite{8892512, 8682266}. The fundamental thrust of the hybrid design is to pre-allocate some of the available sensors  in such a way so as to ensure that all possible correlation lags can be estimated. In this case, the difference between the available sensors and those pre-allocated can  be utilized for maximum  SINR. In essence, the hybrid design locks  few spatial degrees of freedom in an attempt to making the full autocorrelation matrix  available to carry out the array optimization at all times. In that sense, it is a hybrid between structured and non-structured arrays. With pre-allocated sensors, the design approach  offers a simplified antenna switching as the environment changes.   In contrast, the matrix completion-based design   is not tied in to any pre-allocated sensor position and, therefore, has the ability to optimize over all the available sensor locations.  However, low rank matrix completion is a pre-processing step that is required every time we   decide on sensor selection as the environment changes.  This significantly  adds to the overall overhead and  computational complexity. We examine both approaches using estimated autocorrelation function, in lieu of its exact values, and compare their respective performances under different settings and degrees of freedom.

\begin{figure}[!t]
	\centering
	\includegraphics[width=4.47 in, height=2.25 in]{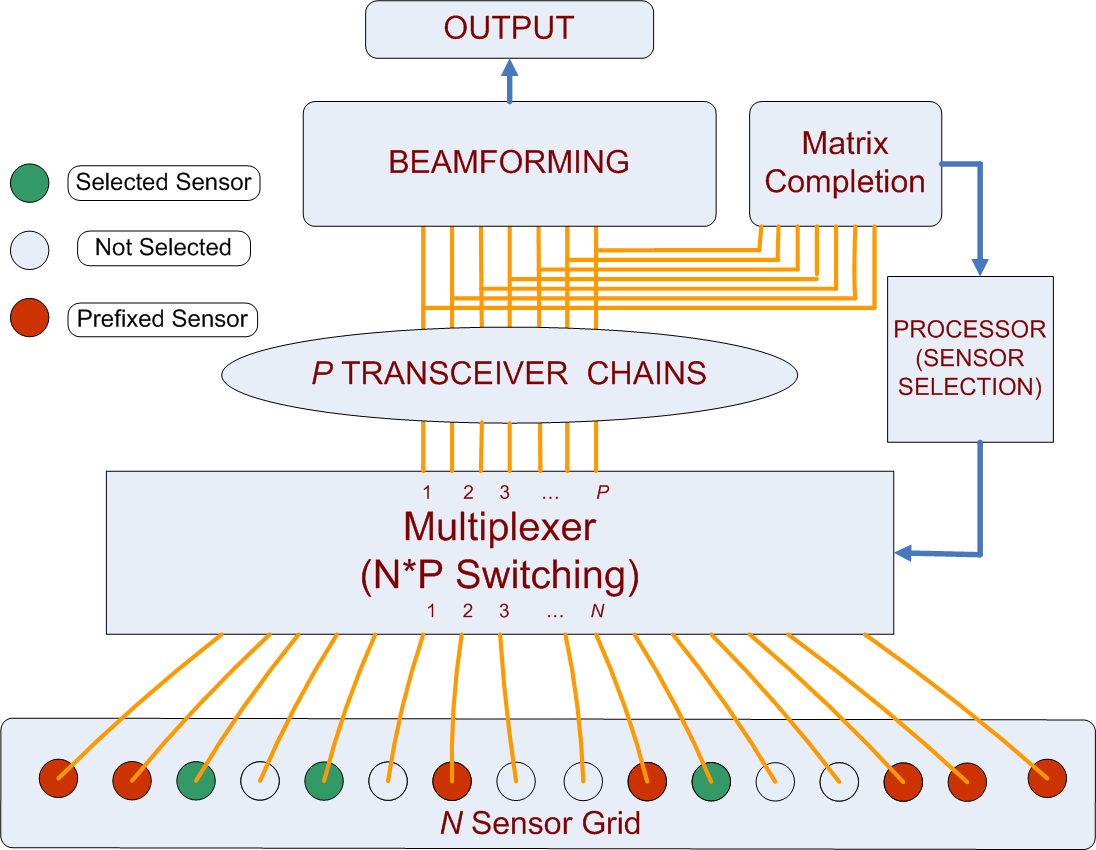}
	\caption{Block diagram implementing adaptive beamforming and antenna switching}
	\label{Block_diagram_ch4}
\end{figure}

It is worth noting that  MaxSINR sparse array design using either  is an entwined optimization problem that  jointly optimizes the beamforming weights and determines the  active sensor locations. The   optimization  is posed as  finding   $P$ sensor positions out of $N$ possible equally spaced grid points for the highest  SINR performance. It is known that maximizing the SINR is equivalent to the problem of  maximizing  the  principal eigenvalue of the product of  the inverse of data correlation matrix and the desired source correlation matrix  \cite{1223538}. However, the maximum eigenvalue problem over all possible sparse topologies  is a combinatorial problem and is challenging to solve in polynomial times. To alleviate the computational complexity of exhaustive combinatorial search,  we pose this problem as successive convex approximation (SCA) with reweighted  $l_1$-norm regularization  to promote sparsity in the final solution.  

To proceed with  the SCA optimization, it is essential to input the algorithm with the full data correlation matrix. For the hybrid design, all the correlation lags are available  and, therefore,  we resort to averaging across the available correlation lags to harness a Toepltiz matrix estimate of the received data.  On the other hand, a sparse array designed freely without preallocating sensor locations necessitates the use of low rank matrix completion to interpolate the missing lags and subsequently applying the SCA optimization \cite{86455521}.  The word `free' implies no preset position of any of the sensors involved. It is shown that the matrix completion  is  an effective approach to accomplish the MaxSINR sparse  design. The performance of matrix completion can potentially surpass the hybrid design at the expense  of more involved sensor switching and additional  computational complexity stemming from the Toeplitz interpolation of the missing correlation lags. The overall system implementation is depicted in  Fig. \ref{Block_diagram_ch4}.

The rest of the chapter is organized as follows:  In the next section, we  state the problem formulation for  maximizing the  output SINR. Section \ref{Optimum sparse array design_ch4} details  the SCA for arrays designed freely  alongside the     hybrid   design approach and the associated modified SCA optimization. Section \ref{sec4_ch4} explains the matrix completion approach.   In section \ref{Simulations_ch4}, with the aid of Monte Carlo simulations,  we compare the performance of  hybrid-designed arrays viz a viz freely-designed arrays in a limited snapshot environment. Concluding remarks follow at the end.

\section{Problem Formulation} \label{Problem Formulation_ch4}
We consider an emitter source in the presence of narrowband  interfering signals. The  signals impinge  on a  uniform grid of $N$ linear elements with the inter-element spacing  of $d$ and the received signal is  given by,
\begin {equation} \label{a_ch4}
\mathbf{x}(n)=    b_s(n) \mathbf{s}( \theta)  + \sum_{k=1}^{I} b _{ik}(n) \mathbf{i}( \theta_k)  + \mathbf{v}(n)
\end {equation}
The sampling instance is  $n$, $I$ denotes the number of interfering sources and ($b_s(n)$, $b _{ik}(n))$ $\in \mathbb{C}$  are the baseband  signals  for the source and interferences, respectively. The  steering vector    corresponding to the  direction  of arrival of desired   source $\mathbf{s} ({\theta})$  $\in \mathbb{C}^N$ is    given by,
\vspace{+2mm}
\begin {equation}  \label{b_ch4}
\mathbf{s} ({\theta})=[1 \,  \,  \, e^{j (2 \pi / \lambda) d cos(\theta)  } \,  . \,   . \,  . \, e^{j (2 \pi / \lambda) d (N-1) cos(\theta)  }]^T
\end {equation}
The interference steering vectors $\mathbf{i} ({\theta_k})$ are similarly defined. The additive  noise  $\mathbf{v}(n)$ $\in \mathbb{C}^N$ is Gaussian with variance    $\sigma_v^2$.
The beamformer processes the received signal  $\mathbf{x}(n)$  linearly   to improve   SINR. The beamformer output $y(n)$ is given by, 
\begin {equation}  \label{c_ch4}
y(n) = \mathbf{w}_o^H \mathbf{x}(n)
\end {equation}
The optimal beamforming weights $\mathbf{w}_o$ that maximizes SINR is given by solving the following optimization problem  \cite{1223538};
\begin{equation} \label{d_ch4}
\begin{aligned}
\underset{\mathbf{w} \in \mathbb{C}^N}{\text{minimize}} & \quad   \mathbf{w}^H\mathbf{R}_{i}\mathbf{w}\\
\text{s.t.} & \quad     \mathbf{ w}^H\mathbf{R}_{s}\mathbf{ w}=1
\end{aligned}
\end{equation}
The  source  correlation matrix,  $\mathbf{R}_s=  \sigma^2 \mathbf{s}( \theta)\mathbf{s}^H( \theta)$, with  power $ \sigma^2 =E\{b_s(n)b_s^H(n)\}$. 
The sum of the  interference and  noise correlation matrix  is $\mathbf{R}_{i}= \sum_{k=1}^{I} \sigma^2_k \mathbf{i}( \theta_k)\mathbf{i}^H( \theta_k)$  $+\sigma_v^2\mathbf{I}_{N\times N}$,    with the $k$th interference power $ \sigma^2_k =E\{b_{ik}(n)b_{ik}^H(n)\}$. Since  $\mathbf{R_{x}}=\mathbf{R}_s+ \mathbf{R}_{i}$,  then formulation (\ref{d_ch4}) can be written as  part of the objective function as follows \cite{1223538},
\begin{equation} \label{e_ch4}
\begin{aligned}
\underset{\mathbf{w} \in \mathbb{C}^N }{\text{minimize}} & \quad   \mathbf{ w}^H\mathbf{R_{x}}\mathbf{ w}\\
\text{s.t.} & \quad     \mathbf{ w}^H\mathbf{R}_{s}\mathbf{ w} \geq 1,
\end{aligned}
\end{equation}
where the equality constraint is relaxed due to the inclusion of the relationship between the data and signal autocorrelation matrices in the cost function. The optimum solution of the above problem only  requires the knowledge of the received data correlation matrix $\mathbf{R_{x}}=E(\mathbf{x}\mathbf{x}^H)$ and the  DOA of the desired source. The former can readily be estimated from the received data vector $\mathbf{x}$ over $T$ snapshots, $\mathbf{\hat{R}_{x}}=\frac{1}{T}\sum_{n=1}^{T} \mathbf{x}(n)\mathbf{x}^H(n)$.

The analytical solution of the optimization problem    is given by $\mathbf{w}_o =  \{  \mathbf{R}_{i}^{-1} \mathbf{s}(\theta)  \}$ with the optimum output SINR$_o$;
\begin{equation}  \label{f_ch4}
\text{SINR}_o=\frac {\mathbf{w}_o^H \mathbf{R}_s \mathbf{w}_o} { \mathbf{w}_o^H \mathbf{R}_{i} \mathbf{w}_o} = \Lambda_{max}\{\mathbf{R}^{-1}_{i} \mathbf{R}_s\},
\end{equation}
which is in fact the maximum eigenvalue ($\Lambda_{max}$) of   the product of   inverse of data correlation matrix  and the desired source correlation matrix. In the next section, the  formulation in (\ref{e_ch4}) is extended to the sparse beamformer design.

\section{Sparse array design through SCA algorithm} \label{Optimum sparse array design_ch4}
The expression in (\ref{f_ch4}) is applicable to any array topology, including uniform and sparse arrays with the respective correlation matrices. To achieve sparse solutions, given the knowledge of full correlation matrix, (\ref{e_ch4}) is introduced with  an additional constraint,
\begin{equation} \label{a2_ch4}
\begin{aligned}
\underset{\mathbf{w \in \mathbb{C}}^{N}}{\text{minimize}} & \quad  \mathbf{ w}^H\mathbf{R_{x}}\mathbf{w}\\
\text{s.t.} & \quad   \mathbf{ w}^H\mathbf{R}_{s}\mathbf{ w}\geq1  \\
& \quad   ||\mathbf{w}||_0=P
\end{aligned}
\end{equation}
The  operator $||.||_0$ denotes the $l_0$ norm which constrains the cardinality of  the weight vector $\mathbf{w}$ to the number of available sensors, $P$.  The problem  in  (\ref{a2_ch4}) is clearly non convex   involving   a hard constraint, rendering the formulation challenging to solve in polynomial time \cite{doi:10.1002/cpa.20132}.

The objective function  and  quadratic constraint  in  (\ref{a2_ch4}) are interchanged,  transforming  into equivalent formulation as follows,
\begin{equation} \label{b2_ch4}
\begin{aligned}
\underset{\mathbf{w \in \mathbb{C}}^{N}}{\text{maximize}} & \quad   \mathbf{ w}^H\mathbf{R}_s\mathbf{ w},\\
\text{s.t.} & \quad     \mathbf{ w}^H\mathbf{R_x}\mathbf{ w} \leq 1 \\
& \quad   ||\mathbf{w}||_0=P
\end{aligned}
\end{equation}
In general, the beamforming weight vector is complex valued, however the quadratic functions are real. The real and imaginary entries of the optimal weight vector are typically decoupled, permitting the involvements of only  real unknowns. This is achieved through concatenating the beamforming weight vector and defining  the respective correlation matrices   \cite{8445845},
\begin{multline} \label{c2_ch4}
\, \, \quad \quad \quad \quad \quad \tilde{\mathbf{R}}_s=\begin{bmatrix}
\text{real}({\mathbf{R}}_s)       & -\text{imag}({\mathbf{R}}_s)   \\
\\
\text{imag}({\mathbf{R}}_s)       & \text{real}({\mathbf{R}}_s)   \\
\end{bmatrix},
\quad \quad \quad  \quad \quad \tilde{\mathbf{w}}=\begin{bmatrix}
\text{real}({\mathbf{w}})       \\
\\
\text{imag}({\mathbf{w}})   \\
\end{bmatrix}
\end{multline}
\begin{align} \label{d2_ch4}
\tilde{\mathbf{R}}_x=\begin{bmatrix}
\text{real}({\mathbf{R}}_x)       & -\text{imag}({\mathbf{R}}_x)   \\
\\
\text{imag}({\mathbf{R}}_x)       & \text{real}({\mathbf{R}}_x)   \\
\end{bmatrix}
\end{align}
Replacing $\mathbf{R_s}$ and $\mathbf{R_x}$ by $\tilde{\mathbf{R}_s}$ and $\tilde{\mathbf{R}_x}$ respectively, (\ref{b2_ch4}) can be expressed in terms of real variables,
\begin{equation} \label{e2_ch4}
\begin{aligned}
\underset{\mathbf{\tilde{w} \in \mathbb{R}}^{2N}}{\text{maximize}} & \quad   \mathbf{\tilde{ w}}^{'}\tilde{\mathbf{R}}_s\mathbf{\tilde{ w}}\\
\text{s.t.} & \quad     \mathbf{\tilde{ w}}^{'}\mathbf{\tilde{R}_x}\mathbf{\tilde{ w}} \leq 1 \\
& \quad   ||\mathbf{w}||_0=P
\end{aligned}
\end{equation}
The quadratic constraint  clearly has the convex feasibility region, however, there is still a  non convex constraint involving the  $l_0$ norm.  In order to realize the convex feasible region,  the  $l_0$ norm is typically relaxed to the $l_1$ norm, which has been effectively used in many sparse recovery applications.  The maximization problem is first transformed to a minimization in order to move the $l_1$ norm constraint to the objective function and realize a sparse solution.    This is achieved   by reversing the sign of the entries of the desired source correlation matrix $\bar{\mathbf{R}}_s=-\mathbf{\tilde R}_s$,
\begin{equation} \label{f2_ch4}
\begin{aligned}
\underset{\mathbf{\tilde{w} \in \mathbb{R}}^{2N}}{\text{minimize}} & \quad   \mathbf{\tilde{ w}}^{'}\bar {\mathbf{R}}_s\mathbf{\tilde{ w}}\\
\text{s.t.} & \quad     \mathbf{\tilde{ w}}^{'}\mathbf{\tilde{R}_x}\mathbf{\tilde{ w}} \leq 1 \\
& \quad   ||\mathbf{w}||_0=P
\end{aligned}
\end{equation}
To convexify the objective function, the concave objective  is  iteratively approximated through successive linear approximation. The approximation coefficients  $\mathbf{m}^i$ and ${b^i}$, are updated iteratively $\mathbf{m}^{i+1}=2\bar{\mathbf{R}}_s\mathbf{\tilde{ w}}^i, b^{i+1}=-\mathbf{\tilde{w}}^i{'}\bar{\mathbf{R}}_s \mathbf{\tilde{w}}^i $ by first order approximation, resulting in the following form,
\begin{equation} \label{m2_ch4}
\begin{aligned}
\underset{\mathbf{\tilde{ w} \in \mathbb{R}}^{2N}}{\text{minimize}} & \quad   \mathbf{m}^i{'}\mathbf{\tilde{ w}}+b^i\\
\text{s.t.} & \quad     \mathbf{\tilde{ w}}^{'}\mathbf{\tilde{R}_x}\mathbf{\tilde{ w}}\leq 1 \\
& \quad   ||\mathbf{w}||_0=P
\end{aligned}
\end{equation}
Finally, the non convex  $l_0$ norm is  relaxed  through minimizing the mixed $l_{1-\infty}$ norm to recover sparse solutions,
\begin{equation} \label{n2_ch4}
\begin{aligned}
\underset{\mathbf{\tilde{w} \in \mathbb{R}}^{2N}}{\text{minimize}} & \quad   \mathbf{m^i{'}}\mathbf{\tilde{ w}}+b^i + \mu(\sum_{k=1}^{N}||\mathbf{\tilde{w}}_k||_\infty)\\
\text{s.t.} & \quad     \mathbf{\tilde{ w}}^{'}\mathbf{\tilde{R}_x}\mathbf{\tilde{ w}}\leq 1
\end{aligned}
\end{equation}
The summation implements the $l_1$ norm that is minimized as a convex  surrogate of $l_0$ norm. The vector $\mathbf{\tilde{w}}_k \in \mathbb{R}^{2}$ has two entries containing  the real and imaginary parts of the beamforming weight corresponding to the $k$th sensor. The $||.||_\infty$ selects  the maximum entry of  $\mathbf{\tilde{w}}_k$ and  discourages the real and imaginary entries concurrently. This is because not selecting a sensor  implies the simultaneous removal of both the real and corresponding imaginary entries in the final solution vector. The sparsity parameter  $\mu$ is set to zero for the first few initial iterations to allow the solution to converge to optimal solution for  the full array elements. The sparsity parameter $\mu$ by itself does not  guarantee   the final solution to be $P$ sparse. To guarantee a $P$ sparse solution, the optimization problem is solved successively against different values of $\mu$. The values of $\mu$  are typically given by a binary search  over the possible upper and lower limit of $\mu$ until the algorithm converges to $P$ sensors \cite{6477161}.  

\begin{table}[t!] 
\noindent\rule{16.5cm}{0.6pt}
	\caption{SCA for  sparse array  beamforming.}
\noindent\rule{16.5cm}{0.6pt}

	\begin{algorithmic}[]
		
		\renewcommand{\algorithmicrequire}{\textbf{Input:}}
		
		\renewcommand{\algorithmicensure}{\textbf{Output:}}
		
		\REQUIRE Received data sparse correlation matrix $\mathbf{R}_P$, look direction DOA $\theta$. \\
		
		\ENSURE  $P$ sensor beamforming weight vector   \\
		
		\textbf{Matrix Completion:} \\
		 Run  Eq. (\ref{a3_ch4}) for free design and Toeplitz averaging for hybrid design to estimate the full correlation matrix.	\\	
		 Set the lowest eigenvalues of $\mathbf{\hat{R}_x}$ corresponding to the noise subspace to be equal to the noise floor.	\\	
		\textbf{Initialization:} \\
		Initialize the beamforming vectors randomly to find $\mathbf{m}$ and $b$. 
		Initialize  $\epsilon$, $\mu=0$.
		\WHILE {(Solution does not converge corresponding to $\mu=0$)}
		\STATE   Run  Eq. (\ref{j2_ch4}).\\ 
		\ENDWHILE \\
		Initialize $\mathbf{h}^{i}$= all ones vector, Binary vector for hybrid design.\\
		Select $\mu$ (Binary search)
		\WHILE {(Beamforming weight vector is not $P$ sparse)}
		\STATE   Run  Eq. (\ref{j2_ch4}) . (for initial iteration use $\mathbf{m}^i$ and $b^i$ from  previous while loop)\\ 
		\STATE   Update the regularization weighting parameter,  $\mathbf{h}^{i+1}(k)=\frac{1}{||\mathbf{\tilde{w}}_k^i||_2+\epsilon}$, Update $\mathbf{m}^i$ and $b^i$ \\
		\ENDWHILE \\
		\STATE After achieving the desired cardinality, analytically solve for  $\mathbf{\tilde{w}}$  corresponding to the selected sensor locations,  yielding, optimal weight vector. 
		
		\RETURN Optimal weight vector $\mathbf{w}_o$ 
	\end{algorithmic}
	\label{algorithm_ch4}
\noindent\rule{16.5cm}{0.6pt}
\end{table}
\subsection{Hybrid sparse array design}
Formulation (\ref{n2_ch4}) penalizes all the sensor weights rather judiciously in an effort to  optimize the objective function. We refer to this approach as free-design. On the other hand, the hybrid sparse array design,  penalizes only some sensor weights, leaving the remaining sensors to assume prefixed positions. These position are chosen to guarantee full augmentatbility of the sparse array, i.e., provide the ability to estimate the autocorrelation at all spatial lags across the array aperture. This provide the means  for thinning the array and carrying out sparse optimization all the times.  In order to discriminate the  prefixed sensors from those which are available for design, the weighted formulation  is adopted, in turn modifying  (\ref{n2_ch4}) as follows, 
\begin{equation} \label{i2_ch4}
\begin{aligned}
\underset{\mathbf{\tilde{w} \in \mathbb{R}}^{2N}}{\text{minimize}} & \quad   \mathbf{m^i{'}}\mathbf{\tilde{ w}}+b^i + \mu(\sum_{k=1}^{N}\mathbf{h}(k)||\mathbf{\tilde{w}}_k||_\infty)\\
\text{s.t.} & \quad     \mathbf{\tilde{ w}}^{'}\mathbf{\tilde{R}_x}\mathbf{\tilde{ w}}\leq 1
\end{aligned}
\end{equation}
The weighting vector $\mathbf{h}$ is a binary vector with $1's$ and $0's$ entries. The entries corresponding to the prefixed sensor locations are set to $0$ while the remaining entries are initialized to $1$. In this way, the partial penalization is implemented in (\ref{i2_ch4}) that ensures  the sparsity is not enforced to the prefixed locations. 
The weighted penalization can easily be extended to the reweighting  formulation which can further promote sparsity and facilitates the  $P$ sparse solution \cite{Candes2008}. This is achieved by iteratively updating weighting vector $\mathbf{h}$ \cite{6663667, article1},
\begin{equation} \label{j2_ch4}
\begin{aligned}
\underset{\mathbf{\tilde{w} \in \mathbb{R}}^{2N}}{\text{minimize}} & \quad   \mathbf{m^i{'}}\mathbf{\tilde{ w}}+b^i + \mu(\sum_{k=1}^{N}\mathbf{h}^i(k)||\mathbf{\tilde{w}}_k||_\infty)\\
\text{s.t.} & \quad     \mathbf{\tilde{ w}}^{'}\mathbf{\tilde{R}_x}\mathbf{\tilde{ w}}\leq 1
\end{aligned}
\end{equation}
The re-weighting vector $\mathbf{h}^i$, at the $i$-th iteration, is updated as an inverse function of the beamforming weights at the present iteration,
\begin{equation} \label{k2_ch4}
\mathbf{h}^{i+1}(k)=\frac{1}{||\mathbf{\tilde{w}}_k^i||_2+\epsilon}
\end{equation}
This relatively suppresses the low magnitude weights in the next iteration to accelerate sparsity. The parameter $\epsilon$ avoids the case of division by zero. The reweighting is applied to both the freely designed  array and the hybrid design. For the former, the vector $\mathbf{h}$ is initialized to all $1's$ vector and updated iteratively. However, to preserve the prefixed sensor locations for the hybrid design, the entries of $\mathbf{h}^i$ corresponding to the prefixed locations must remain zero for all iterations, while the remaining entries are initialized to $1$   and updated as explained above. The procedure is summarized in Table \ref{algorithm_ch4}.

\section{Toeplitz matrix completion and Fully augmentable completion through averaging} \label{sec4_ch4}
The key  concern in the free-design sparse array formulation is the assumption regarding the knowledge of the full array correlation matrix. This is because the data from only $P$ active  sensors is available to  estimate the  correlation matrix.  The full correlation matrix, in this case, is not  readily available and could have many missing correlation lags. Many different approaches for sparse matrix completion, under variant assumptions about the data model, have been considered in the literature including high resolution DOA estimation. We adopt a positive semidefinite Toepltiz  matrix completion scheme that effectively  exploits the structure of the unknown  correlation matrix. It is well known that the  narrowband far field  sources impinging on the ULA resultantly has  the  hermitian positive definite correlation matrix having the Toeplitz structure.    Along with the  Toepltiz  positive definite condition, the trace heuristic is incorporated to interpolate the missing lags. The trace heuristics is successfully used in many areas of control systems and array processing to recover simpler and low rank data models \cite{Recht:2010:GMS:1958515.1958520, Mohan:2012:IRA:2503308.2503351, 5447068}. Moreover, it has been shown  that the trace heuristic is equivalent to the nuclear norm minimization, rendering gridless recovery of the underlying narrowband sources, thus  recovering the missing correlation lags \cite{8474369, 8472789, 7539135, 7935534, 7833203}. The matrix completion problem is, therefore, written as,
\begin{align} \label{a3_ch4}
\underset{l \in \mathbb{C}^{N}}{\text{minimize}} & \quad  ||Toeplitz(l)\odot \mathbf{Z}-\mathbf{R}_P||_F^2 + \zeta \text{Tr}(Toeplitz(l)) \nonumber \\
\text{s.t.} & \quad   Toeplitz(l) \succeq 0  
\end{align}
Here,  $l$ is a complex vector with a real first element,  then $Toeplitz(l)$ returns the symmetric Toeplitz matrix having  $l$ and $l^H$ defining its first row and column  respectively. Matrix $\mathbf{R}_P$ is the received data correlation matrix with missing correlation lags. The entries corresponding to the missing correlation lags  are set to zero. The  symbol `$\odot$' denotes the element wise multiplication and  `$\succeq$' denotes the matrix inequality enforcing the positive  semidefinite constraint. The matrix  $\mathbf{Z}$  is a binary matrix which only fits the non zero elements in $\mathbf{R}_P$ to the unknown Toepltiz matrix.  The function `$||.||_F^2$' is the square of the Frobenius norm of the matrix which seeks to minimize  the sum of error square between the observed correlation values and the corresponding entries of the unknown Toepltiz matrix. The symbol `$\zeta$' gives the trade off  between the denoising  term and the trace heuristic pursuing simpler model. The nominal value of the  parameter  `$\zeta$'   is typically tuned from the numerical experience for the underlying problem.  However, the Toeplitz estimate can potentially be ill conditioned having quite a  few  eigenvalues close to zero.  We utilize the maximum likelihood estimate of the interpolated Toeplitz  correlation matrix by incorporating the knowledge of the noise floor. In  so doing, the eigenvalues corresponding to the noise subspace are set equal to the noise floor. 

Unlike the free-design sparse array, where missing lags manifest themselves as zero values at all entries of some of the autocorrelation matrix sub-diagonals, the hybrid design would ensure that at least one element in each matrix sub-diagonal is available. This facilities the Toeplitz estimation of the received data correlation matrix by averaging the non zero correlation entries across each sub-diagonal. The averaging scheme, however, does not guarantee the positive definiteness of the  Toeplitz estimate \cite{709534}, \cite{Abramovich:1999:PTC:2197953.2201686}. This renders the formulation in (\ref{j2_ch4}) non convex, which essentially requires $\mathbf{R_x}$ to be positive semidefinite. In order to circumvent this issue, we return to  the maximum likelihood estimate adopted for the matrix completion approach to facilitate a positive definite estimate by eliminating the negative eigenvalues typically appearing in the noise subspace. Finally, the estimated data correlation matrix $\mathbf{\hat R_x}=Toeplitz(l)$ is used in lieu of $\mathbf{R_x}$ to carry out the data dependent optimization for MaxSINR.

\section{Simulations} \label{Simulations_ch4}
We show   examples under different design scenarios to access  the  performance of the proposed methodology  achieving  MaxSINR.  We establish two  performance benchmarks in order to examine   the sensitivity of the proposed  algorithm  to  the initial array configuration. This is because the matrix interpolation approach is guided on the    initial configuration that decides the location of the  missing entries in the data correlation matrix. The initial  configuration refers to the $P$-element sparse array topology at the start before commencing of any adaptation process.   In general, the initial configuration could be any random array, or the optimized configuration from the preceding operating conditions.     The first performance benchmark applies the SCA algorithm under the assumption that the   data from  all the perspective  sensor locations  is available. In this way,  the actual  full  correlation matrix utilizing $T$  snapshots is input to the SCA algorithm. Clearly, the performance of the aforementioned  benchmark is not  reliant on the initial configuration  but is  dependent   on the observed data realization and  the number of   snapshots. Another  deterministic performance benchmark  assumes  perfect knowledge of the full correlation matrix, representing the case of  unlimited data snapshots. To draw a proper distinction, the former would be  referred as the `Full correlation statistics-limited snapshots (FCS-LSS),' and the latter is henceforth called the `Full correlation statistics-unlimited snapshots (FCS-USS)'.  
\subsection{Example comparing both designs}

\begin{figure}[!t]
	\centering
	\subfloat[]{\includegraphics[width=0.83\textwidth]{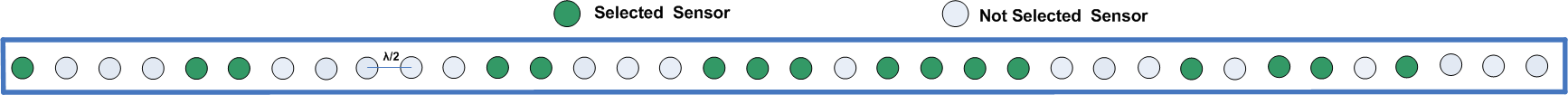}
		\label{best_90_0_ch4}}
		\hfill
	\subfloat[]{\includegraphics[width=0.83\textwidth]{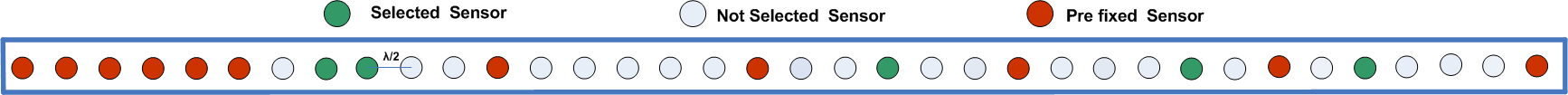}
		\label{worst_90_0_ch4}}
		\hfill
	\subfloat[]{\includegraphics[width=0.83\textwidth]{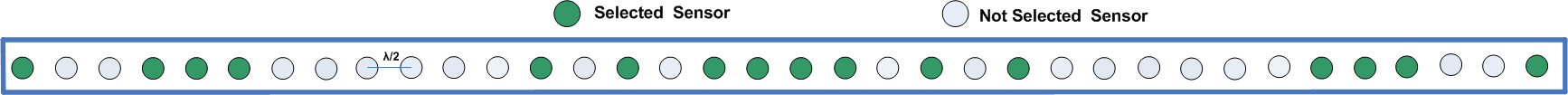}
		\label{op_algo_90_0_ch4}}
		\hfill
	\subfloat[]{\includegraphics[width=0.83\textwidth]{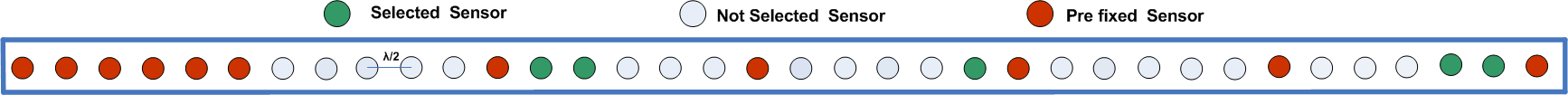}
		\label{op_symalgo_90_0_ch4}}
		\hfill
	\subfloat[]{\includegraphics[width=0.83\textwidth]{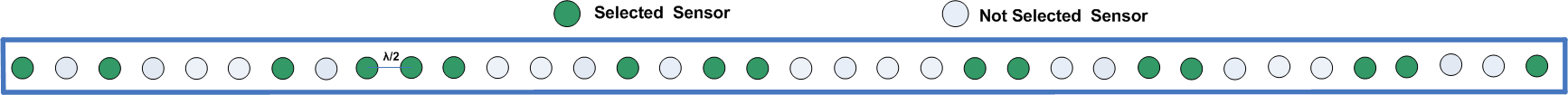}
		\label{best_90_01_ch4}}
		\hfill
	\subfloat[]{\includegraphics[width=0.83\textwidth]{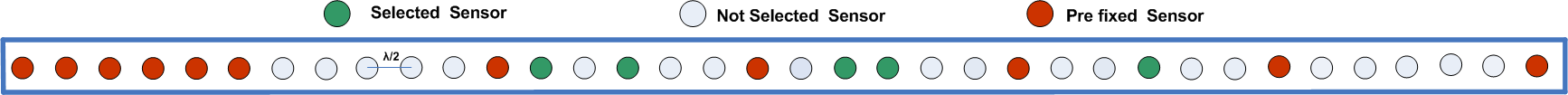}
		\label{worst_90_01_ch4}}
\hfill
	\subfloat[]{\includegraphics[width=0.83\textwidth]{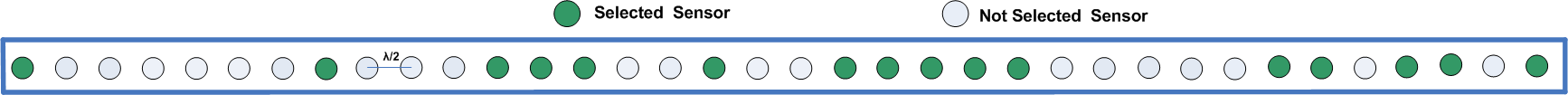}
		\label{op_algo_90_01_ch4}}
		\hfill
	\subfloat[]{\includegraphics[width=0.83\textwidth]{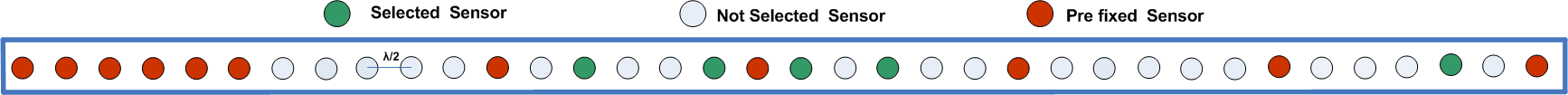}
		\label{op_symalgo_90_01_ch4}}
		\hfill
	\subfloat[]{\includegraphics[width=0.83\textwidth]{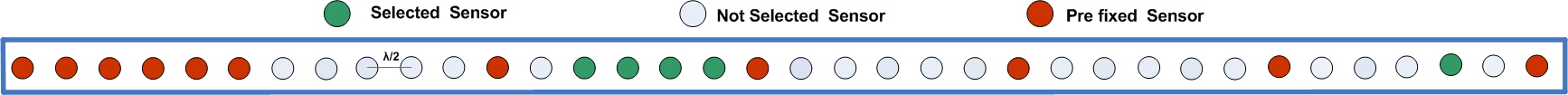}
		\label{op_algo_90_02_ch4}}
		\hfill
	\subfloat[]{\includegraphics[width=0.83\textwidth]{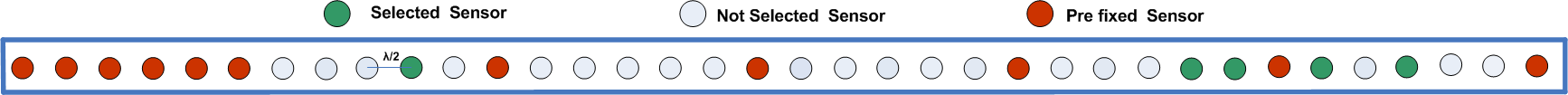}
		\label{op_symalgo_90_02_ch4}}
		\hfill
	\caption{(a) Initial   configuration; randomly selected 16 antennas from 36 (b) Initial   configuration leading to fully augmentable array   (c) Freely designed array (d) Hybrid designed array (e)  Initial random  configuration;  selected 16 antennas from 36 (f) Initial   configuration  leading to  fully augmentable array (g) Freely designed array (h) Hybrid designed array (i) Best performing array configuration  (j) Worst performing array configuration  }
	\label{point_ch4}
\end{figure}
Given $N=36$ perspective   sensor locations placed linearly with an inter-element spacing of $\lambda/2$. Consider selecting $P=16$ sensors among these locations so as to maximize the SINR. A single  source of interest is  operating at   $90^0$, i.e. array broadside. There are also six jammers,   concurrently active at locations  $40^0$, $85^0$, $95^0$, $135^0$, $140^0$  and $160^0$. The SNR of the desired signal is $0$ dB, whereas each  jammer has the interference to noise ratio (INR)  of $20$ dB. The range of binary search for the sparsity parameter $\mu$ is set from $0.01$ to $5$, $\gamma=10^{-3}$ (sparsity threshold)  and $\epsilon=0.05$. The initial $16$-element sparse array configuration to estimate the data correlation matrix is randomly chosen, and shown in the    Fig. \ref{best_90_0_ch4}. This configuration has missing correlation lags and is occupying a fraction of the available aperture. The array collects the data for  $T=1000$ snapshots. The full array Toeplitz estimate is recovered through matrix completion with the regularization parameter $\zeta=0.5$.   The proposed SCA approach  employing  matrix completion renders an array configuration with SINR of $11.73$ dB. It is worth noting that, for the underlying case,  the number of possible array configurations is of  order   $10^9$ which makes the problem infeasible to solve  through exhaustive search. The upper bound of performance, however,  is  $12$ dB which corresponds to the case when interferences are completely canceled in the output. In this regard, the designed array configuration is very close to the performance upper bound. The optimized array configuration is shown in the  Fig. \ref{op_algo_90_0_ch4}. It is noted that this configuration has also missing few correlation lags.

In order to access the performance of the hybrid design approach, we  consider a randomly selected $16$ element fully augmentable array, which is shown in Fig. \ref{worst_90_0_ch4}. The full data correlation matrix is estimated using the same $T=1000$ snapshots and averaging is carried over the available correlation lags to yield a Toepltiz estimate. The SCA approach, in this case, achieves the array design shown in  Fig. \ref{op_symalgo_90_0_ch4} and has a reasonable SINR performance of $10.92$ dB. The designed hybrid array is fully augmentable and involves the  prefixed sensor locations which are arranged in the nested array topology (prefixed configuration shown in red color). The hybrid design is clearly sub optimal as compared to the array designed freely. It is noted that the number of possible  hybrid sparse array configurations associated with the prefixed sensors   is $53130$. Although, the possible fully augmentable configurations are significantly less as compared to $10^9$  possibilities, the maximum SINR hybrid design found through enumeration is $11.93$ dB and is close to the upper performance bound of $12$ dB.  The performance of both designs are compared with the benchmark design initialized with FCS-LSS  estimated from  $T=1000$ samples supposedly collected from all $N$ sensors. The benchmark design yields the freely designed and hybrid sparse configurations  with the  SINR of $11.82$ dB and $11.65$ dB respectively. This performance is  superior to the above mentioned designs that employ the Toeplitz estimation  in lieu of the actual  full correlation matrix. 

It is of interest to analyze the effect of the initial sparse array configuration on the proposed SCA optimization. This time, the data is collected through the initial configurations  depicted in Figs. \ref{best_90_01_ch4} and \ref{worst_90_01_ch4},  instead of the configurations (Figs. \ref{best_90_0_ch4} and \ref{worst_90_0_ch4}) employed for the earlier example. The underlying operating environment and all other parameters remain the same as above. As before, the freely designed array is achieved through matrix completion, whereas the hybrid design involves averaging to estimate the full data correlation matrix.  The free-design and the hybrid design achieve  SINR of $11.82$ dB and $11.65$ dB, respectively. The designed array configurations are shown in the  Figs. \ref{op_algo_90_01_ch4} and \ref{op_symalgo_90_01_ch4}.
These configurations offer superior performances to those optimized earlier, assuming different initial configurations. This  underscores the dependence of sparse array beamforming optimization on the array initial conditions. It is  noted that for the same underlying environment and  initial configuration, the proposed solution   is still not unique and dependent on the random realizations of the received data.  In order to reliably gauge the performance  of the proposed scheme, we report the average results   repeated over $100$ independent trials.  It is found that  under the initial configurations shown in Figs. \ref{best_90_0_ch4} and \ref{worst_90_0_ch4}, the average SINR performances are $11.79$ dB for freely designed SCA and $11.18$ dB for the hybrid design. On the other hand,   the initial configurations, shown in Figs. \ref{best_90_01_ch4} and \ref{worst_90_01_ch4}, yield  the average performances of $11.6$ dB   and $11.54$ dB for the free and hybrid designs, respectively. These performances are compared with the FCS-LSS benchmark.  It is found that  the  FCS-LSS  offers the same performance as is achieved by SCA under initial configurations adopted in Figs. \ref{best_90_01_ch4} and \ref{worst_90_01_ch4}.   We remark that under the initial array configurations shown in Figs. \ref{best_90_0_ch4} and \ref{worst_90_0_ch4}, the SCA-based matrix completion even surpasses the FCS-LSS benchmark, however, it offers slightly lower SINR for the hybrid design ($11.18$ dB as compared to $11.54$ dB). The optimum hybrid array configuration found through enumeration is shown in Fig. \ref{op_algo_90_02_ch4} with an SINR of $11.9$ dB, whereas the worst case hybrid configuration (shown in Fig. \ref{op_symalgo_90_02_ch4}) has an associated SINR of $7.5$ dB which is considerably lower than the above designs.
\subsection{Monte Carlo design for random scenarios}
\begin{figure}[!t]
	\centering
	\includegraphics[width=14.55cm, height=11.50cm]{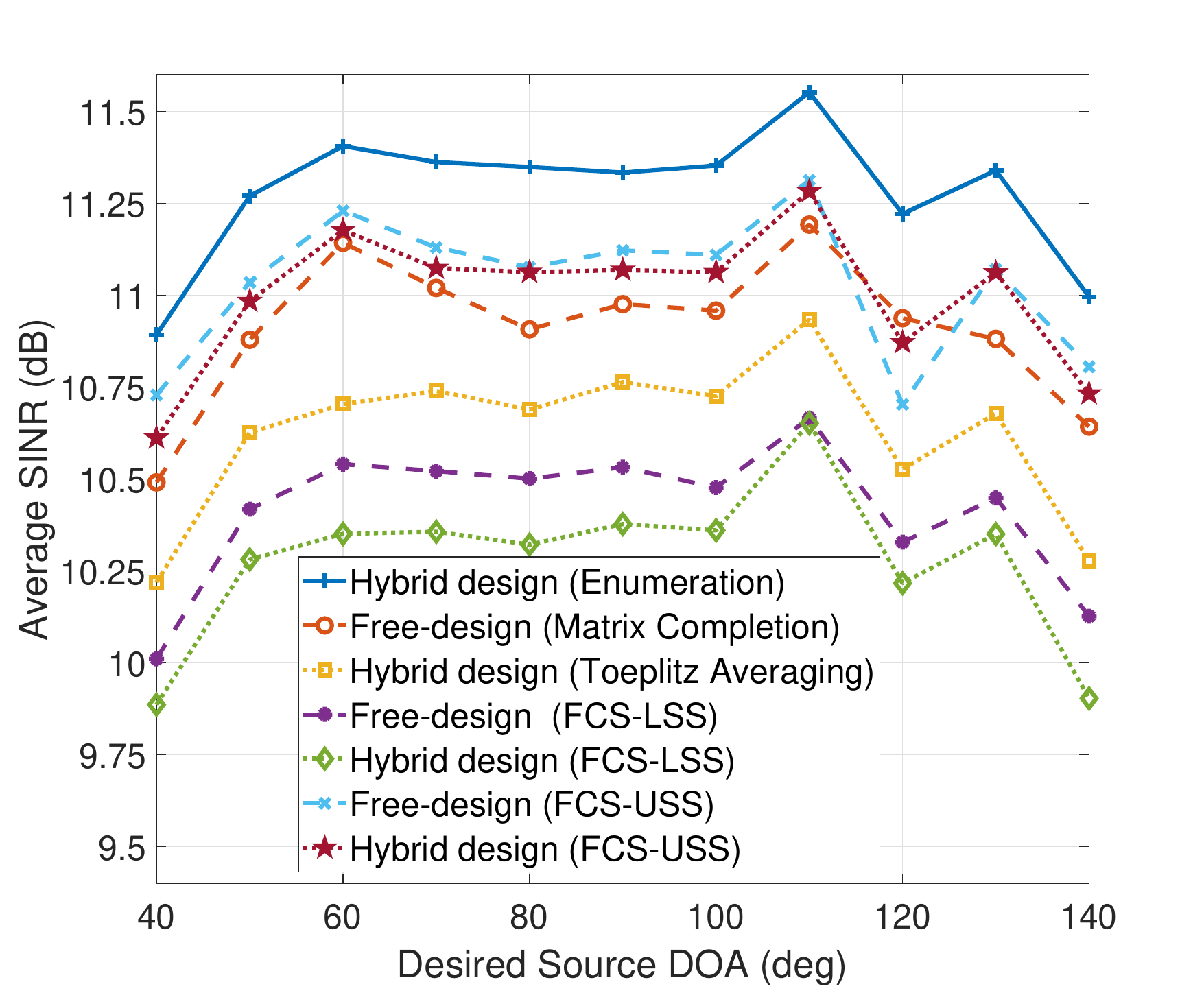}
	\caption{Average SINR performance of various sparse topologies against desired source DOA  for  $T=100$ snapshots.}
	\label{robust_bp1_ch4}
\end{figure}
\begin{figure}[!t]
	\centering
	\includegraphics[width=14.55cm, height=11.50cm]{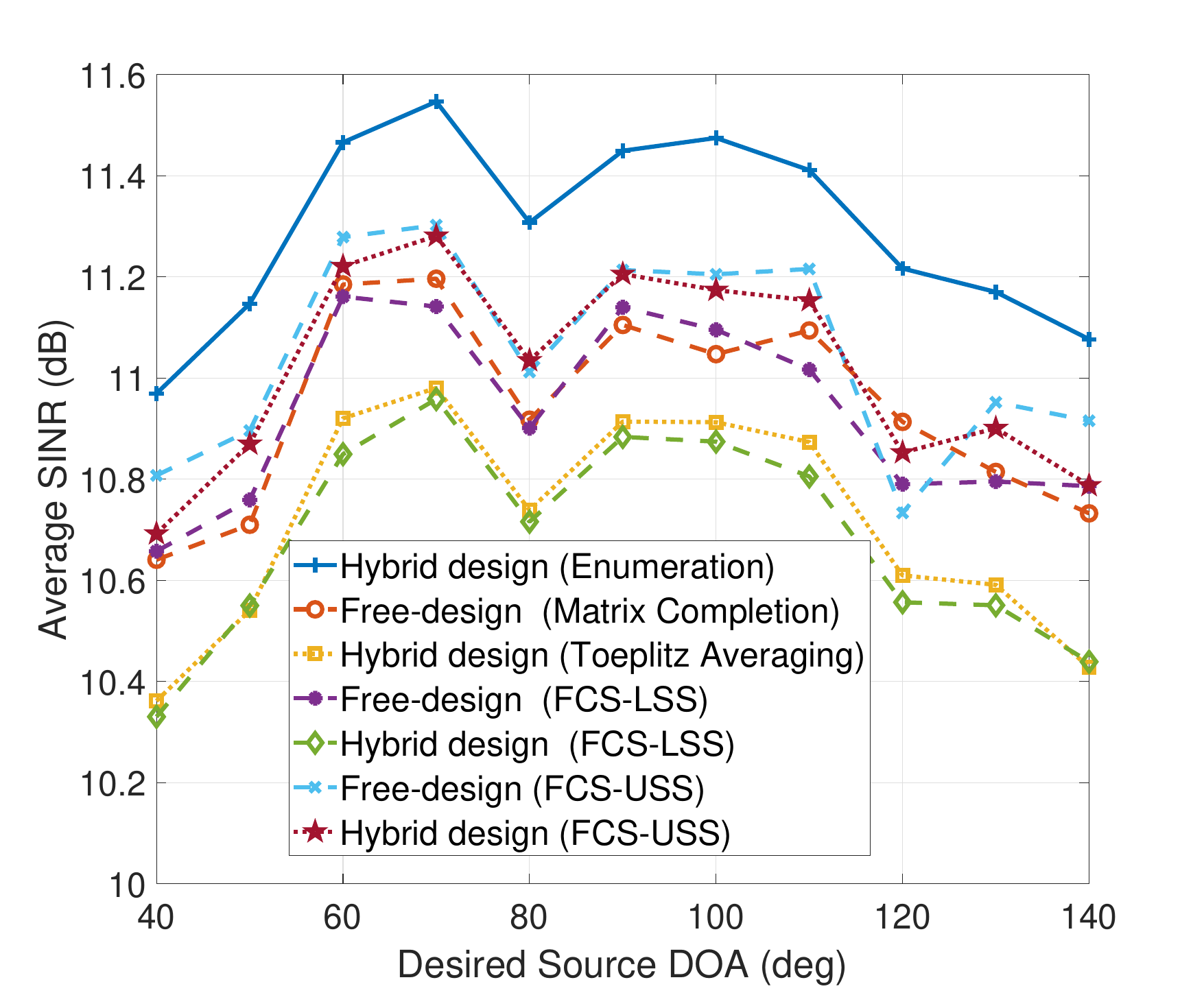}
	\caption{Average SINR performance of various sparse topologies against desired source DOA  for  $T=250$ snapshots.}
	\label{robust_bp2_ch4}
\end{figure}
The above examples tie the performance of the proposed algorithm  not only  to the location of the sources and their respective powers but  also show the dependence on the initial array configuration, the number of snapshots and the observed realization of the received data. In order to provide a more meaningful assessment, the simulation scenarios are designed  keeping the aforementioned variables in perspective. We generate $11$   different scenarios. For each scenario, the desired source DOA is kept fixed, whereas six jammers are randomly placed   anywhere from $30^0$ to $150^0$ with the respective powers uniformly distributed from $5$ to $15$ dB. The experiments are repeated $3000$ times and the initial array configuration is randomly chosen for each experiment. For the freely designed array, the initial array configuration is selected by randomly choosing $16$ sensors from $36$ sensors. However, the initial configuration for  the hybrid design is randomly chosen from all the possible $16$ sensor fully augmentable array configurations associated with the prefixed sensors arranged in nested configuration as depicted in  Fig. \ref{worst_90_0_ch4} (red color sensors). 

Figure \ref{robust_bp1_ch4}  shows the results for $T=100$. The performance curve of the SCA algorithm for the freely designed array incorporating  matrix completion lies in between (for most points) the benchmark designs incorporating   FCS-USS and  FCS-LSS.  That is the matrix completion approach  even outperforms  the benchmark design incorporating the FCS-LSS. This performance is explainable because  matrix completion coupled with the apriori knowledge of noise floor renders a more accurate estimate of the full correlation matrix as compared to  FCS-LSS,  without incorporating knowledge of noise floor, which has high noise variance because of limited snapshots. The performance of the other benchmark incorporating the exact knowledge of the  correlation matrix (FCS-USS) is clearly superior over  matrix completion. The results are fairly similar for the hybrid design, where the performance curve utilizing the Toeplitz averaging is sandwiched between the  benchmark designs incorporating the exact  correlation matrix (FCS-USS) and  the one  utilizing the  presumably observed full data correlation matrix (FCS-LSS). The hybrid designed and freely designed arrays, both demonstrate desirable performances. However, the  matrix completion  marginally outperforms  the hybrid design with an average performance gain of $0.2$ dB.

\begin{figure}[!t]
	\centering
	\includegraphics[width=14.55cm, height=11.50cm]{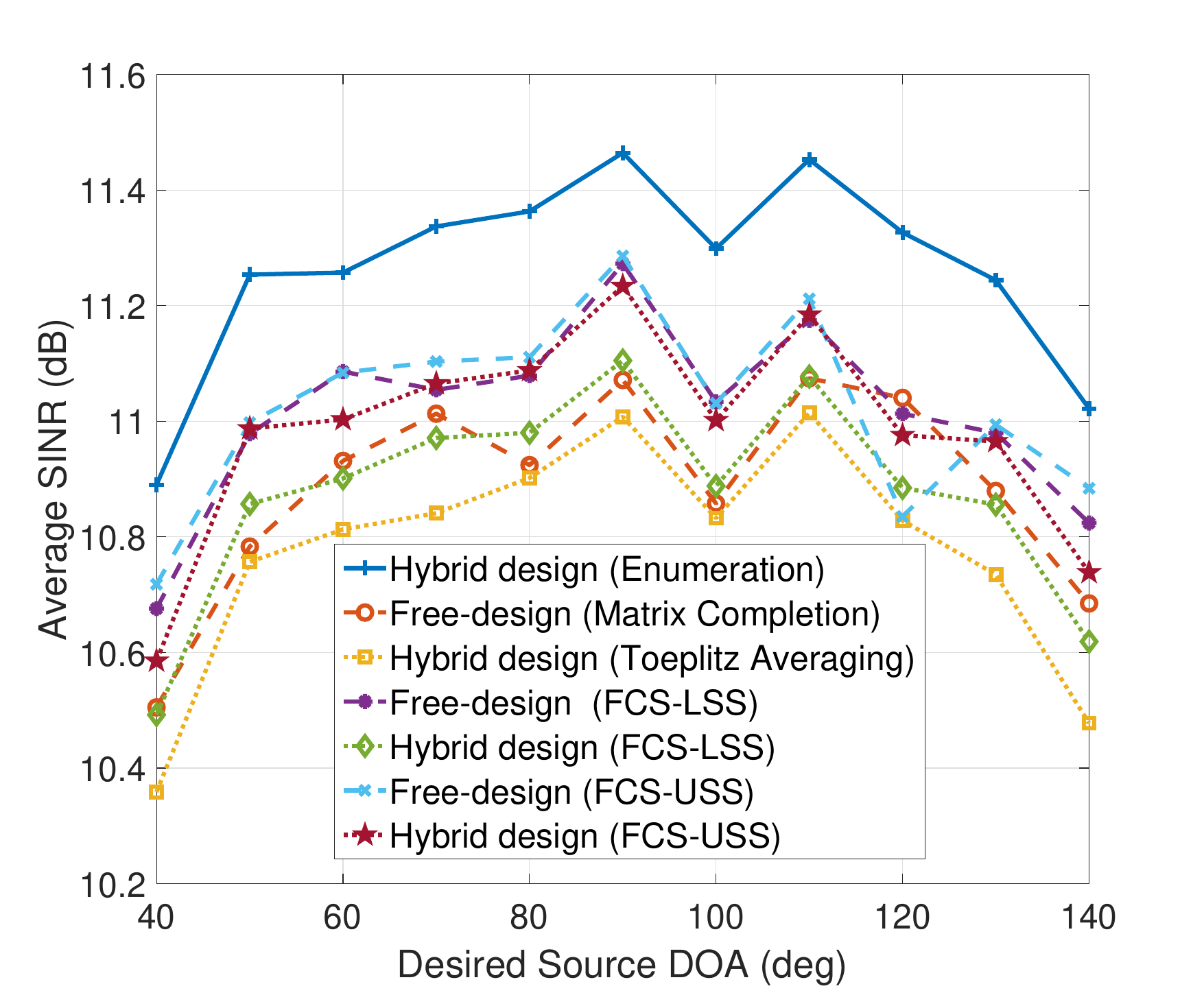}
	\caption{Average SINR performance of various sparse topologies against desired source DOA  for  $T=1000$ snapshots.}
	\label{robust_bp3_ch4}
\end{figure}

\begin{figure}[!t]
	\centering
	\includegraphics[width=13.55cm, height=10.50cm]{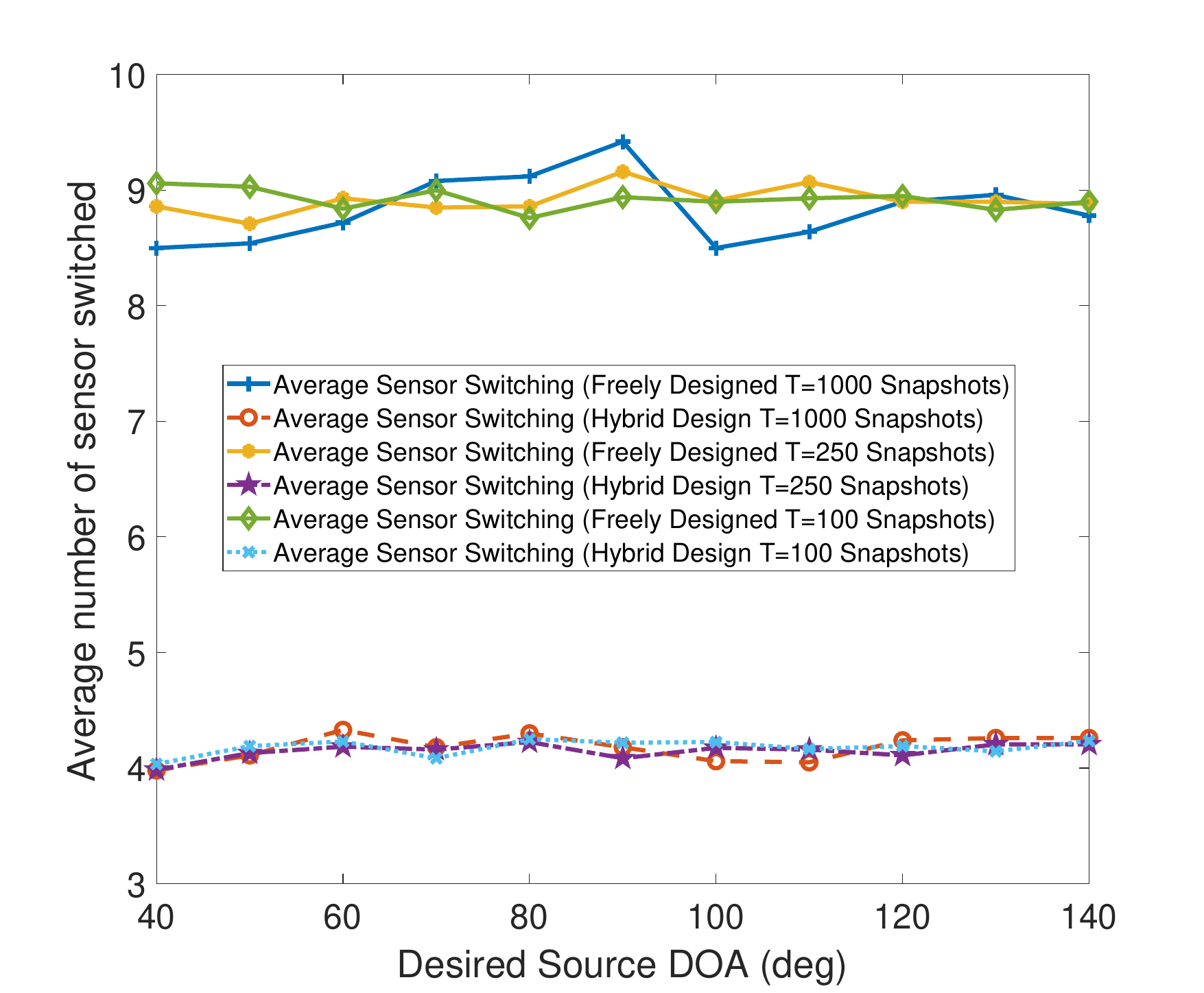}
	\caption{Sensor switching  comparison vs the  free-design and the hybrid design.}
	\label{robust_bp4_ch4}
\end{figure}

The performance curves  are re-evaluated by increasing the snapshots to $T=250$ and $T=1000$, as shown in Figs. \ref{robust_bp2_ch4} and \ref{robust_bp3_ch4}. With such increase, the performances of the proposed SCA using Toeplitz completion move closer to the performances of the FCS-USS benchmark. It is also noted that in contrast to lower snapshots ($T=100$), the  FCS-LSS benchmark  for higher samples ($T=1000$) offers superior average performance over SCA designs incorporating Toeplitz completion.   It is of interest to track the average antenna switching involved per trial for both the free-design and the hybrid design. Fig. \ref{robust_bp4_ch4} shows that freely designed array involves $9$ antenna switching per trial which is more than twice that of  the hybrid design ($4$ antenna switching per trial). It is also  noted that for the hybrid design, the maximum antenna switching is constrained to  $5$ antennas as the rest of $11$ sensors are prefixed. In this regard, the hybrid design    has  more efficient switching as it utilizes 80 percent (4/5) of the DOF as compared to the mere 55 percent (9/16) switching efficiency of freely designed arrays. 

\section{Conclusion}
Sparse array design for maximizing the beamformer output SINR is considered for a single source in an interference active environment. The chapter addressed the problem  that the optimization of the array configuration requires full data correlation matrix which is not readily available in practice. Two different design approaches were considered; one assumes prefixed position of subset of sensors so as to provide full array augmentation, referred to as the hybrid-design approach,  whereas the other, which is referred to as free-design approach,   has no such restriction, and freely allocates all degrees of freedom to maximize the objective function.    It was  shown that the Toeplitz estimation of the autocorrelation at the missing spatial lags has a desirable performance. The SCA was proposed for both the freely designed  and hybrid designed arrays  to achieve MaxSINR  in polynomial run times with a reasonable trade off in  SINR. It was  shown that, in contrast to hybrid design, the matrix completion scheme does not require to pre-allocate sensor resources and, therefore,  offers more design flexibility and better SINR performance. This performance improvement is, however,  at the cost of increased computational complexity and finer parameter tuning as required to accomplish Toepltiz matrix completion. The simulation examples  showed   that the performance of the proposed SCA algorithm incorporating  Toeplitz completion is agreeable with the established benchmark designs.





	\chapter{Sparse Array  Beamforming Design for Wideband Signal Models}
	%
	%
	%


\section{Introduction}  \label{Introduction_ch5}
%
%
%
%

Wideband systems  can deliver accurate target localization   for radar systems \cite{1359140}, provide diversity, reliability and anti jamming capabilities to the wireless communication systems and signal enhancement for microphone arrays \cite{Brandstein2001MicrophoneA, Goldsmith:2005:WC:993515},   whereas the  UWB (Ultra-wideband) systems play a major role in high resolution imagery in medical imaging \cite{2f1ce6c9e62d4b60acddd36683f19924, 5666802}.  

\begin{figure}[!t]
	\centering
	\includegraphics[width=5.28 in, height=2.75 in]{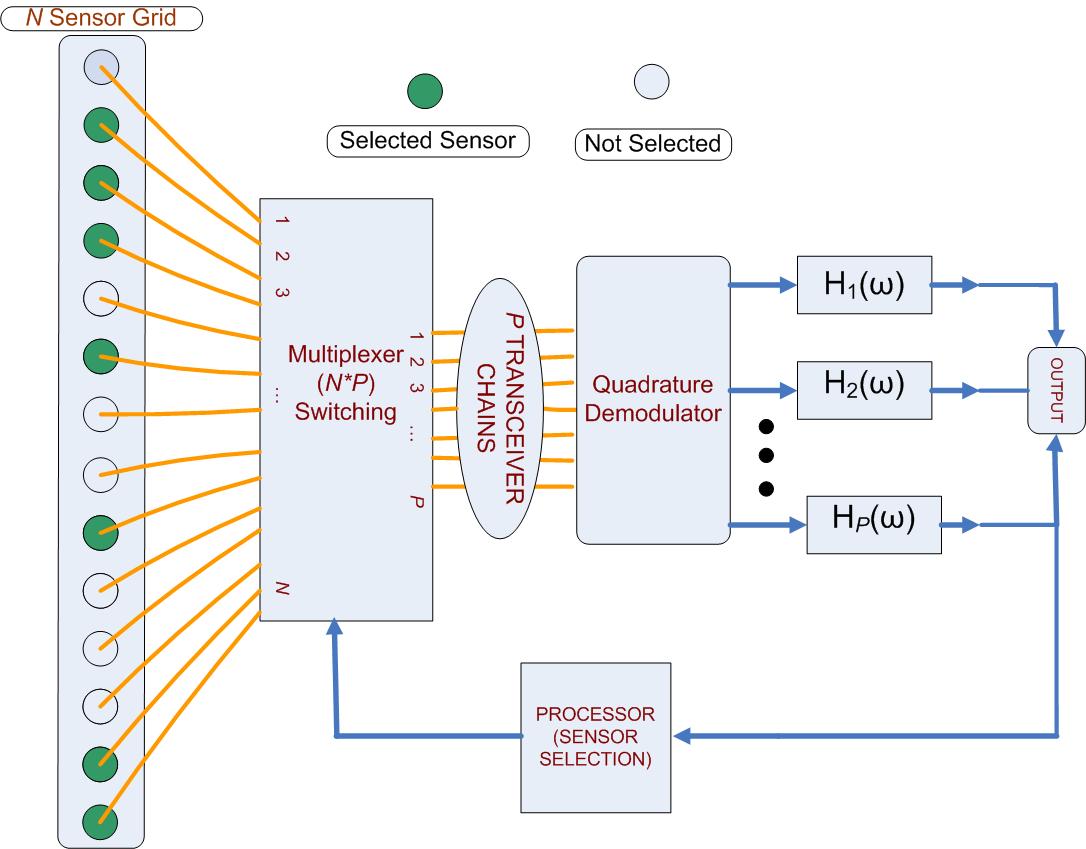}
	\caption{Block Diagram of sparse array wideband processing.}
	\label{labelfig1_ch5}
\end{figure}

 Beamforming techniques for wideband signals  either involve  a fixed design such as a frequency invariant beamformer  or an adaptive design  based on  the linearly constrained minimum variance (LCMV) beamformer \cite{doi:10.1002/0471733482.ch7, Liu2010WidebandBC}.  Irrespective of the design criterion, the wideband beamformers  are typically  implemented jointly in the spatial and temporal domains, as shown in  Fig. \ref{labelfig1_ch5}. This spatio-temporal processing is often realized through   two different  schemes, namely, the tapped delay line  (TDL) filtering  or subband processing like  DFT. For the former, an $L$ TDL filter is assumed for each sensor, and the received data at each sampling instant is processed for all sensors jointly \cite{1165142, 665}.  In the DFT implementation scheme, the data at each sensor is buffered and transformed to the frequency domain by $L$-point DFT. Once expressed in terms of narrowband signals,  optimal beamforming is performed in each DFT bin.  The DFT implementation is computationally more viable \cite{1450747, 1164219, 1143832}. However, the TDL   implementation scheme has an added advantage since buffering is not required and  the spatio-temporal  weight vector can be updated at each sampling instant.   The TDL and DFT beamformers would render  identical output signal  if the corresponding beamformer weights are   related through  a DFT transformation. However, carrying  the beamformer design,  separately in each domain, doesn't  warrant  the beamformer weights forming strictly a  DFT pair.  Resultantly, the output could  differ  slightly for each implementation. To circumvent the  computationally expensive  TDL beamformer design, a DFT beamformer is rather optimized and  the DFT transformation is subsequently used  to obtain  the TDL beamformer. This dual domain  TDL implementation designed primarily in the DFT domain   can  yield adequate output performance in practice \cite{ doi:10.1121/1.413765}.

 Sparse array design strives to optimally deploy  sensors, essentially   achieving desirable beamforming characteristics, lowering  the system hardware costs and  reducing the computational  complexity. Sparse array design is known to yield considerable performance advantages under different design criteria for narrowband signal models.  These   criteria can largely be segregated  into environment-independent  and environment-dependent designs. The minimum redundancy arrays (MRA) and the structured sparse array configurations are cases of the former design. They seek to optimize the environment-blind design criterion to enable the DOA estimation of more sources than physical sensors \cite{5739227,  7012090, 5456168, 7106511}.  More recently, the   switched antenna and beam technologies  have motivated  the design for environment  adaptive sparse arrays. The available sensors are switched  according to  changing environmental conditions, enabling optimum use of expensive transceiver chains  \cite{738234, 1299086, 922993, 967086}. Sparse array design based on Cramer-Rao lower bound (CRLB) minimization criterion was shown effective  for DOA estimation schemes \cite{8386702}.   On the other hand,   beamforming approaches typically implements MaxSINR criterion, yielding efficient  adaptive  performance  that is dependent on the underlying operating environment \cite{6774934, 8036237, 8061016,  663844798989898, 6477161, 1634819, 5680595}. 

 Designing sparse arrays has proved to be particularly advantageous for wideband signal models, as it circumvents the contradicting design requirements posed by the frequency spread of the signal. On the one hand,  the lower frequency end of the spectrum puts a minimum aperture constraint on the array to achieve  certain resolution. On the other hand, the higher end of the spectrum dictates  the minimum separation between consecutive sensor locations so as to avoid spatial aliasing, consequently, resulting in the oversampling of the lower frequency content.   For a limited number of available sensors, wideband sparse array design can, in essence,  yield improved performance in many design applications  by offering more control over beampattern characteristics for all frequencies of interest  \cite{81155, 23483091a3104a7f8bd7d213f8ada7cc, 1071, MURINO1997177}.  Many different  metrics,  such as frequency invariant beampattern synthesis, robust beampattern design   and side-lobe level control   have been proposed for optimal  wideband sparse array beamforming \cite{482017,  doi:10.1121/1.412215, 1071, 968251,  93229}.  For instance,  simulated annealing has been applied in \cite{968251} to achieve the desired peak sidelobe levels, while jointly optimizing  the beamformer weights and sensor locations.   Frequency invariant beampattern optimization  for  wideband sparse array design employing  compressive sensing (CS) technique has  been implemented  in \cite{6822510}.  The authors therein,   invoked sparsity in the temporal and spatial domains, simultaneously,  in an  attempt to decrease the overall processing complexity.  
 
  In this chapter, we examine the Capon based MaxSINR sparse array design from both the DFT  and TDL filtering realization perspectives.   We consider a switched array  adaptive beamformer design  which is   fundamentally different  as compared to the aforementioned wideband  performance metrics that optimize a prefixed receive beampattern for certain sector/frequencies of interest, independent of the received data statistics.      
We examine  environment-dependent sparse arrays that maximize the SINR for  frequency spread  source operating in  wideband jamming environments. The objective of the populated and the sparse wideband beamformers is fundamentally the same;  minimization of noise and interference signals at the array output while simultaneously maintaining a desired response in the direction of interest. We  adopt a Capon based methodology for  enhancing the   signal power for the desired source operating  in an interference active environment. Capon method is a well known linear constraint beamforming approach that rejects the interference and maximizes the output SINR \cite{4101326}. It  provides superior estimation accuracy but  assumes the exact knowledge  or estimated version of the received data correlation matrix across the entire array aperture. The latter is the case in many  key applications in  radar   signal processing and   medical imaging  \cite{Trees:1992:DEM:530789, 1206680, 10.1117/12.2519879, 
8892512, 9257073,  hamza2019sparse}. This assumption, however, cannot be readily made for sparse array configurations.
We pose the design  problem as optimally selecting $P$ sensors out of $N$ possible equally spaced locations. For the scope of this chapter, we ignore any mutual coupling or sensor failure possibilities that could arise due to the closely spaced perspective locations on the grid. Each sensor has an associated  $L$ TDL or $L$ point DFT filtering  to jointly process the  signal in the temporal and spatial  domains. Our approach is a  natural extension of Capon beamforming  at the receiver  and amounts to maximizing the SINR over all possible sparse array configurations. In the case of the TDL realization, we select those sensors that  maximize the principal eigenvalue of the product of  the inverse of received data correlation matrix and the desired source  correlation matrix \cite{1223538}. For the DFT implementation scheme, the  maximization is performed over all DFT bins.  In either case, it   is an NP hard optimization problem. In order to realize convex relaxation  and avoid the computational burden of applying the singular value decomposition (SVD) for each possible configuration, we solve the underlying problem using constrained minimization based approach. We consider two different optimization approaches, namely, the semidefinite relaxation (SDR) and successive convex approximation (SCA). For SDR-based approach, we pose the problem as quadratically constraint quadratic program (QCQP) with weighted $l_{1-\infty}$-norm squared to promote group sparsity. An eigenvector based  iterative methodology is adopted  to promote group sparsity and ensure that only $P$  sensors are finally selected. It is noted that the re-weighted $l_{1-\infty}$-norm squared relaxation is shown to be  effective for reducing the number of sensors in multicast transmit beamforming \cite{6477161}. However, owing to the  computational complexity associated with the SDR approach, we alternatively  pose the  problem as  successive convex relaxation (SCA)   that approximates the problem iteratively by first order gradient approximation \cite{8445845}. The proposed algorithms are  suitable for moderate-size antenna systems to enable real time  applications wherein the environment largely stays stationary relative to the time required for sparse configurability. 

In order to enable a data-dependent design for sparse array wideband beamforming, we require     knowledge of the   received data correlation matrix corresponding to the full array aperture. With only few active sensors at any time instant, it is  infeasible to assume such knowledge due to  missing correlation entries. We circumvent this problem by employing a low rank block Toeplitz matrix completion scheme to interpolate the missing correlation entries. Subsequently,  the interpolated data correlation matrix is input to the proposed sparse optimization algorithms.  We  demonstrate the offerings of the proposed sparse array design utilizing matrix completion under limited data snapshots by comparing its performance with that achieved through enumeration. \par

The rest of the chapter is organized as follows:  In the next section, we  state the problem formulation for  maximizing the  output SINR under wideband source signal model by elucidating the TDL and DFT signal model. Section \ref{Optimum sparse array design_ch5} deals with the optimum sparse array design by  semidefinite relaxation as well as successive convex relaxation to obtain the optimum $P$  sparse array geometry. Section \ref{sec4_ch5} discusses the block Toeplitz matrix completion approach for a conceivable sparse array design from an implementation perspective. Design examples  and conclusion follow at the end.
\section{Problem Formulation} \label{Problem Formulation_ch5}
Consider  a single desired source and $Q$ interfering source signals impinging on a  linear array with $N$ uniformly placed sensors. The   Nyquist sampled received baseband signal $\mathbf{x}(n) \in \mathbb{C}^{N}$   at time instant $n$ is, therefore,  given by, 
\begin{equation} \label{a_ch5}
\mathbf{x}(n)=    \mathbf{s}(n)  + \sum_{k=1}^{Q}  \mathbf{i}_k(n) + \mathbf{v}(n),  
\end{equation}
where $\mathbf{s}(n)\in \mathbb{C}^{N} $ is the contribution from the desired signal located at $\theta_s$, $\mathbf{i}_k(n) $  is the $k$th interfering signal vectors  corresponding to the  respective direction  of arrival, $\theta_{k}$, and $\mathbf{v}(n)$ is the spatially uncorrelated sensor array output noise.
\begin{figure}[!t]
	\centering
	\includegraphics[width=5.28 in, height=2.75 in]{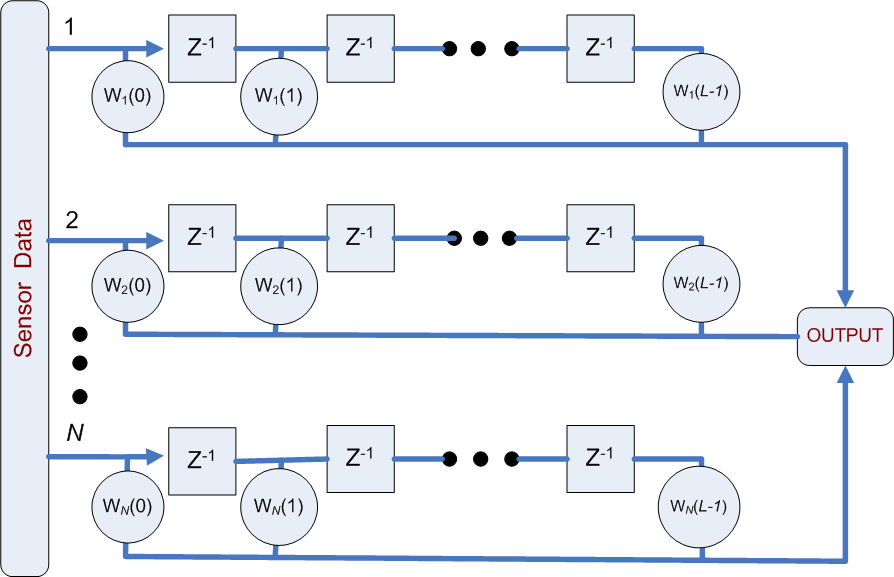}
	\caption{TDL realization of  wideband  beamforming.}
	\label{labelfig2_ch5}
\end{figure}
\subsection{TDL Implementation scheme}
We assume a  TDL of length $L$ associated with each  sensor, as shown in Fig. \ref{labelfig2_ch5}. The symbol z$^{-1}$ denotes the time delay and w$_k(m)$ is the  beamforming weight for the k$th$ sensor at m$th$ sampling instant.   We define a stacked vector $\mathbf{X}=[\mathbf{x}^T(n), \mathbf{x}^T(n-1), ..., \mathbf{x}^T(n-L+1)]^T \in \mathbb{C}^{NL}$ containing the  array data collected  over $L$ sampling instances ((.)$^T$ denotes the transpose).  Rewriting  (\ref{a_ch5}) in a compact way in terms of stacked vectors, we obtain,
\begin{equation} \label{b_ch5}
\mathbf{X}=    \mathbf{S}  + \sum_{k=1}^{Q}  \mathbf{I}_k + \mathbf{V}  
\end{equation}
Here, $\mathbf{S}=[\mathbf{s}^T(n), \mathbf{s}^T(n-1), ..., \mathbf{s}^T(n-L+1)]^T $ and similarly $\mathbf{I}_k$ and  $\mathbf{V}$ are defined, respectively, as interference and noise stacked vectors. The received signal $\mathbf{X}$  is  then  combined linearly  to maximize the output SINR. The output signal $y(n)$ of the optimum beamformer for maximum SINR is given by \cite{1223538}, 
\begin{equation}  \label{c_ch5}
y(n) = \mathbf{w}_o^H \mathbf{X},
\end{equation}
where $\mathbf{w}_o$ is the solution of the following optimization problem,
\begin{equation} \label{d_ch5}
\begin{aligned}
\underset{\mathbf{w} \in \mathbb{C}^{NL}}{\text{minimize}} & \quad   \mathbf{ w}^H\mathbf{R}_{n}\mathbf{w}\\
\text{s.t.} & \quad     \mathbf{w}^H\mathbf{R}_{s}\mathbf{ w}=1
\end{aligned}
\end{equation}
The w$_k(m)$ beamforming weight (shown in Fig. 2) maps to the $(N(m)+k)$th element of the   stacked beamformer $\mathbf{w}$ ($m \in 0, 1, ..., L-1$, $k \in 1, 2, ..., N$). Here, (.)$^H$ denotes Hermitian transpose, $\mathbf{R}_s=E\mathbf{(SS}^H) \in \mathbb{C}^{NL\times NL}$ is the desired signal correlation matrix. Likewise, $\mathbf{R}_{n}$ is the correlation matrix associated with interference and noise stacked vectors. In the case of spatial spread or wideband source signal, the correlation matrix is given by \cite{1165142},

\begin{equation}  \label{e_ch5}
 \mathbf{R}_s =  \int_{B_s}\int_{\Theta_s} \sigma^2_{\theta_s}(\omega)\mathbf{a}(\theta_s,\omega)\mathbf{a}^H(\theta_s,\omega)  d\theta_s d\omega
\end{equation} 
Here, $\sigma^2_{\theta_s}(\omega)$ is the signal power as a function of $\theta_s$ and $\omega$, $\Theta_s$ and $B_s$ are the spatial and spectral supports  of the desired source signal.
We only consider  point sources with no significant  spatial extent, hence rewriting (\ref{e_ch5}) as follows, 
\begin{equation}  \label{f_ch5}
\mathbf{R}_s =  \int_{B_s} \sigma^2_{\theta_s}(\omega)\mathbf{a}(\theta_s,\omega)\mathbf{a}^H(\theta_s,\omega) d\omega
\end{equation}
The space-time steering vector  $\mathbf{a}(\theta_s,\omega) \in \mathbb{C}^{NL}$, corresponding to the  source signal, can be represented as  a Kronecker product ($\otimes$),
\begin{equation}  \label{g_ch5}
\mathbf{a(\theta_s,\omega)} = \pmb{\phi}_{\omega}\otimes \mathbf{a}_{\theta_s}(\omega),
\end{equation}
with,
\begin{equation}  \label{h_ch5}
\pmb{\phi}_{\omega}=[1 \,  \,  \, e^{j (\pi \omega/ \omega_{max})} \,  . \,   . \,  . \, e^{j (\pi \omega/ \omega_{max}) (L-1)}]^T,
\end{equation}
\begin{equation} 
\begin{aligned}  \label{i_ch5}
\mathbf{a}_{\theta_s}(w)  &=[1 \,  \,  \, e^{j (2 \pi / \lambda_\omega) d cos(\theta_s)} \,  . \,   . \,  . \, e^{j (2 \pi / \lambda_\omega) d (N-1) cos(\theta_s)  }]^T\\
 &=[1 \,  \,  \, e^{j \pi( \frac{\omega_c+\omega}{\Omega_{max}} ) cos(\theta_s)} \,  . \,   . \,  . \, e^{j \pi( \frac{\omega_c+\omega}{\Omega_{max}} ) (N-1)cos(\theta_s)}]^T,
\end{aligned}
\end{equation}
where $\lambda_\omega$ is the wavelength corresponding to  $\omega_c+\omega$, $\omega$ and $\omega_{c}$ represent the   baseband source angular frequency and the carrier angular frequency respectively, and $\omega_{max}$ is  the maximum source baseband  angular frequency. The data is sampled temporally  at the Nyquist rate for a given signal bandwidth. Similarly, we set the inter-element spacing   $d=\lambda_{min}/2$  to avoid spatial aliasing corresponding to the highest spatial angular frequency $\Omega_{max}=\omega_c+\omega_{max}$,   where $\lambda_{min}$ is the wavelength corresponding to  $\Omega_{max}$. The correlation matrix $\mathbf{R}_k \in \mathbb{C}^{NL\times NL}$ for the   interferer $\mathbf{i}_k$  is defined according to  (\ref{f_ch5}) with respective to $\theta_k$ and  $B_k$. The sensor noise  correlation matrix, $\mathbf{R}_v=\sigma_v^2\mathbf{I} \in \mathbb{C}^{NL\times NL}$ assumes spatially and  temporally uncorrelated noise $\mathbf{v}(n)$ with variance $\sigma_v^2$.
The problem in  (\ref{d_ch5}) can  be written equivalently by replacing $\mathbf{R}_{n}=\sum_{k=1}^{Q}\mathbf{R}_k+\mathbf{R}_v$ with $\mathbf{R}=\mathbf{R}_s+ \mathbf{R}_{n}$ as follows \cite{1223538},
\begin{figure}[!t]
	\centering
	\includegraphics[width=5.28 in, height=2.75 in]{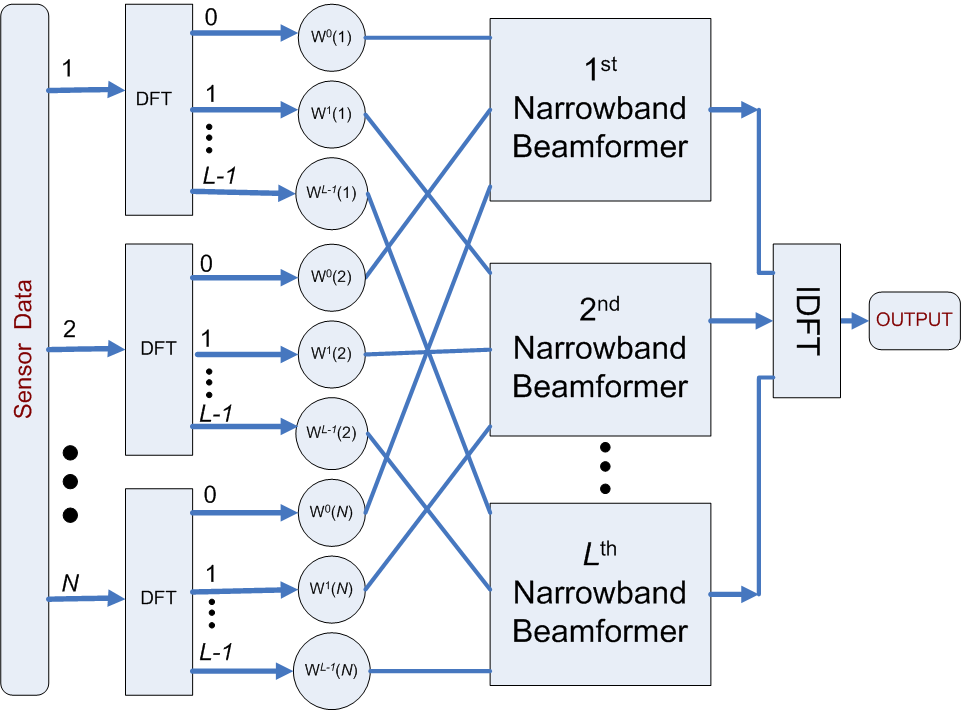}
	\caption{DFT implementation of  wideband  beamforming.}
	\label{labelfig3_ch5}
\end{figure}
\begin{equation} \label{j_ch5}
\begin{aligned}
 \underset{\mathbf{w} \in \mathbb{C}^{NL}}{\text{minimize}} & \quad   \mathbf{ w}^H\mathbf{R}\mathbf{ w}\\
\text{s.t.} & \quad     \mathbf{ w}^H\mathbf{R}_{s}\mathbf{ w} \geq 1
\end{aligned}
\end{equation}
The equality constraint is relaxed in (\ref{j_ch5}) due to the inclusion of the constraint autocorrelation matrix as part of   the objective function, thereby the optimal solution always converges at the equality constraint. The analytical solution of the above  optimization problem exists and  is given by $\mathbf{w}_o = \mathscr{P} \{  \mathbf{R}_{n}^{-1} \mathbf{R}_s  \}=\mathscr{P} \{  \mathbf{R}^{-1} \mathbf{R}_s  \}$. The operator $\mathscr{P} \{. \}$  computes the principal eigenvector of it's argument. Substituting $\mathbf{w}_o= \mathscr{P} \{  \mathbf{R}_{n}^{-1} \mathbf{R}_s  \}  $ into  the SINR formula  yields the corresponding optimum output $SINR_o$ ($\Lambda_{max}$ denotes  the maximum eigenvalue of the matrix);
\begin{equation}  \label{k_ch5}
SINR_o=\frac {\mathbf{w}_o^H \mathbf{R}_s \mathbf{w}_o} { \mathbf{w}_o^H \mathbf{R}_{n} \mathbf{w}_o} = \Lambda_{max}\{\mathbf{R}_{n}^{-1} \mathbf{R}_s\},
\end{equation}
which  shows  that the optimum beamformer for maximizing SINR is directly related to the desired and interference plus noise correlation matrices.  
\subsection{DFT Implementation scheme}
Figure \ref{labelfig3_ch5} shows   the DFT implementation scheme of   wideband array processing. The received signal $\mathbf{x}(n)$ is processed in the spectral domain by taking an $L$ point DFT for the data received by $k$th sensor ${x_k}(n)$, 
\begin{equation} \label{l_ch5}
{X}_k^{(l)} = \sum_{p=0}^{L-1}x_k(n-p)(e^{-j\frac{2\pi}{L}})^{lp}, \, \, \, \, \, \, l \in \{0,1, . . ., \, L-1\}
\end{equation}
Define a vector ${\mathbf{X}}^{(l)} \in \mathbb{C}^N$, containing the  $l$th DFT bin data corresponding to each sensor (superscript $^{(l)}$ denotes the $l$th DFT bin), 
\begin{equation} \label{m_ch5}
{\mathbf{X}}^{(l)}= [{X}_1^{(l)}, {X}_2^{(l)}, . . ., {X}_N^{(l)} ]^T
\end{equation}
These  samples are  then combined linearly by the weight vector $\mathbf{w}^{(l)}  \in \mathbb{C}^N$ such that,
\begin{equation}  \label{n_ch5}
{y}^{(l)} = {\mathbf{w}^{(l)}}^H \mathbf{X}^{(l)},  \, \, \, \, \, \, l \in \{0,1, . . ., \, L-1\}
\end{equation}
Subsequently, the overall beamformer output $y$ is generated by taking the inverse DFT of $y^{(l)}$ across the $L$ beamformers.
 The DFT implementation scheme seeks to maximum the output SINR for each frequency bin, yielding the optimum beamforming weight vector $\mathbf{w}_{o}^{(l)}$ as the solution of the  following optimization problem,
  \begin{equation}  \label{o_ch5}
 \begin{aligned}
 \underset{\mathbf{w}^{(l)}\in \mathbb{C}^N}{\text{minimize}} & \quad   \sum_{l=0}^{L-1}{\mathbf{ w}^{(l)}}^H\mathbf{R}^{(l)}\mathbf{ w}^{(l)}\\
 \text{s.t.} & \quad     {\mathbf{ w}^{(l)}}^H\mathbf{R}_{s}^{(l)}\mathbf{ w}^{(l)} \geq 1.  \quad l \in \{0,1, . . ., \, L-1\}
 \end{aligned}
 \end{equation}
 The correlation matrix $\mathbf{R}^{(l)}=\mathbf{X}^{(l)}{\mathbf{X}^{(l)}}^H$ is the received  correlation matrix for the  $l$th processing bin. Similarly, the source  correlation matrix $\mathbf{R}_{s}^{(l)}$ for the desired source impinging from direction of arrival $\theta_s$ is given by,
 \begin{equation} \label{p_ch5}
\mathbf{R}_{s}^{(l)}=\mathbf{S}^{(l)}{\mathbf{S}^{(l)}}^H={\sigma_{s}^{(l)}}^2\mathbf{a}_{\theta_s}^{(l)}{\mathbf{a}_{\theta_s}^{(l)}}^H 
 \end{equation}
 Here, $\mathbf{S}^{(l)}$ is the received data vector representing the desired source and ${\sigma_{s}^{(l)}}^2 $ denotes the power of this source in the $l$th  bin,   $\mathbf{a}_{\theta_s}^{(l)}$  is  the corresponding  steering vector for the source (DOA $\theta_s$) and is defined as follows,
 \begin{equation} \label{q_ch5}
 \begin{aligned}
 \mathbf{a}_{\theta_s}^{(l)}= {}&[1 \,  \,  \, e^{j \pi( \frac{\Omega_{min}+l\Delta_{\omega}}{\Omega_{max}}) cos(\theta_s)}  {}\\
  &\,  . \,   . \,  . \, \,  \,  \, \,  \,  \, e^{j \pi( \frac{\Omega_{min}+l\Delta_{\omega}}{\Omega_{max}} ) (N-1)cos(\theta_s)}]^T
 \end{aligned}
 \end{equation}
 Eq. (\ref{q_ch5}) models the steering vector for the $l$th DFT bin, where $\Omega_{min}$ is the lower edge of the passband  and $\Delta_{\omega}=\frac{2w_{max}}{L}$ is the spectral resolution. 
The overall output SINR is given by averaging  the SINR over all DFT bins.  Similar to the TDL,  the DFT implementation scheme determines the  optimum sparse array geometry for enhanced MaxSINR performance as explained in the following section.
\section{Optimum sparse array design}  \label{Optimum sparse array design_ch5}
The problem of maximizing the principal eigenvalue for the $P$ sensor selection is a combinatorial optimization problem. In this section, we  assume that the   full array data correlation matrix is known. However,  Section \ref{sec4_ch5} explains the means to   avoid this assumption through fully augmentable sparse array design or utilizing the matrix completion approach \cite{AMIN20171, HAMZA2020102678, 8682266}. We first formulate the sparse array design for maximizing  SINR in the case of wideband beamforming  as an SDR. Owing to the high computational complexity of  SDR, the problem is solved by SCA, for both the TDL and DFT implementation  schemes.
 
 \subsection{Semidefinite relaxation (SDR) for sparse solution}
 \subsubsection{TDL Implementation scheme}
  We assume that the sensor configuration remains the same within the observation time.  In radar, this assumption amounts to selecting the same $P$ sensors over the coherent processing  interval (CPI).  Therefore, the task is to select $P$ entries from the first $N$ elements of $\mathbf{w}$,  and  the same $P$ entries from each subsequent block of $N$ elements (there are $L$ such blocks). Define $\mathbf{w}_k=[\mathbf{w}(k), \mathbf{w}(k+N), \cdot \cdot \cdot, \mathbf{w}(k+N(L-1))] \in  \mathbb{C}^L$ $(k \in \{1, . . ., \, N\})$ as the weights corresponding to TDL of $k$th sensor.   Then,  in seeking sparse solution, the problem  in  (\ref{j_ch5}) can be reformulated as follows,
\begin{equation} \label{a2_ch5}
\begin{aligned}
 \underset{\mathbf{w \in \mathbb{C}}^{NL}}{\text{minimize}} & \quad   \mathbf{ w}^H\mathbf{R}\mathbf{w} + \bar{\mu}(\sum_{k=1}^{N}||\mathbf{w}_k||_q)\\
\text{s.t.} & \quad    \mathbf{ w}^H\mathbf{R}_{s}\mathbf{ w} \geq 1 \\
\end{aligned}
\end{equation}
Here,  $||.||_q$ denotes the $q$-norm of the vector, $\bar{\mu}$ is the sparsity regularization parameter. The mixed $l_{1-q}$ norm regularization  is known to thrive  the group sparsity in the solution. In so doing, the above formulation enables the  decision making on whether to collectively select or discard  all the $L$ sampling instances associated with the $k$th  sensor.  Therefore,  structured group sparsity is essential for wideband beamformer, since the final sparse solution has  to ensure that only $PL$  out of $NL$ spatio-temporal possible sampling instances are chosen through only  $P$ physical  sensors.   The relaxed  problem expressed  in the above equation induces  group sparsity in the optimal weight vector  without placing a hard constraint on the desired cardinality. 
 The  constrained minimization in $(\ref{a2_ch5})$  can be  penalized instead  by the weighted $l_1$-norm function  to further promote   sparse solutions \cite{6663667, article1, Candes2008},
\begin{equation} \label{b2_ch5}
\begin{aligned}
 \underset{\mathbf{w \in \mathbb{C}}^{NL}}{\text{minimize}} & \quad   \mathbf{ w}^H\mathbf{R}\mathbf{w} + \bar{\mu}(\sum_{k=1}^{N}\mathbf{u}^i(k)||\mathbf{w}_k||_q)\\
\text{s.t.} & \quad    \mathbf{ w}^H\mathbf{R}_{s}\mathbf{ w} \geq 1,  \\
\end{aligned}
\end{equation}
where, $\mathbf{u}^i(k)$  is the $k$th element of  re-weighting  vector $\mathbf{u}^i$ at  the $i$th iteration. We choose the  $\infty$-norm  for the  $q$-norm and  replace the weighted $l_1$-norm function   in $(\ref{b2_ch5})$  by the  $l_1$-norm  squared  function  with a modified regularization parameter ($\mu$ instead of $\bar{\mu}$). This change does not affect the regularization property of the  $l_1$-norm  \cite{6477161}. The result is,
\begin{equation} \label{c2_ch5}
\begin{aligned}
 \underset{\mathbf{w \in \mathbb{C}}^{NL}}{\text{minimize}} & \quad   \mathbf{ w}^H\mathbf{R}\mathbf{w} +   \mu(\sum_{k=1}^{N}\mathbf{u}^i(k)||\mathbf{w}_k||_\infty)^2\\
\text{s.t.} & \quad    \mathbf{ w}^H\mathbf{R}_{s}\mathbf{ w} \geq 1 \\
\end{aligned}
\end{equation}
The SDR of the above problem is realized  by substituting $\mathbf{W}=\mathbf{w}\mathbf{w}^H$. The quadratic function, $ \mathbf{ w}^H\mathbf{R}\mathbf{w}= $ Tr$(\mathbf{ w}^H\mathbf{R}\mathbf{w})= $Tr$(\mathbf{R}\mathbf{w}\mathbf{ w}^H)=$ Tr$(\mathbf{R}\mathbf{W})$, where Tr(.) is the trace of the matrix. Similarly, we replace the regularization term in  (\ref{c2_ch5}) by Tr$(\mathbf{U}^i\mathbf{\tilde{W}})$,  with  $\mathbf{U}^i=\mathbf{u}^i(\mathbf{u}^i)^T$  and $\mathbf{\tilde{W}}$ being the auxiliary matrix implementing $\infty-$norm as follows \cite{Bengtsson99optimaldownlink, 5447068, 6477161},
\begin{equation} \label{d2_ch5}
\begin{aligned}
\underset{\mathbf{W \in \mathbb{C}}^{NL\times NL}, \mathbf{\tilde{W} \in \mathbb{R}}^{N\times N}}{\text{minimize}} & \quad   \text{Tr}(\mathbf{R}\mathbf{W}) + \mu \text{Tr}(\mathbf{U}^i\mathbf{\tilde{W}})\\
\text{s.t.} & \quad    \text{Tr}(\mathbf{R}_s\mathbf{W}) \geq 1,  \\
& \quad \mathbf{\tilde{W}} \geq  |\mathbf{W}_{ll}|,  \quad l \in \{0, . . ., \, L-1\}   ,\\
& \quad   \mathbf{W} \succeq 0, \, \text{Rank}(\mathbf{W})=1
\end{aligned}
\end{equation}
\begin{table}[t!] \label{algorithm}
\noindent\rule{16.5cm}{0.6pt}
	\caption{SDR for  the sparse  wideband beamformer}
\noindent\rule{16.5cm}{0.6pt}

	\begin{algorithmic}[]
		
		\renewcommand{\algorithmicrequire}{\textbf{Input:}}
		
		\renewcommand{\algorithmicensure}{\textbf{Output:}}
		
		\REQUIRE Sparse correlation matrices $\overset{\circ}{\mathbf{R}}$ for TDL-SDR ($\overset{\circ}{\mathbf{R}}{^{(l)}}$ for DFT-SDR), $N$, $P$, $L$, ${\mathbf{R}_s}$ for TDL-SDR (${\mathbf{R}_{s}^{(l)}}$ for DFT-SDR). \\
		
	\ENSURE   MaxSINR beamformer against   $P$ active sensors.   \\
		\textbf{Matrix Completion:} \\
		Procedure  in Section \ref{sec4_ch5} is adopted to recover the estimated full data received correlation matrix $\mathbf{\hat{R}}$ for  TDL-SDR  ($\mathbf{\hat{R}}^{(l)}$ for DFT-SDR).\\
		\textbf{Initialization:} \\
		Initialize  $\epsilon$, $\mu_{max}$ (upper limit of binary search), $\mu_{min}$  (lower limit of binary search),  $\mathbf{U}=$ all ones matrix. \\
	
        Appropriate value of  $\mu$ is selected through  the binary search  to ensure $P$ sensor selection.
		\WHILE {(Beamforming weight vector is not $P$ sparse) }
		\STATE Select $\mu=\frac{1}{2}(\mu_{max}+\mu_{min})$ (Binary search).\\
		\STATE   Run the SDR of  (\ref{e2_ch5}) for  TDL-SDR  or (\ref{f2_ch5}) for DFT-SDR (Use either $\mathbf{\hat{R}}$,$\mathbf{\hat{R}}^{(l)}$   in-lieu of $\mathbf{{R}}$,$\mathbf{{R}}^{(l)}$).\\ 
		\STATE   Update  $\mathbf{U}^i$ according to (\ref{h2_ch5}). \\
		Update $\mu_{max}/\mu_{min}$ according to the  binary search. \\
		\ENDWHILE \\
		%
		%
		%
		%
		
		\STATE After achieving the desired cardinality, run SDR for reduced size correlation matrix corresponding to nonzero values of $\mathbf{\tilde{W}}$  and $\mu=0$,  yielding, $\mathbf{w}_o=\mathscr{P} \{  \mathbf{W} \}$ for  TDL-SDR ($\mathbf{w}_{o}^{(l)}=\mathscr{P} \{  \mathbf{W}^{(l)} \}$ for DFT-SDR). 
		
		\RETURN Optimal Beamformer $\mathbf{w}_o$-TDL-SDR ($\mathbf{w}_{o}^{(l)}$-DFT-SDR)
		
		
	\end{algorithmic}
	\label{algorithm1_ch5}
\noindent\rule{16.5cm}{0.6pt}
\end{table}
The operator  `$|.|$' returns the element wise absolute values of the entries of the matrix, `$\geq$' is the element wise comparison and `$\succeq$' represents the generalized matrix inequality, implementing the positive semidefinite constraint, $\mathbf{W}_{ll} \in \mathbb{C}^{N\times N}$ is the $l$th diagonal block matrix  of $\mathbf{W}$.  We note that the solution matrix $\mathbf{W}$ is Hermitian and therefore, it is sufficient to  constrain the upper or lower triangular  entries of $\mathbf{W}_{ll}$ while forcing $\mathbf{\tilde{W}}$ to be symmetric matrix. In so doing, we reduce the inequality constraints and decrease the run time substantially. In addition, we drop  rank constraint in   (\ref{d2_ch5}) which is non convex,    resulting in the following  SDR,
\begin{align} \label{e2_ch5}
\underset{\mathbf{W \in \mathbb{C}}^{NL\times NL}, \mathbf{\tilde{W} \in \mathbb{R}}^{N\times N}}{\text{minimize}} & \quad   \text{Tr}(\mathbf{R}\mathbf{W}) + \mu \text{Tr}(\mathbf{U}^i\mathbf{\tilde{W}}), \nonumber \\
\text{s.t.} & \quad    \text{Tr}(\mathbf{R}_s\mathbf{W}) \geq1, \nonumber \\
& \quad \overset{\bigtriangleup}{\mathbf{\tilde{W}}} \geq  | \overset{\bigtriangleup}{\mathbf{W}}_{ll}|,   \quad l \in \{0,1, . . ., \, L-1\}  , \nonumber\\
& \quad \mathbf{\tilde{W}}=\mathbf{\tilde{W}}^T,  \quad \mathbf{W} \succeq 0.
\end{align}
Here,  $^{\bigtriangleup}$  represents the upper or lower triangular entries of the matrix.
 \subsubsection{DFT Implementation scheme}
 The DFT implementation scheme for sparse array design is achieved by starting from   (\ref{o_ch5}) and following  similar steps of the TDL. The group sparsity is invoked as the data in each DFT bin requires the underlying array configuration to remain the same for processing over  all DFT bins.  The SDR is finally realized as follows, 
 \begin{align}\label{f2_ch5}
 \underset{\mathbf{W}^{(l)} \mathbf{\in \mathbb{C}}^{N\times N}, \mathbf{\tilde{W} \in \mathbb{R}}^{N\times N}}{\text{minimize}} &    \sum_{l=0}^{L-1}\text{Tr}(\mathbf{R}^{(l)}\mathbf{W}^{(l)}) + \mu \text{Tr}(\mathbf{U}^i\mathbf{\tilde{W}}) \nonumber\\
 \text{s.t.} &  \, \,  \text{Tr}(\mathbf{R}_{s}^{(l)}\mathbf{W}^{(l)}) \geq1, \, \,l \in \{0,1, . . ., \, L-1\}, \nonumber\\  
 &  \, \,  \overset{\bigtriangleup}{\mathbf{\tilde{W}}}{^{(l)}} \geq  |\overset{\bigtriangleup}{\mathbf{W}}{^{(l)}}|, \quad l \in \{0,1, . . ., \, L-1\}, \nonumber\\
&   \, \, \mathbf{\tilde{W}}=\mathbf{\tilde{W}}^T,  \quad \mathbf{W} \succeq 0.
 \end{align}
 In general, QCQP of order $M$ with $T$ quadratic constraints can   be solved to an arbitrary small accuracy  $\zeta$ by employing  interior point methods  involving   the   worst case complexity of  $\mathcal{O}\{$max$(T,M)^4M^{(1/2)}\log(1/\zeta)\}$   \cite{5447068}. It is apparent from   (\ref{e2_ch5}) and (\ref{f2_ch5})  that the dimensionality of the TDL implementation scheme is $NL\times NL$, whereas that for the DFT approach involves $L$ unknown variables of size $N\times N$. Therefore, the computational complexity for TDL implementation scheme is  $\mathcal{O}\{(NL)^9\log(1/\zeta)\}$, and is  $\mathcal{O}\{(N^9L^{(9/2)})\log(1/\zeta)\}$ for the DFT implementation scheme, which renders the DFT implementation scheme  computationally more  viable. 
 \subsubsection{Unit rank promoting iteration}  \label{Unit rank promoting iteration_ch5}
 Promoting sparse solutions iteratively would depend on   careful selection of the regularization weighting matrix at each iteration. As suggested in \cite{Candes2008, 6477161}, the weighting matrix  $\mathbf{U}^i$ in  (\ref{e2_ch5}) and (\ref{f2_ch5})  is initialized unweighted, i.e., a matrix of all ones. Afterwards, this matrix is iteratively updated in an inverse relationship to $\mathbf{\tilde{W}}$, which is related to    the solution matrix $\mathbf{W}$ from the previous iteration. This enables the beamforming weights  having relatively lower magnitude to be   penalized aggressively, hence encouraging them to go to zero in an iterative fashion. The parameter $\epsilon$ prevents  the unwanted case of division by zero and also avoids the solution to converge to  local minima as follows,
\begin{equation} \label{g2_ch5}
\mathbf{U}^{i+1}(m,n)=\frac{1}{\mathbf{\tilde{W}}^i(m,n)+\epsilon}
\end{equation}
However,   the solution matrix $\mathbf{W}$ in the case of the TDL implementation scheme is not exactly rank  one  at initial iterations. The problem aggravates when the  weight matrix is  updated according to    the above equation, inadvertently pushing the  rank of $\mathbf{W}$ to build up with each iteration. In this respect, updating $\mathbf{U}$ according to (\ref{g2_ch5}) favors solutions of higher ranks, and, as such, fails to yield  sparse configurations iteratively. To mitigate this problem, we  propose a modified re-weighting scheme that relies on updating the regularization weighting matrix  that depends on the inverse of a  rank $1$ matrix $\mathbf{Y}$ instead of   $\mathbf{\tilde{W}}$ as follows, 
\begin{equation} \label{h2_ch5}
\mathbf{U}^{i+1}(m,n)=\frac{1}{\mathbf{Y}^i(m,n)+\epsilon},
\end{equation}
where, $\mathbf{Y}^i=\mathbf{y}^i(\mathbf{y}^i)^T,$ for $\mathbf{y}^i=\frac{1}{L}\sum_{l=1}^{L}(\mathbf{|\mathscr{P}\{W}_{ll}^i\}|)^2 $.  Clearly, $\mathbf{Y}^i$ is a rank one matrix that is synthesized from the rank one approximation of each block diagonal matrix $\mathbf{{W}}_{ll}^i$. It is noted that  $\mathbf{{W}}_{ll}^i$ are the only entries of $\mathbf{{W}}$ that are constrained  by the SDR formulated in (\ref{e2_ch5}). Since  $\mathbf{{W}}_{ll}^i$ are the diagonal block matrices of  the solution matrix $\mathbf{{W}}$, then sparsity is implicitly  encouraged in $\mathbf{{W}}$ by unit rank penalization.    This modified re-weighting approach  given by  (\ref{h2_ch5}) effectively solves the  optimum sparse array  selection problem for the wideband beamforming.  It is noted that the arbitrarily chosen sparsity parameter $\mu$ does not ensure  the   final solution to be exactly $P$ sparse. In order to do so, the optimization problem should be solved  against various values of $\mu$. This is typically achieved by successively running the optimization and updating the  values of $\mu$  through a binary search  over the possible upper and lower limit of $\mu_{max}/\mu_{min}$, until the solution converges to $P$ sensors \cite{6477161}.   The proposed algorithm for achieving  the sparse  optimal weight vector  under the TDL and DFT implementation schemes is summarized in  Table \ref{algorithm1_ch5}. 
\begin{table}[t!] 
\noindent\rule{16.5cm}{0.6pt}
	\caption{SCA   for the sparse wideband beamformer}
\noindent\rule{16.5cm}{0.6pt}

	\begin{algorithmic}[]
		
		\renewcommand{\algorithmicrequire}{\textbf{Input:}}
		
		\renewcommand{\algorithmicensure}{\textbf{Output:}}
		
		\REQUIRE Sparse correlation matrices $\overset{\circ}{\mathbf{R}}$  for  TDL-SCA ($\overset{\circ}{\mathbf{R}}{^{(l)}}$ for DFT-SCA), $N$, $P$, $L$, ${\mathbf{R}_s}$ for TDL-SCA (${\mathbf{R}_{s}^{(l)}}$ for DFT-SCA). \\
	
	\ENSURE   MaxSINR beamformer against   $P$ active sensors.   \\
	\textbf{Matrix Completion:} \\
	Procedure  in Section \ref{sec4_ch5} is adopted to recover the estimated full data received correlation matrix $\mathbf{\hat{R}}$ for  TDL-SCA  ($\mathbf{\hat{R}}^{(l)}$ for DFT-SCA).\\
	\textbf{Initialization:} \\
		Initialize the beamforming vectors randomly to find $\mathbf{m}$ and $b$. 
		Initialize  $\epsilon$, $\mu_{max}$ (upper limit of binary search), $\mu_{min}$  (lower limit of binary search), $\mu=0$.
		\WHILE {(Solution does not converge corresponding to $\mu=0$)}
		
		\STATE   Run   (\ref{n2_ch5}) for  TDL-SCA or (\ref{o2_ch5}) for DFT-SCA.\\ 
		\ENDWHILE \\
		Initialize $\mathbf{u}^{i}$= all ones vector. 
		\WHILE {(Beamforming weight vector is not $P$ sparse)}
		\STATE Select $\mu=\frac{1}{2}(\mu_{max}+\mu_{min})$ (Binary search).\\
		\STATE   Run   (\ref{n2_ch5}) for  TDL-SCA or (\ref{o2_ch5}) for DFT-SCA (Use  $\mathbf{\hat{R}}$ or $\mathbf{\hat{R}}^{(l)}$ in-lieu of $\mathbf{{R}}$,$\mathbf{{R}}^{(l)}$ to synthesize $\mathbf{\tilde{R}}$, $\mathbf{\tilde{R}}^{(l)}$).\\  
		(Also use the optimal non sparse solution from  the previous while loop for  $\mathbf{m}$ and $b$).\\ 
		\STATE   Update the regularization weighting parameter $\mathbf{u}^{i+1}(k)=\frac{1}{||\mathbf{\tilde{w}}_k^i||_2+\epsilon}$.\\
		Update $\mu_{max}/\mu_{min}$ according to the  binary search. \\
		\ENDWHILE \\
		%
		%
		%
		%
		
		\STATE After achieving the desired cardinality, run  (\ref{n2_ch5}) for  TDL-SCA or (\ref{o2_ch5}) for DFT-SCA, with reduced dimension  corresponding to nonzero values of $\mathbf{\tilde{w}}$  and $\mu=0$,  yielding, optimal weight vector. 
		
		\RETURN Optimal Beamformer $\mathbf{w}_o$-TDL-SCA ($\mathbf{w}_{o}^{(l)}$-DFT-SCA)
	\end{algorithmic}
	\label{algorithm2_ch5}
\noindent\rule{16.5cm}{0.6pt}
\end{table}\vspace{-3mm}
\subsection{Successive convex approximation (SCA)}
\subsubsection{TDL Implementation scheme}
The problem in (\ref{j_ch5}) can equivalently be rewritten by swapping the objective and constraint functions as follows,
\begin{equation} \label{i2_ch5}
\begin{aligned}
 \underset{\mathbf{w \in \mathbb{C}}^{NL}}{\text{maximize}} & \quad   \mathbf{ w}^H\mathbf{R}_s\mathbf{ w}\\
\text{s.t.} & \quad     \mathbf{w}^H\mathbf{R}\mathbf{ w} \leq 1
\end{aligned}
\end{equation}
Although this swapping operation allows the associated constraint to be convex, however it renders the objective function non convex. We note that the formulation in (\ref{i2_ch5}) doesn't have a trivial solution $\mathbf{w} =0$, as the objective and constraint are coupled due to $\mathbf{R}_s$  being part of $\mathbf{R}$.   The maximization of the convex function  is replaced by the minimization of the concave function. The transformation to the minimization problem will later enable carrying out the minimization based on  $P$ sparse solution. Rewriting (\ref{i2_ch5}) by reversing the sign of the desired source correlation matrix  $\bar{\mathbf{R}}_s=-\mathbf{R}_s$ as follows, 
\begin{equation} \label{j2_ch5}
\begin{aligned}
 \underset{\mathbf{w \in \mathbb{C}}^{NL}}{\text{minimize}} & \quad   \mathbf{ w}^H\bar{\mathbf{R}}_s\mathbf{ w}\\
\text{s.t.} & \quad     \mathbf{ w}^H\mathbf{R}\mathbf{ w} \leq 1
\end{aligned}
\end{equation}
The beamforming weight vectors are generally complexed valued, whereas the  quadratic functions are real. This observation  allows expressing the problem with  only real variables and  is typically accomplished by  replacing the correlation matrix $\bar{\mathbf{R}}_s$ by $\tilde{\mathbf{R}}_s$  and concatenating the beamforming weight vector accordingly \cite{8445845}, \vspace{-2mm}
\begin{multline} \label{k2_ch5}
\quad \quad \quad \quad \quad \tilde{\mathbf{R}}_s=\begin{bmatrix}
\text{real}(\bar{\mathbf{R}}_s)       & -\text{imag}(\bar{\mathbf{R}}_s)   \\
\\
\text{imag}(\bar{\mathbf{R}}_s)       & \text{real}(\bar{\mathbf{R}}_s)   \\
\end{bmatrix},
\quad \quad \quad \quad \quad \quad \tilde{\mathbf{w}}=\begin{bmatrix}
\text{real}({\mathbf{w}})       \\
\\
\text{imag}({\mathbf{w}})   \\
\end{bmatrix}
\end{multline}
Similarly, the received data correlation matrix $\mathbf{R}$ is replaced by $\tilde{\mathbf{R}}$. The problem in  (\ref{j2_ch5}) then  becomes,
\begin{equation} \label{l2_ch5}
\begin{aligned}
 \underset{\mathbf{\tilde{w} \in \mathbb{R}}^{2NL}}{\text{minimize}} & \quad   \mathbf{\tilde{ w}}^{T}\tilde{\mathbf{R}}_s\mathbf{\tilde{ w}}\\
\text{s.t.} & \quad     \mathbf{\tilde{ w}}^{T}\mathbf{\tilde{R}}\mathbf{\tilde{w}} \leq 1
\end{aligned}
\end{equation}
We can then proceed to convexify the objective function  by utilizing  the first order approximation  iteratively,
 \begin{equation} \label{m2_ch5}
 \begin{aligned}
 \underset{\mathbf{\tilde{ w} \in \mathbb{R}}^{2NL}}{\text{minimize}} & \quad   {\mathbf{m}^i}^{T}\mathbf{\tilde{ w}}+b^i\\
 \text{s.t.} & \quad     \mathbf{\tilde{ w}}^{T}\mathbf{\tilde{R}}\mathbf{\tilde{w}} \leq 1,
 \end{aligned}
 \end{equation}
The linearization coefficients  $\mathbf{m}^i$ and ${b^i}$  are updated as, $\mathbf{m}^{i+1}=2\tilde{\mathbf{R}}_s\mathbf{\tilde{ w}}^i$, and  $b^{i+1}=-{{\mathbf{\tilde{w}}^i}}^{T}\tilde{\mathbf{R}}_s \mathbf{\tilde{w}}^i $ (superscript $^i$ denotes the iteration number). Finally, to invoke sparsity in the beamforming weight vector, the re-weighted mixed $l_{1-2}$ norm is adopted primarily for promoting group sparsity,  \vspace{-3mm}
 \begin{equation} \label{n2_ch5}
 \begin{aligned}
  \underset{\mathbf{\tilde{w} \in \mathbb{R}}^{2NL}}{\text{minimize}} & \quad   {\mathbf{m}^i}^{T}\mathbf{\tilde{ w}}+b^i + \mu(\sum_{k=1}^{N}\mathbf{u}^i(k)||\mathbf{\tilde{w}}_k||_2)\\
 \text{s.t.} & \quad     \mathbf{\tilde{ w}}^{T}\mathbf{\tilde{R}}\mathbf{\tilde{ w}}\leq 1
 \end{aligned}
 \end{equation}
 Here, $\mathbf{\tilde{w}}_k \in \mathbb{R}^{2L}$ are the beamforming weights corresponding to TDL of $k$th sensor. Discouraging a sensor ($||.||_2$ denotes the $l_2$ norm) implies a simultaneous removal of both the real and corresponding imaginary entries of all beamforming weights  associated with the removed sensor \cite{8445845}. 
\subsubsection{DFT implementation scheme}
The above formulation can be  extended  for the DFT  implementation scheme as follows:\vspace{-3mm}
\begin{equation} \label{o2_ch5}
\begin{aligned}
 \underset{\mathbf{\tilde{w}}^{(l)} \mathbf{\in \mathbb{R}}^{2N}}{\text{minimize}} &    \sum_{l=0}^{L-1}( {{\{\mathbf{m}^{(l)}}\}^i}^{T}\mathbf{\tilde{ w}}^{(l)}+\{b^{(l)}\}^i) + \mu(\sum_{k=1}^{N}\mathbf{u}^i(k)||\mathbf{\tilde{w}}_k||_2)\\
\text{s.t.} & \quad     {\mathbf{\tilde{ w}}^{(l)}}^{T}\mathbf{\tilde{R}}^{(l)}\mathbf{\tilde{ w}}^{(l)} \leq 1, \quad   l \in \,\, \{0, 1, . . .  L-1\}, 
\end{aligned}
\end{equation}
where $\mathbf{\tilde{w}}_k \in R^{2L}$ contains  the $L$ DFT bins data for the $k$th sensor, $\{{\mathbf{m}^{(l)}}\}{^i}$ and $\{{b^{(l)}\}{^i}}$ are the approximation coefficients at the $i$th iteration for the desired source correlation matrix in the $l$th bin, with $\{{\mathbf{m}^{(l)}}\}{^i}=2\tilde{\mathbf{R}}_{s}^{(l)}\{{\mathbf{\tilde{ w}}^{(l)}}\}{^{i}},  \{{b^{(l)}\}{^i}}=-{\{\mathbf{\tilde{w}}^{(l)}\}^i}^{T}\tilde{\mathbf{R}}_{s}^{(l)}\{\mathbf{\tilde{w}}^{(l)}\}^i$.
The initial estimates $\{{\mathbf{m}^{(l)}}\}{^i}$ and $\{{b^{(l)}\}{^i}}$  are calculated for the optimal non sparse solution. These parameters can be found by setting the sparsity parameter  $\mu$ to zero. In  so doing,  the solution and the  corresponding parameters $\{{\mathbf{m}^{(l)}}\}{^i}$ and $\{{b^{(l)}\}{^i}}$    converge to the optimal value against  the full array elements. Using these values as initial conditions  has proven  appropriate in our design for  recovering effective sparse solutions.  The sparsity parameter $\mu$ is  chosen according to the binary search over the possible range of $\mu$ to warrant the desired cardinality of the beamforming weight vector as explained before in section \ref{Unit rank promoting iteration_ch5}. The $k$th entry of re-weighting vector $\mathbf{u}^i(k)$ is updated according to $\mathbf{u}^{i+1}(k)=\frac{1}{||\mathbf{\tilde{w}}_k^i||_2+\epsilon}$. The computational complexity  involving SCA is considerably lower as compared to the SDR formulation, as the latter  intrinsically  squares the number of   variables involved, essentially exacerbating the  runtime \cite{8445845}. The SCA for sparse array design for wideband beamforming is summarized in  Table \ref{algorithm2_ch5}.
\section{Sparse  matrix completion of block Toeplitz matrices} \label{sec4_ch5}
The aforementioned  sparse array design formulations   require    the received data correlation matrix corresponding to the full array aperture. This  is a  rather stringent requirement  in an adaptive switching environment  where the data is fetched  from  only $P$ active  sensor locations over a given observation period.  The received data correlation matrix for the sparse array design using TDL implementation scheme has $L^2$($N^2-P^2$) missing correlation entries, whereas there  are $L$($N^2-P^2$) missing correlation values for the DFT implementation scheme. Clearly, for the large values of $L$, the TDL implementation scheme has significantly higher number of missing correlation entries as compared to the DFT implementation scheme. 

Recently, the hybrid sparse design for the  narrowband beamforming was introduced to alleviate the issue of  missing correlation lags in the received data correlation matrix \cite{8892512}.  This is primarily  achieved by pre-allocating few sensors  to guarantee a fully augmentable sparse array, while engaging the remaining degrees of freedom (DOF) to maximize the SINR.  However, locking in few DOFs to ensure the array full augmentability can lead to suboptimal sparse beamformers. Alternatively, the matrix completion    approach can be used to provide the missing lags \cite{Abramovich:1999:PTC:2197953.2201686, 709534, 7935534}.  We propose, herein,  sparse matrix completion to efficiently exploit the structure of the data correlation matrix to recover the missing correlation values. The received data correlation matrix for the TDL implementation scheme is a Hermitian positive definite matrix but it also follows a block Toeplitz formation, as shown below,
\begin{multline} \label{a4_ch5}
\quad \quad \quad \quad \quad \quad \mathbf{R}=\begin{bmatrix}
\mathbb{T}_0       & {\mathbb{T}}_1  & {\mathbb{T}}_2 & \ldots & {\mathbb{T}}_{L-1} \\
\\
\mathbb{T}_{-1}       & {\mathbb{T}}_0    & {\mathbb{T}}_1  & \ldots   & {\mathbb{T}}_{L-2}  \\
\\
{\mathbb{T}}_{-2}       & {\mathbb{T}}_{-1}    & {\mathbb{T}}_0  & \ldots  & {\mathbb{T}}_{L-3}  \\
\\
\vdots       & \vdots    & \vdots  & \ddots  & \vdots   \\
\\
{\mathbb{T}}_{-(L-1)}       & {\mathbb{T}}_{-(L-2)}    & {\mathbb{T}}_{-(L-3)}  & \ldots  & {\mathbb{T}}_0  \\
\end{bmatrix} \quad \quad \quad \quad \quad \quad
\end{multline}
By definition, block Toeplitz matrices  doesn't necessitate each comprising block  to be Toeplitz within itself. Therefore, matrix $\mathbf{R}$ in (\ref{a4_ch5}) represents a special case of block Toeplitz matrices, where the Toeplitz structure  is also preserved for each constituent block $\mathbb{T}_k \in \mathbb{C}^{N\times N}$. Because of  the matrix Hermitian symmetry, we also have $\mathbb{T}_k^{H}=\mathbb{T}_{-k} (\text{for} \, \, \, k\neq 0 $). 
Instead of recovering $\mathbf{R}$ as a single unit, we focus on  completing  the constituent blocks and then synthesizing the full correlation matrix $\mathbf{R}$. This approach  potentially caps the computational expenses considerably but also efficiently exploits the formation of $\mathbf{R}$.

There is an important distinction between the constituent blocks $\mathbb{T}_0$ and $\mathbb{T}_k \, \, \,  (\text{for} \, \, \, k\neq 0 $). It is noted that $\mathbb{T}_0$ is positive definite Hermitian Toeplitz matrix, whereas $\mathbb{T}_k$ $ \, \, \,  (\text{for} \, \, \, k\neq 0 $) are indefinite Toeplitz matrices which  are not necessarily Hermitian. Therefore, we  resort to  two different ways with regards to our treatment of $\mathbb{T}_0$ and $\mathbb{T}_k$ while adopting a  Toepltiz  matrix completion scheme  under  the low rank assumption. It is  known that the correlation matrix $\mathbf{R}$  for the  wideband far field  sources impinging on the ULA resultantly follows the structure in (\ref{a4_ch5}) and can be represented effectively with  a relatively  low rank approximation depending on the observed source  time-bandwidth product \cite{1143832}.   The  trace heuristic is a well known approach which is  adopted generally as a convex surrogate in recovering low rank matrices. This approach has been successfully used in many areas of control systems and array processing to recover simpler and low rank data models \cite{Recht:2010:GMS:1958515.1958520, Mohan:2012:IRA:2503308.2503351, 5447068}. Moreover, it has been shown  that the trace heuristic is equivalent to the nuclear norm minimization in recovering positive semidefinite correlation matrices\cite{8474369,  7539135, 7935534, 7833203}. The low rank positive semidefinite Toeplitz matrix completion problem has been proposed in \cite{8472789} for interpolating missing correlation lags in coprime array configuration and can be adopted to interpolate $\mathbb{T}_0$ as follows,
\begin{equation} \label{b4_ch5}
\begin{aligned}
\underset{l \in \mathbb{C}^{N}}{\text{minimize}} & \quad  ||\mathcal{T}(l)\odot \mathbf{Z}-\overset{\circ}{\mathbb{T}}_0  ||_F^2 + \zeta \text{Tr}(\mathcal{T}(l))\\
\text{s.t.} & \quad   \mathcal{T}(l) \succeq 0
\end{aligned}
\end{equation}
Here,  the unknown Hermitian Toeplitz matrix $\mathcal{T}(l)$, can uniquely be defined by a single  vector $l$ representing  the  first row of  $\mathcal{T}(l)$, and $l^{H}$ denoting the matrix first  column.  Matrix $\overset{\circ}{\mathbb{T}}_0 $ is the received data correlation matrix with the missing correlation values set equal  to zero. The   element wise product is denoted by symbol `$\odot$' and  `$\succeq$' implements the positive  semidefinite constraint. The objective function attempts to minimize  the  error  between the observed correlation values and the corresponding entries of $\mathcal{T}(l)$ implemented through the Frobenius norm of the error matrix (The function `$||.||_F^2$',  represents the square of the Frobenius norm of matrix which returns the sum of square of it's entries).   The parameter `$\zeta$' pursues  the trade off   between the denoising  term and the trace heuristic to recover a simpler low rank model. The nominal value of the  parameter  `$\zeta$'   is challenging to locate and is typically gleaned from the numerical experience. In order to do away with the nuisance parameter  `$\zeta$',   we adopt a fusion based approach more suited to our application. We note that the non zero elements in $\overset{\circ}{\mathbb{T}}_0 $ can be segregated into two classes. With regards to the sparse entries in $\overset{\circ}{\mathbb{T}}_0 $,  either we have the  whole sub-diagonal entries  missing in $\overset{\circ}{\mathbb{T}}_0 $ or the sub-diagonals are sparse. The former situation arises if there is missing correlation lag in $\overset{\circ}{\mathbb{T}}_0 $, whereas the latter arises when the corresponding correlation lag is present but lacking the intrinsic redundancy corresponding to the compact sensor grid. The observed correlation lags are averaged across the sub diagonal to filter the sensor noise as follows, 
 \begin{equation} \label{c4_ch5}
  \doublehat{l}^i(k)=(\frac{1}{c_k}) \sum_{\forall \, m-n=k}\overset{\circ}{\mathbb{T}}^i_0 (m,n)    
 \end{equation}
 Here,   $\overset{\circ}{\mathbb{T}}^i_0$  is the $\mathbf{R}(i,i)$ entry in (\ref{a4_ch5}) denoting the estimate of $\overset{\circ}{\mathbb{T}}^i_0$ at $i$th sampling instance, $k \in 0,1, \cdots, N-1$ represents the respective lag or the sub diagonal of $\overset{\circ}{\mathbb{T}}^i_0$ and  $c_k$ is the observed redundancy of the $k$th lag. As evident in (\ref{a4_ch5}), there are $L$ copies of  $\overset{\circ}{\mathbb{T}}_0$ corresponding to $L$ sampling instances. Hence, $\doublehat{l}^i(k)$ is averaged over $L$ blocks to yield an estimate of the given lag  $\hat{l}(k)=\frac{1}{L}\sum_{i=0}^{L-1}\doublehat{l}^i(k)$. The fused matrix completion formulation, therefore, substitutes the sparse sub-diagonals with the estimated average value $\hat{l}(k)$, whereas the completely missing sub-diagonals are interpolated   as follows,
\begin{equation} \label{d4_ch5}
\begin{aligned}
\underset{l \in \mathbb{C}^{N}}{\text{minimize}} & \quad \text{Tr}(\mathcal{T}(l))\\
\text{s.t.} & \quad   l(\text{lag present}) = \hat{l}(k),\\
 & \quad   \mathcal{T}(l) \succeq 0  
\end{aligned}
\end{equation}
The above formulation relies on fairly accurate estimate of the observed correlation lag which  not only involves averaging over the corresponding sub diagonal but also  across $L$ sampling instances. Such accuracy can  become   challenging to meet the semidefinite constraint if the available degrees of freedom are few. To circumvent this problem, the $0$ lag is removed from the constraint which gives the additional degree of freedom to the  algorithm  to set aside an  appropriate loading factor to make the problem feasible. It is noted that the formulation in (\ref{d4_ch5}) would choose the  minimum possible  diagonal loading as it requires to minimize the trace heuristic and hence strives to maximize the sparse  eigenvalues.

The trace heuristic can also be extended for indefinite matrices to recover sparse models \cite{Candes2009}. We  couple this observation along with the above discussion  to perform low rank matrix completion for $\mathbb{T}_k \, \, \,  (\text{for} \, \, \, k\neq 0 $) as follows, 
\begin{equation} \label{e4_ch5}
\begin{aligned}
\underset{l_r, l_c \in \mathbb{C}^{N}}{\text{minimize}} & \quad \text{Tr}(\mathbf{W}_1) +  \text{Tr}(\mathbf{W}_2)\\
\text{s.t.} & \quad   l_r(\text{lag present}) = \hat{l_r}(k),\\
\text{s.t.} & \quad   l_c(\text{lag present}) = \hat{l_c}(k),\\
     & \quad    
\begin{bmatrix}%
\mathbf{W}_1 & \mathcal{T}(l_r,l_c) \\
\mathcal{T}(l_r,l_c) & \mathbf{W}_2
\end{bmatrix}\succeq 0
\end{aligned} 
\end{equation}
Here, $\mathbf{W}_1$ and $\mathbf{W}_2$ are auxiliary matrices implementing trace heuristic to recover low rank indefinite matrices.  The function  $Toeplitz(l_r,l_c)$ returns the Toeplitz matrix with $l_r$ and $l_c$ being the first row and the first column, respectively. This distinction is important as in general $\mathbb{T}_k \, \, \,  (\text{for} \, \, \, k\neq 0 $) is not a Hermitian Toeplitz matrix. The formulation in (\ref{e4_ch5}) is repeated to yield an estimate for all constituent Toeplitz blocks. 

 Upon performing matrix completion for each constituent Toeplitz block, the individual Toeplitz blocks can be plugged back into  (\ref{a4_ch5}) to yield an estimate  $\mathbf{\doublehat{{R}}}$. We can improve the estimate  $\mathbf{\doublehat{{R}}}$ by incorporating the   noise variance which  is generally known or estimated apriori. This is  achieved through the maximum likelihood estimate (MLE) approach where the eigenvalues corresponding to the  eigenvectors of the  noise subspace are set equal to the noise floor while the remaining eigenvalues are kept the same. The MLE of $\mathbf{\doublehat{{R}}}$ is donated by $\mathbf{\hat{R}}$ and is given by the outer product of the original eigenvectors of $\mathbf{\doublehat{{R}}}$ reweighted by the modified eigenvalues.  However, it is noted that in practice  the number of  eigenvectors associated with the noise floor are not exactly known. Nevertheless, we only reset those  eigenvalues which are less than the noise floor. Finally, the maximum likelihood estimate  $\mathbf{\hat{R}}$, is used in lieu of $\mathbf{\doublehat{{R}}}$, to carry out the data dependent optimization for MaxSINR. It is also noted that unlike $\mathbf{\doublehat{{R}}}$, the  matrix $\mathbf{\hat{R}}$ is no longer block Toeplitz   but is now guaranteed to be positive definite as is strictly required to implement the proposed sparse optimization algorithms. The formulation in (\ref{d4_ch5}) is sufficient for the DFT implementation scheme  which only involves $L$ received data matrices corresponding to the $L$ DFT bins.  
 \section{Simulations} \label{Simulations_ch5}
The effectiveness of  the sparse array design for  MaxSINR  beamforming is demonstrated by  design examples considering a wideband source operating in the presence of a mix of narrowband and wideband jammers. The MATLAB-based CVX modeling tool  is used for convex optimization. The  importance  of   sparse array design for MaxSINR is further emphasized  by comparing the  optimum  design with the sub optimum  array configurations under the TDL  and DFT implementation schemes. The simulation results are presented for perspective linear sensor locations, nevertheless, the proposed algorithms are applicable to  rectangular grid points or arbitrary placed sensors on  3D surfaces.
\subsection{Example 1}
The task is to select $P=14$ sensors from $N=20$  possible equally spaced locations with  inter-element spacing of $\lambda_{min}/2$.  We consider an  $8$ delay line filter taps associated  with each selected  sensor ($8$ DFT bins for DFT implementation scheme) implying $L=8$. 
A wideband  desired point source   impinges on a linear array from  DOA $40^0$. Five strong wideband jammers with uniform PSDs (power spectral density) are located at angles $30^0$, $45^0$, $50^0$, $140^0$ and $150^0$. The PSD  of all  frequency spread sources are uniform with the fractional bandwidth of  $0.22$. A narrowband jammer locked onto the source carrier frequency is active at $130^0$.    The SNR of the desired signal is $0$ dB, and the INR of  each interfering signals is set to $30$ dB. Thereby, the input SINR is $-37$ dB. 
\begin{figure}[!t]
	\centering
	\includegraphics[width=5.78 in, height=2.90 in]{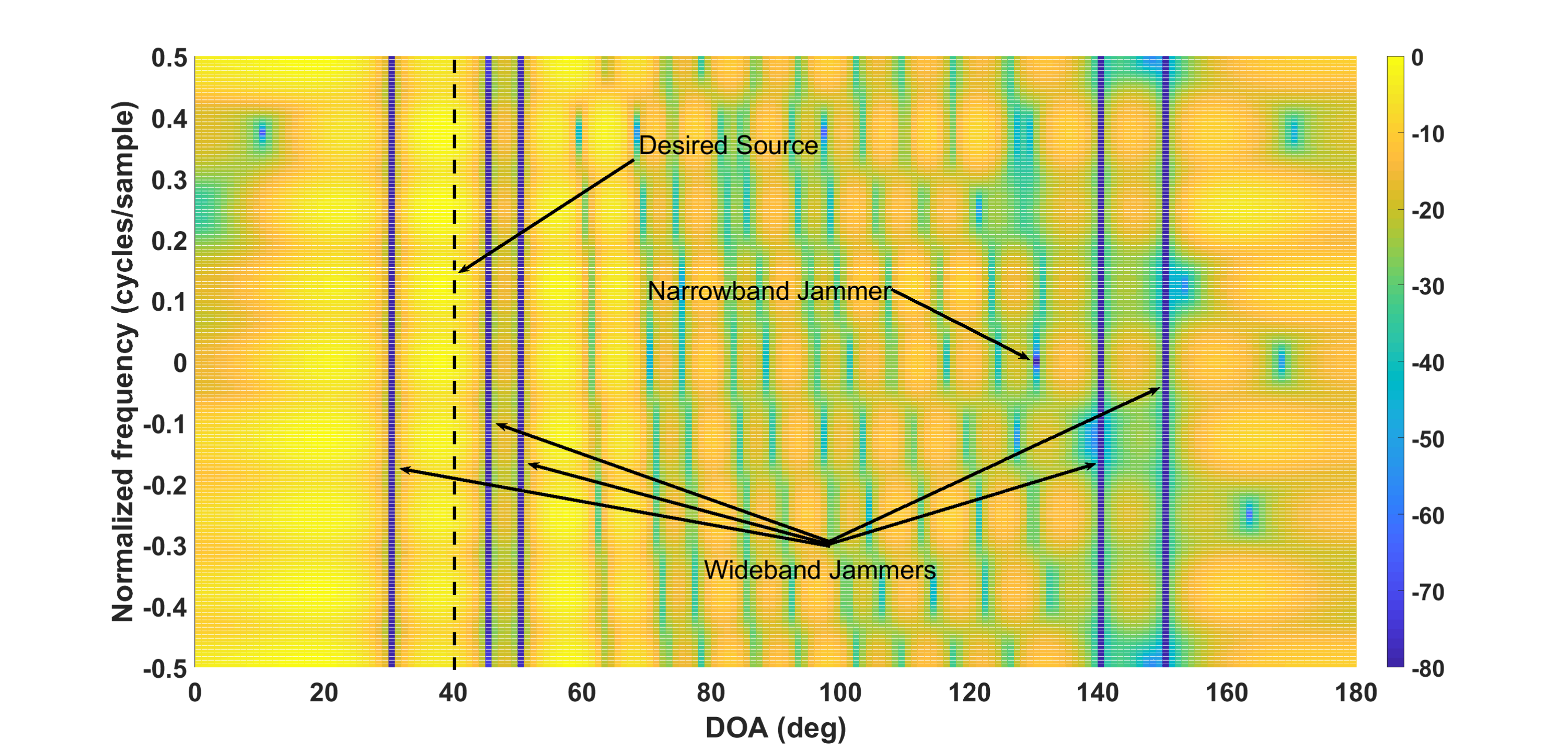}
	\caption{Frequency dependent beampattern for the  array configuration recovered through convex relaxation. }
	\label{labelfig4_ch5}
\end{figure}

Figure \ref{labelfig4_ch5} shows the frequency dependent beampattern  for the   array configuration recovered through TDL-SDR. It is evident from the beampattern that a  high gain is maintained throughout the band of interest for the desired source signal while alleviating  the interferers from respective  DOAs and  frequency bands. These beampattern characteristics   translate to an output SINR of  $7.05$ dB. The TDL-SDR performs  close to the optimum array found by enumeration which has an output SINR of $7.83$ dB. The corresponding sparse array configurations obtained by enumeration and TDL-SDR are shown in the Fig. \ref{op_enumer_608_ch5} and \ref{op_algo_2027_ch5}, respectively (the green color denotes corresponding sensor location selected, whereas the gray color shows the sensor location not selected). It is important to mention  that  the optimum array found by enumeration requires a search over $38760$ possible sparse array configurations, which has a prohibitive  computational cost attributed to expensive singular value decomposition (SVD) for each enumeration.  The optimum sparse array design achieved through TDL-SCA is shown in  Fig. \ref{op_notap_775_ch5}. This array configuration is capable of delivering output SINR of $7.7$ dB with the use of optimal beamforming weights. It  performs very close to the optimum sparse array found through enumeration with the performance degradation of around $0.1$ dB. This performance is superior to that of the sparse array found through TDL-SDR by around $0.65$ dB with  less computational complexity.

In general, for any given  array configuration,  the TDL implementation scheme yields marginally higher SINR as compared to the DFT implementation scheme. Owing to the reduced computational cost of the DFT sparse design, and relatively better performance of the TDL sparse design, one can entwine TDL and DFT beamformers to capitalize on both merits. That is, we proceed to find the optimum TDL beamformer weights based on the sensor array configuration that results from the optimum DFT-based implementation scheme.  This way, optimum sparse array design would involve  both implementation schemes. This is referred as the dual-domain design implementation scheme  since it considers both time and frequency domains in generating the optimum receiver design. This design has slightly elevated  computational expense over the  DFT design, as it  involves calculating the  optimum TDL beamformer weights corresponding to the DFT optimized configuration.  For the underlying problem, the dual-domain design, through either the DFT-SDR or DFT-SCA gives an output SINR of $7.76$~dB  which is close to the   maximum possible SINR of $7.83$~dB.  The DFT-SDR and DFT-SCA both render an identical sparse configurations,  which is shown in  Fig. \ref{wo_enu_924_ch5}. Therefore, we remark that for the  example considered, the dual domain design underlying the DFT implementation scheme, performs slightly better  than the design carried out exclusively by the TDL implementation scheme.  
 \begin{figure}[!t]
	\centering
	\includegraphics[width=5.78 in, height=2.90 in]{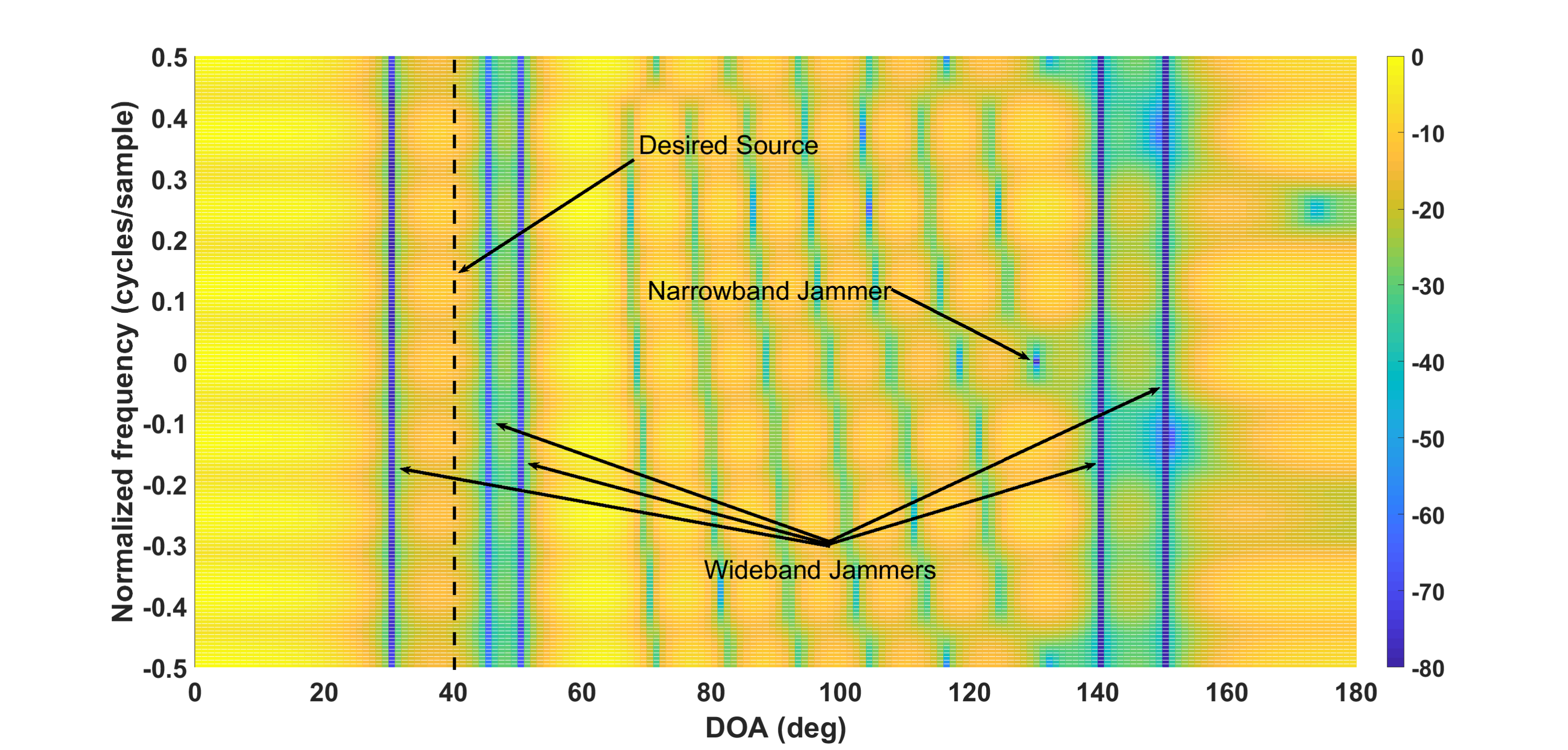}
	\caption{Frequecy dependent beampattern for the compact ULA (Fig. \ref{wo_enu_ula_ch5}).}
	\label{labelfig5_ch5}
\end{figure}
\begin{figure}[!t]
	\centering
	\subfloat[]{\includegraphics[width=0.83\textwidth]{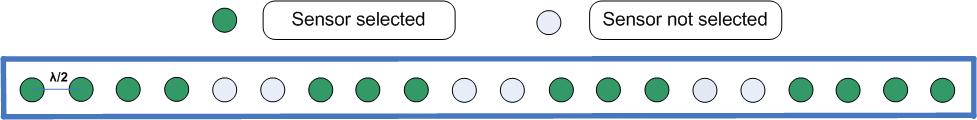}
		\label{op_enumer_608_ch5}}
		\hfill
\subfloat[]{\includegraphics[width=0.83\textwidth]{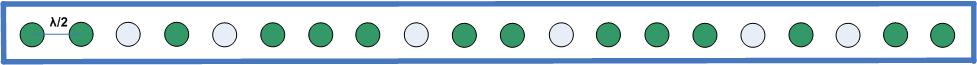}
		\label{op_algo_2027_ch5}}
		\hfill
\subfloat[]{\includegraphics[width=0.83\textwidth]{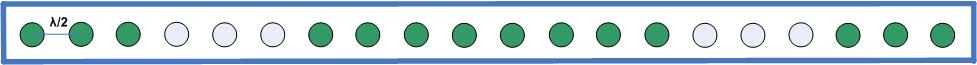}
		\label{op_notap_775_ch5}}
		\hfill
\subfloat[]{\includegraphics[width=0.83\textwidth]{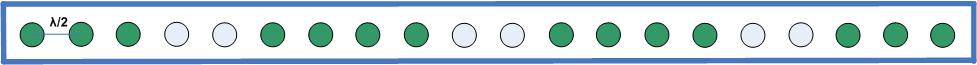}
		\label{wo_enu_924_ch5}}
		\hfill
\subfloat[]{\includegraphics[width=0.83\textwidth]{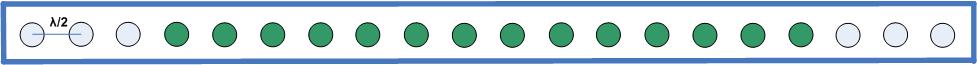}
		\label{wo_enu_ula_ch5}}
		\hfill
	\caption{Example 1 - (a) Optimum  array TDL implementation scheme (Enumeration)  (b)  TDL-SDR  (c) TDL-SCA  (d) DFT-SDR, DFT-SCA (e)  $14$ sensor compact ULA }
	\label{op_c_co_c_ch5}
\end{figure}
\begin{figure}[!t]
	\centering
\subfloat[]{\includegraphics[width=0.83\textwidth]{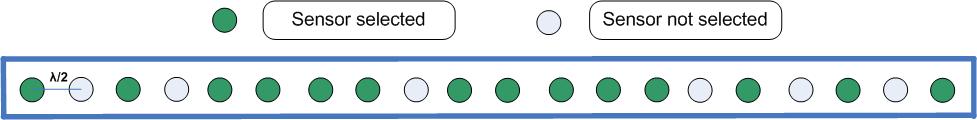}
		\label{op_enumer_6082_ch5}}
		\hfill
\subfloat[]{\includegraphics[width=0.83\textwidth]{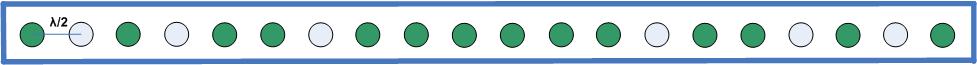}
		\label{op_notap_7752_ch5}}
		\hfill
\subfloat[]{\includegraphics[width=0.83\textwidth]{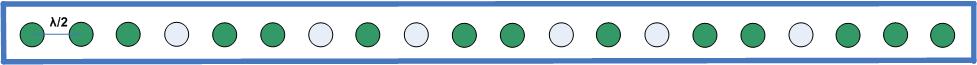}
		\label{op_algo_20272_ch5}}
		\hfill
\subfloat[]{\includegraphics[width=0.83\textwidth]{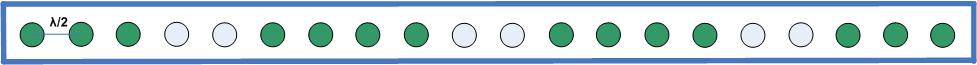}
		\label{wo_enu_9262_ch5}}
		\hfill
\subfloat[]{\includegraphics[width=0.83\textwidth]{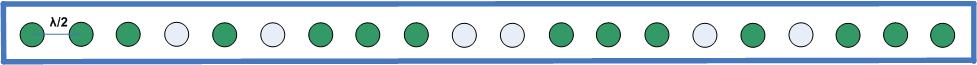}
		\label{wo_enu_9252_ch5}}
		\hfill
\subfloat[]{\includegraphics[width=0.83\textwidth]{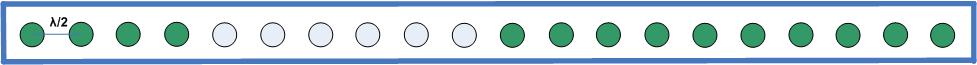}
		\label{wo_enu_9272_ch5}}
		\hfill
	\caption{Example 2 -  (a) Optimum  array TDL implementation scheme (Enumeration)  (b) TDL-SCA    (c) TDL-SDR   (d) DFT-SCA  (e) DFT-SDR (f) Worst case   array (TDL, DFT)}
	\label{op_c_co_c2_ch5}
\end{figure}

Fig. \ref{labelfig5_ch5} shows the beampattern for the sparse array design resulting in the worst case output SINR. In this case, the worst array configuration  turns out to be the compact ULA (Fig. \ref{wo_enu_ula_ch5}). This configuration  could not deliver the  resolution required to mitigate the closely located jammers. It can be observed from the beampattern that the compact ULA, in an effort to remove the strong interfering signals, failed to provide maximum gain towards the source of interest.  This results in a considerable low output SINR that is equal to $-4.5$ dB. Therefore, a sparse array design which optimally and efficiently utilizes the array aperture and the  additional degrees of freedom made available by the switching   transceiver chains  yields  an  SINR improvement of more than $12$ dB. 
\subsection{Example 2}
Consider a  wideband source of interest  at $45^0$  and the wideband jammers located at  $35^0$, $55^0$, $60^0$, $145^0$ and $155^0$. A narrowband jammer is located at $135^0$ at an operating frequency of $f_c$, all other parameters are the same as in  Example $1$.  The SINR of the  optimum array for the TDL implementation scheme  (Fig. \ref{op_enumer_6082_ch5}) is $11.32$ dB (found through enumeration). Optimization performed using SCA yields the array in Fig. \ref{op_notap_7752_ch5}, with respective  SINR of $11.2$ dB, whereas that performed by SDR yields the array shown in Fig. \ref{op_algo_20272_ch5} and corresponding SINR of  $10.9$ dB.   For the dual domain design, the optimum sparse arrays found through DFT-SCA (Fig. \ref{wo_enu_9262_ch5})  and through the  DFT-SDR algorithm  (Fig. \ref{wo_enu_9252_ch5})  deliver an approximately similar output SINR, around $11.02$ dB. This  is inferior  to the exclusive TDL design.     It is important to  note that the array  configuration resulting in the worst case performance,   shown in  Fig. \ref{wo_enu_9272_ch5}, spans  the full  aperture as  the optimum array, yet it offers an output SINR of only $7.35$ dB, underscoring the importance of carefully performing sensor selection for the sparse array design. 
\subsection{Comparison of SDR and SCA under both models }
The design examples discussed thus far   show  amicable  performance of the proposed algorithms under the assumption of the knowledge of full data correlation matrix. The results clearly tie the performance to the location of the sources and their respective powers. However,  evaluating the performance under  matrix completion involves analysis for additional variables, namely, the initial sparse array configuration prior to optimization and the number of snapshots. The performance is, therefore,  dependent on  the observed realization of the received data. In order to have a holistic  assessment of the proposed algorithms,  Monte Carlo simulations   are  generated. We select $P=8$ locations out of  $N=16$ available locations. For  specified DOA of the  desired source,  trials are generated involving six jammers occupying  random locations from $30^0$ to $150^0$.  The SNR of the desired source is $0$ dB, while the  powers  of the jammers are uniformly distributed from $10$ to $20$ dB. The simulation is repeated at $11$ different desired source DOAs, and the average SINR computed. In total, $1500$ experiments are conducted.    For each trial, a random $P$-sparse  array topology serves as an initial  array configuration. This   configuration  could be an optimized configuration from the preceding operating conditions.  The sparse data correlation matrix is estimated based on sensor locations in the initial configuration before performing matrix completion and subsequent optimization process.   The  binary search for the sparsity parameter $\mu$ ranges from $0.01$ to $3$, sparsity threshold $\gamma=10^{-3}$  and $\epsilon=0.05$, relative signal bandwidths and other parameters are the same as given in Example 1.   

\begin{figure}[!t]
	\centering
	\includegraphics[height=4.35in,width=5.68in]{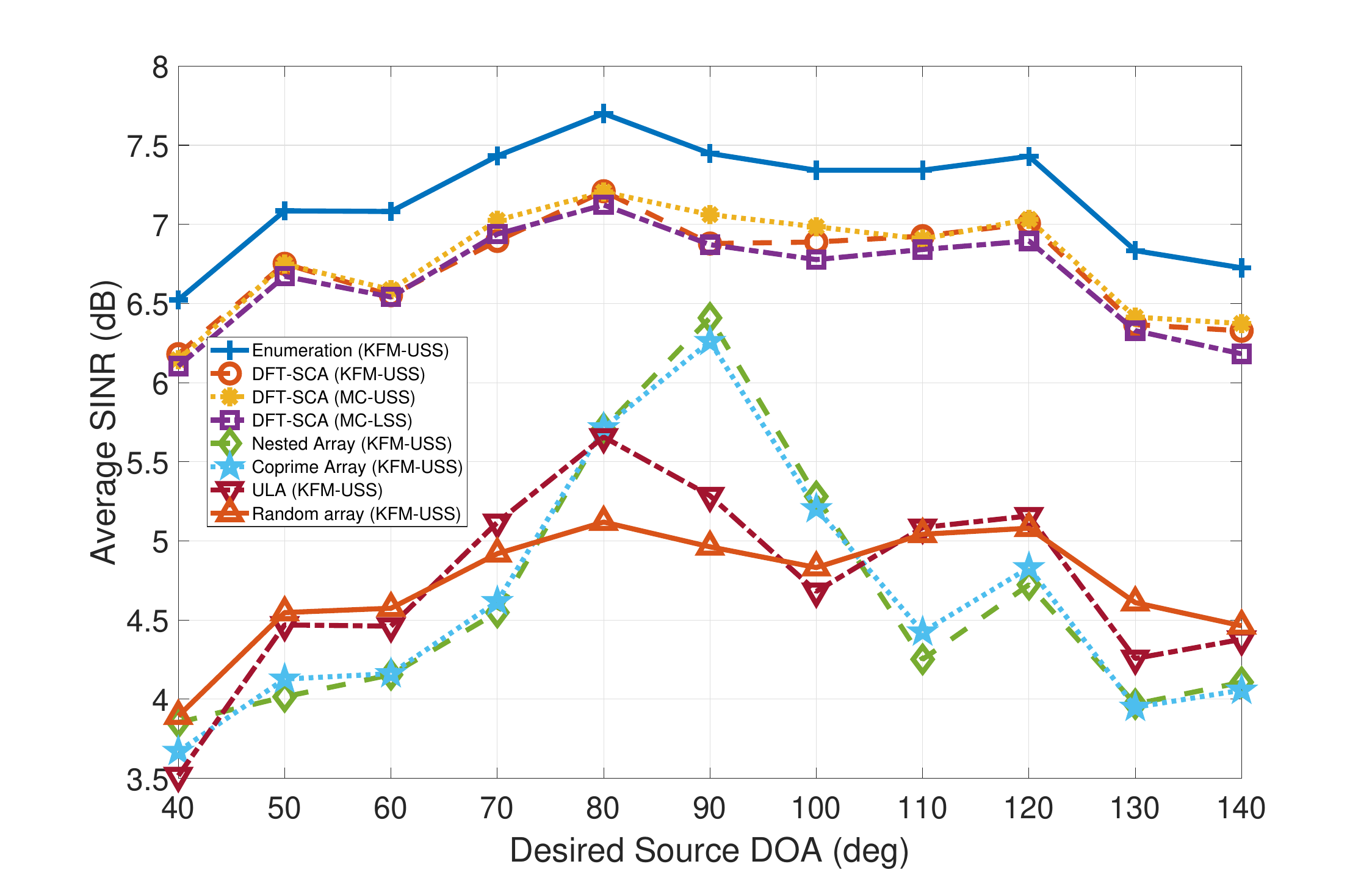}
	\caption{Performance comparisons of SCA under DFT model.}
	\label{labelfig8_ch5}
\end{figure}

Three  benchmarks are established to access the performance of the proposed algorithm under limited snapshots and lack of knowledge of the full data correlation matrix. The first benchmark applies the enumeration technique for MaxSINR design under the assumption that the   data from  all the perspective  sensor locations  is available and  accurate knowledge of the data correlation matrix, i.e., assuming unlimited number of snapshots. This benchmark is referred as ``Knowledge of Full Correlation Matrix-Unlimited Snapshots (KFM-USS)''. Other benchmarks utilize   matrix completion  to recover the missing lags (corresponding to $N-P$ missing sensors).  We refer to these benchmarks as "Matrix Completion-Unlimited Snapshots (MC-USS)" and "Matrix Completion-Limited Snap Shots (MC-LSS)," depending on whether the correlation values are  accurately known through unlimited snapshots or estimated from the limited snapshots. The evaluation  under limited snapshots considers $T=500$.    The performance of the SCA algorithm for the DFT implementation scheme is shown in the Fig. \ref{labelfig8_ch5}. The  performance upper bound  is given by the MaxSINR design   evaluated through enumeration (DFT-Enumeration (KFM-USS)). In this case, the  average performance over all the desired source DOAs is  $7.24$ dB.  The proposed  DFT-SCA algorithm under the KFM-USS benchmark  offers an average SINR  of $6.63$ dB. This performance is also comparable to the one achieved through the proposed  matrix completion, as is evident in Fig. \ref{labelfig8_ch5}. However,  the DFT-SCA design incorporating the MC-LSS benchmark ($T=500$) has a slight performance trade off of $0.14$ dB w.r.t. the DFT-SCA MC-USS design.  The aforementioned robustness of the MaxSINR design under limited snapshots is partially attributable  to a rather accurate full  matrix estimate achieved by incorporating the apriori knowledge of noise floor. 
\begin{figure}[!t]
	\centering
	\includegraphics[height=4.35in,width=5.68in]{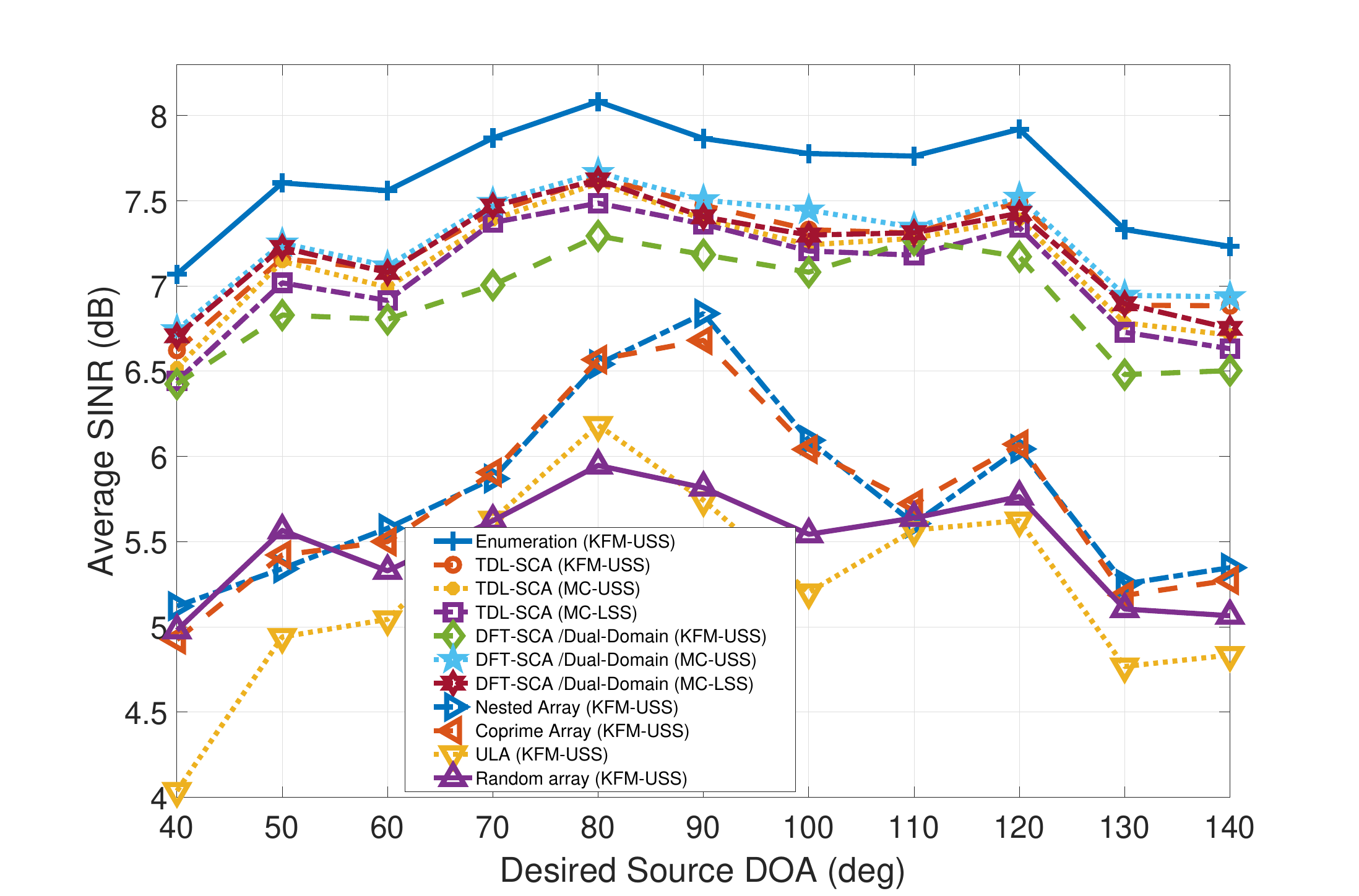}
	\caption{Performance comparisons of SCA under TDL model.}
	\label{labelfig9_ch5}
\end{figure}
The performance of the SCA under the TDL model is evaluated based on the aforementioned benchmarks, as  depicted in Fig. \ref{labelfig9_ch5}. The performance  trends are similar, however, the average SINR offered by the TDL implementation scheme is slightly superior to the DFT implementation scheme which is consistent with the literature on wideband beamforming for compact arrays \cite{doi:10.1121/1.413765}. Moreover, it is noted that the DFT-SCA dual domain design achieves comparable performance  to the TDL-SCA under all design benchmarks. This demonstrates the potential of the dual design in achieving an effective  MaxSINR beamformer with reduced complexity.  Figs. \ref{labelfig10_ch5} and   \ref{labelfig11_ch5} depict the Monte Carlo performance results analyzing the  proposed SDR. It shows that the SDR offers comparable performance to the  SCA technique, but involves  a heavy computational overhead.      It is also clear from the plots that the optimized array configurations offer  a consequential advantage over both the compact ULA and the high resolution structured arrays, such as coprime and nested arrays \cite{5739227,  5456168}, and the randomly selected $P$ sparse array configuration, each employing their respective optimal beamforming weights. The average worst case SINR is, however, reduced significantly to only $1.1$ dB. The performance is also re-evaluated at  varying number of snapshots with  consistent results.  The  performances of the proposed algorithms under MC-LSS inch closer to the MC-USS benchmark with increased data.
\begin{figure}[!t]
	\centering
	\includegraphics[height=4.35in,width=5.68in]{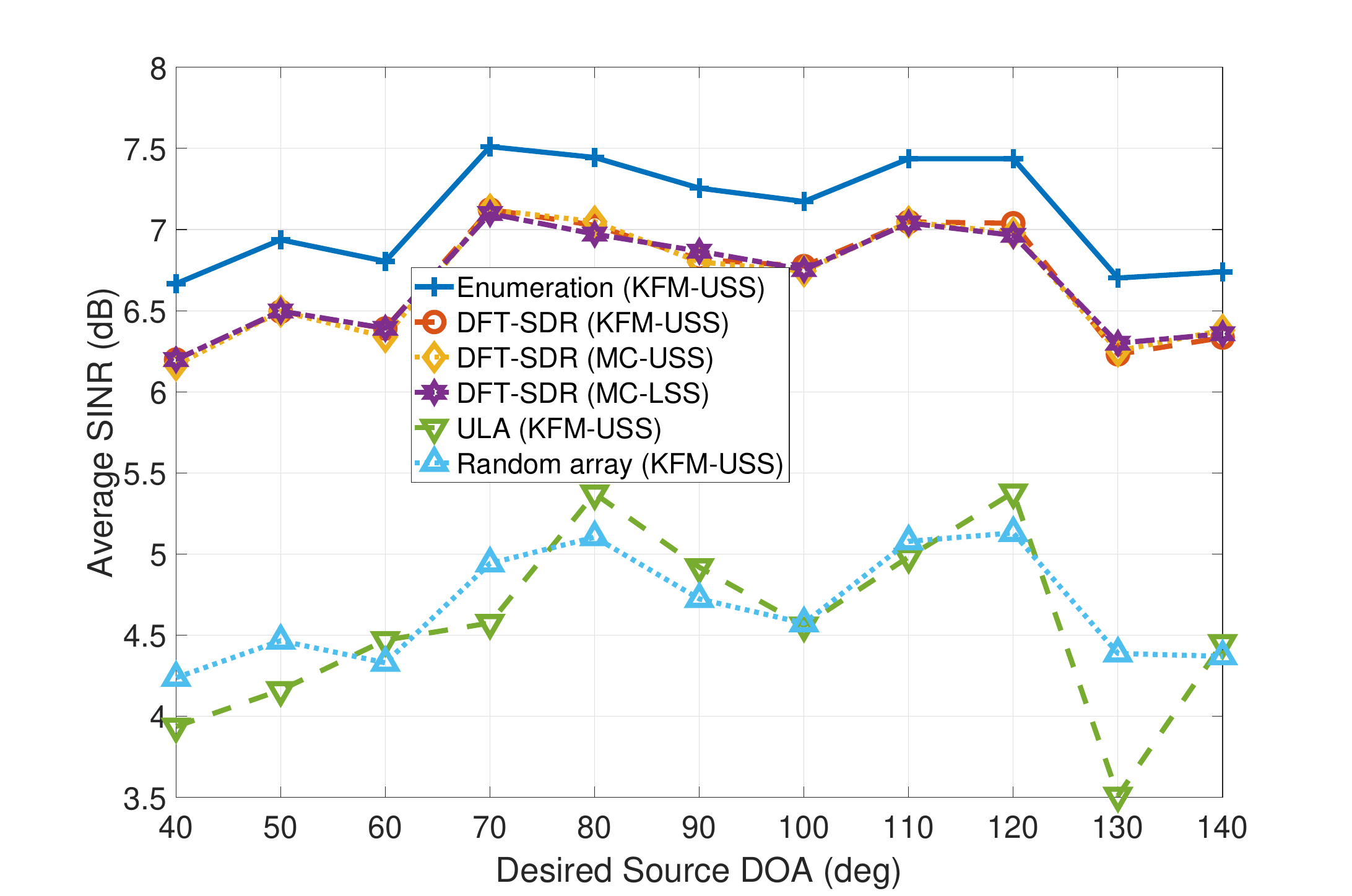}
	\caption{Performance comparisons of SDR under DFT model.}
	\label{labelfig10_ch5}
\end{figure}
\begin{figure}[!t]
	\centering
	\includegraphics[height=4.35in,width=5.68in]{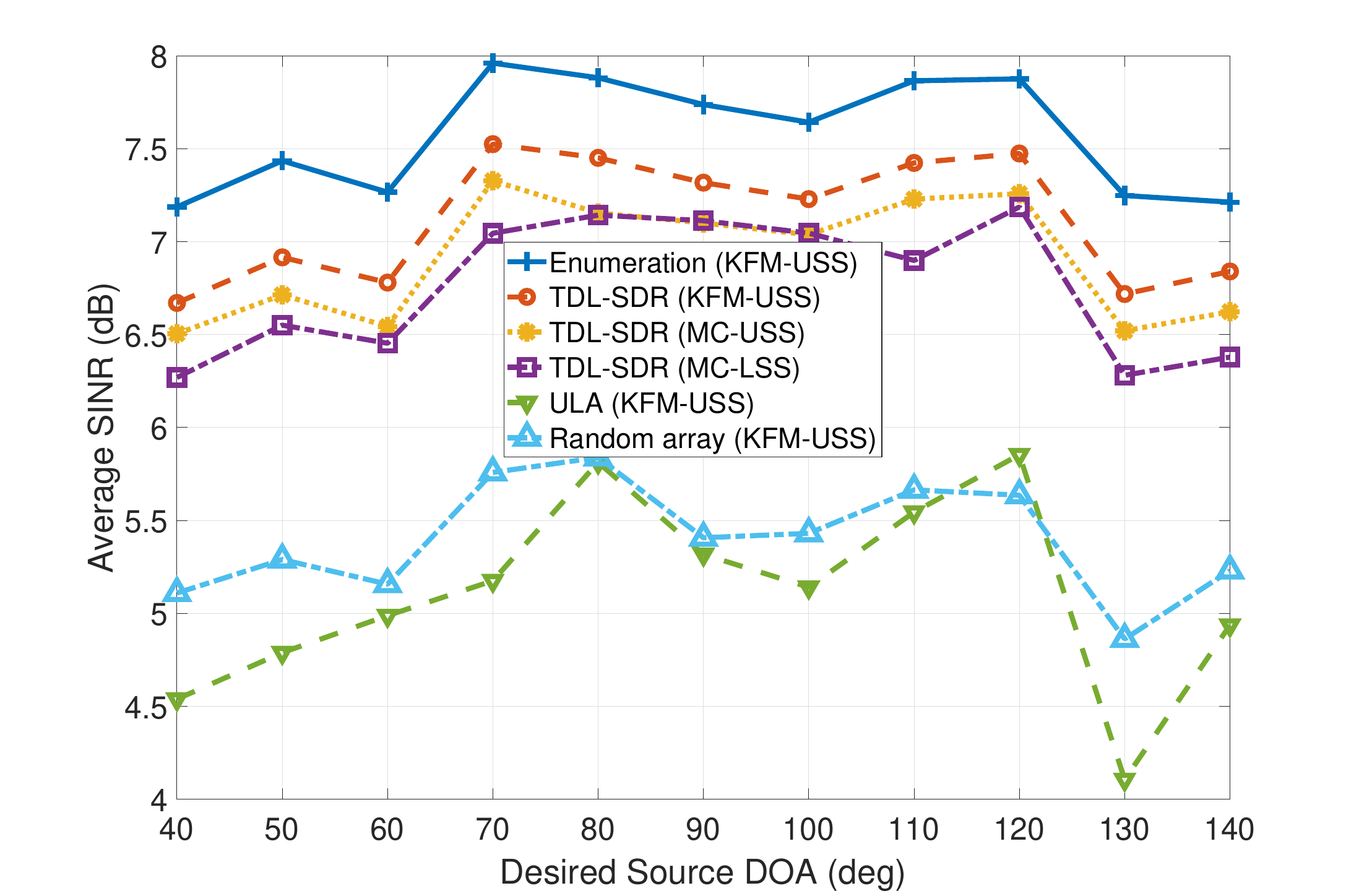}
	\caption{Performance comparisons of SDR under TDL model.}
	\label{labelfig11_ch5}
\end{figure}
\begin{figure}[!t]
	\centering
	\includegraphics[height=4.05in,width=5.68in]{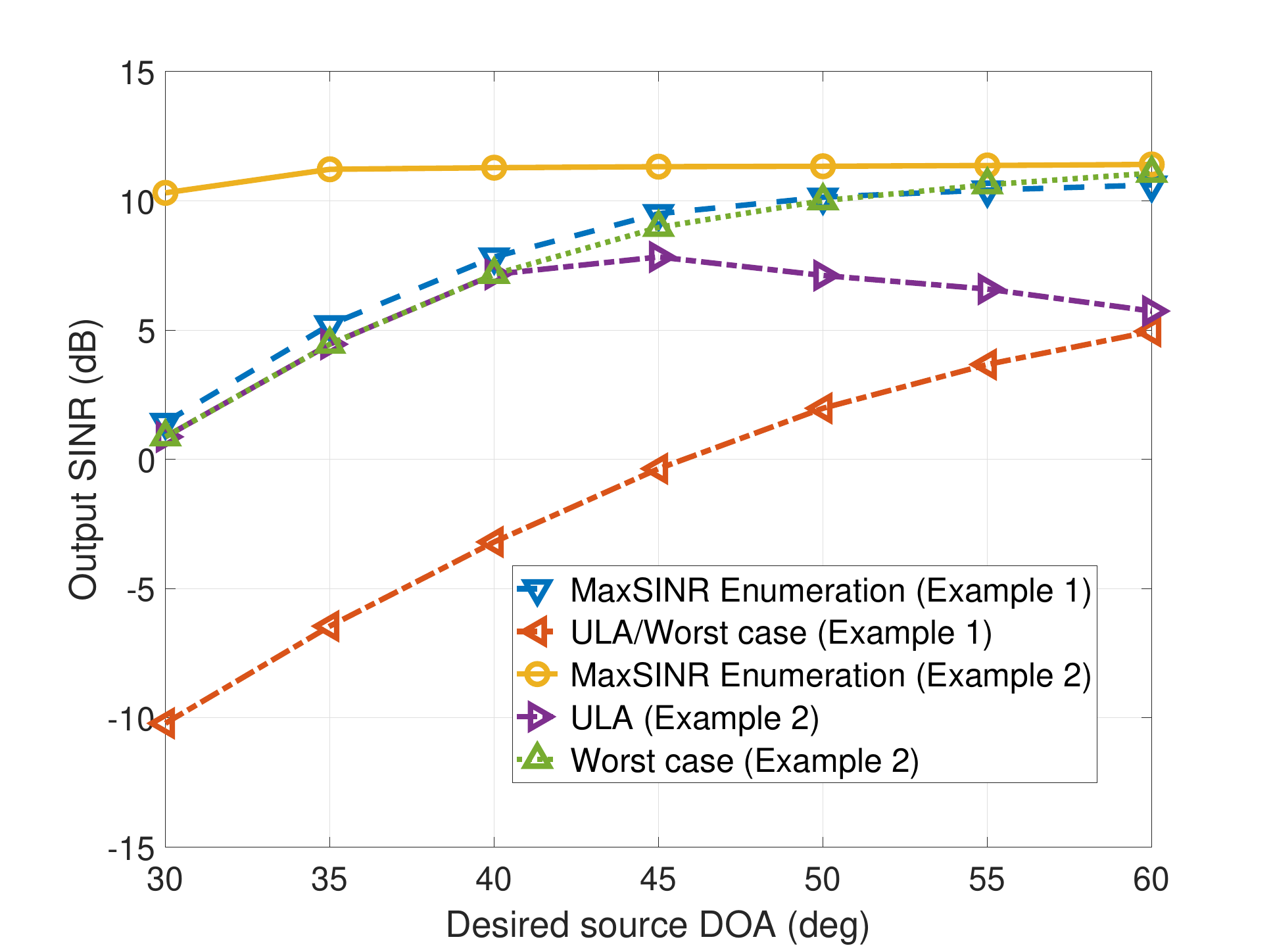}
	\caption{Performance comparisons of the optimum sparse array, the worst performing array and the compact ULA (TDL implementation scheme).}
	\label{labelfig12_ch5}
\end{figure}

\subsection{Practical Considerations for sparse array design} \label{Practical Considerations_ch5}
To assess  the  SINR  advantage of  the optimum sparse array design, we consider the effect of   two important environment dependent parameters, namely, the DOA  of the desired source and the relative locations of the jammers w.r.t. the desired source.    To demonstrate this effect, the desired source DOA in the above examples is  changed  in   steps of $5^0$, with the relative locations of the jammers remaining the same with respect to   the desired source. For example, when the desired source is  at  $50^0$ instead of $45^0$, the corresponding jammer locations shift  by  $5^0$. Figure \ref{labelfig12_ch5} compares the performance of the  optimal configuration, the worst performing array, and the compact ULA, with the desired source DOA varying from $30^0$ to $60^0$ under Example 1 and 2. It is evident from  Fig. \ref{labelfig12_ch5} that  under all  scenarios generated in Example 1, the  compact ULA delivers the worst performance, irrespective of  the  desired source DOA. This is because the  jammers are located closer to the source of interest and the compact ULA  lacks the resolution due to it's  limited array aperture.  On the other hand, for Example 2, the jammers are comparatively widely spaced and as such, the compact ULA  has a satisfactory performance that is close to the optimum sparse arrays, especially when the source DOAs are near the  array broadside. The performance degradation of the ULA near end-fire is due to the increasing  overlap between   the desired signal subspace and the interference plus noise subspace \cite{17564},   therefore lowering SINR performance. In such  scenarios, the sparse array design efficiently utilizes its degrees of freedom to improve SINR by  increasing the separation between the two subspaces. These examples show that the sparse array design is most critical when the underlying jamming environment calls for  additional degrees of freedom to engage the available array aperture more efficiently and to fulfill the  resolution requirements posed by the closer proximity of jammers and  the desired source DOA. 

\section{Conclusion}
This chapter considered optimum sparse array design  for maximizing the beamformer output SINR for the case of wideband signal models. Two different implementations, namely the TDL and the DFT implementation schemes, were presented for the optimum sparse array design. The DFT implementation scheme reduces the MaxSINR sparse array design  problem to the lower dimensional space, thereby reducing computational cost. The sparse array configuration optimized in the DFT domain  and later  imported to the TDL implementation scheme is analyzed to alleviate the  computational cost  of the TDL sparse implementation.  It was  shown that the  imported design can possibly yield comparable performance to  the design carried out exclusively through the TDL implementation scheme. For both approaches, we solved the problem using the iterative unit rank promoting SDR algorithm  and a simplified implementation using SCA.  The parameter-free block Toeplitz matrix completion was proposed to realize the data dependent design. It was shown that the SDR and SCA formulation  perform reasonably close to the optimum sparse array design achieved through enumeration under limited data snapshots. The  MaxSINR optimum sparse array yielded considerable performance improvement over suboptimal  sparse arrays and compact ULA for the underlying sensing scenarios. 

	\chapter{Sparse Array Design  for Transmit Beamforming}

\section{Introduction}
Low latency sensor selection technology enables cost effective sparse array design that can be readily configured to meet environment-dependent performance metrics. In this case, the system cost can be considerably reduced by multiplexing the expensive transmission chains to serve many more perspective sensor locations. However, at any given time, only a few sensor locations are operational which correspond to the active connections to the transmission chains. This approach is fundamentally different from the environment-independent sparse array design which seeks to maximize the number of spatial autocorrelation lags by producing a hole-free coarray aperture for a given number of sensors. There, the main task is to enhance source identifiability for DOA estimation and be able to handle more sources than sensors  \cite{1139138, 5456168, AMIN20171, 7012090, 8683315}.  This task can also be achieved for wideband or multiple frequency sources \cite{7106511}. Similarly, the environment-independent fixed beamformer design optimizes the receive beampattern characteristics such as desirable main lobe width and minimum sidelobe levels as well as  frequency invariant beampattern design for wideband sources  \cite{81155, 23483091a3104a7f8bd7d213f8ada7cc, 1071, MURINO1997177}.   

Environment-dependent sparse array optimum configuration design has primarily been considered from the perspective of receive beampattern characteristics. The optimality criteria such as maximizing the signal-to-interference plus noise ratio (MaxSINR) at the receiver can potentially harness effective array configurations for enhancing target parameter estimation accuracy  \cite{663844798989898, 6774934, 7944303, 86455521, 10.1117/12.2557923,
9114655}.    In this chapter, we consider environment-dependent sparse array design for transmit beamforming. Transmit bempattern design is critical to implementing an efficient receiver design for adaptive radar operations \cite{7746569, 8828023, 9287795}. A desirable transmit design focuses the transmitted power towards the perceived target locations and, as such, suppress the transmission towards undesired directions. In so doing, it can steer clear of heavy clutter environment or avoid probing towards an adversary location in covert situations. Another critical design objective for the transmitter is to minimize the cross correlation from the  returns from different targets to enable effective adaptive receiver processing, as high target cross correlation  can severely deteriorate performance.


The transmit beamforming design, in essence, is realized by designing the waveforms launched across all array sensors. This is in contrast to receive beamforming which optimizes the weights to yield a desirable output of a linear combiner. The transmit design can be simplified into two steps, one pertains to designing the correlation matrix of the transmitted waveform and the other involves synthesizing the actual transmitted sequence from the optimized correlation matrix     \cite{4276989}. For the scope of this chapter, we consider the transmit waveform correlation matrix optimization. There are also additional constraints that arise in transmit waveform design \textit{vis-a-vis} receiver design. The former typically requires a total transmit power constraint as well as a unit modulus or equal power constraint for each sensor location so as to utilize system resources judiciously.

In this chapter, we propose sparse array design for transmit beamforming which is formulated as maximizing the transmit signal power at the target locations w.r.t. to the power transmitted to the undesired locations while minimizing the cross correlation among the transmitted signals towards target locations. Our approach extends the sparse array transmit design proposed in  \cite{5680595} which optimizes the transmit waveform correlation matrix to maximize the power towards the desired targets for a given total power constraint. The methodology therein neither incorporates the cross correlation term nor proposes a method to suppress transmission towards certain unwanted directions.

We pose the active sensing design problem as optimally selecting $P$ antennas out of $N$ possible equally spaced locations. The optimum sparse transmit array beamformer, in this case, is the one that achieves the design objective, considering all possible sparse array configurations that stem from different arrangements of the available sensors. It is important to note that for the compact arrays (ULAs), the transmit beampattern design only involves optimizing the transmitted waveform sequence  \cite{1223538}. However, for the underlying problem, sparse array design involves optimization over two sets of variables, namely, sensor placements and transmitted waveforms. Selecting sensor positions for transmit beamformer is a combinatorial optimization problem and is, therefore, NP hard. In order to avoid the extensive computational burden for enumerating each possible array configuration, we solve the problem by convex approximation. The design problem at hand is posed as QCQP with weighted mixed  $l_{1-\infty}$-norm  penalization. We propose an approach for transmit beampattern design to handle the cross correlation term rather implicitly and best suited for the sparse transmitter design. The proposed formulation involves an $L$ rank waveform correlation matrix design, synthesized by  $L$   unit rank correlation matrices corresponding to each target location. As a result, the proposed  sparse array design promotes group sparsity harmoniously across $L$ constituent correlation matrices to ensure that the designed correlation matrix correspond to $P$ sparse solution. The QCQP is solved by the SDR approach through convex relaxation and, therefore, can be solved in polynomial time.   

%
The rest of the chapter is organized as follows:  In the next section, we  state the problem formulation for  transmit beamformer design in case of a pre-specified array configuration. Section \ref{Optimum sparse array design_ch6} elaborates the  sparse array design by  semidefinite relaxation  to find the optimum $P$   sparse transmit array geometry. In the subsequent section,  simulations are presented to demonstrate the offerings of the proposed sparse array transmit design, followed by the  concluding remarks.

\section{Problem Formulation} \label{Problem Formulation_ch6}
We consider $L$  target locations at $\{\theta_l\}_1^L$  and $Q$ known adversary or undesired locations for transmission specified by $\{\theta_q\}_1^Q$.   The received baseband probing signal  ${x}_l(n)$    at target  locations $\theta_l$ at  time instant $n$  is given by,
\begin {equation} \label{a_ch6}
{x}_l(n)=    \mathbf{s}^H(n) \mathbf{a}_l
\end {equation}
 The signal vector $\mathbf{s}(n)$ is launched from  a linear array with $N$ uniformly placed sensors with an inter-element spacing of $d$. Assuming   a narrowband signal model with  targets at far field, the steering vector $\mathbf{a}_l \in \mathbb{C}^{N}$ is  given by, 
\begin {equation}  \label{b_ch6}
\mathbf{a}_l=[1 \,  \,  \, e^{j (2 \pi / \lambda) d cos(\theta_l)  } \,  . \,   . \,  . \, e^{j (2 \pi / \lambda) d (N-1) cos(\theta_l)  }]^T
\end {equation}
Likewise, the    baseband signal ${x_q}(n) \in \mathbb{C}$ at locations $\theta_q$ is defined. The expression of the signal power at locations $\theta_l$  is  given as follows,
\begin {equation} \label{c_ch6}
P_l=    \mathbf{a}^H_l \mathbf{R} \mathbf{a}_l 
\end {equation}
The  matrix $\mathbf{R}=E(\mathbf{s}(n)\mathbf{s}(n)^H)$ is the transmitted  waveform correlation matrix. The symbol $E$ denotes the expectation operator.  The problem of maximizing the signal power at the target locations w.r.t. the unwanted locations can then be formulated as  seeking to maximize the directed power  for each target location, yielding the following optimization problem,
  \begin {equation}  \label{d_ch6}
 \begin{aligned}
 \underset{\mathbf{R}}{\text{minimize}} & \quad   \sum_{q=1}^{Q}w_q(\mathbf{a}_{q}^H\mathbf{R}\mathbf{a}_{q})\\
 \text{s.t.} & \quad    \mathbf{a}_l^H\mathbf{R}\mathbf{a}_l=1, \quad   l \in \,\, \{ 1, . . .  L\}
 \end{aligned}
 \end {equation}
  The weighting coefficients $w_q$ determines the relative preference of minimizing one signal over another. We incorporate the cross-correlations of any  two target signals located at $\theta_{l^{'}}$ and $\theta_l$, which   is given by,
\begin {equation} \label{e_ch6}
P_{l,{l^{'}}}=    |\mathbf{a}_l^H \mathbf{R}\mathbf{a}_{l^{'}}|,
\end {equation}
 where `$|.|$' is the modulus operator. The objective function in (\ref{d_ch6}) can then be rewritten  as follows \cite{4276989},
 \begin{align} \label{f_ch6}
 	\underset{\mathbf{R}}{\text{minimize}} & \quad   \sum_{q=1}^{Q}w_q(\mathbf{a}_{q}^H\mathbf{R}\mathbf{a}_{q}) + \sum_{l=1}^{L}\sum_{{l^{'}}=l+1}^{L}w_{ll^{'}}|\mathbf{a}_{l}^H\mathbf{R}\mathbf{a}_{l^{'}}|\nonumber\\
 	\text{s.t.} & \quad    \mathbf{a}_l^H\mathbf{R}\mathbf{a}_l=1, \quad   l \in \,\, \{ 1, . . .  L\}
 \end{align}
 The weighting coefficients $w_{ll^{'}}$ determines the relative trade off in minimizing the cross correlation terms. We adopt  a method to handle the cross correlation term in a more implicit way. Consider the correlation matrix $\mathbf{R}$ decomposed into $L$ matrices such that  $\mathbf{R}=\sum_{l=1}^{L}\mathbf{R}_{l}$. Matrix $\mathbf{R}_{l}$ is chosen such that the associated cross correlation term is zero, i.e.,  $\mathbf{a}_{l}^H\mathbf{R}_{l}\mathbf{a}_{l^{'}}=0$ for $l\neq l^{'}$. This is possible, for instance, if $\mathbf{a}_{l^{'}} \in \mathcal{N}(\mathbf{R}_{l}) \, \, \, \forall \, \, l\neq l^{'}$. The symbol $\mathcal{N}$ denotes the null space of the input matrix. Assuming the aforementioned conditions are satisfied, then (\ref{f_ch6}) can be rewritten as,
  \begin {equation}  \label{g_ch6}
  \begin{aligned}
 	\underset{\mathbf{R}_1, \mathbf{R}_2, ..., \mathbf{R}_{L}}{\text{minimize}} & \quad   \sum_{l=1}^{L}\sum_{q=1}^{Q}w_q(\mathbf{a}_{q}^H\mathbf{R}_{l}\mathbf{a}_{q})\\
 	\text{s.t.} & \quad    \sum_{l=1}^{L}\mathbf{a}_{l^{'}}^H\mathbf{R}_l\mathbf{a}_{l^{'}}=1, \quad   {l^{'}} \in \,\, \{ 1, . . .  L\}
 \end{aligned}
 \end {equation}
It is noted that the  constraints in the above equation can also be simplified owing to the aforementioned assumptions on   $\mathbf{a}_{l^{'}} \in \mathcal{N}(\mathbf{R}_{l}) \, \, \, \forall \, \, l\neq l^{'}$. We rewrite  (\ref{g_ch6}) as,
  \begin {equation}  \label{h_ch6}
\begin{aligned}
	\underset{\mathbf{R}_1, \mathbf{R}_2, ..., \mathbf{R}_{L}}{\text{minimize}} & \quad   \sum_{l=1}^{L}\sum_{q=1}^{Q}w_q(\mathbf{a}_{q}^H\mathbf{R}_{l}\mathbf{a}_{q})\\
	\text{s.t.} & \quad    \mathbf{a}_l^H\mathbf{R}_l\mathbf{a}_l=1, \quad   l \in \,\, \{ 1, . . .  L\}
\end{aligned}
\end {equation} 
The  zero cross-correlation conditions  can be    achieved  approximately by introducing additional terms as part of the objective function, that is,  
   \begin {equation}  \label{i_ch6}
 \begin{aligned}
 	\underset{\mathbf{R}_1, \mathbf{R}_2, ..., \mathbf{R}_{L}}{\text{minimize}} & \quad   \sum_{l=1}^{L}\sum_{q=1}^{Q}w_q(\mathbf{a}_{q}^H\mathbf{R}_{l}\mathbf{a}_{q}) + \sum_{l=1}^{L}\sum_{l^{'}\neq l}w_{l^{'}}(\mathbf{a}_{l^{'}}^H\mathbf{R}_{l}\mathbf{a}_{l^{'}})\\
 	\text{s.t.} & \quad    \mathbf{a}_l^H\mathbf{R}_l\mathbf{a}_l=1, \quad   l \in \,\, \{ 1, . . .  L\}
 \end{aligned}
 \end {equation}
Here, we have utilized  the fact that since $\mathbf{a}_{l^{'}}^H\mathbf{R}_{l}\mathbf{a}_{l^{'}}$ bounds the cross-term $|\mathbf{a}_{l}^H\mathbf{R}_{l}\mathbf{a}_{l^{'}}|$, therefore  minimizing the terms $\mathbf{a}_{l^{'}}^H\mathbf{R}_{l}\mathbf{a}_{l^{'}}$ for all $l\neq l^{'}$ also minimizes  the cross terms $|\mathbf{a}_{l}^H\mathbf{R}_{l}\mathbf{a}_{l^{'}}|$ for  all $l\neq l^{'}$.  The weighting coefficients $w_{l^{'}}$ controls the  the relative cross correlations among the  target locations. Therefore, instead of  enforcing  exact zero value for the cross-terms, we resort to minimizing the  augmented term in (\ref{i_ch6}), leading to, \vspace{-1mm}
\begin{align} \label{k_ch6} 
	\underset{\mathbf{R}_1, \mathbf{R}_2, ..., \mathbf{R}_{L}}{\text{minimize}} & \quad   \sum_{l=1}^{L}(\sum_{q=1}^{Q}w_q(\mathbf{a}_{q}^H\mathbf{R}_{l}\mathbf{a}_{q}) + \sum_{l^{'}\neq l}w_{l^{'}}(\mathbf{a}_{l^{'}}^H\mathbf{R}_{l}\mathbf{a}_{l^{'}})) \nonumber\\
	\text{s.t.} & \quad    \mathbf{a}_l^H\mathbf{R}_l\mathbf{a}_l=1, \quad   l \in \,\, \{ 1, . . .  L\}
\end{align}
In essence, the  proposed formulation in (\ref{k_ch6}) limits the cross-terms inherently by   treating the $Q$ unwanted signals somewhat similarly to  the  other target locations except the $l$th target.    We define the manifold matrix corresponding to the unwanted locations $\theta_q$, $\mathbf{V}=[\mathbf{a}_1, \mathbf{a}_2, ..., \mathbf{a}_Q]$. Let $\mathbf{S}_l=[\mathbf{a}_1, \mathbf{a}_2, ..., \mathbf{a}_{L-1}]$ incorporate all target steering vectors except the $l$th target.   Define $\mathbf{B}_l=[\mathbf{V} \, \, \, \, \mathbf{S}_{l}]$. Lumping the weighting coefficients in a diagonal matrix $\mathbf{W}_l$ other than the $l$th target allows rewriting (\ref{k_ch6}) in a compact manner as, 
 \begin {equation}  \label{l_ch6}
 \begin{aligned}
 	\underset{\mathbf{R}_1, \mathbf{R}_2, ..., \mathbf{R}_{L}}{\text{minimize}} & \quad   \sum_{l=1}^{L}(\mathbf{W}_l\mathbf{B}_l^H\mathbf{R}_{l}\mathbf{B}_l +\rho_l\text{Tr}(\mathbf{R}_{l}))\\
 	\text{s.t.} & \quad    \mathbf{a}_l^H\mathbf{R}_l\mathbf{a}_l=1, \quad   l \in \,\, \{ 1, . . .  L\}
 \end{aligned}
 \end {equation}
The additional term involving Tr($\mathbf{R}_{l}$) is introduced in (\ref{l_ch6}), where the symbol `Tr(.)' represents the trace of the matrix and  its minimization limits the norm of  $\mathbf{R}_{l}$. This additional regularization term ($\rho_l$ is the regularization parameter) can also help to alleviate high sidelobe levels in the transmit beampattern \cite{1206680}.  The above quadratic formulation requires the  information of the target locations as well as the undesired locations, but it  holds true irrespective of the array configuration and is equally valid for  compact arrays as well as  sparse arrays. The sparse optimization of the above formulation is explained in the next section. 
\section{Sparse array design}  \label{Optimum sparse array design_ch6} 
The above quadratic problem for $P$ antenna selection is a combinatorial optimization problem and can't be solved in polynomial times.  We  formulate the sparse array design  as a rank relaxed  semidefinite program (SDR) with the mixed norm regularization to recover group sparse solutions. 
 \subsection{Sparse solution through SDR}
 Invoking sparsity in the designed correlation matrix $\mathbf{R}_{l}$ of  the   transmitted waveform  essentially implies that a missing sensor translates to the sparsity in the  corresponding row and column of $\mathbf{R}_{l}$. Furthermore,  $\mathbf{R}=\sum_{l=1}^{L}\mathbf{R}_{l}$ implies that the sparse rows and columns should be consistent across all $L$ correlation matrices.  In order to do so, and invoke sparsity in a systematic manner, the positive semidefinite correlation matrices  $\mathbf{R}_{l}'s$ are assumed to be unit rank and expressed as an outer product, $\mathbf{R}_{l}=\mathbf{r}_l\mathbf{r}_l^H$.     Define $\mathbf{r}^{(k)}=[\mathbf{r}_1(k), \mathbf{r}_2(k), ..., \mathbf{r}_L(k) ] \in \mathbb{C}^L$ as the $k$th entry  for all the $L$ vectors $\mathbf{r}_l$'s which is associated   with the $k$th transmitter.   The problem formulated in (\ref{l_ch6}) can be rewritten   with an additional regularization term as,
 \vspace{-1mm}
\begin{align} \label{a2_ch6}
 \underset{\mathbf{R}_l \mathbf{\in \mathbb{C}}^{N}}{\text{minimize}} & \quad   \sum_{l=1}^{L}(\mathbf{W}_l\mathbf{B}_l^H\mathbf{R}_{l}\mathbf{B}_l +\rho_l\text{Tr}(\mathbf{R}_{l})) + \mu(\sum_{k=1}^{N}||\mathbf{r}^{(k)}||_q) \nonumber\\
\text{s.t.} & \quad    \mathbf{a}_l^H\mathbf{R}_{l}\mathbf{a}_l=1, \quad   l \in \,\, \{ 1, . . .  L\}  
\end{align}
Here,  $||.||_q$ denotes the $q$-norm of the vector. The mixed $l_{1-q}$ norm regularization  is known to thrive  the group sparsity in the solution. This ensures that the identical $P$ transmitters are selected for  each constituent correlation matrix $\mathbf{R}_{l}$, and without  specifying the cardinality of active transmitters. 
 For the underlying design problem, the  re-weighted $l_1$-norm is adopted    which is a well known regularizer for  promoting sparse solutions iteratively \cite{Candes2008}, \vspace{-2mm}
\begin{align*} 
\underset{\mathbf{R}_l \mathbf{\in \mathbb{C}}^{N}}{\text{minimize}} & \quad   \sum_{l=1}^{L}(\mathbf{W}_l\mathbf{B}_l^H\mathbf{R}_{l}\mathbf{B}_l +\rho_l\text{Tr}(\mathbf{R}_{l})) 
 \end{align*}\vspace{-7mm}
\begin{align*} 
 & \quad \quad \quad \quad \qquad \qquad \qquad \qquad \qquad \qquad   + \mu(\sum_{k=1}^{N}\mathbf{u}^i(k)||\mathbf{r}^{(k)}||_q) 
 \end{align*}\vspace{-7mm}
 \begin{align} \label{b2_ch6}
\qquad \qquad \qquad  \text{s.t.} & \quad    \sum_{l=1}^{L} \mathbf{a}_l^H\mathbf{R}_{l}\mathbf{a}_l=1, \quad l \in \,\, \{ 1, . . .  L\}  
\end{align}
The choice of the  $k$th element of the re-weighting  vector $\mathbf{u}^i(k)$   at  the $i$th iteration is later discussed. Implementing the group sparsity through the   $\infty$-norm  and choosing   $l_1$-norm  squared  function instead of the  $l_1$-norm  can facilitate the semidefinite realization \cite{6477161}, \vspace{-4mm}
\begin{align*}
 \underset{\mathbf{R}_l \mathbf{\in \mathbb{C}}^{N}}{\text{minimize}} & \quad   \sum_{l=1}^{L}(\mathbf{W}_l\mathbf{B}_l^H\mathbf{R}_{l}\mathbf{B}_l +\rho_l\text{Tr}(\mathbf{R}_{l})) 
\end{align*}\vspace{-8mm}
\begin{align*} 
& \quad \quad \quad \quad \quad \qquad \qquad \qquad \qquad \qquad   +
\mu(\sum_{k=1}^{N}\mathbf{u}^i(k)||\mathbf{r}^{(k)}||_\infty)^2 
\end{align*}\vspace{-7mm}
\begin{align} \label{c2_ch6}
\qquad \qquad \qquad  \text{s.t.} & \quad    \sum_{l=1}^{L} \mathbf{a}_l^H\mathbf{R}_{sl}\mathbf{a}_l=1, \quad l \in \,\, \{ 1, . . .  L\}  
\end{align}
The  above problem can then be posed as  semidefinite programming  by equivalently expressing the quadratic functions, $ \mathbf{W}_l\mathbf{B}_l^H\mathbf{R}_l\mathbf{B}_l= $ Tr$(\mathbf{W}_l\mathbf{B}_l^H\mathbf{R}_l\mathbf{B}_l)= $Tr$(\mathbf{R}_l\mathbf{W}_l\mathbf{B}_l\mathbf{B}_l^H)=$ Tr$(\mathbf{R}_l\mathbf{\bar{B}}_l)$. Here, $\mathbf{\bar{B}}_l=\mathbf{W}_l\mathbf{B}_l\mathbf{B}_l^H$ and similarly    $\mathbf{{A}}_l=\mathbf{a}_l\mathbf{a}_l^H$ is the outer product of $l$th target steering vector. The formulation in (\ref{c2_ch6}) takes  the following form \cite{Bengtsson99optimaldownlink, 5447068,  8682266, 8378759, 6477161},
\begin{align}\label{e2_ch6}
\underset{\mathbf{R}_l \mathbf{\in \mathbb{C}}^{N.N}, \mathbf{\tilde{R} \in \mathbb{R}}^{N.N}}{\text{minimize}} &
 \quad   \sum_{l=1}^{L}\text{Tr}(\mathbf{R}_l(\mathbf{\bar{B}}_l+\rho_l\mathbf{I})) + \mu \text{Tr}(\mathbf{U}^i\mathbf{\tilde{R}})  \nonumber \\
\text{s.t.} & \quad    \text{Tr}(\mathbf{R}_{l}\mathbf{A}_l) \geq1, \quad l \in \,\, \{ 1, . . .  L\}, \nonumber\\
& \quad \mathbf{\tilde{R}} \geq  |\mathbf{R}_{l}|  \quad l \in \,\, \{ 1, . . .  L\}, \nonumber \\
& \quad   \mathbf{R}_l \succeq 0, \, \text{Rank}(\mathbf{R}_l)=1
\end{align}
Here `$\geq$' is the element wise comparison and `$\succeq$' denotes the positive semidefinite constraint, and $\mathbf{U}^i=\mathbf{u}^i(\mathbf{u}^i)^T$ ((.)$^T$ denotes the transpose). The  SDR realization of the problem in (\ref{e2_ch6}) is pursued  by dropping the non-convex rank limiting constraint,
\begin{align} \label{f2_ch6}
\underset{\mathbf{R}_l \mathbf{\in \mathbb{C}}^{N.N}, \mathbf{\tilde{R} \in \mathbb{R}}^{N.N}}{\text{minimize}} &
\quad   \sum_{l=1}^{L}\text{Tr}(\mathbf{R}_l(\mathbf{\bar{B}}_l+\rho_l\mathbf{I})) + \mu \text{Tr}(\mathbf{U}^i\mathbf{\tilde{R}})  \nonumber \\
\text{s.t.} & \quad    \text{Tr}(\mathbf{R}_{l}\mathbf{A}_l) \geq1, \quad l \in \,\, \{ 1, . . .  L\}, \nonumber \\
& \quad \mathbf{\tilde{R}} \geq  |\mathbf{R}_{l}|  \quad l \in \,\, \{ 1, . . .  L\}, \nonumber\\
& \quad   \mathbf{R}_l \succeq 0, \quad l \in \,\, \{ 1, . . .  L\}
\end{align}
Although the formulation in (\ref{f2_ch6}) doesn't enforce the rank constraint, it has been observed, through extensive numerical simulations, that it usually renders  rank 1 solutions and, therefore, approximates the original problem fairly accurately. 
\begin{table}[t!] 
\noindent\rule{16.5cm}{0.6pt}
	\caption{SDR-Transmit Beamformer Design}
\noindent\rule{16.5cm}{0.6pt}

	\begin{algorithmic}[]
		
		\renewcommand{\algorithmicrequire}{\textbf{Input:}}
		
		\renewcommand{\algorithmicensure}{\textbf{Output:}}
		
		\REQUIRE  $N$, $P$, $L$, Look directions $\theta_q$ and $\theta_l$. \\
		
		\ENSURE  $\mathbf{R}=\sum_{l=1}^{L}\mathbf{R}_l$  transmit waveform correlation matrix.   \\
		
		\textbf{Initialization:} \\
		
		Initialize $\gamma$, $\epsilon$, $\mathbf{W}_l$, $\rho_l$, $\mathbf{U}$ as all ones matrix.
		
%
		Bisection search for desired cardinality $P$\\
		$l=\mu_{lower}$, $u=\mu_{upper}$ (Initializing lower and upper limits of sparsity parameter range)\\
		\WHILE {(Cardinality of $\mathbf{r}_l\text{'s}$ $\neq$ $P$)}
		\STATE  $\mu=[(l+u)/2]$
		\STATE   Run the  SDR of Eq. (16). \\
				\STATE   Update  $\mathbf{U}$ according to Eq. (17). \\ 
		\STATE  Update $\mu$ through bisection method.
		
		\ENDWHILE
		
		\STATE After achieving the desired cardinality, run SDR for reduced size correlation matrix corresponding  to nonzero values of $\mathbf{\tilde{R}}$  and $\mu=0$,  yielding, $\mathbf{R}=\sum_{l=1}^{L}\mathbf{R}_l$. 
		
		\RETURN $\mathbf{R}$
		
	\end{algorithmic}
	\label{algorithm_ch6}
\noindent\rule{16.5cm}{0.6pt}

\end{table}
\subsection{Reweighting sparsity}
The reweighting matrix  $\mathbf{U}^i$ is typically initialized unweighted with an all ones matrix. It is subsequently  updated with an inverse function of the absolute values of the entries of the  correlation matrix   as follows \cite{Candes2008},
\begin{equation} \label{g2_ch6}
\mathbf{U}_{m,n}^{i+1}=\frac{1}{\mathbf{\tilde{R}}^i(m,n)+\epsilon}
\end{equation}
The $m,n$th entry of $\mathbf{\tilde{R}}$ is given by $\mathbf{\tilde{R}}^i(m,n)$. The parameter $\epsilon$ prevents  the solution to converge to  local minima and avoids the unwanted case of division by zero. Choosing the randomly selected  sparsity parameter $\mu$ does not exactly correspond to  $P$ sparse solution. In order to trace $\mu$ corresponding to $P$ sparse solution, the optimization problem is typically solved  for different values of $\mu$ found through a bisection search, until the solution converges to $P$ sensors \cite{6477161}.  The proposed algorithm for controlling the sparsity of the transmitted waveform correlation   matrix is elaborated further  in  Table \ref{algorithm_ch6}.
  \subsection{Transmit power constraint}
It is  noted that the formulation adopted so far doesn't explicitly account for the total transmit power constraint which is critical for a pragmatic design. However, the proposed approach  can inherently accommodate this constraint as it is primarily based on relative target power maximization w.r.t. the undesired locations. This renders the design invariant to the scaling factor. Therefore, for a given transit power of $P_t$ and the designed transmit correlation matrix $\mathbf{R}$, the total transmit power constraint is realized by a scaling factor $\alpha$  such that $\alpha$Tr$(\mathbf{R})=P_t$. For the scope this chapter, we don't consider the uniform elemental power constraint. 

 \section{Simulations} \label{Simulations_ch6}
In this section,  we demonstrate the effectiveness of the active sparse array design from the perspective of the transmit beampattern.  The performance   of  sparse array design  is  compared with the   compact ULA and the structured sparse array design. 
\begin{figure}[!t]
	\centering
\subfloat[]{\includegraphics[width=0.83\textwidth]{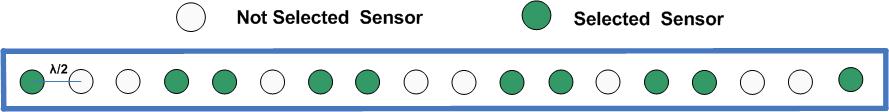}
		\label{1234a_ch6}}
		\hfill
\subfloat[]{\includegraphics[width=0.83\textwidth]{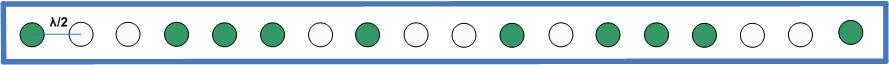}
		\label{1234b_ch6}}
		\hfill
\subfloat[]{\includegraphics[width=0.83\textwidth]{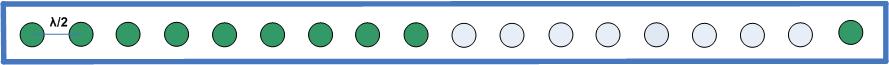}
		\label{1234c_ch6}}
		\hfill
\subfloat[]{\includegraphics[width=0.83\textwidth]{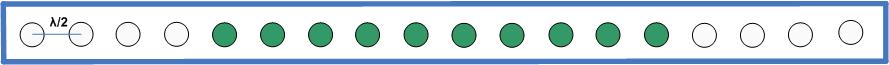}
		\label{1234d_ch6}}
		\hfill
	\caption{Various  array configurations}
\end{figure}
\subsection{Example}
We consider  $N=18$    equally spaced locations are potentially available for sensor selection. The minimum spacing  among the  sensors is $d=\lambda/2$, and $P=10$ sensor locations can be chosen concurrently.  
The three   targets locations, $L=3$, namely  Targets 1,  2 and  3 are perceived at $40^0$, $50^0$ and $65^0$, respectively.   Four unwanted directions are located at $25^0$, $60^0$, $110^0$ and $120^0$.  All the weighting coefficients corresponding to the targets and unwanted locations are set to one ($\rho_l=1)$. The  sparsity parameter $\mu$ is found through bisection  search  ranging from $0.01$ to $3$, the sensor locations corresponding to the  correlation values of less than  the sparsity threshold $\gamma=10^{-5}$ are declared sparse  and $\epsilon=0.05$ is chosen for the underlying problem.

\begin{figure}[!t]
 	\centering
 	\includegraphics[height=3.15in, width=4.75in]{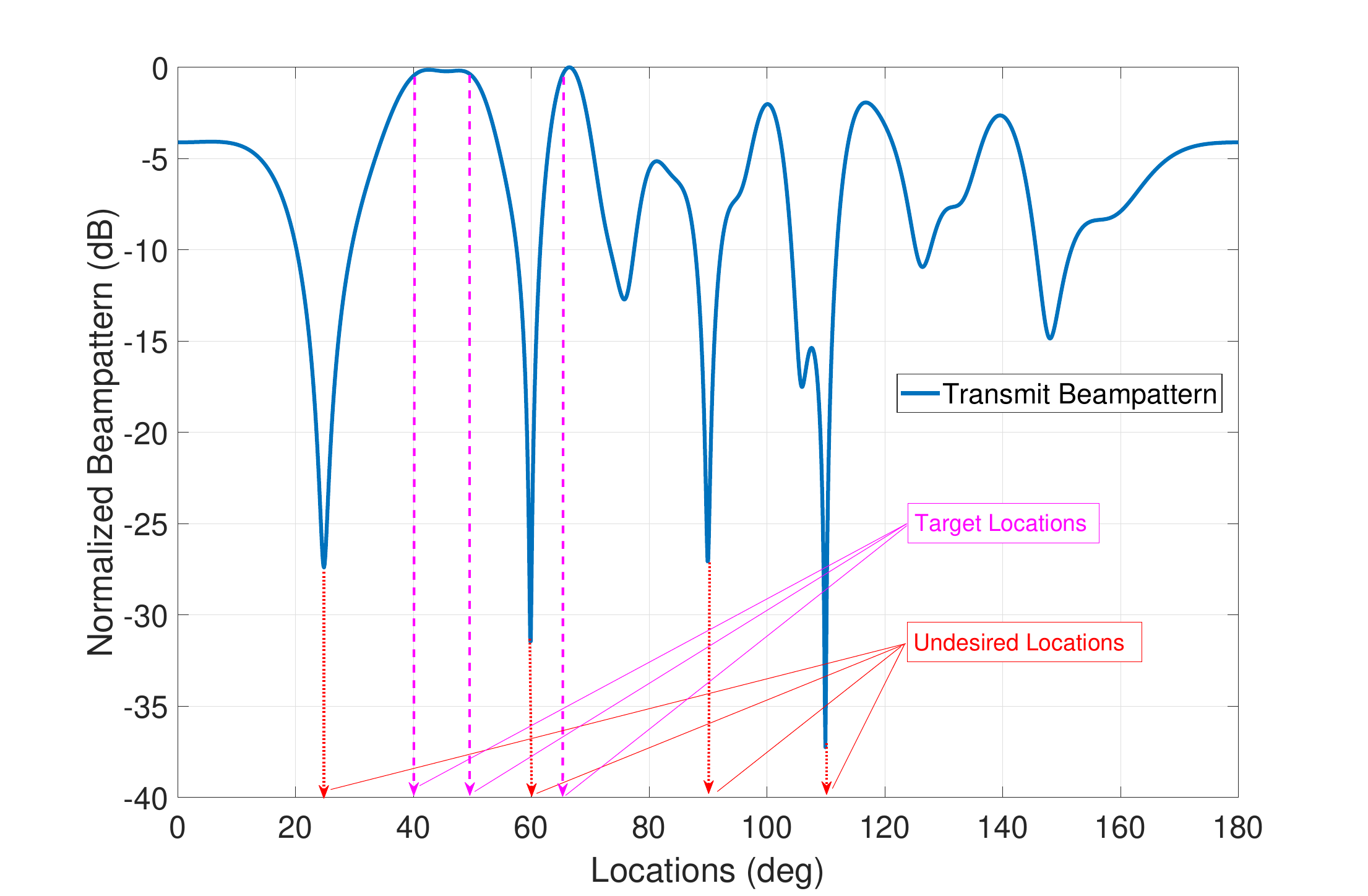}
 	\caption{Transmit beampattern for the SDR-optimized configuration.}
 	\label{op_algo_2027_8_ch6}
 \end{figure}

\begin{figure}[!t]
 	\centering
 	\includegraphics[height=3.15in, width=4.75in]{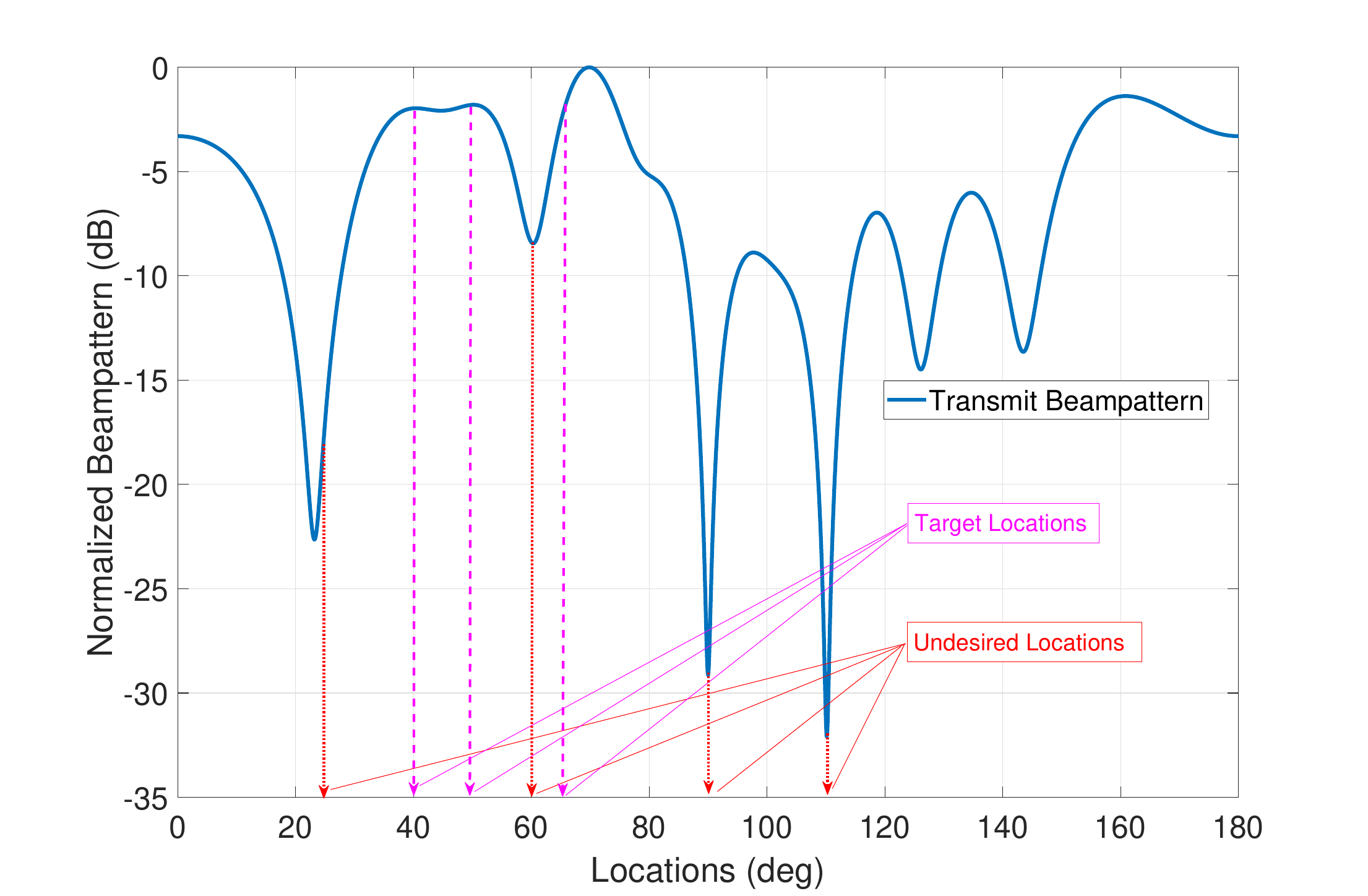}
 	\caption{Transmit beampattern for the nested array configuration.}
 	\label{uni_uni_1_81_ch6}
 \end{figure}

\begin{figure}[!t]
	\centering
	\includegraphics[height=3.15in, width=4.75in]{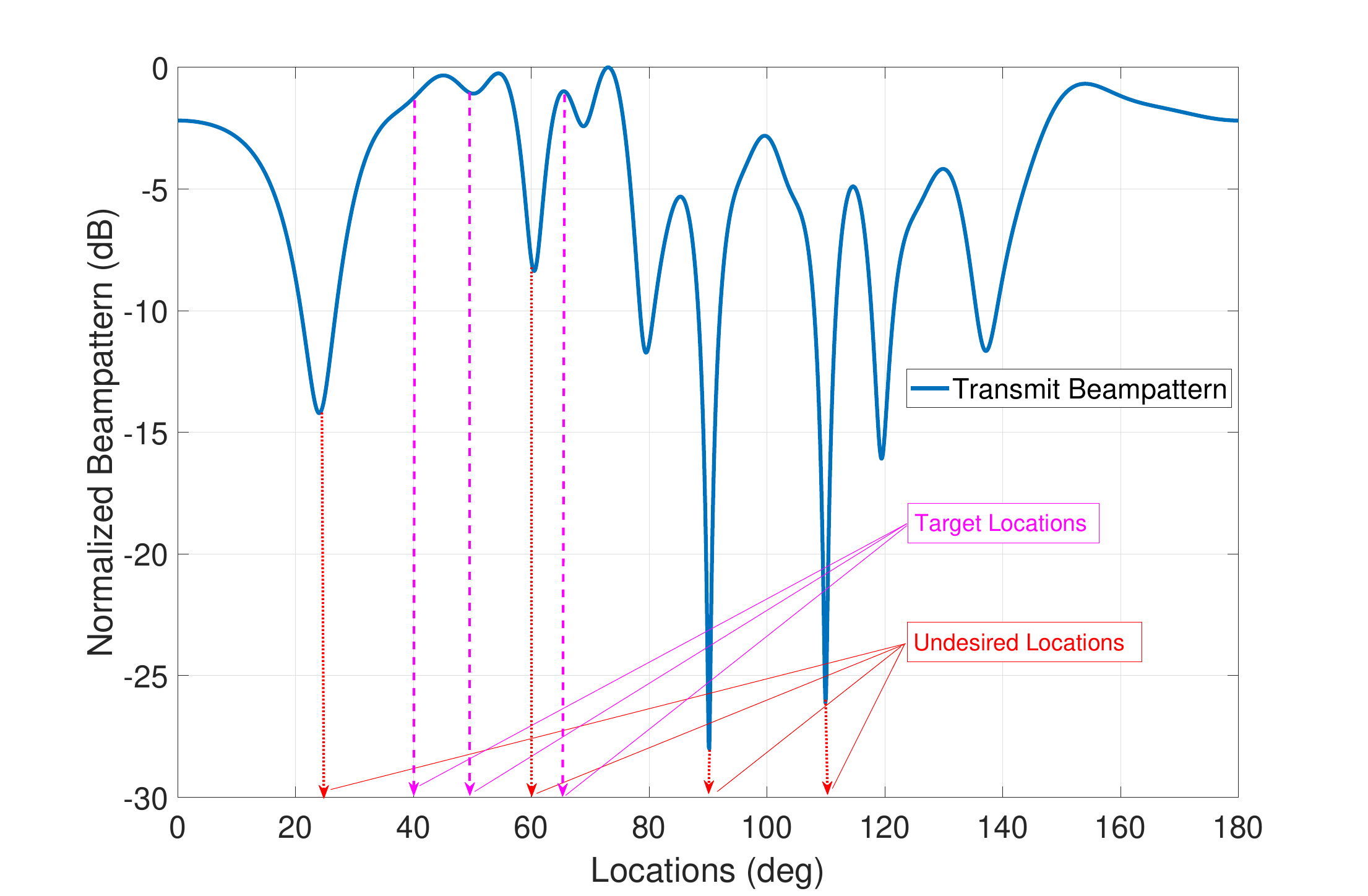}
	\caption{Transmit beampattern for the compact ULA.}
	\label{uni_uni_1_82_ch6}
\end{figure}

\begin{figure}[!t]
	\centering
	\includegraphics[height=3.3in, width=4.75in]{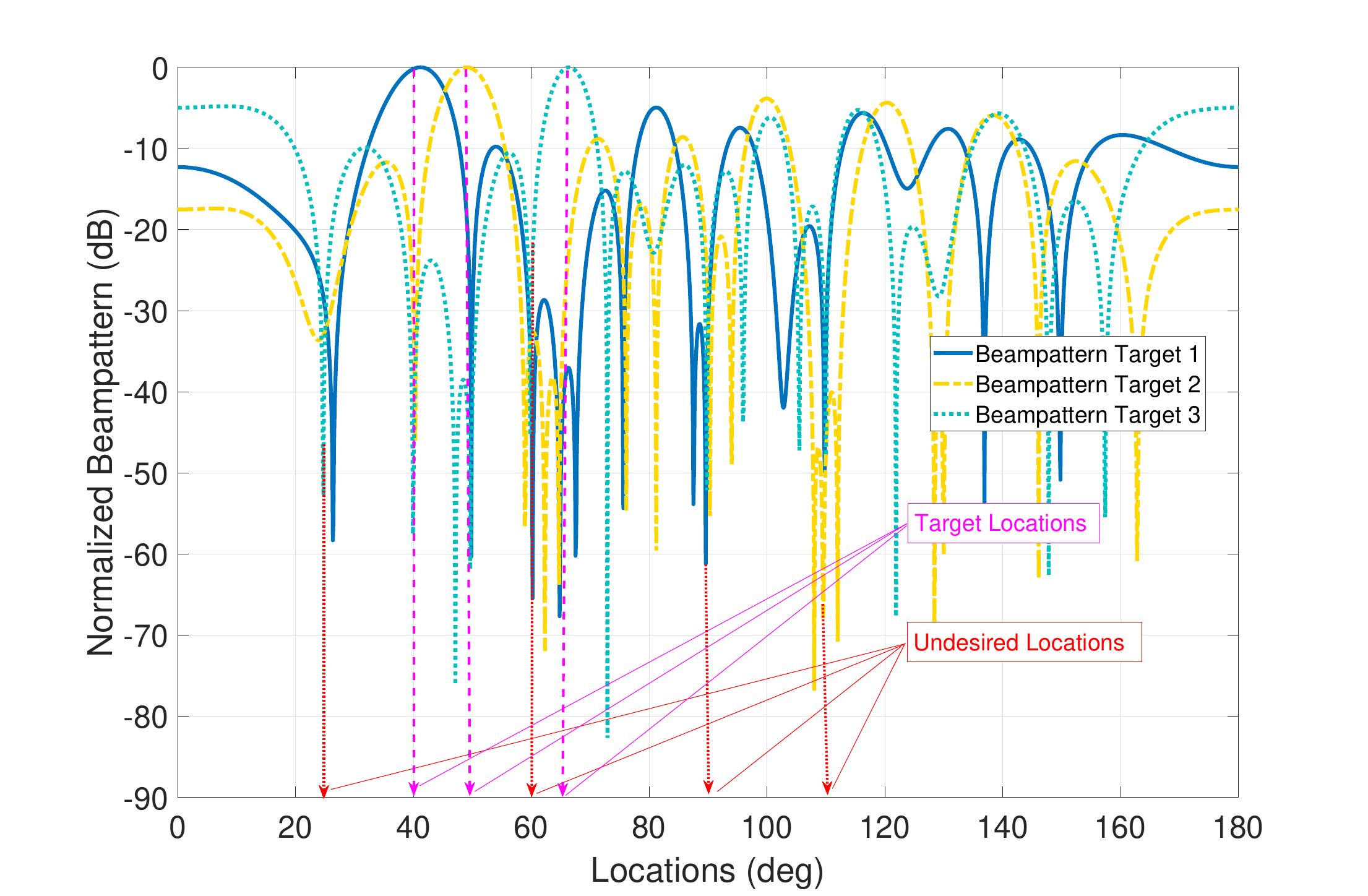}
	\caption{Constituent transmit beampatterns for the SDR-optimized configuration.}
	\label{uni_uni_1_83_ch6}
\end{figure}

 Figure (\ref{1234a_ch6}) shows the sparse array found through enumeration that optimizes the formulated objective function over all possible configurations. This sparse array configuration has a  performance that is comparable  to that  of the array optimized through the proposed SDR algorithm. The SDR-optimized topology is shown in (\ref{1234b_ch6}), and  has the performance trade off less than 0.23 dB as compared to the enumerated design. The transmit beampattern of the optimized topology is  depicted in Fig.  (\ref{op_algo_2027_8_ch6}). It is observed form the normalized beampattern that the designed configuration fairly attempts to  maximize the transmitted power towards the target locations while mitigating the power towards the unwanted directions. In  so doing, the  average  directed  gain  towards the target locations  is less than 0.5 dB down as compared to the maximum gain. On the other hand, the nested array configuration and the compact ULA offers  an average gain towards the  targets  which is $2.1$ dB and $2.45$ dB down, respectively,  as compared to the   corresponding  locations with the maximum transmitted power. The nested array and the compact ULA are shown on Figs. (\ref{1234c_ch6}) and (\ref{1234d_ch6}), with the respective beampatterns depicted in Figs. (\ref{uni_uni_1_81_ch6})  and (\ref{uni_uni_1_82_ch6}) , respectively. 
 It is noted from the depicted  beampatterns  that the proposed methodology, in an attempt to minimize the cross correlation among the target signals, renders a synthesized beampattern whose maxima are precisely towards the  target locations.  This is also  evident from the beampattern  of the SDR-optimized array as  shown in Fig. (\ref{op_algo_2027_8_ch6}), where the maxima is slightly skewed from the target locations. This effect can be explained from the  transmit beampatterns of the constituent correlation matrices shown in  Fig. (\ref{uni_uni_1_83_ch6}). It is clear that the individual constituent beampatterns have the maxima at the respective target locations. However,  since the  transmit  beampattern is a cumulative pattern found by the superposition of constituent beampatterns, the overall maxima can drift    from the proximity of the target  locations. This effect doesn't pose a major drawback to the proposed design as it is evident from the beampattern of Fig. (\ref{op_algo_2027_8_ch6}), where the beampattern maxima is located somewhere in the middle of the closely located targets at $40^0$, $50^0$ and is merely $0.5$ dB strong as compared to the target gains. In this case, still significantly higher power is directed to  the target locations relative to the total average transmitted power.  

\begin{figure}[!t]
	\centering
	\includegraphics[height=2.91in, width=4.75in]{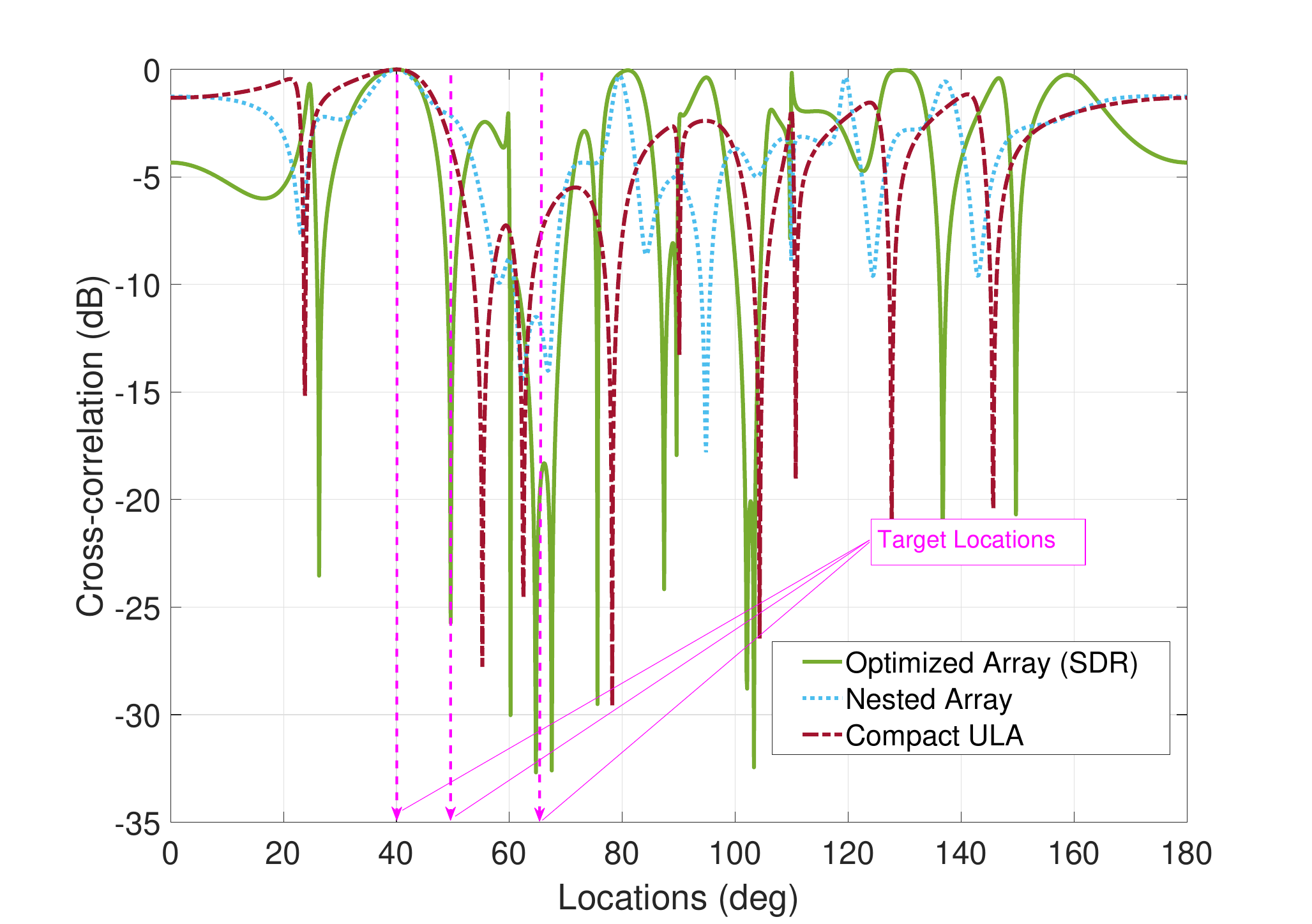}
	\caption{Cross-correlation pattern against the Target 1 for various sparse configurations.}
	\label{uni_uni_1_84_ch6}
\end{figure}

\begin{figure}[!t]
	\centering
	\includegraphics[height=2.91in, width=4.75in]{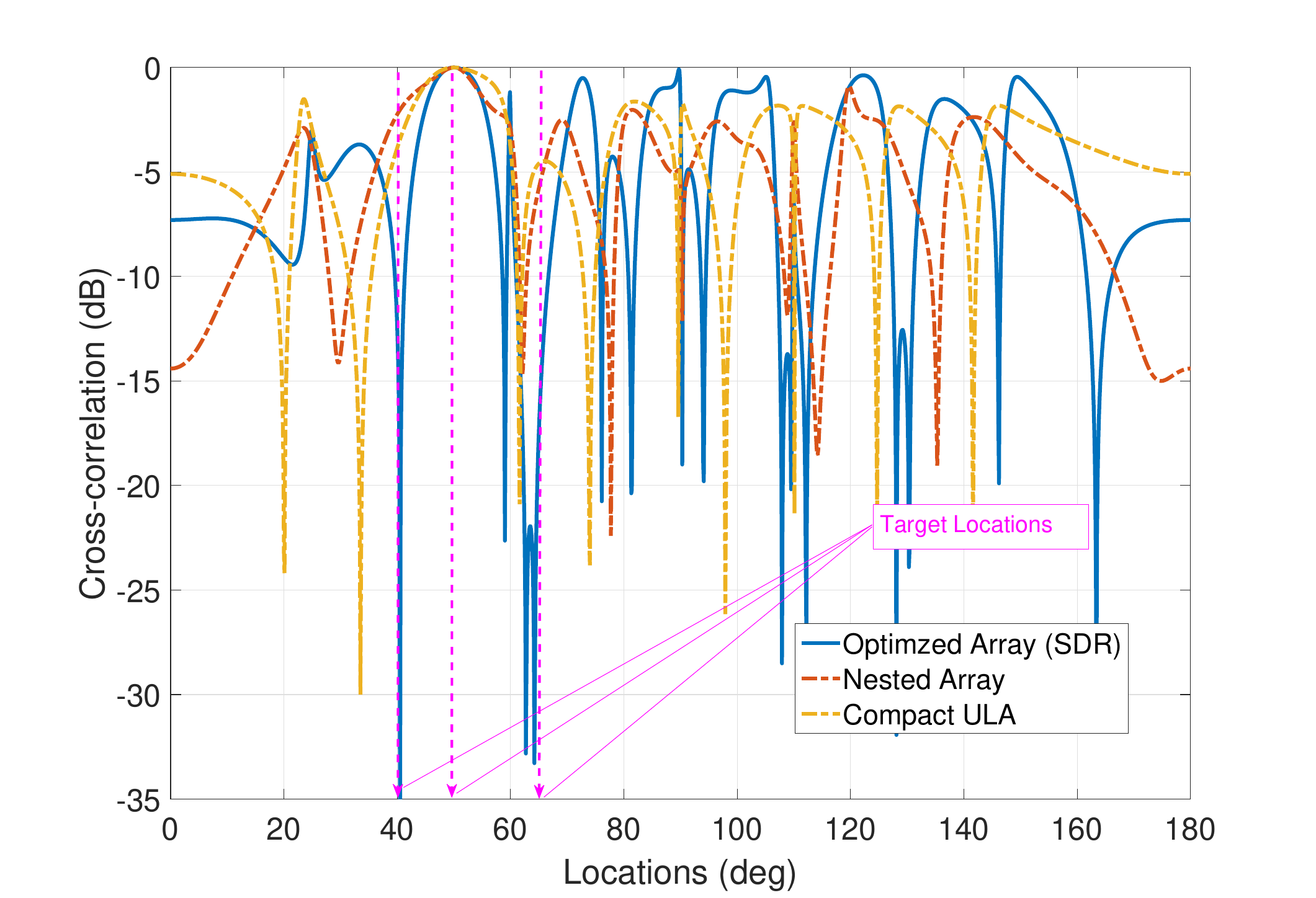}
	\caption{Cross-correlation pattern against the Target 2 for various sparse configurations.}
	\label{uni_uni_1_85_ch6}
\end{figure}

 To analyze the  cross correlation term, we plotted the following normalized cross-term $\hat{P}_{l,{l^{'}}}$ for each target location,
 \begin {equation*} \label{a3_ch6}
 P_{l,{l^{'}}}=    \frac{|\mathbf{a}^H(\theta_l) \mathbf{R}\mathbf{a}(\theta_{l^{'}})|}{(\mathbf{a}^H(\theta_l) \mathbf{R}\mathbf{a}(\theta_{l}))^{\frac{1}{2}}(\mathbf{a}^H(\theta_{l^{'}}) \mathbf{R}\mathbf{a}(\theta_{l^{'}}))^{\frac{1}{2}}}
 \end {equation*}
Fig. (\ref{uni_uni_1_84_ch6}) shows the cross-correlation pattern w.r.t. to the Target 1. It is clear that the  cross-correlation contributions of the  Target 1 to the target locations 2 and 3 are negligible and are less than $20$ dB.    In contrast, the cross correlation  coefficients of the Target 1 to Target 2, in case of  the nested array and the compact ULA are $0.46$ and $0.53$ which are significantly higher. However, the cross correlations of the Target 1 to Target 3 are substantially lower for the nested array and the compact ULA as Target 3 is sufficiently farther away  from Target 1 location. This also holds true for the  cross correlations of the Target 2 to Target 3 owing to the spacing of both target locations as shown in  Fig. (\ref{uni_uni_1_85_ch6}).
\section{Conclusion}
This chapter considered active sensing using sparse array configuration. It sought optimum transmit beamforming for radar applications.   The problem was formulated as optimizing the transmit waveform correlation matrix and solved through   convex relaxation  with low  computational complexity. It was shown that the active sparse array design can achieve desirable beampattern characteristics such as directing a high proportion of the transmitted power towards the perspective target locations with minimal cross-correlation among  target signals. We showed the effectiveness of the proposed approach by comparing  its performance with that of compact ULA, a structured array, and the design achieved   through  enumeration.


\chapter{Sparse Array Capon Beamformer Design Availing Deep Learning Approach \, \,}

\section{Introduction}
Sparse array design reduces system transceiver costs  by reducing the hardware  and processing complexity through sensor selection. It is useful in multitude of sensor signal processing tasks for MIMO communications, radar/sonar, satellite navigation,  radio telescopes, speech enhancement  and medical imaging applications \cite{710573, 6477161, 1428743, 4058261,  4663892,   6031934}. The performance gains in using sparse arrays stem from their inherent ability of tending the additional degrees of freedom to accomplish certain pre-defined performance metrics. Several different performance metrics have been proposed in the literature, and can generally be categorized  into environment-independent or environment-dependent design. In the latter case, the receiver performance then  largely depends on the operating environment, which may change according to the source and interference signals and locations, implemented via reconfigurable sensor arrays.   This is in contrast to environment-independent sparse arrays whose configurations follow certain formulas and seek to attain structured sparse configurations with  extended aperture co-arrays. The driving objective, in this case,  is to enable direction of arrival (DOA) estimation of more sources than the available physical sensors. Common examples of structured sparse arrays are the nested and coprime arrays \cite{1139138, 5456168, 7012090}.

Reliably extracting a desired  signal waveform  by enhancing SINR has a direct bearing on improving target detection and localization for radar   signal processing,   increasing   throughput or  channel capacity for MIMO wireless communication systems, and enhancing   resolution capability in  medical imaging \cite{Goldsmith:2005:WC:993515, Trees:1992:DEM:530789, 1206680}.  Maximizing  signal to noise ratio (MaxSNR) and MaxSINR criteria have been shown to yield significantly efficient beamforming performance and interference  mitigation. For sparse array design, the  MaxSINR beamforming performance depends mainly on the selected positions of the sensors as well as the locations of sources in the field of view (FOV) \cite{6774934,  1634819, 6714077, 8036237, 8061016, 8892512}.  It is noted that with sparse arrays, the commonly used Capon beamforming must not only find the optimum weights but also the optimum array configuration.  This is clearly an entwined optimization problem and requires attaining maximum SINR over all possible sparse array configurations.    \par

\begin{figure}[!t]
	\centering
	\includegraphics[width=5.35 in, height=3.2 in]{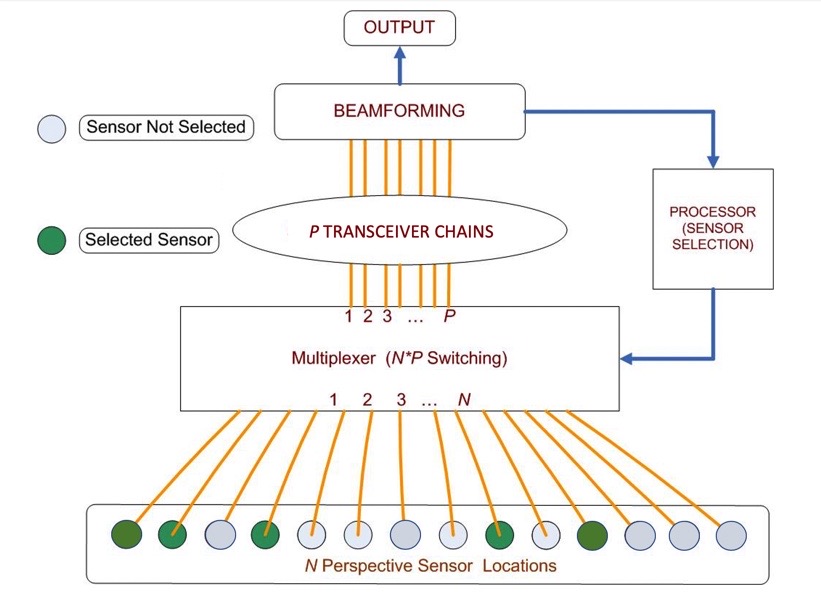}
	\caption{Block diagram of adaptive switched sensor beamformer}
	\label{Block_diagram1_ch8}
\end{figure}

Sparse array design typically involves the selection of a subset of uniform grid points for sensor placements. For a given number of sensors, it is assumed that the number of perspective grid points, spaced by half wavelength, is limited due to  a size constraint on the physical aperture. For environment-dependent sparse arrays, the antenna positions are selected from uniformly spaced locations that are served by a limited number of transceiver chains. The environment-sensitive array design objectives have recently become more realizable due to advances in efficient sensor switching technologies that readily activates a subset of sensors on a predefined grid points resulting in rapid array reconfigurability. Thereby, the system cost can significantly be reduced by limiting the number of expensive transceivers chains as shown in Fig. \ref{Block_diagram1_ch8} \cite{738234, 967086, 922993, 1299086}.

Environment-dependent sparse array design algorithms generally follow two different approaches. The designs based on prior knowledge of interfernce parameters, essentially require that the real time interference parameters such as DOAs and respective SINRs are either known or are estimated apriori in real time. The other approach doesn’t explicitly require the information of interfering environment which is the case in the Capon beamforming formulation. In both cases, several iterative algorithms have been proposed to optimize the sparse array beamformer design. Although, convex based optimization algorithms, such as semidefinite relaxation (SDR) and successive convex approximation (SCA) have been developed to yield sparse configurations with superior beamforming performances \cite{8061016, 8892512}, the real time implementation of these algorithms remains limited due to the relatively high computation cost. The latter is undesirable  in rapidly changing environments stemming from temporal and spatial nonstationary behaviors of the sources in the field of view.

\begin{figure}[!t]
	\centering
	\includegraphics[width=5.55 in, height=3.3 in]{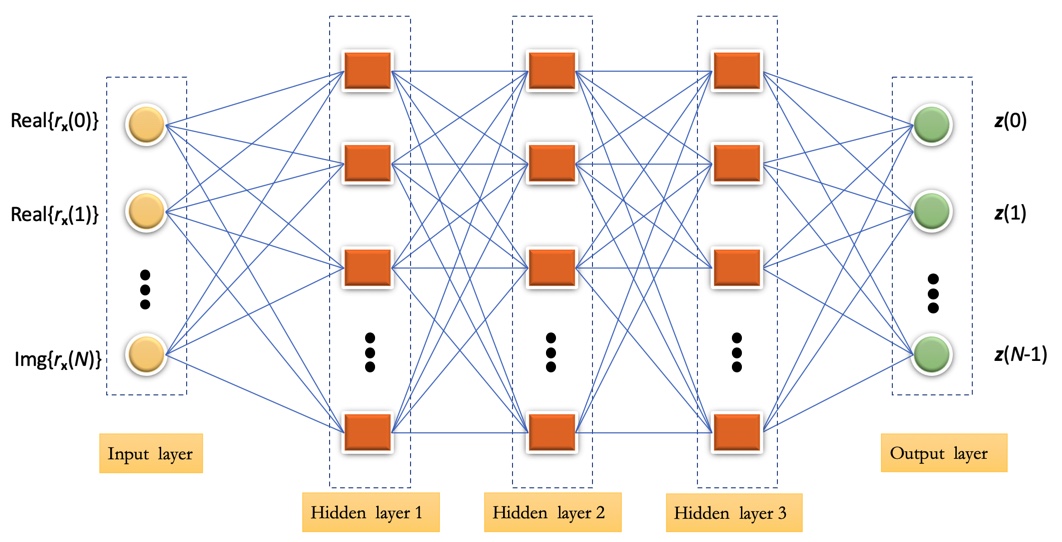}
	\caption{Architecture of Deep Neural Network (DNN)}
	\label{DNN_diagram2_ch8}
\end{figure}

In this chapter, we propose a sparse beamformer methodology implementing  data-driven array design by training the DNN to learn and mimic the sparse beamforming design algorithms \cite{9266359,
10.1117/12.2561526}. Towards this goal, the training scenarios are simulated in two different ways. In the first approach, we use enumeration technique to generate the training labels for any given received sensor correlation function. This is achieved by finding the MaxSINR configuration by sifting through all possible sparse configurations and choosing the best performing topology. Although in practice, the training data is generated offline,  it becomes infeasible to obtain optimum configuration even for a moderate size arrays due to the enormous sensor permutations. In order to circumvent this problem, we propose a  new technique to expedite the generation of large number to training data labels to input the DNN. For a given environment in the training set, this technique considers the array spatial spectrum and incorporates  the lag redundancy in determining desired array structures. Aside from efficient generation of DNN training data, the same technique can be used as a stand alone method to determine the best array configuration if prior information of the interference parameter is provided.   The DNN approximates the unknown mapping from the input correlation matrix to the output of sensor placements (Fig. \ref{DNN_diagram2_ch8}). It is shown that DNN effectively learns the optimum array structure. This makes DNN suitable for  real-time implementation  as DNN output prediction only requires a fewer number of trivial operations.

\textit{Prior work:} DNNs have shown great potential  due to its automatic feature selection with demonstrated ability of effective feature learning  in many applications including computer vision, speech recognition, and natural language processing. Recently, 'learn to optimize' approaches has been proposed to improve and  automate the implementation of DNNs iteslf, which  largely require  the laborious task  of designing the learning algorithms alongside  model and hyperparameter selection that needs to be manually  tailored from task to task.   In these methods, an additional DNN is trained  (called the meta-DNN) to   ensure better optimization of the original DNN (called the base-DNN)  and  generalize  over   many tasks, hence avoiding to redesign algorithm for  closely related   tasks \cite{DBLP:conf/iclr/LiM17, 10.5555/3157382.3157543, oshea2016recurrent}. 

 Depending on the task at hand, the  DNN employed to learn  can either be  trained by reinforcement learning or supervised learning. It has been shown that the reinforcement learning is a reliable approach, in case, the training examples are not independent and identically distributed (i.i.d.). This is precisely the case in optimizer learning  because the step vector towards the optimizer,  for any given iteration, affects the gradients at all subsequent iterations. 
On the other hand, the DNN  designs based on standard  supervised learning approach has been shown  to realize computationally efficient implementation of  iterative signal processing algorithms \cite{8444648,  DBLP:journals/twc/ShenSZL20}. These computationally intensive algorithms,  take a given set of design parameters and produce the “optimized” solutions as their outputs. The DNN learns from the training examples that are generated by running  these algorithms offline. Also, the parameters of the network are learned entirely offline, whereas efficient online implementation is realized by  passing the input through this pre-trained DNN which  requires a small number of simple operations to yield the optimized output.

In this chapter, we develop sparse array beamformer design  implementation using supervised training. Learning  sparse optimization techniques has been studied before in the contest of developing sparse representations and seeking simpler  models \cite{articlesparseli}. Unlike,  sparse representation approaches the sparse beamformer optimization refers to  'sparsity in the sensing' rather than 'sparsity in the sensed' and has important distinctions. Learning sparse algorithms, thus far, are mainly focused   on iterative “unfolding” concept implemented by a single layer of the network \cite{DBLP:journals/corr/OSheaEC17, hershey2014deep, doi:10.1137/080716542, 10.5555/3104322.3104374, articlesparseli}. The proposed approach  approximate rather simple algorithms implemented through iterative soft-thresholding such as  ISTA algorithm for sparse optimization. The sparse beamformer design, on the other hand, involves intricate  operations such as singular value decomposition and matrix inversion. Also, the  sparse beamfomer designs mainly  implemented through convex relaxation  are based on SDR and SCA algorithms involving sparsity promoting regularizers and are very expensive to implement in real time. 

\textit{Main Contributions:} The main contributions of this chapter can be summarized as follows,

1) Sparse beamformer spectral analysis (SBSA) algorithm is proposed which provides an insightful perspective for MaxSINR beamformer design with reduced computational complexity and superior performance. The design elucidates the concept of ’inherently amenable’ sparse configurations for interference mitigation.

 2) The DNN based approach is developed, for the first time, to configure a Capon based data driven sparse beamfomer by learning the enumerated algorithm as well as SBSA design. The design is achieved through a direct mapping of the received correlation matrix to the optimum sparse configuration for a given `look direction'. The proposed methodology combines the merits of the data dependent designs and the designs assuming prior information of the interfering environment.
 
3) The proposed design is shown to be robust in the case of limited data snapshots and can easily be extended to robust adaptive beamforming to cater the uncertainty regarding the look direction DOA as well as array calibration errors.

The rest of the chapter is organized as follows: In the next section, we state the problem formulation for maximizing the output SINR. Section 7.3 deals with the optimum sparse array design by SBSA algorithm and DNN based Capon implementation. In section 7.4, with the aid of number of design examples, we demonstrate the usefulness of proposed algorithms in achieving MaxSINR sparse array design. Concluding remarks follow at the end.

\section{Problem Formulation} \label{Problem Formulation}
Consider  a desired source and $L$ independent interfering source signals impinging on a  linear array with $N$ uniformly placed sensors. The baseband signal received at the array  at  time instant $t$ is then given by; 
\begin {equation} \label{a_ch8}
\mathbf{x}(t)=   (\alpha(t)) \mathbf{s}( \theta)  + \sum_{l=1}^{L} (\beta _l(t)) \mathbf{v}( \theta_l)  + \mathbf{n}(t),
\end {equation}
where, $\mathbf{s} ({\theta_k})$  and $\mathbf{v} ({\theta_l})$ $\in \mathbb{C}^N$ are  the  steering vectors   corresponding to the direction  of arrival, $\theta$ or $\theta_l$, and are defined  as follows;
\vspace{+2mm}
\begin {equation}  \label{b_ch8}
\mathbf{s} ({\theta})=[1 \,  \,  \, e^{j (2 \pi / \lambda) d cos(\theta)  } \,  . \,   . \,  . \, e^{j (2 \pi / \lambda) d (N-1) cos(\theta)  }]^T.
\end {equation}
where $d$ is the inter-element spacing  and ($\alpha (t)$, $\beta _l(t))$ $\in \mathbb{C}$  are the complex amplitudes of the incoming baseband signals \cite{trove.nla.gov.au/work/15617720}. The additive Gaussian noise $\mathbf{n}(t)$ $\in \mathbb{C}^N$ is of   variance of  $\sigma_n^2$.
The received signal vector $\mathbf{x}(t)$   is   combined linearly by the $N$-sensor  beamformer that strives to maximize the output SINR. The output signal $y(t)$ of the optimum beamformer for maximum SINR is given by \cite{1223538}, 
\begin {equation}  \label{c_ch8}
y(t) = \mathbf{w}_o^H \mathbf{x}(t),
\end {equation}
where $\mathbf{w}_o$ is the solution of the  optimization problem given by,
\begin{equation} \label{d_ch8}
\begin{aligned}
\underset{\mathbf{w} \in \mathbb{C}^N}{\text{minimize}} & \quad   \mathbf{w}^H\mathbf{R}_{s^{'}}\mathbf{w},\\
\text{s.t.} & \quad     \mathbf{ w}^H\mathbf{R}_{s}\mathbf{ w}=1.
\end{aligned}
\end{equation}
For statistically independent signals, the desired source correlation matrix is $\mathbf{R}_s= \sigma^2 \mathbf{s}( \theta)\mathbf{s}^H( \theta)$, where $ \sigma^2 =E\{\alpha (t)\alpha ^H(t)\}$. Likewise,  the  interference and noise correlation matrix, $\mathbf{R}_{s^{'}}= \sum_{l=1}^{L} (\sigma^2_l \mathbf{v}( \theta_l)\mathbf{v}^H( \theta_l)$) + $\sigma_n^2\mathbf{I}_{N\times N}$,    with $ \sigma^2_l =E\{\beta _l(t)\beta_l^H(t)\}$ being the power of the $l$th interfering source. The problem in (\ref{d_ch8}) can  be written equivalently by replacing $\mathbf{R}_{s^{'}}$ with the received data covariance matrix, $\mathbf{R_{xx}}=\mathbf{R}_s+ \mathbf{R}_{s^{'}}$  as follows \cite{1223538},
\begin{equation} \label{e_ch8}
\begin{aligned}
\underset{\mathbf{w} \in \mathbb{C}^N }{\text{minimize}} & \quad   \mathbf{ w}^H\mathbf{R_{xx}}\mathbf{ w},\\
\text{s.t.} & \quad     \mathbf{ w}^H\mathbf{R}_{s}\mathbf{ w} \geq 1.
\end{aligned}
\end{equation}

\begin{figure*}[!t]
	\centering
	\includegraphics[width=6.55 in, height=3.3 in]{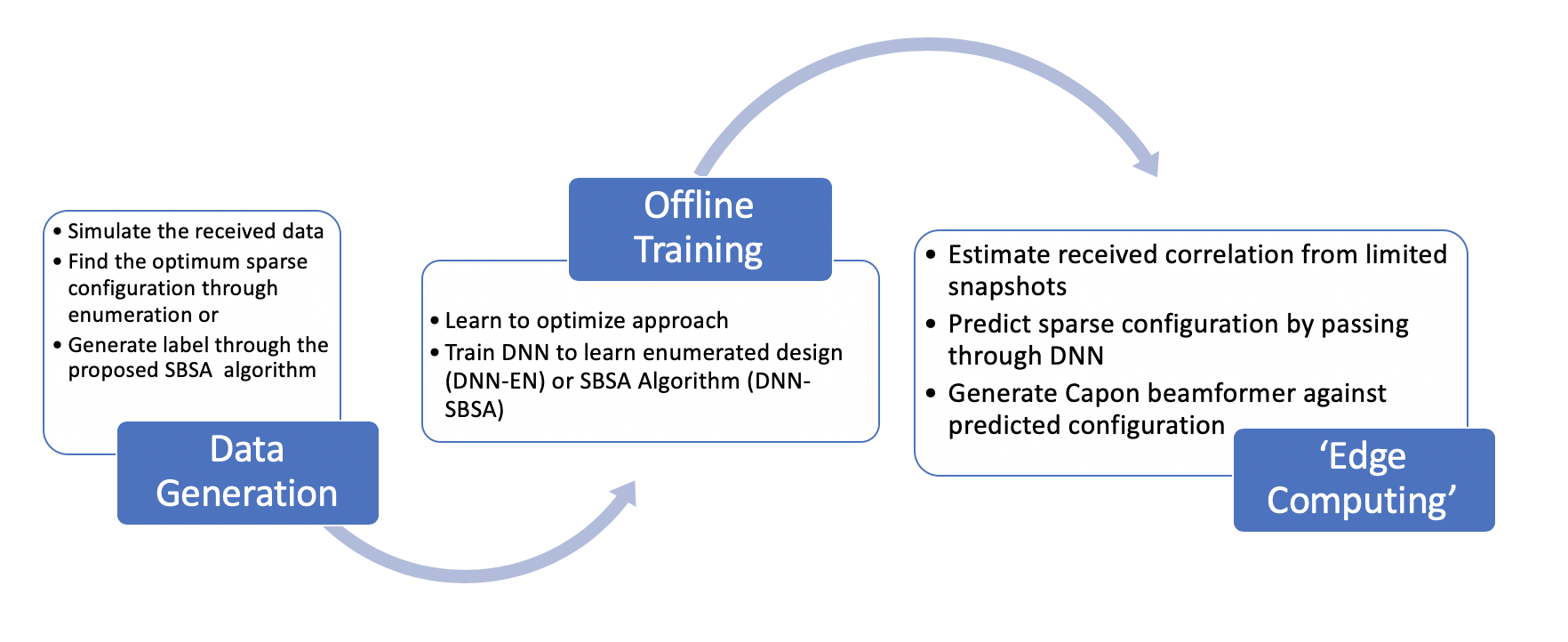}
	\caption{Overview of the proposed  approach using Deep Neural Network (DNN)}
	\label{Smart_diagram_ch8}
\end{figure*}
It is noted that the  equality constraint in (\ref{d_ch8}) is  relaxed in (\ref{e_ch8}) due to  the inclusion of  the constraint as part of the objective function, and as such, (\ref{e_ch8})  converges to  the equality constraint. Additionally, the optimal solution in (\ref{e_ch8}) is invariant up to  uncertainty in  the absolute power of the source of interest.   However, in practice, these assumptions can deviate from the actual received data statistics and hence the discrepancy is typically mitigated, to an extent,  by preprocessing  the received data correlation matrix through diagonal loading or tapering the  correlation matrix \cite{1206680}.

     The closed form solution of the above  optimization problem exists  and  is given by $\mathbf{w}_o = \mathscr{P} \{  \mathbf{R}_{s^{'}}^{-1} \mathbf{R}_s  \}=\mathscr{P} \{  \mathbf{R_{xx}}^{-1} \mathbf{R}_s  \}$. The operator $\mathscr{P} \{. \}$  computes the principal eigenvector of the input matrix. Substituting $\mathbf{w}_o$ into  (\ref{c_ch8})  yields the corresponding optimum output SINR$_o$;
\begin{equation}  \label{f_ch8}
\text{SINR}_o=\frac {\mathbf{w}_o^H \mathbf{R}_s \mathbf{w}_o} { \mathbf{w}_o^H \mathbf{R}_{s^{'}} \mathbf{w}_o} = \Lambda_{max}\{\mathbf{R}^{-1}_{s^{'}} \mathbf{R}_s\}.
\end{equation}
 This shows that the optimum output SINR$_o$ is given by the maximum eigenvalue ($\Lambda_{max}$) associated with   the product of  the inverse of interference plus noise correlation matrix  and the desired source correlation matrix. Therefore, the performance of the optimum beamformer for maximizing the output SINR is directly related to the desired and interference plus noise correlation matrices. 

\section{Optimum sparse array design}  \label{Optimum sparse array design}
 The  constraint optimization (\ref{e_ch8}) can be re-formulated for optimum sparse array design by incorporating  an additional constraint on the cardinality of  the weight vector;   
\begin{equation} \label{a2_ch8}
\begin{aligned}
\underset{\mathbf{w \in \mathbb{C}}^{N}}{\text{minimize}} & \quad  \mathbf{ w}^H\mathbf{R_{xx}}\mathbf{w},\\
\text{s.t.} & \quad   \mathbf{ w}^H\mathbf{R}_{s}\mathbf{ w}\geq1,  \\
& \quad   ||\mathbf{w}||_0=P.
\end{aligned}
\end{equation}
Here,  $||.||_0$ determines the cardinality of the weight   vector $\mathbf{w}$.  This is a combinatorial optimization problem and can be solved optimally by enumerating over all possible locations.   Several different approaches have been developed to mitigate the computational expense of the   combinatorial search either by exploiting the  information of the interference parameters  or employing  data-dependent approaches realized through the SDR and SCA algorithms.  These algorithms have high computational costs, impeding  real time implementations, especially in applications involving rapidly changing environments. In the context of designing sparse arrays using machine learning, the optimum sparse arrays for different operating environments constitute labels for the network training data. These labels can be generated off-line  through enumerations. Although, this procedure is off line, accounting for all possible source and interference scenarios leads to a formidable computational  problem rendering solutions for large arrays and  a large number of training samples infeasible. In order to simplify the problem, we propose a  sparse beamformer spectral analysis (SBSA) design technique to efficiently train the DNN.  This technique  is detailed below, and  has improved  performance  and yields lower computational complexity as compared to the state-of-the-art. The subsequent subsection presents  the implementation of the SBSA and enumerated design algorithms  using the DNN approach, this further reduces the implementation time and also paves the way to data dependent design (refer to Fig. \ref{Smart_diagram_ch8} for an overview). 
  
\subsection{Sparse beamformer spectral analysis (SBSA) design}
 \begin{figure*}[!t]
	\centering
	\includegraphics[width=5.48 in, height=0.55 in]{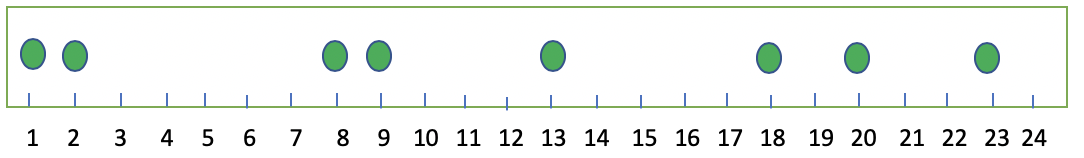}
	\caption{Eight element sparse array configuration}
	\label{mvrk-1_ch8}
\end{figure*}
\begin{figure*}[!t]
	\centering
	\includegraphics[width=6.48 in, height=2.55 in]{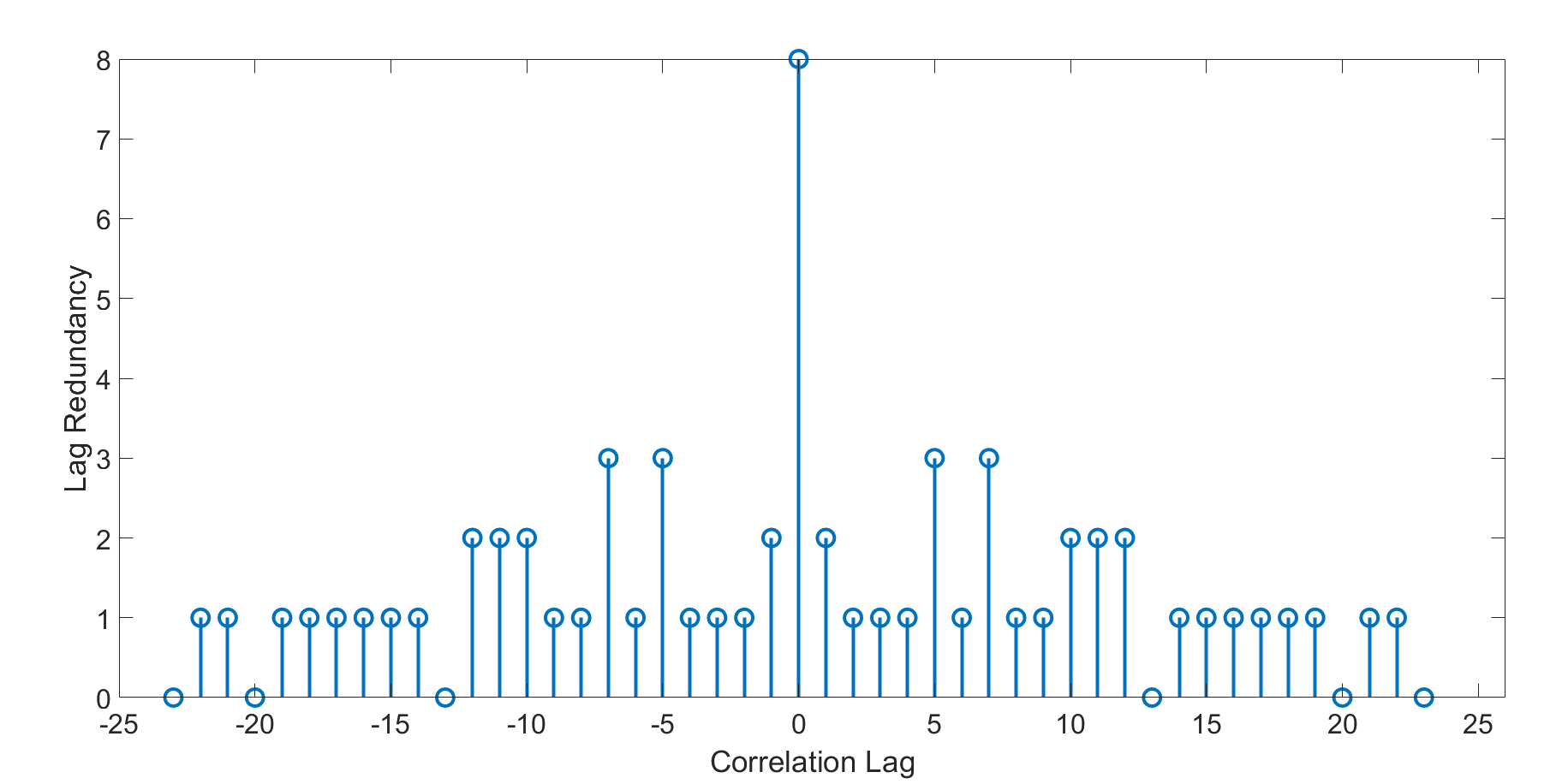}
	\caption{Lag Redundancy  of the sparse array shown in Fig. \ref{mvrk-1_ch8}}
	\label{mvrk-2_ch8}
\end{figure*}
\begin{figure*}[!t]
	\centering
	\includegraphics[width=6.48 in, height=2.55 in]{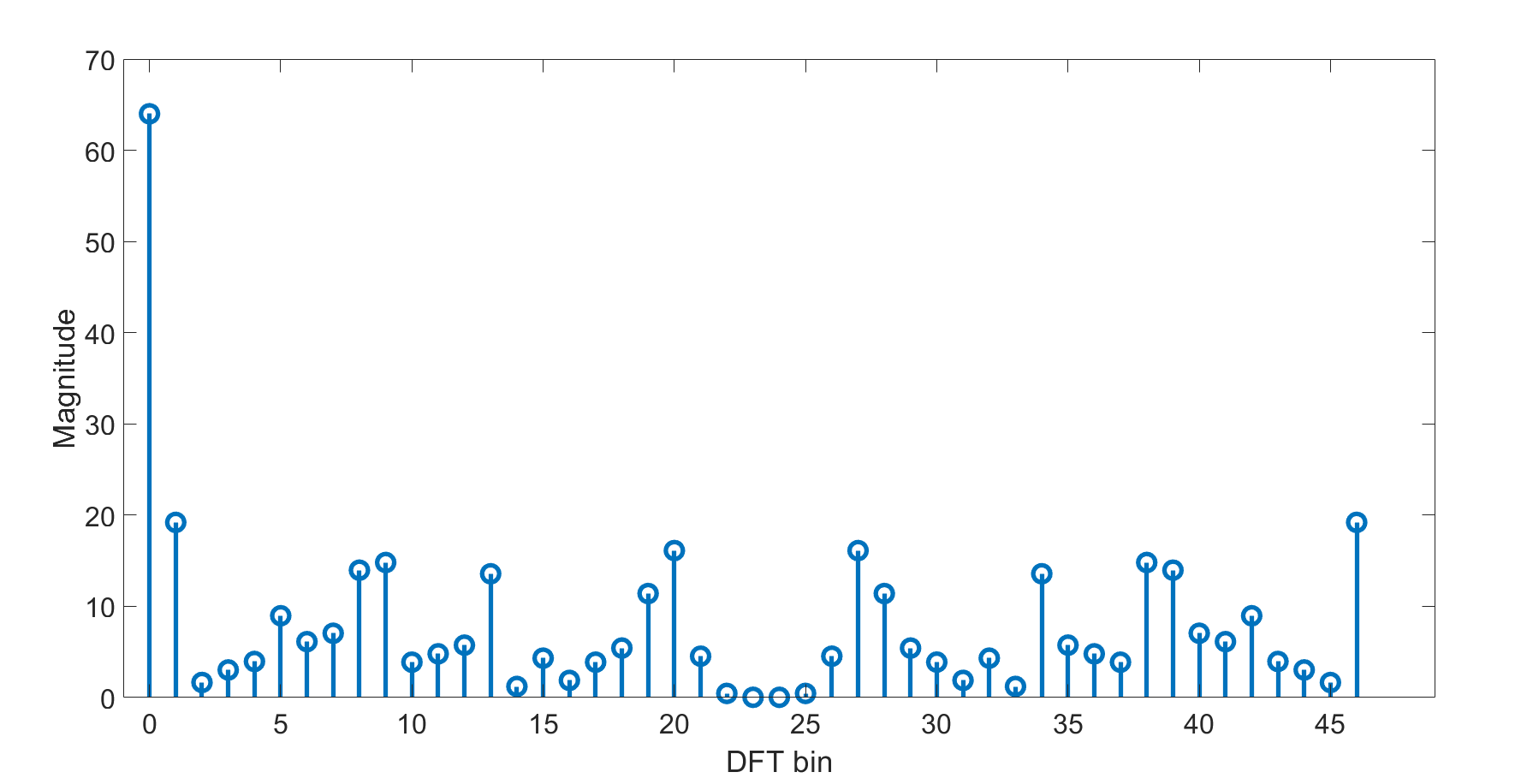}
	\caption{DFT  of the lag redundancy }
	\label{mvrk-3_ch8}
\end{figure*}
\begin{figure*}[!t]
	\centering
	\includegraphics[width=6.28 in, height=2.55 in]{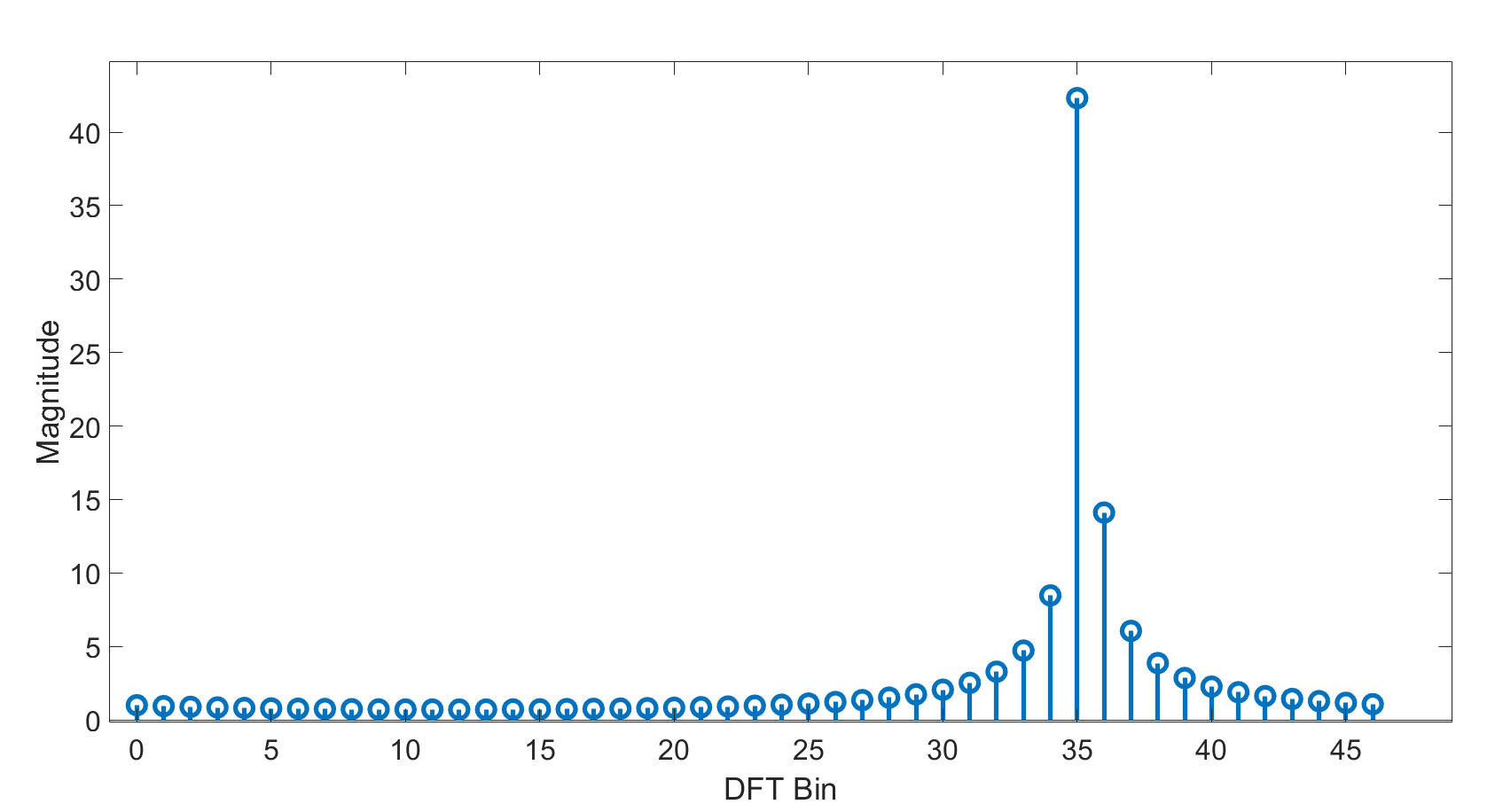}
	\caption{Power spectrum  of the desired signal}
	\label{mvrk-4_ch8}
\end{figure*}
\begin{figure*}[!t]
	\centering
	\includegraphics[width=6.78 in, height=2.75 in]{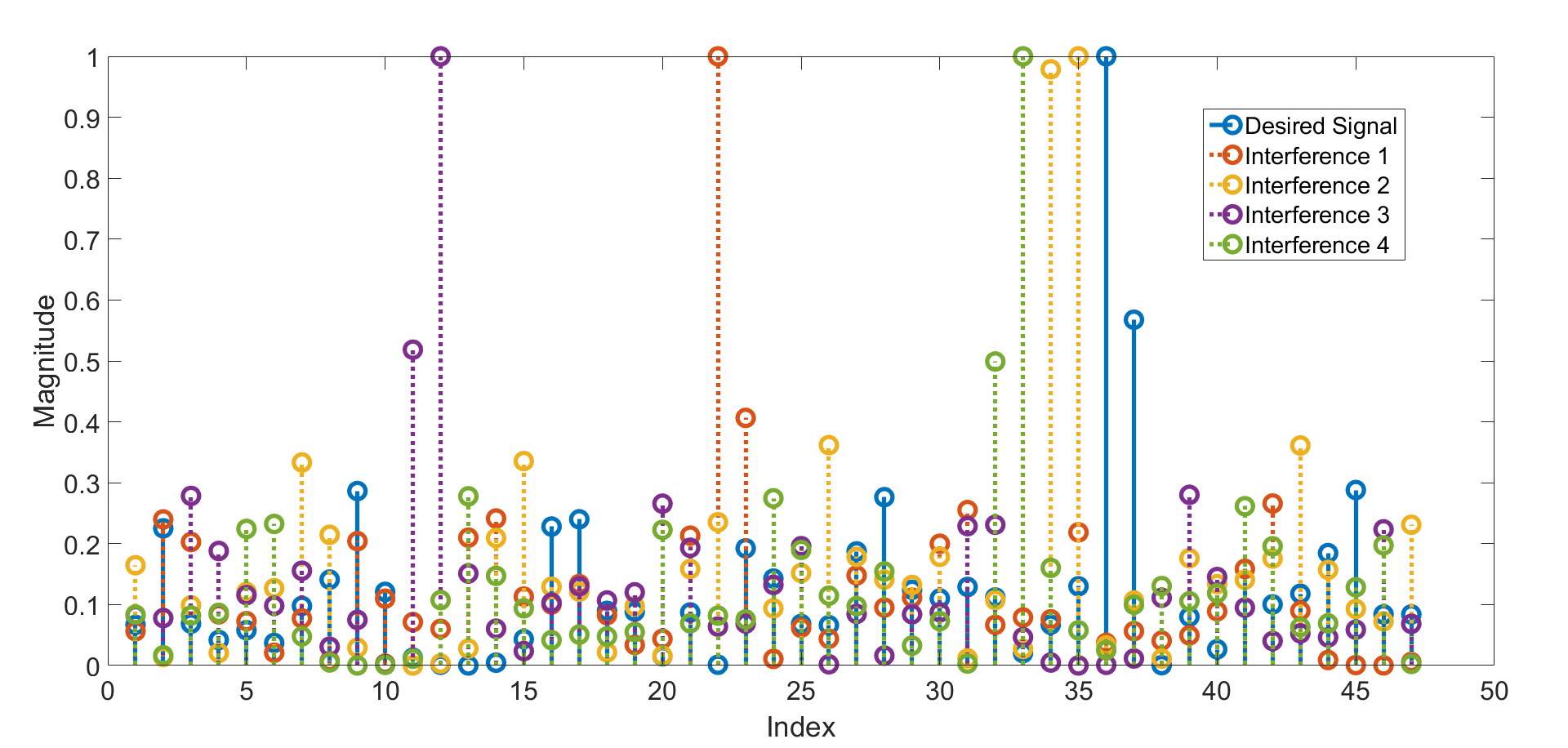}
	\caption{Explanation of the proposed objective criterion  for the optimum array configuration shown in Fig. \ref{mvrk-1_ch8}}
	\label{mvrk-5_ch8}
\end{figure*}
\begin{figure*}[!t]
	\centering
	\includegraphics[width=6.48 in, height=2.75 in]{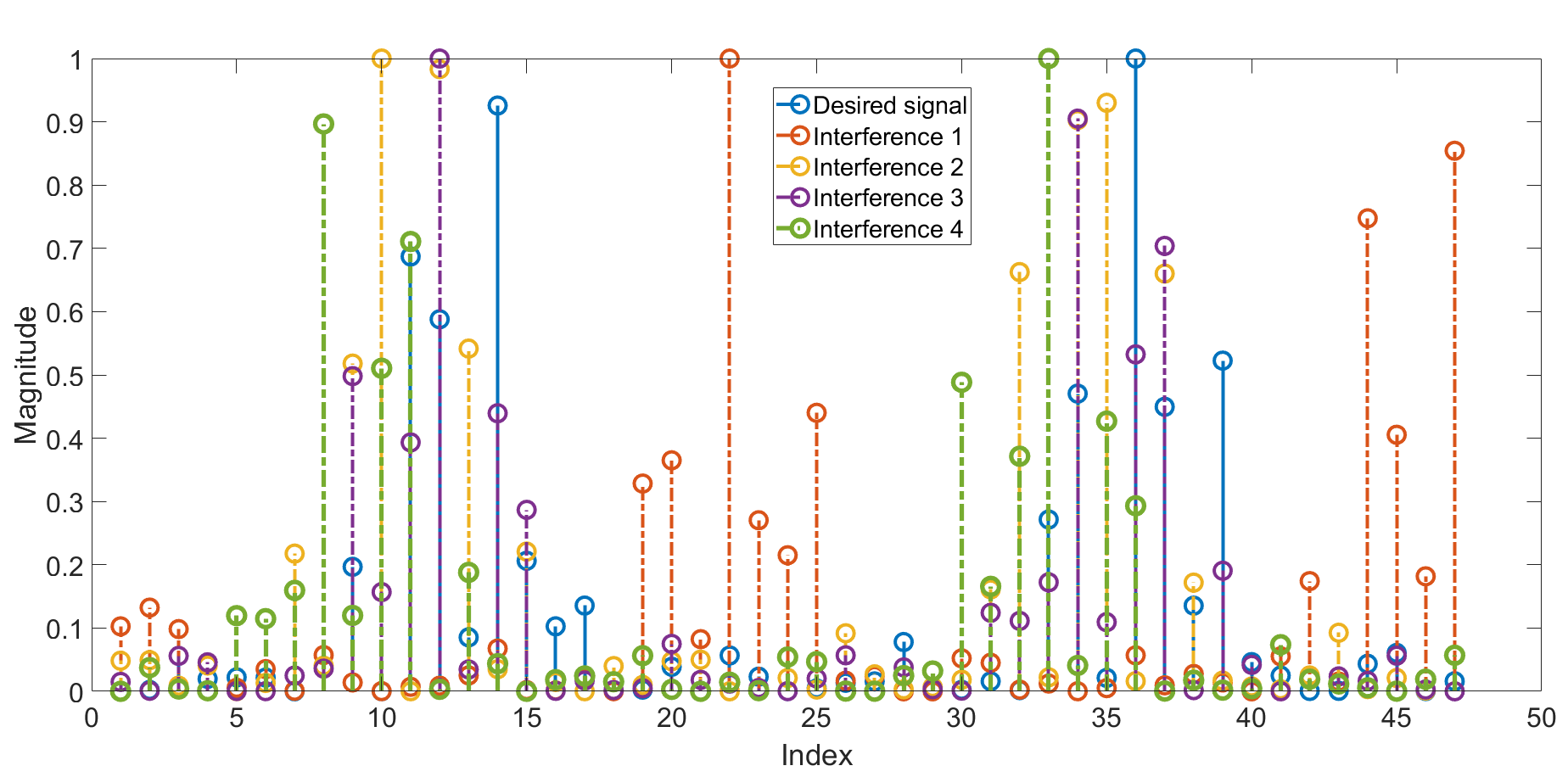}
	\caption{ Explanation of the proposed objective criterion for the worst possible array configuration}
	\label{mvrk-6_ch8}
\end{figure*}
\begin{figure*}[!t]
	\centering
	\includegraphics[width=6.48 in, height=3.15 in]{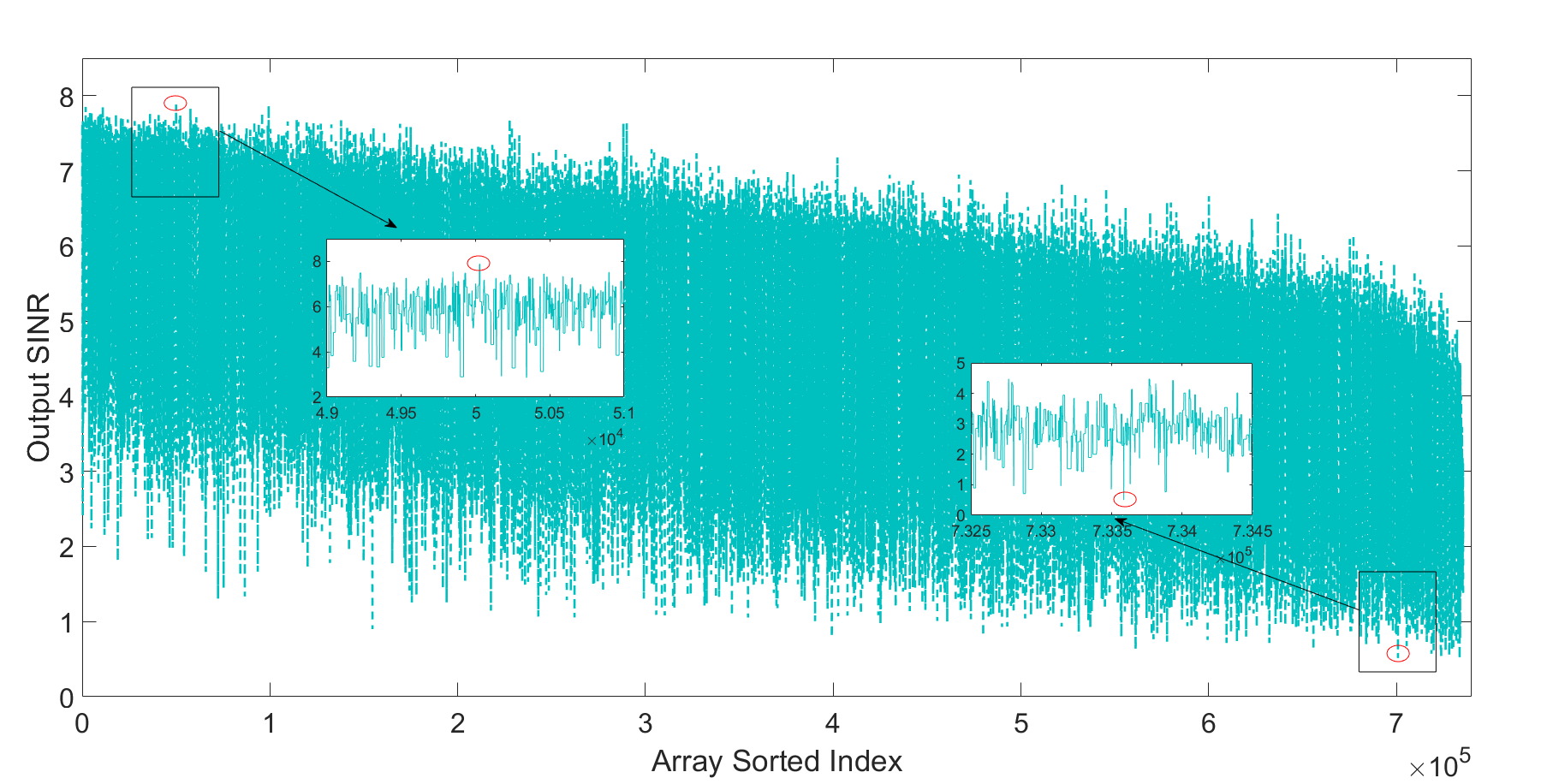}
	\caption{Plot of the proposed objective criterion   in ascending order}
	\label{mvrk-7_ch8}
\end{figure*}
 As shown in  Eq. \ref{f_ch8}, the  sparse array configuration for MaxSINR design depends on the beamforming weights and also on the  sparse array configuration. Therefore, the problem of interference mitigation is not  a cascade design but is an entwined task, calling for the simultaneous optimization for the beamforming weights and the sparse array configuration. We adopt a frequency domain approach offering a unique insight to the problem and addressing the suitability of a given sparse configuration, irrespective of the beamforming weights, in canceling the interfering signals. 
 
 The problem formulation developed in the previous section is valid irrespective of the array configuration and holds true for the compact ULA or any  generic sparse configuration. The beamformer output signal, $y(t) = {\mathbf{w}}^H \mathbf{x}$,  for a sparse beamfromer can  be rewritten as $\overset{\circ}{y}(t) = \overset{\circ}{\mathbf{w}}^H \mathbf{x}$. The typeset `${{\circ}}$' indicates that the corresponding vector is sparse with few zero entries. The sparse beamformer $\overset{\circ}{y}(t)$ can also be rewritten, equivalently, as $\overset{\circ}{y}(t) = \overset{\circ}{\mathbf{w}}^H \{\overset{\circ}{\mathbf{z}}\odot \mathbf{x}\}$. Here, the point-wise multiplication ($\odot$) of  the   received vector $\mathbf{x}$  with a sparse selection vector $\overset{\circ}{\mathbf{z}}$ sets the entries of the received signal to zero which corresponds to the  zero beamforming weights of $\overset{\circ}{\mathbf{w}}$. The entries of $\overset{\circ}{\mathbf{z}} \in \mathbb{R}^N$ are either 1's or 0's depending on whether the corresponding sensor location is active or inactive respectively. 
 
 
 The above formulation shows that the sparse beamforming filter $\overset{\circ}{\mathbf{w}}$ can also be viewed as pre-processing the  received signal by point-wise multiplication,   prior to applying the beamforming filter. We analyze how the point-wise multiplication of received signal with the selection vector can either help or hinder the performance of the subsequent beamforming filter $\overset{\circ}{\mathbf{w}}$.  Denote the DFT of the input data as $\mathbf{X}= \mathscr{F}(\mathbf{x})$ and $\mathbf{Z}= \mathscr{F}(\overset{\circ}{\mathbf{z}})$, respectively. Then, the DFT of the pointwise multiplication of these two vectors is given by the circular convolution of the correspond DFTs,  $\mathscr{F}(\overset{\circ}{\mathbf{z}}\odot \mathbf{x})=\mathbf{X} \circledast \mathbf{Z}$, where  $\circledast$ denotes the circular convolution.  Furthermore,  because the received data vector  can be written as $\mathbf{x}=   \alpha \mathbf{s}( \theta_k)  + \sum_{l=1}^{L} \beta _l\mathbf{v}( \theta_l)  + \mathbf{n}$, then by invoking the linearity of the DFT transform, we obtain  $\mathscr{F}(\overset{\circ}{\mathbf{z}}\odot \mathbf{x})=\alpha \mathbf{S} \circledast \mathbf{Z}+ \sum_{l=1}^{L}\beta_l \mathbf{V}_l \circledast \mathbf{Z}+\mathbf{N}.  \circledast \mathbf{Z}$.  For an effective sparse beamformer, it is desirable to minimize the  overlapping of the spatial spectrum of the desired  source and the spectra of interfering signals.  This action is poised to  aid the subsequent beamformer filtering  to effectively suppress  the  unwanted signals. We propose  a design metric $\Omega(\mathbf{Z})$ based on  weighted sum of the   spatial spectrum of the individual interfering signals scaled by the desired signal  spatial spectrum as follows,
 \begin {equation}  \label{c3_ch8}
\Omega(\mathbf{Z}) = \sum_{l=1}^{L} \{\alpha^2 |\mathbf{S}|^2 \circledast |\mathbf{Z}|^2\} \odot\{\beta_l^2 |\mathbf{V}_l|^2 \circledast |\mathbf{Z}|^2\}
\end {equation}
 It is clear that  Eq. \ref{c3_ch8} performs  element wise scaling of  the interfering powers in the DFT domain.  Therefore, if the interfering signal power,  after convolution through the selection filter, is concentrated in the    DFT bins different from   those  occupied primarily by the desired signal, then the pointwise product is always lower. Conversely, if there is a significant overlap between the results of the two convolutions, then the objective function  $\Omega(\mathbf{Z})$ is significantly higher.
 
 The spatial spectrum can be estimated  by computing the  DFT of the autocorrelation function of the corresponding signal. It is worth noting that for a given sparse configuration, the autocorrelation function of the selection vector is given by the corresponding  redundancy  of the autocorrelation lags. Therefore, unlike the structured sparse array design which seeks to maximize the contiguous correlation lags, the MaxSINR sparse design is directly linked   to the DFT of the autocorrelation sequence of the  lag redundancy.

 We   illustrate the proposed approach with the help of the following example.
  Consider an 8-element sparse array on the 24 point   equally spaced grid locations that are potentially available for sensor selections. The minimum spacing  among the  sensors is $d=\lambda/2$. Consider a source signal  located at $60^0$ and  four unwanted intereferneces  located at $154^0$, $55^0$, $117^0$ and $50^0$ with the  INRs ranging from $10-20$ dB. The sparse array configuration achieving the best performance is found through enumeration and shown in  Fig. \ref{mvrk-1_ch8}. The associated correlation lag redundancy of this configuration and the corresponding spatial spectrum is depicted in Figs.  \ref{mvrk-2_ch8} and \ref{mvrk-3_ch8}, respectively. The spatial spectrum  of the desired source at $60^0$  is depicted in the Fig. \ref{mvrk-4_ch8}. The resultant normalized spectrum of the two convolved  spectra in Figs \ref{mvrk-3_ch8},\ref{mvrk-4_ch8} is shown as  solid lines (blue) in the Fig. \ref{mvrk-5_ch8}. The normalized  spectrum for each interfering signal is shown as dotted lines in the same figure. Fig. \ref{mvrk-6_ch8} plots the normalized spectrum for the sparse array worst case scenario for comparison purposes. Note the maximum of the convolved desired signal spectrum  is at 35th DFT bin. For the best case scenario,  all the convolved interfering signals assume minimum power at the aforementioned DFT position. This is in contrast to the  worst case scenario depicted in  Fig. (\ref{mvrk-6_ch8}) where  the convolved interfering signals have considerable power at the maxima of the desired source. Apart form the maxima location, it is noted that for the best case design, there is  minimum overlapping between the desired signal and interfering signals at the   DFT bin locations which is clearly in contrast to the worst case design.
 
 In order to further understand the effectiveness of the proposed approach, Fig. (\ref{mvrk-7_ch8}) plots the SINR performance of all  possible sparse configurations  after sorting the array topologies in ascending order  of the output of the proposed objective function. It is clear that the average SINR in the plot is  higher and more desirable towards the left side where the objective function is minimum  vis a vis the right side where the objective function  is high. It is also noted that  the best enumerated result of  MaxSINR performance does not correspond to  the array configuration with the smallest objective function. This is because, the optimum sparse configuration also depends on  the  beamformer weights to  minimize the interfering signals. This is also clear from the high variance of the curve. Similarly, the worst performing array is very close to the right side of the curve where the proposed objective function is high due to the significant overlap of the desired and undesired signal spectra. The significantly reduced  performance  towards the right side of the plot can   now be  explained  intuitively  in terms of the inherent inability of these configurations to mitigate the significant overlap in the desired and interfering signals which is an artifact of the sparse beamformer explained in the frequency domain.

 Therefore, in  light of the above discussions, it becomes prudent to seek sensor arrangements  that  minimize the proposed objective function in an attempt to find more suitable  sparse configuration for MaxSINR design.   Towards this end, we propose an iterative algorithm that minimize the proposed objective function  by successive sensor selection, hence deciding on  one senor  location at a time.  For the  initial iteration, a  sensor location is chosen randomly on an $N $ grid points.  For  the $i$th subsequent iteration, the proposed objective is evaluated at the remaining $N-i$ locations and then selecting  the  sensor location that yields the minimum objective function. The procedure is iterated $P$ times for selecting the  $P$ locations from $N$ possible locations.  Due to  high variance of the curve  in the above figure, it is best to initialize  the algorithm with  different sensor location or DFT size and find the corresponding configurations for each initialization, and eventually consider the one with the  best SINR performance.  Steps of the algorithm are detailed in  Table \ref{algorithm_ch8}.

\begin{table}[t!] 
\noindent\rule{16.5cm}{0.6pt}
	\caption{SBSA Algorithm}
\noindent\rule{16.5cm}{0.6pt}

	\begin{algorithmic}
		
		\renewcommand{\algorithmicrequire}{\textbf{Input:}}
		
		\renewcommand{\algorithmicensure}{\textbf{Output:}}
		
		\REQUIRE  $N$, $P$, Look direction DOA $\theta$, Interfernce DOAs, SNR and INRs  \\
		
		\ENSURE  Sparse beamformer $\mathbf{w}_o$
				       	\STATE Initialize  $\mathbf{z}$=$[0 \, \, 1 \, \, 0\,  ... \, 0]$ where  all entries of $\mathbf{z}$ are zero except an arbitrarily selected  entry \\
				     	\STATE Compute the spatial spectrum of desired source and interfering signals \\
	\STATE 	\textbf{For} (j=1 to $P$-1)
    \STATE 	\textbf{For} (i=1 to $P$-1-j)
        	\STATE  Select the i\textit{th} sensor  from $P$-1-j remaining locations
   	\STATE Compute the lag redundancy of this  sparse array consisting of  j+1  sensor
   	\STATE Compute the spatial spectrum of the j+1 sensor sparse array
   	\STATE Convolve the spatial spectrum of the j+1 sensor sparse array with the spectrum of the desired source and the interfering sources
     	\STATE Compute the overlapping power in the spatial spectra by computing the proposed metric in Eq. \ref{c3_ch8} \\
                    \textbf{EndFor}\\
		    	\STATE Select the $i$th sensor  from the inner for loop which  results in the minimum overlapping power computed by Eq. \ref{c3_ch8} \\
	\STATE Update $\mathbf{z}$  by setting the j\textit{th}  location in $\mathbf{z}$ to 1 \\ 
                    \textbf{EndFor} \\
			\STATE  After finding the sparse configuration find $\mathbf{w}_o$ by running Eq. \ref{e_ch8} for reduced size correlation matrix while ignoring the sensor locations corresponding to $\mathbf{z}$ 
	\end{algorithmic}
	\label{algorithm_ch8}
\noindent\rule{16.5cm}{0.6pt}

\end{table}

\subsection{DNN based learning of the SBSA and enumerated  designs }

Modeling  the behaviour of optimization algorithms by training   a DNN is justifiable,  as DNNs are universal  approximators for arbitrary continuous functions and have the required capacity to accurately map almost  any  input/output formula. For effective learning, it makes sense that the DNN can generalize to a more broader class that is represented by the finite number of drawn training examples. From Capon beamforming perspective, it is instructive to realize that  a given arrangement of a desired source direction, interference DOAs and respectives SNR/INRs constitute one particular example. The class, in this case, would constitute any arbitrary permutation of the interference DOAs and respective powers while keeping the desired  source DOA fixed.  The DNN task is, therefore, to learn from a dataset, characterized by a set of different training examples and corresponding  optimum sparse array predictions. 

Here, an important question arises that whether we could aim for a more stronger notion of generalization, i.e., instead of training for a fixed desired source, is it possible to  generalize over all possible desired source DOAs. To answer this query, we note that  for a given desired  source and interference setting, the received signal remains the same even if one of the interfering  signal is assumed as a desired signal and the desired signal is treated as an interference. In this case, although the received signal is identical in both scenarios yet the Capon beamformer task is flipped and is now required to be pointing to an entirely different direction. In so doing, the corresponding  optimum sparse configuration and beamformer weights could be very  different than before. Therefore, instead of relying entirely on the information of the received data correlation it is imperative to incorporate the knowledge of the desired source or look direction,   which is always assumed in Capon beamforming formulation. For DNN learning, this information can either be incorporated by exclusively training the DNN for each desired source DOA or the desired source DOA can be incorporated as an additional input feature to DNN. In this chapter we adopt the former approach.

Our proposed approach uses a fully connected neural network with the input layer of size $2N-1$ and the output layer of size $N$. Although there are $N$ unique correlation lags corresponding to the $N$ sensor locations on the grid,  the dimensionality of the input layer is $2N-1$. This is due to concatenating the real and imaginary entries of the generally complex valued  correlation lags except the  zeroth lag.  We have 3 hidden layers with 450, 250 and 80 nodes, respectively. The ReLU activation function is used for all  hidden layers activation. The network is input the correlation  values of the received data assuming a stationary environment. The network output is a binary vector such that 1 indicate sensor selection and 0 indicates the absence of the corresponding sensor location. 
For the scope of this chapter, we  assume that we have an estimate of all the filled co-array  correlation lags corresponding to the correlation matrix of the  full  aperture array.    

The received data is generated in the following manner. For a given desired source location, the $i$th training  realization   is simulated by randomly selecting $L$ interfering signals from a DOA grid spanning the range of  $10^0$  to $170^0$, with a grid spacing of $1^0$. The interfering jammers are allocated random powers uniformly distributed with INR from 10 dB to 20 dB. For this given scenario, the  received correlation function, which includes the desired source signal, is calculated corresponding to the full sensor configuration. The corresponding optimum configuration is found through  enumeration and also through the proposed SBSA algorithm.  The process is repeated  30000 times against a given desired source DOA to complete the training data set.  Similarly,  a small sized validation data set is generated for hyperparamter tuning for model selection, minibatch size and learning rate. 

For the training stage,  the   weights of the neural network are optimized to minimize  the mean squared error between the label and the output of the network. The ADAM algorithm is used to carry out parameter optimization. This algorithm  used  an efficient implementation of
mini-batch stochastic gradient descent called the ADAM algorithm \cite{DBLP:journals/corr/KingmaB14}, that improves  learning as a function of the most recent gradient values. The  learning rate is set to 0.001 and dropout regularization with the keep probability of 0.9 is used. The initial weights are initialized using the Xavier initialization. 

The robustness of the learned models is demonstrated by  generating the testing data  that is different from the training stage, and is generated by  assuming the DOA of the interfering signals off grid. This is simulated by adding the Gaussian noise to the interference DOAs on the grid. We also present  the results under  limited and unlimited data snapshots. Additionally, to ensure the selection of $P$ antenna locations   at the output of DNN, we  declare the $P$ highest values in the output as the selected sensor locations. Therefore, generalization in this context means that the learned DNN works on different interference settings which can change according to changing environment conditions.

\section{Simulations} \label{Simulations}
In this section, we show the effectiveness of the proposed approach for sparse array design achieving  MaxSINR. The   results are examined first by training the DNN to learn  the enumerated optimum array configurations. Then,  we  demonstrate in the follow-on examples  the effectiveness  of the DNN when trained by the labels drawn from SBSA algorithm. 

\subsection{Enumerated design} \label{Single point source}

In this example, we pose the problem as selecting $P=6$ antennas from $N=12$  possible equally spaced locations with  inter-element spacing of $\lambda/2$. For all  numerical results, we use a network with three hidden layers, one input layer, and one output layer. Accordingly, the input to the network is of size 23 and output is size 12.

\begin{figure}[!t]
	\centering
	\includegraphics[width=5.6 in, height=3.6 in]{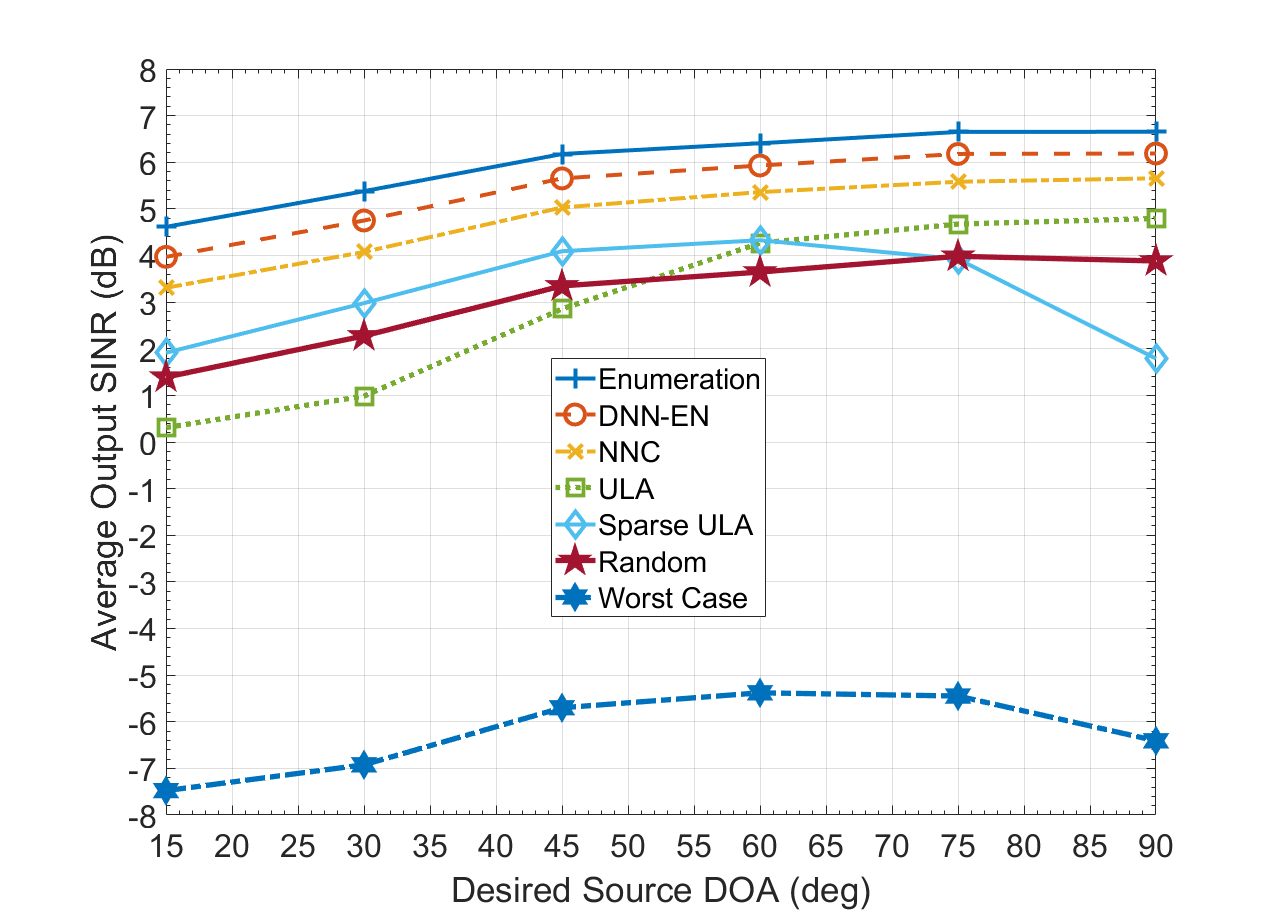}
	\caption{Performance comparison enumerated design under unlimited snapshots}
	\label{DNN-1_ch8}
\end{figure}

Figure \ref{DNN-1_ch8} shows the output SINR performance comparisons for different  array configurations.  The horizontal axis is the DOA of the desired point source, and the performance is computed at six different source DOAs varying from   $15^0$ to $90^0$ in  steps of $15^0$.  The SNR of the desired signal is $0$ dB, and the  interference-to-noise-ratio (INR) for each interference is chosen randomly between $10-20$ dB for a given realization.  The results presented in  Fig. \ref{DNN-1_ch8} are obtained by using  unlimited number of data snapshots (USS) and employing  enumerated labels to train the DNN.  The network performance is reported by averaging over  $900$ testing scenarios for each desired source DOA. It is evident that the DNN-EN design (prediction using the enumerated labels)   performs  close (0.45 dB trade off) to the performance of the optimum array found by enumeration ($980$ possible configurations). The latter  gives the highest SINR performance but involves expensive singular value decomposition (SVD) for each enumeration and is also not scale-able with the problem size, facing the curse of dimensionality. 

The DNN-EN design performance is also compared with the  NNC (nearest neighbour correlation) design which returns the label corresponding to the input nearest neighbour correlation function (in terms of mean square error). NNC design is simply a lookup table such that for a given test data it  returns the  label of the closest training example by sifting over the entire training set. It is  noted that the DNN-EN design outperforms the NNC design with an average performance gain of 0.5 dB. The former approach not only offers superior performance but also is more economically viable from `Edge computing' perspective as compared to the nearest neighbour design which requires to maintain   a large dictionary and run  exhaustive search over the entire  training set  for  each prediction. Similar results are obtained for the NNC  design obtained by minimizing the   mean absolute error instead of the mean square error. For the underlying case, the DNN has around 88 $\%$ accuracy for the training data, and has around 54$\%$ accuracy on the test data (meaning that all sensor locations are correctly predicted). This implies that the superior performance of the DNN is not  simply because it memorizes the optimum but   due to the ability of DNN to generalize the learning as applied to the test set which is not seen before by the DNN. The superior performance of DNN  over NNC design again signifies that  the DNN   doesn't  memorize a lookup table implementation   to locate the nearest training data point and output  the corresponding memorized  optimal sparse configuration.  It is also clear that the proposed design yields significant gains over a compact ULA, sparse ULA and randomly selected array topology. The utility of the effective sparse design is also evident from the worst case performance which exhibits  significantly low output SINR of around -5 dB on average.

\begin{figure}[!t]
	\centering
	\includegraphics[width=5.6 in, height=3.6 in]{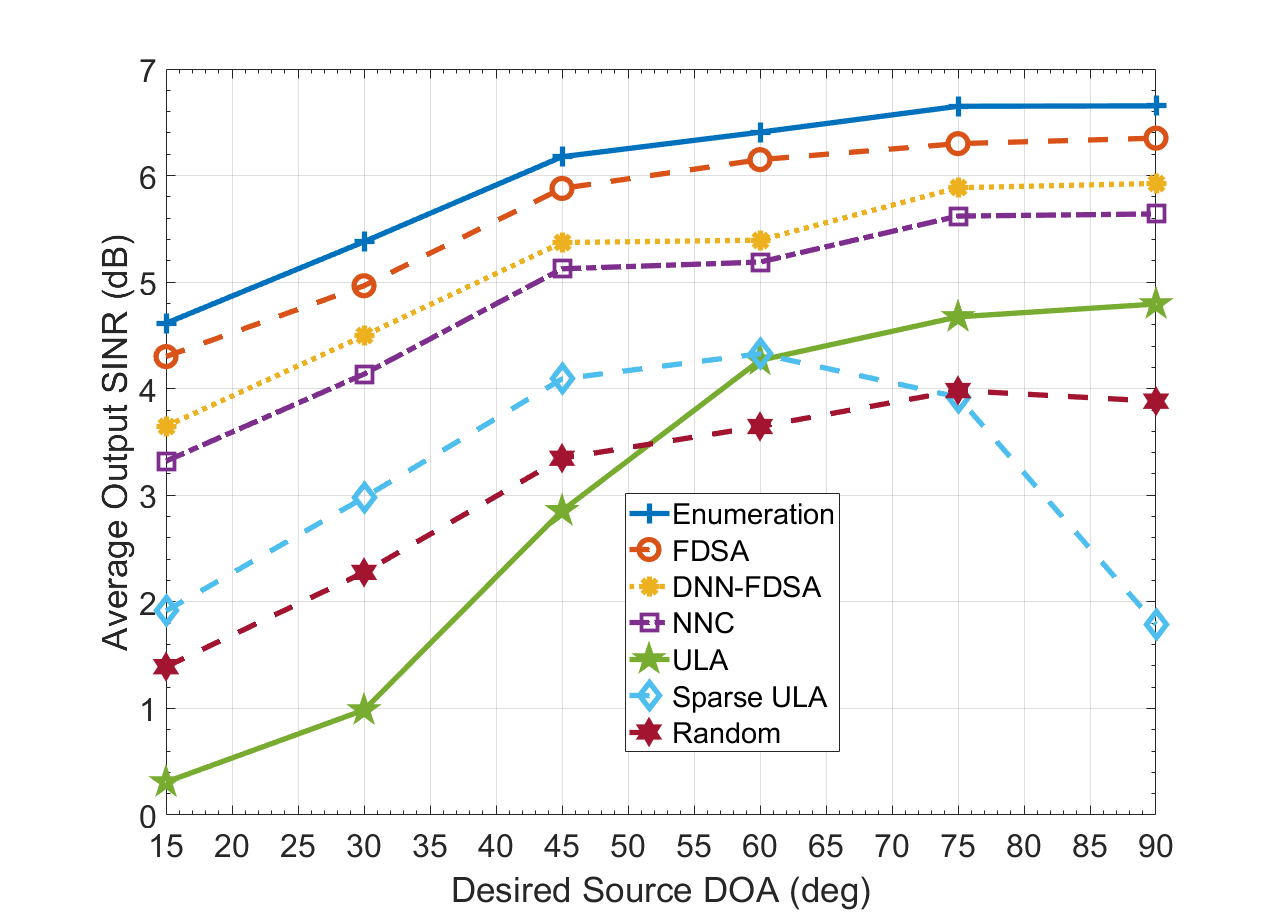}
	\caption{Performance comparison SBSA design under unlimited snapshots}
	\label{DNN-2_ch8}
\end{figure}
\subsection{DNN based SBSA design} \label{Single point source}

Although the training phase is entirely offline, it is infeasible to train the DNN while relying on the enumerated results. This is because  the number of possible sparse solutions can be considerably larger even for the modest size problem. For instance, choosing 14 elements out of 24 possible locations results in the order of  $10^6$ candidate spare configurations  for each training data, i.e., for each environment scenario.  In order to circumvent this problem and generate a large amount of training  data labels, we resort to  the proposed SBSA design.  Fig. \ref{DNN-2_ch8} shows the performance of SBSA design which is merely 0.3 dB down than the design obtained through enumeration.  Quite similar to  the DNN-EN design, the DNN-SBSA design is around 0.4 dB down from DNN-SBSA design. However,  this places the DNN-SBSA design further 0.8 dB suboptimal to the enumerated design. This is still a reasonable performance yielding significant dividends over the commonly used compact ULA, sparse ULA and random sparse topology. 

\subsection{Robust design} \label{Single point source}

\begin{figure}[!t]
	\centering
	\includegraphics[width=5.6 in, height=3.6 in]{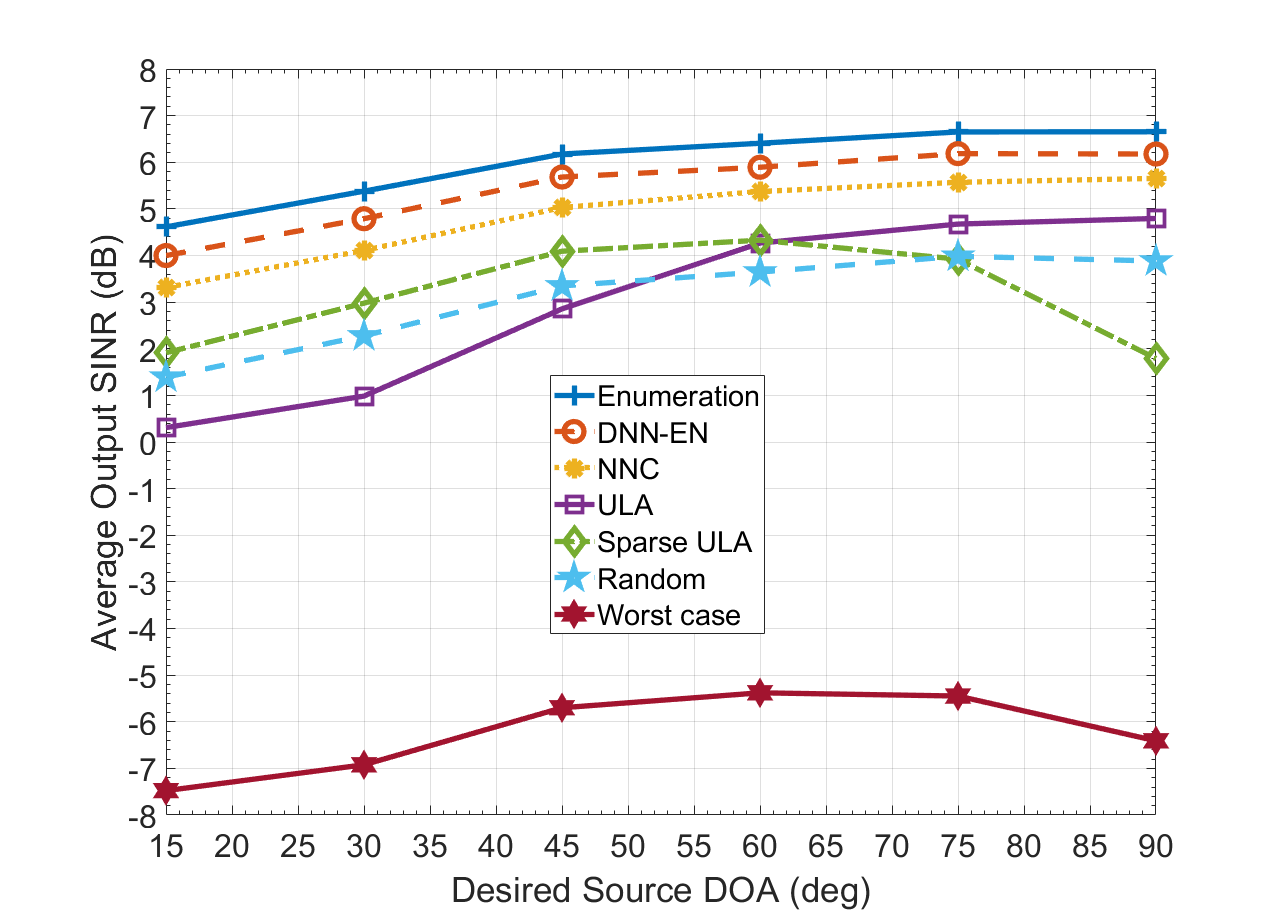}
	\caption{Performance comparison enumerated design under 1000 snapshots}
	\label{DNN-3_ch8}
\end{figure}

\begin{figure}[!t]
	\centering
	\includegraphics[width=5.6 in, height=3.6 in]{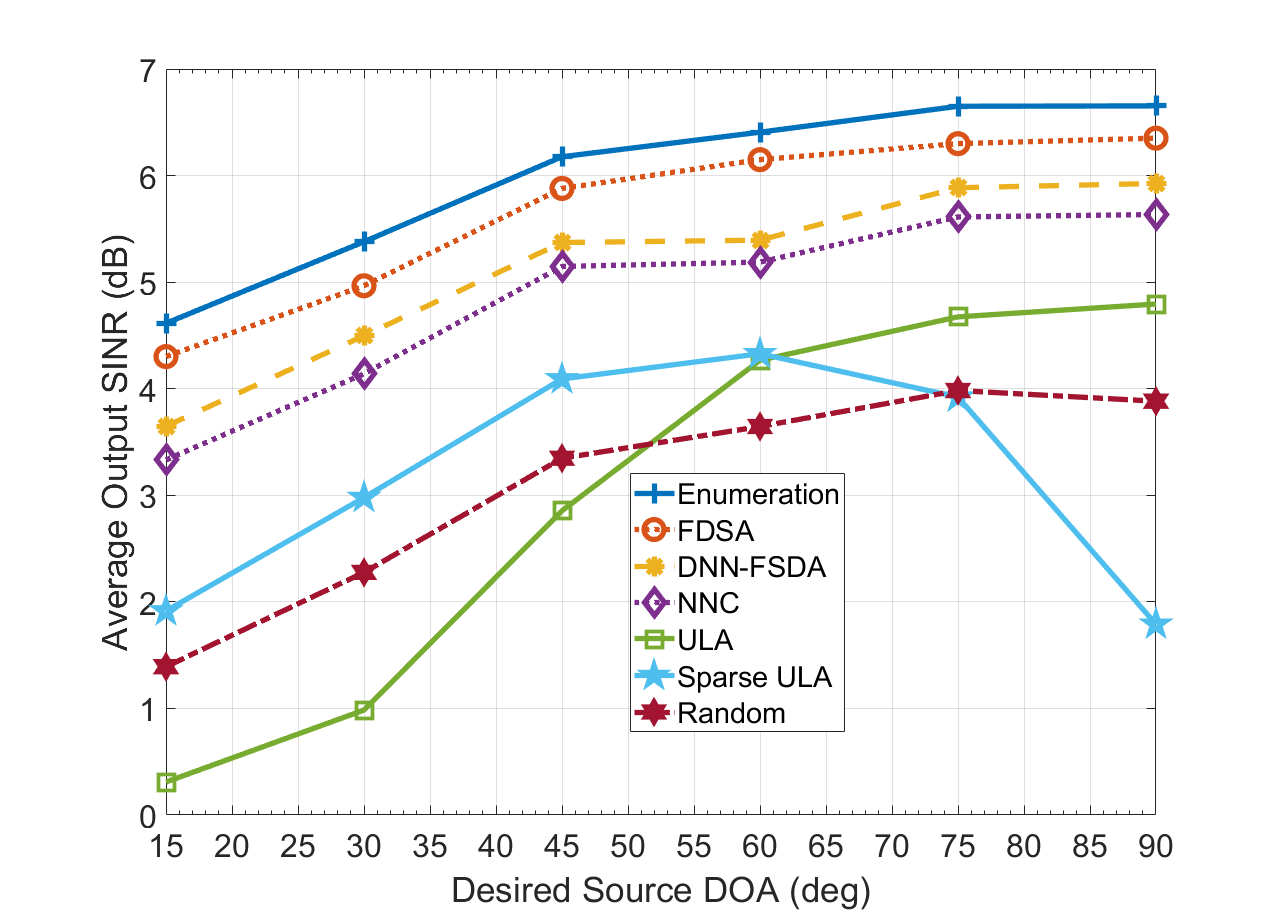}
	\caption{Performance comparison SBSA design under 1000 snapshots}
	\label{DNN-4_ch8}
\end{figure}

In order to gauge the robustness of the DNN based scheme, the performance is evaluated under a limited number of data  snapshots. Also, the desired source DOA is perturbed with Gaussian noise of 0.25$^0$ variance, in generating the test data to account of possible uncertainty around the desired source DOA. For simulating the limited snapshot  scenario, 1000 snapshots are generated assuming  the incoming signals (source and interfering signals) are independent BPSK  signals in the presence of Gaussian noise. The correlation matrix under limited snapshots doesn't follow the Toeplitz structure. Therefore, we average along the diagonal entries of the correlation matrix  to calculate the average correlation values.   Figs. \ref{DNN-3_ch8} and  \ref{DNN-4_ch8} show the performance of DNN-EN and DNN-SBSA designs under the limited data snapshots. It is clear from the figures that the performance is largely preserved with an SINR discrepancy of less than 0.01 dB demonstrating the robustness of the proposed scheme. The  NNC design, in this case is suboptimal with more than 0.3 dB additional performance loss.

\subsection{Performance comparisions with state-of-the-art}

The performance of the proposed SBSA, DNN-EN and DNN-SBSA are compared with  existing work on sparse array design which is based on SDR and SCA approaches \cite{8061016, 8892512}. It is clear from Fig. 15 that the SBSA algorithm outperforms the other designs and is also more than 100X  computationally efficient as compared to the SDR and SCA (Wang etal. \cite{8061016}) approach. However, it is only fair to compare the SBSA design with the SCA (Wang etal.) approach because  both incorporate the apriori knowledge of interference parameters. Therefore, in comparing  the data dependent designs, it is found that  SDR design (also the SDR-symmetric \cite{8892512}) is comparable to the  DNN-EN design, whereas the  DNN-SBSA is marginally suboptimal with the average performance degradation of 0.37 dB.  This slight performance trade off is readily leveraged by the real time implementation of the DNN-SBSA algorithm implementing the Capon beamformer in time frames of the order of few milli-seconds.

\label{Single point source}

\section{Conclusion}
This chapter considered  sparse array design for maximizing the beamformer output SINR for a desired source in an interference active scenario. A sparse beamformer spectral analysis (SBSA) algorithm was proposed which provided an insightful perspective for MaxSINR beamformer design. Also a DNN based approach was developed to configure a data driven sparse beamfomer by learning the enumerated algorithm as well as SBSA design.  The proposed methodology combined the merits of the data dependent designs and the designs assuming prior information of interference parameters. It was shown through the design examples that the  proposed schemes  are robust in the case of limited data snapshots and showed promising performance with reduced computational complexity.

\begin{figure}[!t]
	\centering
	\includegraphics[width=5.6 in, height=4.6 in]{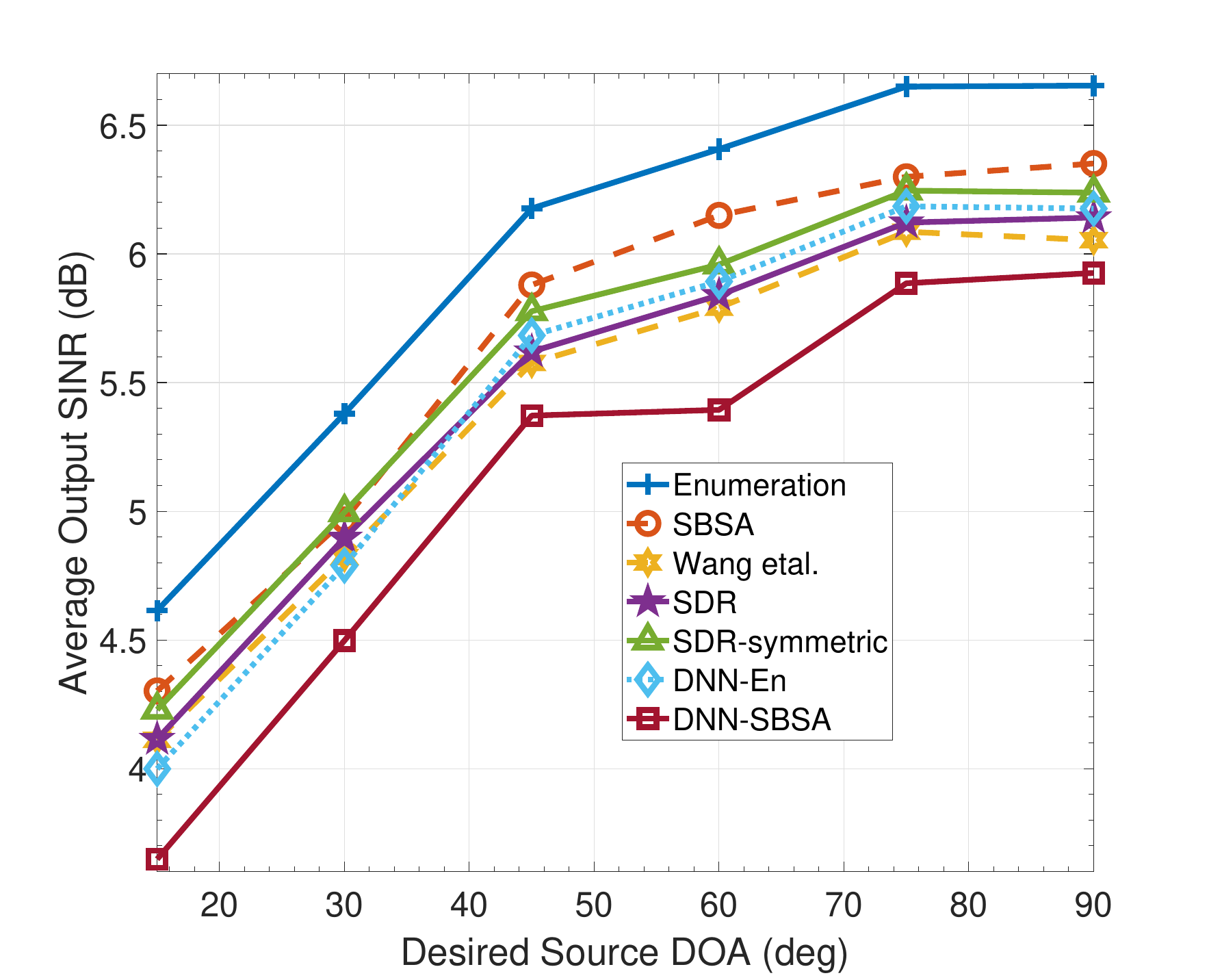}
	\caption{Performance comparisons with the state-of-the-art}
	\label{DNN-5_ch8}
\end{figure}

\chapter{CONCLUSIONS AND RECOMMENDATIONS}

In this work, sparse array design was proposed to improve active and passive sensing for radar and communication applications. Novel algorithms for sparse beamfomer design were presented for both receive and transmit signal platforms. The  proposed approaches were analyzed analytically as well as with the help of realistically simulated data.  
 
  The thesis addressed the problem  that the optimization of the array configuration requires full data correlation matrix which is not readily available in practice. Two different design approaches were considered; one assumed prefixed position of subset of sensors so as to provide full array augmentation, referred to as the hybrid-design approach,  whereas the other used the  Toeplitz estimation of the autocorrelation at the missing spatial lags.  We also considered maximizing the beamformer output SINR for the case of wideband signal models. Two different implementations, namely the TDL and the DFT implementation schemes, were presented for the wideband sparse array design. The sparse array configuration optimized in the DFT domain  and later  imported to the TDL implementation scheme was proposed, yielding promising results.  In all cases,  the  MaxSINR optimum sparse array delivered  considerable  dividends over other suboptimal  sparse arrays and compact uniform array. 
 
 A frequency domain sparse array  algorithm was proposed which provided an insightful perspective for MaxSINR beamformer design. Also a DNN based approach was developed to configure a data driven sparse beamfomer by learning various sparse design algorithms to enable real time implementation.  We also considered active sensing using sparse array configuration and sought optimum transmit beamforming for radar applications.    It was shown that the active sparse array design can achieve desirable beampattern characteristics such as directing a high proportion of the transmitted power towards the perspective target locations with minimal cross-correlation among  target signals. 
      
Future work will use the findings of this thesis to address some of the practical issues for sparse beamformer design such as evaluating the performance on real data and  incorporating  physical constraints such as antenna radiation pattern  and mutual coupling effects \cite{7027190, BOUDAHER20173}.

\bibliographystyle{IEEEtran}
\bibliography{IEEEabrv,main}
\end{document}